\documentclass[sn-mathphys,Numbered]{sn-jnl}

\usepackage{graphicx}%
\usepackage{multirow}%
\usepackage{amsmath,amssymb,amsfonts}%
\usepackage{amsthm}%
\usepackage{mathrsfs}%
\usepackage[title]{appendix}%
\usepackage{xcolor}%
\usepackage{textcomp}%
\usepackage{manyfoot}%
\usepackage{booktabs}%
\usepackage{algorithm}%
\usepackage{algorithmicx}%
\usepackage{algpseudocode}%
\usepackage{listings}%

\usepackage{tabularx}
\usepackage{bm}

\makeatletter
\newcommand*{\transpose}{\bgroup\@transpose}
\newcommand*{\@transpose}[1][0]{\mathpalette\@@transpose{#1}\egroup}
\newcommand*{\@@transpose}[2]{\setbox0=\hbox{\m@th$#1\mkern-#2mu\intercal$}\raise\dp0\box0}
\makeatother

\newcommand{\lm}{\lambda}
\newcommand{\lmd}{\lambda}
\newcommand{\clm}{\Lambda}
\newcommand{\Lmd}{\Lambda}
\newcommand{\Gma}{\Gamma}
\newcommand{\Omg}{\Omega}

\newcommand{\p}{\partial}
\newcommand{\al}{\alpha}
\newcommand{\afa}{\alpha}

\newcommand{\ve}{\varepsilon}
\newcommand{\veps}{\varepsilon}
\newcommand{\fai}{\varphi}

\newcommand{\mcalC}{\mathcal{C}}
\newcommand{\mcalD}{\mathcal{D}}
\newcommand{\mcalF}{\mathcal{F}}

\newcommand{\mcalI}{\mathcal{I}}

\newcommand{\mcalL}{\mathcal{L}}
\newcommand{\mcalP}{\mathcal{P}}

\newcommand{\mcalR}{\mathcal{R}}
\newcommand{\mcalU}{\mathcal{U}}
\newcommand{\mcalUtil}{\widetilde{\mathcal{U}}}

\newcommand{\bbC}{\mathbb{C}}
\newcommand{\bbZ}{\mathbb{Z}}
\newcommand{\bft}{\bm{t}}

\newcommand{\muhat}{\widehat{\mu}}
\newcommand{\Omghat}{\widehat{\Omega}}
\newcommand{\Omgtilhat}{\widetilde{\widehat{\Omega}}}
\newcommand{\Thetahat}{\widehat{\Theta}}
\newcommand{\Bhat}{\widehat{B}}
\newcommand{\Rhat}{\widehat{R}}
\newcommand{\ftil}{\tilde{f}}
\newcommand{\gtil}{\tilde{g}}
\newcommand{\ttil}{\tilde{t}}
\newcommand{\ctil}{\tilde{c}}
\newcommand{\etil}{\tilde{e}}
\newcommand{\ptil}{\tilde{p}}
\newcommand{\vtil}{\tilde{v}}
\newcommand{\Etil}{\widetilde{E}}
\newcommand{\Ftil}{\widetilde{F}}
\newcommand{\Gtil}{\widetilde{G}}
\newcommand{\Ltil}{\tilde{L}}
\newcommand{\Mtil}{\widetilde{M}}
\newcommand{\Rtil}{\tilde{R}}
\newcommand{\Stil}{\widetilde{S}}
\newcommand{\nablatil}{\widetilde{\nabla}}
\newcommand{\thetatil}{\tilde{\theta}}
\newcommand{\etatil}{\tilde{\eta}}
\newcommand{\mutil}{\tilde{\mu}}

\newcommand{\rme}{\mathrm{e}}
\newcommand{\rmi}{\mathrm{i}}
\newcommand{\rmD}{\mathrm{D}}
\newcommand{\bfa}{\bm{a}}

\newcommand*{\pp}[1]
  {\frac{\partial   }
        {\partial #1}
  }

\newcommand*{\pfrac}[2]
  {\frac{\partial #1}
        {\partial #2}
  }

\newcommand*{\pair}[2]
  {
    \left\langle
      #1,#2
    \right\rangle
  }

\newcommand*{\Bigset}[2]
  {
   \left\{ #1 \,\middle|\, #2 \right\}
  }

\newcommand{\beq}{\begin{equation}}
\newcommand{\eeq}{\end{equation}}

\newcommand{\dtafrac}[2]{\frac{\delta#1}{\delta #2}}

\DeclareMathOperator{\res}{Res}

\DeclareMathOperator{\nd}{d\!}
\DeclareMathOperator{\td}{d\!}

\theoremstyle{thmstyleone}%
\newtheorem{thm}{Theorem}[section]
\newtheorem{cor}[thm]{Corollary}
\newtheorem{lem}[thm]{Lemma}
\newtheorem{prop}[thm]{Proposition}
\newtheorem{conj}[thm]{Conjecture}

\theoremstyle{thmstyletwo}%
\newtheorem{rmk}{Remark}[section]

\theoremstyle{definition}
\newtheorem{ex}{Example}[section]

\theoremstyle{thmstylethree}%
\newtheorem{defn}{Definition}[section]

\numberwithin{equation}{section}

\allowdisplaybreaks[4]

\begin{document}
\title[Generalized Frobenius Manifolds and Integrable Hierarchies]{Generalized Frobenius Manifolds with Non-flat Unity and Integrable Hierarchies}

\author[1]{\fnm{Si-Qi} \sur{Liu}}\email{liusq@tsinghua.edu.cn}
\equalcont{These authors contributed equally to this work.}

\author[2]{\fnm{Haonan} \sur{Qu}}\email{qhn1121@pku.edu.cn}
\equalcont{These authors contributed equally to this work.}

\author*[1]{\fnm{Youjin} \sur{Zhang}}\email{youjin@tsinghua.edu.cn}
\equalcont{These authors contributed equally to this work.}

\affil[1]{\orgdiv{Department of Mathematical Sciences}, \orgname{Tsinghua University}, \orgaddress{\city{Beijing}, \postcode{100084}, \country{ P.\,R.~China}}}
\affil[2]{\orgdiv{School of Mathematical Sciences},
\orgname{Peking University}, \orgaddress{\city{Beijing}, \postcode{100871}, \country{P.\,R. China}}}

\abstract{
For any generalized Frobenius manifold with non-flat unity, we construct a bihamiltonian integrable hierarchy of hydrodynamic type which is an analogue of the Principal Hierarchy of a Frobenius manifold. We show that such an integrable hierarchy, which we also call the Principal Hierarchy, possesses Virasoro symmetries and a tau structure, and the Virasoro symmetries can be lifted to symmetries of the tau-cover of the integrable hierarchy. We derive the loop equation from the condition of linearization of actions of the Virasoro symmetries on the tau function, and construct the topological deformation of the Principal Hierarchy of a semisimple generalized Frobenius manifold with non-flat unity. We also give two examples of generalized Frobenius manifolds with non-flat unity and show that they are closely related to the well-known integrable hierarchies: the Volterra hierarchy, the \textit{q}-deformed KdV hierarchy and the Ablowitz-Ladik hierarchy.}

\keywords{Generalized Frobenius manifold; Principal Hierarchy; Tau structure; Virasoro symmetry; Loop equation}

\maketitle

\section{Introduction}

The notion of Frobenius manifold was introduced by Dubrovin in \cite{Dubrovin-Frob-mfd} as a coordinate free description of the Witten-Dijkgraaf-Verlinde-Verlinde (WDVV) associativity equations satisfied by the primary free energies of 2D topological field theories (TFTs) \cite{DVV, Witten-1, Witten-2}. It has played a central role in the study of Gromov-Witten theory, singularity theory, mirror symmetry, integrable systems and some other research subjects of mathematical physics in the past thirty years. The main motivation for the study of Frobenius manifolds in \cite{Dubrovin-Frob-mfd} is to reconstruct a complete 2D TFT starting from the corresponding Frobenius manifold, i.e., starting from a solution of the WDVV associativity equations.
The reconstruction of
the genus zero potential of a 2D TFT
was given in \cite{Dubrovin-92}.
In this reconstruction, a bihamiltonian integrable hierarchy of hydrodynamic type, known as the Principal Hierarchy of the corresponding Frobenius manifold, plays a crucial role. This integrable hierarchy possesses a tau structure which enables one to assign a tau function to any of its solution, and the logarithm of the tau function of a particular solution selected by the string equation yields the genus zero free energy of the 2D TFT.
The reconstruction of
the full genera potential of a 2D TFT
was given in \cite{DZ-cmp-98, normal-form} under the assumption that the corresponding Frobenius manifold is semisimple. It is achieved by performing a certain quasi-Miura transformation to the Principal Hierarchy, which linearizes the action of the Virasoro symmetries of the Principal Hierarchy on the tau function of the resulting deformed integrable hierarchy. Such a deformed integrable hierarchy is called the topological deformation of the Principal Hierarchy, and its tau function which is selected by the string equation yields the partition function of the 2D TFT.

The topological deformation of the Principal Hierarchy of a semisimple Frobenius manifold is a bihamiltonian integrable hierarchy \cite{normal-form, LWZ-1}. For the one dimensional case,  the associated integrable hierarchy is the KdV hierarchy which controls the 2D topological gravity as we know from Witten's conjecture and its proof by Kontsevich \cite{Kontsevich, Witten-2}. Other well-known bihamiltonian integrable hierarchies that are topological deformations of the Principal Hierarchies of certain semisimple Frobenius manifolds include the extended Toda hierarchy \cite{CDZ} and the Drinfeld-Sokolov hierarchies associated with untwisted affine Kac-Moody algebras of ADE type \cite{DS}, they correspond to the Frobenius manifolds that are associated respectively with the Gromov-Witten invariants of the projective line \cite{DZ-cmp-2004, Getzler, OP, Z-2002}, and the FJRW invariants of simple singularities \cite{FJRW, Givental-Milanov}.

In this paper, we study the relationship of a class of generalized Frobenius manifold with bihamiltonian integrable hierarchies. By Dubrovin's definition \cite{Dubrovin-2DTFT}, on a Frobenius manifold $M$ there is defined a flat metric, and on each of its tangent space there is a commutative associative algebra structure with unity, and moreover the unit vector field is required to be flat w.r.t. the flat metric. In applications of the notion of Frobenius manifold to the study of singularity theory, Gromov-Witten theory, quantum K-theory and the theory of integrable systems, certain generalized Frobenius manifold structures
also arise and play important roles. One class of generalized Frobenius manifolds, which are called F-manifolds, was introduced and studied by Hertling and Manin in \cite{Manin-Hertling}. On such a generalized Frobenius manifold the existence of a flat metric is replaced by a certain weaker condition imposed on the commutative associative algebra structure of the tangent spaces. By further imposing a certain compatible flat connection and a flat unity on an F-manifold, one obtains the notion of flat F-manifold \cite{Getzler-04, Manin}, whose relationship with integrable hierarchies of hydrodynamic type and their deformations are studied in \cite{Lorenzoni-3, Lorenzoni-2, Buryak-Rossi,Lorenzoni-1}.
The generalized Frobenius manifolds that we study in this paper are defined by Dubrovin's definition of Frobenius manifold but without the flatness condition imposed on the unit vector fields, and they will be called generalized Frobenius manifolds with non-flat unity or simply generalized Frobenius manifolds.

Our study of generalized Frobenius manifolds with non-flat unity is mainly motivated by the relationship of the Ablowitz-Ladik hierarchy, a well-known bihamiltonian integrable hierarchy, with the equivariant Gromov-Witten invariants of the resolved conifold, as it was revealed by Brini in \cite{Brini-1}. Brini conjectured in \cite{Brini-1} that a particular tau function of the Ablowitz-Ladik hierarchy yields the generating function of the equivariant Gromov-Witten invariants of the resolved conifold with anti-diagonal action, and verified his conjecture at the genus one approximation. In \cite{Brini-2, Brini-3} Brini and his collaborators showed that the dispersionless limit of the Ablowitz-Ladik hierarchy belongs to the Principal Hierarchy of a generalized Frobenius manifold with non-flat unity,
and that the almost dual of this generalized Frobenius manifold coincides with the non-conformal Frobenius manifold associated with the equivariant Gromov-Witten invariants of the resolved conifold with anti-diagonal action. It is natural to ask the question: is there an analogue of the construction for the topological deformation of the Principal Hierarchy of a usual semisimple Frobenius manifold for the one of a semisimple generalized Frobenius manifold with non-flat unity? Such a construction could lead to a quasi-Miura transformation which relates the
dispersionless Ablowitz-Ladik hierarchy with the dispersionfull one, and provides support of Brini's conjecture of the relation of this integrable hierarchy with the equivariant Gromov-Witten invariants of the resolved conifold.

Another main motivation of our study comes from the relationship between certain special cubic Hodge integrals and the Volterra hierarchy (also called the discrete KdV hierarchy) that is established in \cite{DLYZ-1, DLYZ-2}. This relationship is given by the identification of the generating function of these Hodge integrals with a certain tau function of the Volterra hierarchy. The Volterra hierarchy has a bihamiltonian structure, from the hydrodynamic limit of which we can obtain a 1-dimensional generalized Frobenius manifold with non-flat unity. The reconstruction of the Volterra hierarchy from this generalized Frobenius manifold would help us to have a deeper understanding of the above-mentioned relationship between Hodge integrals and integrable hierarchies.

For a generalized Frobenius manifold with non-flat unity, we can
construct its Principal Hierarchy as it is done for a usual Frobenius manifold, and we can prove that the Principal Hierarchy possesses Virasoro symmetries and a tau structure. To construct the topological deformation of the Principal Hierarchy by using the condition of linearization of the actions of the Virasoro symmetries on the tau function
of the deformed integrable hierarchy, we need to lift the Virasoro symmetries of the Principal Hierarchy to symmetries of its tau-cover and to represent the actions of these symmetries on the tau function in terms of certain linear differential operators. However, due to the non-flatness of the unit vector field we can not achieve this in a direct way. One important point in our solution of this problem is the introduction of infinitely many additional commuting flows to the Principal Hierarchy. These additional flows include the one given by the translation along the spatial variable $x$, and they commute with each other and with the original flows of the Principal Hierarchy. Then we show that the the Virasoro symmetries can be lifted to the tau-cover of the extended integrable hierarchy which we still call the Principal Hierarchy of the generalized Frobenius manifold, and their actions on the tau function can be represented in terms of certain linear differential operators which satisfy the Virasoro commutation relations. Notably, in our study of properties of the Virasoro symmetries of the Principal Hierarchy of an $n$-dimensional generalized Frobenius manifold with non-flat unity, we need to use the structure of a special $(n+2)$-dimensional Frobenius manifold which has a flat unity and has the original generalized Frobenius manifold as its generalized Frobenius submanifold.

We derive the loop equation of a generalized Frobenius manifold with non-flat unity by using the condition of linearization of the actions of the Virasoro symmetries on the tau functions of deformations of the Principal Hierarchy of certain class. A solution of the loop equation of a generalized Frobenius manifold with non-flat unity gives a quasi-Miura transformation which yields a deformation of the Principal Hierarchy. For the two examples of generalized Frobenius manifolds that are associated with special cubic Hodge integrals and the equivariant Gromov-Witten invariants of the resolved conifold with anti-diagonal action which we mentioned above, we solve their loop equations at the genus two approximation, and we verify that the deformations of the Principal Hierarchies obtained by the unique solutions of these loop equations
contain the first two flows of the Volterra hierarchy and the Ablowitz-Ladik hierarchy. This result is in agreement with the relationship between the special cubic Hodge integrals and the Volterra hierarchy, and the conjectural relationship between the equivariant Gromov-Witten invariants of the resolved conifold with anti-diagonal action and the Ablowitz-Ladik hierarchy. So for a generalized Frobenius manifold with non-flat unity, we call the deformation of its Principal Hierarchy that is given by a solution of the loop equation a topological deformation. For a semisimple generalized Frobenius manifold with non-flat unity, the existence and uniqueness of the solution of its loop equation is shown in \cite{LWZ-2}, so in such a case we have a unique topological deformation of the associated Principal Hierarchy.

The paper is organized as follows.
In Sect.\,2, we present some properties of an $n$-dimensional generalized Frobenius manifold $M$ with non-flat unity. In Sect.\,3, we give the definition of the Principal Hierarchy of $M$.
In Sect.\,4, we construct the tau-cover of the Principal Hierarchy. In Sect.\,5, we introduce an $(n + 2)$-dimensional Frobenius manifold associated with $M$. In Sect.\, 6--8, we introduce the notion of periods of $M$ and in terms of which we present the Virasoro symmetries of the tau-cover of $M$. In Sect.\,9 and 10, we derive the loop equation of $M$. In Sect.\,11, we present two examples of generalized Frobenius manifolds with non-flat unity, and relate the topological deformations of their Principal Hierarchies to the well-known integrable hierarchies: the Volterra hierarchy, the \textit{q}-deformed KdV hierarchy and the Ablowitz-Ladik hierarchy. In Sect.\,12, we give some concluding remarks. In Appendix \ref{Appendix-A} and Appendix \ref{Appendix-B}, we give the proofs of Theorem \ref{thm:G-nu as Gram},
Lemma \ref{lemma: Laplace-Bilinear identity} and Theorem \ref{main-lemma-orig}.

\section{Generalized Frobenius manifolds}\label{zh-24}

Let us begin with the description of the class of generalized Frobenius manifolds that we are to study following Dubrovin's definition of Frobenius manifold \cite{Dubrovin-2DTFT}.
\begin{defn}\label{zh-38}
A smooth or analytic manifold $M$ is called a generalized Frobenius manifold with non-flat unity if
each of its tangent spaces is endowed with
a Frobenius algebra structure $A=\left(\cdot, \langle\ ,\, \rangle, e\right)$, i.e., a commutative associative algebra (over $\mathbb{R}$ or $\mathbb{C}$) with multiplication $\cdot$ and unity $e$, and with a nondegenerate symmetric bilinear form $\langle\ ,\, \rangle$ which is invariant w.r.t. the multiplication; this Frobenius algebra structure depends smoothly or analytically on the point of $M$, and is required to satisfy the following conditions.
\begin{enumerate}
\item[1.] The bilinear form $\langle\,\, ,\, \rangle$ yields a flat metric $\eta$ on $M$, and $\nabla e$ does not vanish identically on $M$, where $\nabla$ is the Levi-Civita connection of $\eta$.
\item[2.] Define a 3-tensor $c$ on $M$ by
$c(u,v,w)=\langle u\cdot v, w\rangle$,
then the 4-tensor
\[\nabla c: (z,u,v,w)\mapsto\nabla_zc(u,v,w)\]
is symmetric, here $z,u,v,w\in\mathrm{Vect}(M)$.
\item[3.] There exists an Euler vector field $E$ on $M$ such that $\nabla\nabla E=0$, $\nabla E$ is diagonalizable, and
\begin{align}
    \mcalL_E(a\cdot b)&=\mcalL_Ea\cdot b+a\cdot\mcalL_Eb+a\cdot b
    \label{FM4-c}\\
    \mcalL_E\langle a,b\rangle&=
    \langle\mcalL_Ea,b\rangle+
    \langle a,\mcalL_Eb\rangle
    +(2-d)\langle a,b\rangle.
    \label{FM4-eta}
  \end{align}
  The constant $d$ is called the charge of $M$.
\end{enumerate}
\end{defn}

Notably, in the above definition we only replace the flatness condition $\nabla e=0$ of the unit vector field that is given in Dubrovin's original definition of a Frobenius manifold by the condition that $\nabla e$ does not vanish identically on $M$. In what follows, we will call a generalized Frobenius manifold with non-flat unity a generalized Frobenius manifold.

Let us first recall some basic properties of a Frobenius manifold that are still possessed by a generalized one \cite{Dubrovin-2DTFT}. We fix a system of flat coordinates $v^1,\dots, v^n$ of the flat metric in a neighborhood $U$ of a certain point $v_0\in M$, and denote
\[\eta_{\al\beta}=\left\langle\p_\al,\p_\beta\right\rangle,\quad (\eta^{\al\beta})=(\eta_{\al\beta})^{-1},\]
where $\p_\al=\frac{\p}{\p v^\al}$. The structure constants of the Frobenius algebras $c_{\al\beta}^\gamma=c_{\al\beta}^\gamma(v)$ are defined by
\[\p_\al\cdot\p_\beta=c^\gamma_{\al\beta}\p_\gamma,\quad \al,\beta=1,\dots, n.\]
Here and in what follows summation over repeated upper and lower Greek indices ranging from $1$ to $n$ is assumed. From the definition of a generalized Frobenius manifold it follows the existence of a function $F=F(v)$, called the potential of the Frobenius manifold, such that
\[
\frac{\p^3F}{\p v^\afa\p v^\beta \p v^{\gamma}}=\eta_{\gamma\xi}c^\xi_{\al\beta},\quad \p_E F=(3-d) F+\frac12 A_{\al\beta} v^\al v^\beta+B_\al v^\al+C,\]
here $A_{\al\beta}, B_\al, C$ are certain constants, and the functions $c^\gamma_{\al\beta}$ satisfy the associativity equations
\begin{equation}\label{zh-5}
c^\xi_{\al\beta}c^\delta_{\xi\gamma}
=c^\xi_{\gamma\beta}c^\delta_{\xi\al},\quad  \al,\beta,\gamma,\delta=1,\dots,n.
\end{equation}

Denote
\begin{equation}\label{zh-2}
\mu=\frac{2-d}{2}-\nabla E,
\end{equation}
then from \eqref{FM4-eta} we have
\begin{equation}\label{zh-1}
\mu\eta+\eta\mu=0.
\end{equation}
Since $\nabla E$ is diagonalizable, we can choose the flat coordinates $v^1,\dots, v^n$ such that
\begin{equation}\label{q and mu}
  \mu:=\mathrm{diag}(\mu_1,\dots,\mu_n),
\end{equation}
and the Euler vector field $E$ can be represented in the form
\begin{equation}\label{Euler field}
  E=E^\al\p_\al=\sum_{\afa=1}^{n}
    \left[
      \left(
        1-\frac d2-\mu_\afa
      \right)v^\afa+r^\afa
    \right]\p_\afa,
\end{equation}
here the constants $r^\al\ne 0$ only if $\mu_\al=1-\frac{d}2$.

As for a usual Frobenius manifold, we define the intersection form $g$ of $M$ by \cite{Dubrovin-2DTFT}
\begin{equation}\label{zh-27}
(\omega_1, \omega_2)=\iota_E (\omega_1\cdot \omega_2),\quad
\omega_1, \omega_2\in T^*_vM,
\end{equation}
where the multiplication on $T^*_v M$ is defined via the isomorphism
between $T_v M$ and $T^*_v M$ induced by the bilinear form $\langle\ ,\, \rangle$. In the flat coordinates we have
\begin{equation}\label{zh-19}
g^{\al\beta}(v)=(d v^\al, d v^\beta)=E^\gamma c_{\gamma}^{\al\beta},
\end{equation}
and the contravariant coefficients of the Levi-Civita connection of $g$ can be represented in the form
\begin{equation}\label{connection Gamma}
  \Gma^{\afa\beta}_\gamma
 =\left(
   \frac12-\mu_{\beta}
 \right)c^{\afa\beta}_\gamma,
\end{equation}
where $c^{\afa\beta}_\gamma=\eta^{\afa\lmd}c^\beta_{\lmd\gamma}$.
It can be verified that
the intersection form $g$ and the flat metric $\eta$ yield a flat pencil
of metrics $g^{\al\beta}-\lm \eta^{\al\beta}$ on $M$
by using the same method as for a usual Frobenius manifold \cite{Dubrovin-2DTFT}.

\begin{prop}\label{prop existence of theta 00}
On a generalized Frobenius manifold $M$ there locally
exists a function $\fai$
such that its gradients w.r.t. the metrics $\eta$ and $g$ yield the unity vector field $e$ and the Euler vector field $E$ respectively, i.e., it  satisfies the relations
\begin{equation}\label{e=nabla fai}
  e^\afa = \eta^{\afa\beta}\pfrac{\fai}{v^\beta},\quad
  E^\afa = g^{\afa\beta}\pfrac{\fai}{v^\beta},\quad \al=1,\dots,n.
\end{equation}
Here $e^\al, E^\al$ are the coefficients of the unit vector field $e=e^\al \p_\al$ and the Euler vector field $E=E^\al\p_\al$.
\end{prop}

\begin{proof}
To prove the existence of a function $\fai$ satisfying the first relation of \eqref{e=nabla fai},
it suffices to show that
  \begin{equation}\label{existence of fxx}
    \p^\afa e^\beta=\p^\beta e^\afa,
  \end{equation}
  where $\p^\afa=\eta^{\afa\beta}\p_\beta$.
  In fact, from $e\cdot \p_\afa=\p_\afa$ it follows that
  $
    e^\gamma c^\beta_{\gamma\xi}=\delta^\beta_\xi,
  $
  so we have
  \[
    0=e^\gamma\p^\al\left(
      e^\xi c^\beta_{\gamma\xi}
    \right)
   =e^\gamma
    \left(\left(\p^\al e^\xi\right)
    c^\beta_{\gamma \xi}
    + e^\xi c^{\al\beta}_{\gamma\xi}\right)
   =
     \p^\al e^\beta+ e^\gamma e^\xi c^{\al\beta}_{\gamma\xi},
  \]
  which yields the required relation
  \begin{equation}\label{p-afa-e-beta}
  \p^\al e^\beta=-e^\gamma e^\xi c^{\al\beta}_{\gamma\xi}=\p^\beta e^\al.
  \end{equation}
Here we use the notation $c^{\al\beta}_{\gamma\xi}=\p_\xi c^{\al\beta}_{\gamma}$.
The validity of the second relation of \eqref{e=nabla fai} follows from the fact that
  \[
    g^{\afa\beta}\pfrac{\fai}{v^{\beta}}
   =E^\xi c^\afa_{\xi\gamma}\eta^{\gamma\beta}
    \pfrac{\fai}{v^{\beta}}
   =
     E^\xi c_{\xi\gamma}^\afa e^\gamma
   =
    E^\xi\delta^\afa_\xi = E^\afa.
  \]
The proposition is proved.
  \end{proof}

\begin{prop} For a generalized Frobenius manifold the following identities
hold true:
\begin{equation}\label{existence of fxx-2}
  [E,e]=-e,\quad  \nabla\langle E,e\rangle=(1-d)e.
\end{equation}
\end{prop}

\begin{proof}
The first identity of \eqref{existence of fxx-2} follows from \eqref{FM4-c}.
By using this identity and \eqref{existence of fxx}
we obtain
\begin{align}
 &\eta_{\afa\beta}E^\afa\p^\gamma e^\beta
=
  \eta_{\afa\beta}E^\afa\p^\beta e^\gamma
 =
  -e^\gamma+e^\afa \p_\afa E^\gamma\notag\\
&\quad =
  -e^\gamma+e^\gamma
  \left(
    \frac{2-d}{2}-\mu_\gamma
  \right)
 =
  -\left(\frac d2+\mu_\gamma\right)e^\gamma.\label{zh-8}
\end{align}
Thus from the relation \eqref{zh-1} it follows that
\begin{align*}
  &\p^\gamma\langle E,e\rangle
=
  (\p^\gamma E^\afa)e^\beta\eta_{\afa\beta}
 +E^\afa(\p^\gamma e^\beta)\eta_{\afa\beta}
=
  \eta^{\gamma\xi}
  (\p_\xi E^\afa)e^\beta\eta_{\afa\beta}
 -\left(\frac d2+\mu_\gamma\right)e^\gamma  \\
&\quad =
  \left[
    \left(
      \frac{2-d}{2}+\mu_\gamma
    \right)
   -\left(
    \frac d2+\mu_\gamma
   \right)
  \right]e^\gamma
=
  (1-d)e^\gamma,
\end{align*}
hence the second identity of \eqref{existence of fxx-2} also holds true. The proposition is proved.
\end{proof}

From the second identity of \eqref{existence of fxx-2} we see that the function $\fai$ that is given in Proposition \ref{prop existence of theta 00} can be taken as
\[\fai=\frac{\langle E,e\rangle}{1-d}\]
when $d\ne 1$.

As for a usual Frobenius manifold, on a generalized Frobenius manifold $M$ we also have the deformed flat connection $\tilde{\nabla}$ defined by
\[\nablatil_uv=\nabla_uv+zu\cdot v,\quad u,v\in\mathrm{Vect}(M).\]
It can be extended to a flat affine connection on $M\times \mathbb{C}^*$ by defining
\[
   \nablatil_u\frac{\td}{\td z}=0,
   \quad \nablatil_{\frac{\td}{\td z}}\frac{\td}{\td z}=0,
   \quad \nablatil_{\frac{\td}{\td z}}v=\p_zv+E\cdot v-\frac1z\mu v,
\]
here $u, v$ are vector fields on $M\times\mathbb{C}^*$ with zero components
along $\mathbb{C}^*$, and $z$ is the coordinate on $\mathbb{C}^*$.
We can choose a system of deformed flat coordinates
$\tilde{v}^1(v;z),\dots,\tilde{v}^n(v;z)$ of $M$ which have the form 
  \begin{equation}\label{levelt system}
  (\vtil_1( v,z),\dots,\vtil_n( v,z))
 =(\theta_1( v,z),\dots,\theta_n( v,z))z^{\mu}z^R,
\end{equation}
and such that
\[\zeta_\al=\frac{\p \vtil_\al}{\p v^\gamma} \td v^\gamma,\quad
\al=1,\dots,n\]
yield a basis of solutions of the system $\nablatil \zeta=0$.
Here the constant matrices $\mu, R$ are the monodromy data of $M$ at
$z=0$, with $\mu$ given by \eqref{zh-2} and $R=R_1+\dots+R_m$
(for a certain $m\in\bbZ_+$)
satisfies the relations
\begin{align}
&(R_s)^\afa_\beta=0 \quad\text{if}\quad
  \mu_{\afa}-\mu_\beta\neq s, \label{monodromy-R}\\
&R_s^{\transpose}\eta=(-1)^{s+1}\eta R_s,\quad s\geq 1. \label{zh-3}
\end{align}
The functions $\theta_{\afa}(v;z)$ are analytic at $z=0$, and can be represented in the form
\[\theta_\al(v;z)=\sum_{p\ge 0}\theta_{\al,p}(v) z^p,\quad \al=1,\dots,n.\]
The proofs of the above assertions are the same as for a usual Frobenius manifold, the details of which can be found, for example, in Lecture 2 of \cite{Dubrovin-painleve}.

Denote
\begin{equation}\label{zh-10}
\xi_{\al,p}=\xi^\gamma_{\al,p}\p_\gamma=\nabla\theta_{\al,p},\quad \al=1,\dots,n,\, p\ge 0,
\end{equation}
then these vector fields satisfy the recursion relations
\begin{equation}\label{zh-11}
\p_\afa\xi_{\beta,p+1}^\gamma = c^\gamma_{\afa\sigma}\xi^\sigma_{\beta,p}
\end{equation}
and the quasi-homogeneity condition
\begin{equation}\label{zh-12}
\p_E\xi_{\al,p}=(p+\mu_\al-\mu)\xi_{\al,p}+\sum_{s=1}^p (R_s)^\gamma_\al\xi_{\gamma,p-s}.
\end{equation}
We also impose, as for a usual Frobenius manifold \cite{Dubrovin-2DTFT},
the following normalization conditions on the deformed flat coordinates:
\begin{align}
&\xi_{\al}(v;0)=\xi_{\al,0}=\p_\al,\label{zh-15}\\
&
\!\!\pair{\xi_{\afa}(v;-z)}{\xi_{\beta}(v;z)}
 = \eta_{\afa\beta},\label{nabla-theta-ortho}
\end{align}
where
\[\xi_\al(v;z)=\sum_{p\ge 0} \xi_{\al,p} z^p,\quad \al=1,\dots, n.\]

In general, the equations \eqref{zh-11}--\eqref{nabla-theta-ortho} do not uniquely determine the vector fields $\xi_{\al,p}$, since we have the following ambiguity in determining $\xi_{\al,p+1}$ from $\xi_{\al, 0}$,\dots, $\xi_{\al,p}$ for $\al=1,\dots,n$ by using the recursion relations \eqref{zh-11}:
\[\xi_{\al,p+1}\mapsto \xi_{\al,p+1}+a_{\al,p+1},\]
where the constant vectors  $a_{\al,p+1}=a_{\al,p+1}^\gamma\p_\gamma$ satisfy the conditions
\[(p+1+\mu_\al-\mu_\gamma) a^\gamma_{\al,p+1}=0,
\quad a_{\al,p+1}^\gamma \eta_{\gamma\beta}=(-1)^p a_{\beta,p+1}^\gamma \eta_{\gamma\al}.\]
\textit{In what follows, we will fix a system of solutions $\xi_{\al,p}$ of the equations
\eqref{zh-11}--\eqref{nabla-theta-ortho}, from which we will also fix a choice of the functions $\theta_{\al,p}$ satisfying \eqref{zh-10} in the next section.}

\section{The Principal Hierarchy}

For a usual Frobenius manifold $M$ with flat unity, the functions $\theta_{\al,p}$
yield a hierarchy of integrable Hamiltonian systems
\begin{equation}\label{zh-4}
  \pfrac{v^\al}{t^{\beta, q}}
  =\eta^{\al\gamma}\frac{\p}{\p x}\left(\frac{\p\theta_{\beta,q+1}}{\p v^\gamma}\right)=\eta^{\al\gamma}\frac{\p^2\theta_{\beta,q+1}}{\p v^\gamma\p v^\xi} v^\xi_x,\quad 1\leq\al,\beta\leq n,\, q\geq 0
\end{equation}
defined on the jet space $J^\infty(M)$, and it is called the Principal Hierarchy of the Frobenius manifold. For a generalized Frobenius manifold we also have such an integrable hierarchy which possesses a bihamiltonian structure, a tau structure and an infinite number of Virasoro symmetries, just as for a Frobenius manifold with flat unity. However, due to the non-flatness of the unit vector field, the flow given by the translation along the spatial variable $x$ does not belong to this hierarchy, and we are no longer able to lift the actions of the Virasoro symmetries to the tau function of this integrable hierarchy. In order to solve this problem, we propose to introduce an additional set of commuting flows to this integrable hierarchy, and we call the integrable hierarchy \eqref{zh-4} together
with these additional flows the Principal Hierarchy of the generalized Frobenius manifold. We construct these additional flows in this section.

In this and the subsequent sections, we will denote by $M$ an $n$-dimensional generalized Frobenius manifold.

\begin{lem}\label{lemma-existence of theta-0-p}
 There exist $\xi_{0,p}\in\mathrm{Vect}(M)$ for $p\geq 0$,
  and constants $r_p^\al\in\bbC$ for $p\geq 1,\, \al = 1,2,\dots,n$
  such that
  \begin{align}
\xi_{0,0} &= e,\quad \xi_{0,1}= v^\afa\p_\afa,\label{zh-6}\\
\p_\afa\xi_{0,p}^\gamma &= c_{\afa\beta}^\gamma\xi^\beta_{0,p-1},\quad \al,\gamma=1,\dots,n,\, p\geq 1.
  \label{zh-7}\\
\p_E\xi_{0,p}
    &=\left(p-\frac d2-\mu\right)\xi_{0,p}
    +
      \sum_{s=1}^p r_s^\lm\xi_{\lm, p-s}
    ,\quad p\geq 0,\label{theta0p-quasi-homog}
\end{align}
and the constants $r_p^\al$ satisfy the conditions
\begin{align}
   r_1^\al &= r^\al,\\
   r^\al_p&\neq 0\quad \textrm{only if}\quad
    \mu_\al+\frac d2 = p.\label{r^veps_p condition}
    \end{align}
  Here $r^\al$ are defined by the expression \eqref{Euler field} of the Euler vector field $E$.
\end{lem}

\begin{proof}
Let us prove the existence of the vector fields $\xi_{0,p}$ and the constants $r_p^\al$ by induction on $p$. From \eqref{existence of fxx-2} and the expression \eqref{Euler field} of the Euler vector field it follows that \eqref{zh-6}--\eqref{r^veps_p condition} hold true for $p=0, 1$. Assume that we have already found the vector fields $\xi_{0,p}$ and the constants $r_p^\al$ for $1\le p\le k$ and $1\le \al\le n$ satisfying the relations \eqref{zh-6}--\eqref{r^veps_p condition}, then by using the associativity equations \eqref{zh-5} we know the existence of a vector field $\xi_{0,k+1}=\xi_{0,k+1}^\gamma\p_\gamma$ satisfying the recursion relation \eqref{zh-7}
for $p=k+1$.
From the relation
\begin{align*}
&  \p_E\left(
    \p_\afa\cdot\xi_{0,k}
  \right)^\lmd
=
  \left(
    \p_Ec^\lmd_{\afa\beta}
  \right)
  \xi_{0,k}^\beta
 +c_{\afa\beta}^\lmd\p_E\xi_{0,k}^\beta\\
&\quad =
  \left(
    \mu_\afa+\mu_\beta-\mu_\lmd+\frac d2
  \right)c^\lmd_{\afa\beta}\xi_{0,k}^\beta
 +c^\lmd_{\afa\beta}
  \left(
    \left(
      k-\frac d2-\mu_\beta
    \right)\xi_{0,k}^\beta
   +\sum_{s=1}^k r^\veps_s\xi_{\veps,k-s}^\beta
  \right) \\
&\quad =
  (k+\mu_\afa-\mu_\lmd)c^\lmd_{\afa\beta}
  \xi_{0,k}^\beta
  +\sum_{s=1}^k
    r^\veps_s c^\lmd_{\afa\beta}\xi_{\veps,k-s}^\beta\\
&\quad =
  (k+\mu_\afa-\mu_\lmd)\p_\afa\xi^\lmd_{0,k+1}
  +\sum_{s=1}^k
    r^\veps_s \p_\afa\xi^\lmd_{\veps,k+1-s},
\end{align*}
it follows that
\begin{align*}
 & \p_\afa\left(
    \p_E\xi_{0,k+1}
  \right)
=
  \left(
    1-\frac{d}{2}-\mu_\afa
  \right)\p_\afa\xi_{0,k+1}
 +\p_E\left(
   \p_\afa\cdot\xi_{0,k}
 \right)\\
 &\quad =
  \left(
   k+1-\frac d2-\mu
 \right)\p_\afa\xi_{0,k+1}
+
 \sum_{s=1}^k r^\veps_s\p_\afa\xi_{\veps,k+1-s}.
\end{align*}
So there exist constants $r_{k+1}^\al$ such that
\begin{align*}
 & \p_E\xi_{0,k+1}
=
  \left(
   k+1-\frac d2-\mu
 \right)\xi_{0,k+1}
+
  \sum_{s=1}^{k}r^\veps_s \xi_{\veps,k+1-s}
+
  r_{k+1}^\veps\p_\veps\\
&\quad =
  \left(
   k+1-\frac d2-\mu
 \right)\xi_{0,p+1}
+
  \sum_{s=1}^{k+1}r^\veps_s\xi_{\veps,k+1-s}.
\end{align*}
The last identity can be written in the form
\[
  \p_E\xi^\al_{0,k+1}
=
   \left(
    k+1-\frac d2-\mu_\afa
  \right)\xi^\al_{0,k+1}
 +\sum_{s=1}^{k}r^\veps_s\xi^\al_{\veps, k+1-s}
 +r^\afa_{k+1},
\]
so we can adjust $\xi^\al_{0,k+1}$, if needed, by adding a constant term so that $r_{k+1}^\al$ satisfies the
condition \eqref{r^veps_p condition} with $p=k+1$. The lemma is proved.
\end{proof}

When we use the recursion relation \eqref{zh-7} to find $\xi_{0,p}^\al$ from $\xi_{0,p-1}^\al$, there is the ambiguity
\[\xi_{0,p}\mapsto \xi_{0,p}+a_{0,p},\]
here the constant vector fields $a_{0,p}$ satisfy the condition
\[\left(p-\frac{d}2-\mu\right) a_{0,p}=0.\]
In what follows we will fix a choice of the vector fields $\xi_{0,p}$ for $p\ge 0$ that satisfy the requirement of the above lemma.

\begin{rmk}
For a Frobenius manifold with flat unity, one can choose the flat coordinates such that $e=\pp{v^1}$,
and $r^\afa = (R_1)^\afa_1$.
In such a case , one can choose
\[
  \xi_{0,p}(v):=\xi_{1,p}(v),\quad p\geq 0,
\]
and the constants $r_k^\veps := (R_k)^\veps_1$.
\end{rmk}

\begin{lem}\label{thm-existence of theta-0,-p}
  Let $\xi_{0,-p}=\xi_{0,-p}^\gamma\p_\gamma$ for $p\geq 1$ be vector fields on $M$ defined recursively by
\begin{equation}\label{zh-13}
 \xi_{0,0}=e,\quad \xi_{0,-p}=\p_e\xi_{0,-p+1},\quad p\geq 1.
\end{equation}
Then the following relations hold true:
\begin{align}
  \p_\afa\xi_{0,-p}^\gamma&=
    c_{\afa\beta}^\gamma\xi^\beta_{0,-p-1},\quad \al,\gamma=1,\dots,n,\,p\geq 0,\label{theta0,-p-recur}\\
    \p_E\xi_{0,-p}
    &=\left(-p-\frac d2-\mu\right)\xi_{0,-p},\quad p\geq 0.\label{theta0,-p-quasi-homog}
 \end{align}
\end{lem}

\begin{proof}
Let us prove the validity of \eqref{theta0,-p-recur} and \eqref{theta0,-p-quasi-homog} by induction on $p\ge 0$.
By using \eqref{existence of fxx}--\eqref{existence of fxx-2} we have
\begin{align*}
c_{\al\beta}^\gamma\xi^\beta_{0,-1}&=c_{\al\beta}^\gamma\p_e\xi_{0,0}^\beta=c_{\al\beta}^\gamma\p_e e^\beta
=c_{\al\beta}^\gamma e^\lmd \p_\lmd e^\beta\\
&=\p_\lmd\left(e^\beta c_{\al\beta}^\gamma\right) e^\lmd-e^\lmd e^\beta c_{\al\beta\lmd}^\gamma\\
&=-e^\lmd e^\beta c_{\al\beta\lmd}^\gamma=\p_\al e^\gamma
=\p_\al\xi_{0,0}^\gamma,
\end{align*}
thus \eqref{theta0,-p-recur} holds true for $p=0$.
The validity of \eqref{theta0,-p-quasi-homog} for $p=0$ follows from  \eqref{theta0p-quasi-homog}.
Now we assume that  \eqref{theta0,-p-recur} and \eqref{theta0,-p-quasi-homog} also hold true for $0\le p\le k-1$, then we have
\begin{align*}
  &\p_\afa\xi^\gamma_{0,-k}
=
  \p_\afa\p_e\xi^\gamma_{0,-k+1}
 =
  \p_\afa\left(
    e^\lmd\p_\lmd\xi_{0,-k+1}^\gamma
  \right)\\
&\quad =
  (\p_\afa e^\lmd)
  \p_\lmd\xi_{0,-k+1}^\gamma
 +e^\lmd\p_\lmd\p_\afa\xi_{0,-k+1}^\gamma\\
&\quad=
   (\p_\afa e^\lmd)
  c^\gamma_{\lmd\beta}\xi_{0,-k}^\beta
 +e^\lmd\p_\lmd
 \left(
 c^\gamma_{\afa\beta}\xi_{0,-k}^\beta
 \right)\\
&\quad =
   (\p_\afa e^\lmd)
  c^\gamma_{\lmd\beta}\xi_{0,-k}^\beta
 +e^\lmd\left(\p_\lmd
 c^\gamma_{\afa\beta}\right)\xi_{0,-k}^\beta
+c^\gamma_{\afa\beta}\p_e\xi_{0,-k}^\beta
\\
&\quad =
  (\p_\afa e^\lmd)
  c^\gamma_{\lmd\beta}\xi_{0,-k}^\beta
  +e^\lmd\left(\p_\lmd
 c^\gamma_{\afa\beta}\right)\xi_{0,-k}^\beta
 +c^\gamma_{\afa\beta}\xi_{0,-k-1}^\beta
\\
&\quad =
     (\p_\afa e^\lmd)
  c^\gamma_{\lmd\beta}\xi_{0,-k}^\beta
  +\left(\p_\al(e^\lmd c_{\lmd\beta}^\gamma)-(\p_\al e^\lmd) c_{\lmd\beta}^\gamma\right)\xi_{0,-k}^\beta+c^\gamma_{\afa\beta}\xi_{0,-k-1}^\beta
\\
&\quad =
 c^\gamma_{\afa\beta}\xi_{0,-k-1}^\beta.
\end{align*}
On the other hand, by using \eqref{existence of fxx-2} we have
\begin{align*}
 & \p_E\xi_{0,-k}
=
  \p_E\p_e\xi_{0,-k+1}
=
  \p_{[E,e]}\xi_{0,-k+1}
 +\p_e\p_E\xi_{0,-k+1}\\
&\quad =
  -\p_e\xi_{0,-k+1}
  +\p_e\left(-k+1-\frac d2-\mu\right)\xi_{0,-k+1}
 = \left(
    -k-\frac d2-\mu
  \right)\xi_{0,-k},
\end{align*}
hence \eqref{theta0,-p-recur} and \eqref{theta0,-p-quasi-homog} also hold true when $p=k$.
The lemma is proved.
\end{proof}

\begin{rmk}
  For a Frobenius manifold with flat unity, the vector fields $\xi_{0,-p}$ are trivial.
\end{rmk}

Now let us fix a choice of the functions $\theta_{\al,p}$ for $\al=1,\dots,n$, $p\ge 0$ that satisfy \eqref{zh-10} by the following proposition.
\begin{prop}\label{zh-16}
Let $\xi_{\al,p}\in\mathrm{Vect}(M)$, $\al=1,\dots, n, p\ge 0$ be a fixed solution of the equations
\eqref{zh-11}--\eqref{nabla-theta-ortho}, and $\xi_{0, p}\in\mathrm{Vect}(M)$, $p\ge 0$ satisfy the conditions of Lemma \ref{lemma-existence of theta-0-p},
then the functions $\theta_{\al,p}$ defined by
\begin{equation}\label{real-theta-afa,p}
  \theta_{\afa, p}=
  \sum_{k=0}^{p}
    (-1)^k
    \pair{\xi_{0,k+1}}{\xi_{\afa,p-k}},\quad
  1\leq\afa\leq n,\, p\geq 0
\end{equation}
satisfy the relation \eqref{zh-10}, and they also satisfy the quasi-homogeneity condition
\begin{equation}\label{real D-E-theta}
\p_E\theta_{\afa, p}
=\left(
  p+1-\frac d2+\mu_\afa
\right)\theta_{\afa, p}
+\sum_{s=1}^p
  \theta_{\veps, p-s}(R_s)^\veps_\afa
  +
  (-1)^p r^\veps_{p+1}\eta_{\veps\afa},
\end{equation}
where the constants $r^\veps_{p}$
are given by Lemma \ref{lemma-existence of theta-0-p}.
\end{prop}

\begin{proof}
By using the recursion relations \eqref{zh-11} and \eqref{zh-7} we obtain
\begin{align*}
\p_\beta\theta_{\al,p}&=
    \sum_{k=0}^{p}(-1)^k
    \Big(
      \pair{\p_\beta\cdot\xi_{0,k}}{\xi_{\afa,p-k}}
     +\pair{\xi_{0,k+1}}{\p_\beta\cdot\xi_{\afa,p-k-1}}
    \Big)\\
  &=
    \pair{\p_\beta\cdot\xi_{0,0}}{\xi_{\afa,p}}
   =
     \pair{\p_\beta}{\xi_{\afa,p}}
    =\eta_{\beta\gamma}\xi^\gamma_{\al,p},
     \end{align*}
 so \eqref{zh-10} holds true. Here and in what follows we define  $\xi_{\al,-k}:=0$ for $k\ge 1$. To check the quasi-homogeneity condition, we use \eqref{zh-1}, \eqref{zh-12}, \eqref{nabla-theta-ortho}, and \eqref{theta0p-quasi-homog} to arrive at
\begin{align*}
  \p_E\theta_{\afa,p}
  =&\,
    \sum_{k=0}^{p}
      (-1)^k
      \Big(
        \pair{\p_E\xi_{0,k+1}}{\xi_{\afa,p-k}}
       +\pair{\xi_{0,k+1}}{\p_E\xi_{\afa,p-k}}
      \Big)\\
   =&\,
    \sum_{k=0}^{p}
      (-1)^k
      \left(
        \pair{(k+1-\frac d2-\mu)\xi_{0,k+1}}{\xi_{\afa,p-k}}
       +
       \sum_{s=1}^{k+1}
       r^\veps_s\pair{\xi_{\veps,k+1-s}}{\xi_{\afa,p-k}}
      \right)\\
  &\,
    +\sum_{k=0}^{p}
      (-1)^k
      \left(
        \pair{\xi_{0,k+1}}{(p-k+\mu_\afa-\mu)\xi_{\afa,p-k}}
       +\sum_{s=1}^{p-k}(R_s)^\veps_\afa
         \pair{\xi_{0,k+1}}{\xi_{\veps, p-k-s}}
      \right)\\
 =&\,
    \left(p+1-\frac d2+\mu_\afa\right)
    \sum_{k=0}^{p}
      (-1)^k
      \pair{\xi_{0,k+1}}{\xi_{\afa,p-k}}\\
  &\,
    + \sum_{s=1}^{p+1}
    r^\veps_s \sum_{k=s-1}^{p}
      (-1)^k
      \pair{\xi_{\veps,k+1-s}}{\xi_{\afa,p-k}}
    +\sum_{s=1}^p
     (R_s)^\veps_\afa
     \sum_{k=0}^{p-s}
     (-1)^k
     \pair{\xi_{0,k+1}}{\xi_{\veps,p-k-s}}\\
  =&\,
  \left(
  p+1-\frac d2+\mu_\afa
\right)\theta_{\afa, p}
+\sum_{s=1}^p
 (R_s)^\veps_\afa \theta_{\veps, p-s}
+
  (-1)^pr^\veps_{p+1}\eta_{\veps\afa}.
  \end{align*}
  The proposition is proved.
\end{proof}

\begin{prop}\label{zh-17}
 There exist functions $\{\theta_{0,p}\}_{p\in\bbZ}$ on $M$ which
  satisfy the relation
  \begin{equation}\label{zh-14}
  \nabla\theta_{0,p}=\xi_{0,p},\quad p\in\mathbb{Z}
  \end{equation}
  and the quasi-homogeneity condition
  \begin{equation}\label{def of c-p}
    \p_E\theta_{0,p}=(p-d+1)\theta_{0,p}
    +\sum_{s=1}^p r^\veps_s \theta_{\veps,p-s}+c_p,
    \quad p\in\bbZ,
  \end{equation}
  where we define $\theta_{\veps,p}:=0$ if $p<0$ and $\veps\neq 0$,
  and $c_p$ are certain constants which are possibly non-zero only if
  $p=d-1\in\bbZ$ and $p$ is even.
 \end{prop}

\begin{proof}
 The existence of functions $\theta_{0,p}$ for $p\in\mathbb{Z}$ satisfying \eqref{zh-14} and \eqref{def of c-p} follows from the recursion relations \eqref{zh-7}, \eqref{theta0,-p-recur} and the associativity equations \eqref{zh-5}. In particular, when $p=2q+1$ we can take
 \begin{equation}\label{real theta0,2p+1}
  \theta_{0,p}=
  \begin{cases}
      \frac12\sum_{k=0}^{p-1}
        (-1)^k
        \pair{\xi_{0,p-k}}{\xi_{0,k+1}},
   & p>0,
  \\[5pt]
    \frac12\sum_{k=0}^{-p-1}
    (-1)^k
    \pair{\xi_{0,p+1+k}}{\xi_{0,-k}},
   & p<0,
  \end{cases}
  \end{equation}
then it follows from \eqref{theta0p-quasi-homog} and \eqref{theta0,-p-quasi-homog} that $c_p=0$. When $p$ is even and $p\ne d-1$, we can adjust $\theta_{0,p}$ by adding a constant so that $c_p$ vanishes in the relation \eqref{def of c-p}. The proposition is proved.
\end{proof}

From \eqref{real-theta-afa,p}, \eqref{real theta0,2p+1} and \eqref{zh-15}, \eqref{zh-6}, \eqref{zh-13} we know that
\begin{equation}\label{zh-18}
\theta_{0,0}=\fai,\quad \theta_{0,1}=\frac12 v^\gamma v_\gamma,\quad \theta_{0,-1}=\frac12 e^\gamma e_\gamma,\quad \theta_{\al,0}=v_\al,
\quad \al=1,\dots,n.
\end{equation}
Here $\fai$ is defined in Proposition \ref{prop existence of theta 00}.
When $d\ne 1$, we can adjust the choice of $\fai$ by adding to it a constant so that $\theta_{0,0}=\fai$ satisfies the quasi-homogeneity condition \eqref{def of c-p} with $c_0=0$.

Now we are ready to introduce the Principal Hierarchy of the generalized Frobenius manifold $M$.
\begin{defn}
  The Principal Hierarchy of $M$
  is given by the following family of evolutionary PDEs:
  \begin{equation}\label{principal hierarchy}
   \begin{cases}
    \pfrac{v^\afa}{t^{\beta,q}}
    =\eta^{\al\gamma}\frac{\p}{\p x}\left(\frac{\p\theta_{\beta,q+1}}{\p v^\gamma}\right),& 1\leq\beta\leq n,\, q\geq 0,\\
    \pfrac{v^\afa}{t^{0,q}}
    =\eta^{\al\gamma}\frac{\p}{\p x}\left(\frac{\p\theta_{0,q+1}}{\p v^\gamma}\right), & q\in\bbZ,
   \end{cases}
  \end{equation}
  where $\theta_{\afa, p\geq 0}$, $\theta_{0,q\in\bbZ}$ are defined in Propositions \ref{zh-16}, \ref{zh-17}.
\end{defn}

From the associativity equations \eqref{zh-5} it follows that the flows of the Principal Hierarchy pairwise commute, and from \eqref{zh-18} we know that the Principal Hierarchy contains the flow given by the
translation along the spatial variable $x$, i.e.,
\[
  \pfrac{v^\afa}{t^{0,0}}
 =v^\afa_x,
\]
so in what follows we will identify the time variable $t^{0,0}$ with the
spatial variable $x$, i.e.,
\[t^{0,0}=x.\]

We note that the Principal Hierarchy can also be written in the form
\begin{equation}
\begin{cases}
\pfrac{v}{t^{\afa,p}}=\p_x\nabla\theta_{\afa,p+1}=
\nabla\theta_{\afa,p}\cdot v_x, & 1\leq\afa\leq n,\, p\geq 0,\\[3pt]
\pfrac{v}{t^{0,q}}=\p_x\nabla\theta_{0,p+1}=
\nabla\theta_{0,q}\cdot v_x, & q\in\bbZ,
\end{cases}
\end{equation}
where $v:=(v^1,\dots,v^n)^{\transpose}$, and $v_x:=(v^1_x,\dots,v^n_x)^{\transpose}$.

As for a Frobenius manifold with flat-unity \cite{Dubrovin-2DTFT,normal-form},
on the jet space $J^\infty(M)$ we also have a bihamiltonian structure given by the compatible Hamiltonian operators $\mathcal{P}_a=(\mathcal{P}_a^{\al\beta})$, $a=1,2$ with
\begin{equation}\label{Bihamiltonian structure}
\mathcal{P}_1^{\al\beta}=\eta^{\afa\beta}\p_x,\quad
  \mathcal{P}_2^{\al\beta} = g^{\afa\beta}\p_x+\Gamma^{\afa\beta}_\gamma v^\gamma_x,
\end{equation}
here the intersection form $g^{\afa\beta}$ and
the contravariant coefficients of the Levi-Civita connection of $g$ are given in \eqref{zh-19} and \eqref{connection Gamma}.
The Principal Hierarchy \eqref{principal hierarchy} is a bihamiltonian hierarchy with respect to this bihamiltonian structure, i.e., its flows are Hamiltonian systems of the form
\[\frac{\p v^\al}{\p t^{j,q}}=\mathcal{P}_1^{\al\gamma}\frac{\delta H_{j,q}}{\delta v^\gamma},\quad \al=1,\dots,n,\, (j,q)\in\mcalI,\]
and they satisfy the
bihamiltonian
recursion relations
\begin{align*}
\mathcal{P}_2^{\al\gamma}\frac{\delta H_{\beta,q-1}}{\delta v^\gamma}&=\left(q+\frac12+\mu_\beta\right)\mathcal{P}_1^{\al\gamma}\frac{\delta H_{\beta,q}}{\delta v^\gamma}+\sum_{s=1}^q (R_s)^\lmd_\beta \mathcal{P}_1^{\al\gamma}\frac{\delta H_{\lmd,q-s}}{\delta v^\gamma},\\
\mathcal{P}_2^{\al\gamma}\frac{\delta H_{0,p-1}}{\delta v^\gamma}&=\left(p+\frac12-\frac{d}2\right)\mathcal{P}_1^{\al\gamma}\frac{\delta H_{0,p}}{\delta v^\gamma}+\sum_{s=1}^p r_s^\lmd \mathcal{P}_1^{\al\gamma}\frac{\delta H_{\lmd,p-s}}{\delta v^\gamma},
\end{align*}
for $\al, \beta=1,\dots, n$, $q\ge 0$, $p\in\mathbb{Z}$. Here the index set $\mcalI$ is defined by
\begin{equation}\label{zh-23}
  \mcalI= \Big(\{1,\dots,n\}\times\bbZ_{\geq 0}\Big)\cup
  \Big(\{0\}\times\bbZ\Big),
\end{equation}
and the Hamiltonians are given by
\[H_{j,q}=\int \theta_{j,q+1}(v(x)) \td x.\]

\section{The tau-cover of the Principal Hierarchy}
In this section, we introduce the \textit{tau structure} and \textit{tau-cover} of the Principal Hierarchy of the generalized Frobenius manifold $M$ following
\cite{Dubrovin-2DTFT, normal-form}.

Let $\Bigset{\theta_{i,p}}{(i,p)\in\mcalI}$ be the set of functions on $M$ given by Propositions \ref{zh-16}, \ref{zh-17}, where the index set $\mcalI$ is defined by \eqref{zh-23}. We introduce a set of functions $\{\Omega_{i,p;j,q}\mid (i,p), (j,q)\in\mcalI\}$ on $M$,
which is called the \textit{tau structure} of the Principal Hierarchy \eqref{principal hierarchy}, as follows:
  \begin{align}
  \Omg_{\afa,p;\beta,q}&=
    \sum_{k=0}^{q}
    (-1)^k
    \langle
      \nabla\theta_{\afa,p+k+1},
      \nabla\theta_{\beta,q-k}
    \rangle,
  \label{Omg afa p beta q}
  \\
  \Omg_{0,\ell;\beta ,q}
  &=
    \sum_{k=0}^{q}
      (-1)^k
      \pair{\nabla\theta_{0,\ell+1+k}}{\nabla\theta_{\beta,q-k}},
  \label{Omg 0 p beta q}
  \\
  \Omg_{\afa,p; 0,m}
  &=
    \sum_{k=0}^{p}
      (-1)^k
      \pair{\nabla\theta_{\afa, p-k}}{\nabla\theta_{0,m+1+k}},
  \label{Omg afa p 0 q}
  \\
  \Omg_{0,\ell; 0,m}
  &=
    \begin{cases}
      \sum_{k=0}^{\ell-1}
        (-1)^k
        \pair{\nabla\theta_{0,\ell-k}}{\nabla\theta_{0,m+1+k}}+(-1)^{\ell}\theta_{0,\ell+m}, & \ell\geq 0,\\[4pt]
      \sum_{k=0}^{-\ell-1}
        (-1)^k
        \pair{\nabla\theta_{0,\ell+1+k}}{\nabla\theta_{0,m-k}}+(-1)^{\ell}\theta_{0,\ell+m}, & \ell<0.\\
    \end{cases}
  \label{Omg 0 p 0 q}
  \end{align}
Here $\afa,\beta=1,\dots,n$, $p,q\in\bbZ_{\geq 0}$ and $\ell, m\in\bbZ$.
We also define $\Omg_{i,p; j,q}:=0$ if $i\neq 0, p<0$
or $j\neq 0, q<0$.

\begin{prop}
The functions $\Omg_{i,p;j,q}$ defined above
satisfy the identities
\begin{align}
\Omg_{0,0; i,p}&=\theta_{i,p},\quad
\Omg_{i,p; j,q}=\Omg_{j,q; i,p},\quad
\nabla\Omg_{i,p;j,q}=\nabla\theta_{i,p}\cdot\nabla\theta_{j,q},\label{Omg prop 4}
\end{align}
and for any solution $v(\bft)=(v^1(\bft),\dots,v^n(\bft))$ of the Principal  Hierarchy \eqref{principal hierarchy}, we have
\begin{equation}
\pfrac{\theta_{i,p}}{t^{j,q}}=\p_x\Omg_{i,p; j,q},\label{Omg prop 3}
\end{equation}
here the independent variables $v^\al$ of the functions $\theta_{i,p}$ and $\Omega_{i,p;j,q}$ are understood to be substituted
by $v^\al(\bft)$, and we denote $\bft=(t^{i,p})_{(i,p)\in\mcalI}$.
\end{prop}

\begin{proof}
  The first two identities in \eqref{Omg prop 4}
  can be verified straightforwardly from \eqref{nabla-theta-ortho}, \eqref{real theta0,2p+1} and
  \eqref{Omg afa p beta q}--\eqref{Omg 0 p 0 q}.
  The last identity in \eqref{Omg prop 4} follows from the relations
  \eqref{zh-11}, \eqref{zh-7}, \eqref{theta0,-p-recur}  and \eqref{real-theta-afa,p}.
  For example, for integers $\ell\geq 0$ and $m\in\bbZ$, we have
  \begin{align*}
   & \p_\afa\Omg_{0,\ell;0,m} \\
  &\quad=
    \sum_{k=0}^{\ell-1} (-1)^k
      \Big(
        \pair{\p_\afa\cdot\nabla\theta_{0,\ell-1-k}}{\nabla\theta_{0,m+1+k}}
       +\pair{\nabla\theta_{0,\ell-k}}{\p_\afa\cdot\nabla\theta_{0,m+k}}
      \Big)\\
    &\qquad+(-1)^\ell \p_\afa\theta_{0,\ell+m} \\
  &\quad=
    \left(
      \sum_{k=0}^{\ell-1}
     -\sum_{k=1}^{\ell}
    \right)(-1)^k
    \pair{\nabla\theta_{0,\ell-k}\cdot\nabla\theta_{0,m+k}}{\p_\afa}
   +(-1)^{\ell}\pair{\nabla\theta_{0,\ell+m}}{\p_\afa} \\
  &\quad=
    \pair{\nabla\theta_{0,\ell}\cdot\nabla\theta_{0,m}}{\p_\afa}
   -(-1)^\ell
   \pair{\nabla\theta_{0,0}\cdot\nabla\theta_{0,m+\ell}}{\p_\afa}
   +(-1)^{\ell}\pair{\nabla\theta_{0,\ell+m}}{\p_\afa}  \\
  &\quad=
    \pair{\nabla\theta_{0,\ell}\cdot\nabla\theta_{0,m}}{\p_\afa}
  =
    \eta_{\afa\beta}
    \big(
      \nabla\theta_{0,\ell}\cdot\nabla\theta_{0,m}
    \big)^\beta,
  \end{align*}
  which implies $\nabla\Omg_{0,\ell;0,m} = \nabla\theta_{0,\ell}\cdot\nabla\theta_{0,m}$
  for $\ell\geq 0$ and $m\in\bbZ$. In the same way, it can be verified that
  $\nabla\Omg_{i,p;j,q} = \nabla\theta_{i,p}\cdot\nabla\theta_{j,q}$ hold true for all $(i,p), (j,q)\in\mcalI$, we omit
  the details here.

  The equation \eqref{Omg prop 3} can be derived from the last identity in \eqref{Omg prop 4} as follows:
  \begin{align*}
    \pfrac{\theta_{i,p}}{t^{j,q}}
  &=\,
    \p_\afa\theta_{i,p}
    \eta^{\afa\beta}\p_x\p_\beta\theta_{j,q+1}
  =
    (\p^\beta\theta_{i,p}) c_{\beta\lmd}^\delta
    (\p^\lmd\theta_{j,q})\eta_{\delta\gamma}v^\gamma_x \\
  &=\,
    \pair{\nabla\theta_{i,p}\cdot\nabla{\theta_{j,q}}}{v_x}
  =
    \pair{\nabla\Omg_{i,p;j,q}}{v_x}
  =
    \p_x\Omg_{i,p;j,q}.
  \end{align*}
The proposition is proved.
\end{proof}

We note that the functions $\Omg_{\afa,p; \beta,q}$ coincide with the ones defined by the following formulae of \cite{Dubrovin-2DTFT}:
\begin{equation}\label{Omega afa beta series}
\Omg_{\afa\beta}(z,w)
 =\frac{\langle\nabla\theta_\afa(z), \nabla\theta_\beta(w)\rangle-\eta_{\afa\beta}}
       {z+w},
\end{equation}
where the generating functions $\Omg_{\afa\beta}(z,w)$ are defined by
\begin{equation}
  \Omg_{\afa\beta}(z,w):=\sum_{p,q\geq 0}\Omg_{\afa,p;\beta,q}z^pw^q,
  \quad
  1\leq \afa,\beta\leq n.
\end{equation}

Now let us introduce the tau-cover of the Principal Hierarchy of $M$ following \cite{Biham coh 2, normal-form}.

\begin{defn} The system of PDEs
  \begin{equation}\label{tau-cover}
        \pfrac{f}{t^{j,q}}=f_{j,q},\quad
    \pfrac{f_{i,p}}{t^{j,q}}=\Omg_{i,p;j,q},\quad
    \pfrac{v^\afa}{t^{j,q}}=\eta^{\al\gamma}\p_x\Omg_{\gamma,0;j,q}
  \end{equation}
 is called the tau-cover of the Principal Hierarchy \eqref{principal hierarchy} of $M$, where $(i,p), (j,q)\in\mcalI$ and $\al=1,\dots, n$.
\end{defn}

For a given solution
\[\{f(\bft), f_{i,p}(\bft), v^\al(\bft)\mid (i,p)\in\mcalI,\al=1,\dots,n\}\]
of \eqref{tau-cover}, we call the function $\tau^{[0]}=\exp(f)$ the tau function of the solution
\[v(\bft)=(v^1(\bft),\dots,v^n(\bft))\]
of the Principal Hierarchy \eqref{principal hierarchy}.
We have the following relation of the tau function with the associated solution of the Principal Hierarchy:
\begin{equation}
  v^\afa=\eta^{\afa\gamma}\frac{\p^2 \log\tau^{[0]}}{\p t^{0,0}\p t^{\gamma,0}}=\eta^{\afa\gamma}
  \frac{\p^2 \log\tau^{[0]}}{\p x\p t^{\gamma,0}}.
\end{equation}

We can specify a particular class of solutions of the tau-cover \eqref{tau-cover} of the Principal Hierarchy following the approach of
\cite{Dubrovin-2DTFT, normal-form}. To this end, we fix a point $v_0\in M$ and a set of
constants $c^{i,p}$, $(i,p)\in\mcalI$ with only finitely many of them being nonzero. These constants are also required to satisfy the conditions that
\[c^{0,0} e^\al(v_0)+c^{\al,0}=-\sum_{(i,p)\in\mcalI,\, p\ne 0} c^{i,p}\nabla^\al\theta_{i,p}(v_0)\]
and the matrix
\[\left(\sum_{(i,p)\in\mcalI} c^{i,p}\frac{\p^2\theta_{i,p}}{\p v^\al\p v^\beta}(v_0)\right)\]
is invertible. Then from \cite{Dubrovin-2DTFT, normal-form} we know that the Euler-Lagrange equation
\begin{equation}\label{EL equation}
  \sum_{(i,p)\in\mcalI}
  \tilde{t}^{i,p}\nabla\theta_{i,p}(v(\bft))=0
\end{equation}
yields a solution $v(\bft)$ of the Principal Hierarchy \eqref{principal hierarchy} with
\[v(\bft)|_{\bft=0}=v_0.\]
Here we denote
\[\tilde{t}^{i,p}=t^{i,p}-c^{i,p},\quad (i,p)\in\mcalI,\]
and we identify $t^{0,0}$ with the spatial variable $x$.
For such a solution $v(\bft)$ of the Principal Hierarchy, let us introduce, following \cite{Dubrovin-2DTFT}, the \textit{genus zero free energy}
\begin{equation}\label{genus 0 free energy}
  \mcalF_0(\bft)
 :=\frac12\sum_{(i,p), (j,q)\in\mcalI}
  \tilde{t}^{i,p}
  \tilde{t}^{j,q}
  \Omg_{i,p; j,q}(v(\bft)),
\end{equation}
then we have a solution of the tau-cover \eqref{tau-cover} of the Principal Hierarchy given by $v(\bft)$ and
\[f(\bft)=\mcalF_0(\bft),\quad  f_{i,p}(\bft)=\sum_{(j,q)\in\mcalI}\tilde{t}^{j,q}\Omg_{i,p;j,q}(v(\bft)).\]

\section{The associated $(n+2)$-dimensional Frobenius manifold}

In this section, we are to associate the generalized Frobenius manifold $M$ with an $(n+2)$-dimensional Frobenius manifold $\Mtil$
with flat unity following the construction of Mironov and Taimanov given in \cite{algebraic examples}. This Frobenius manifold structure will be used in our consideration of the Virasoro symmetries of the Principal Hierarchy of $M$ and their linearization.

\begin{lem}[see Lemma 3 of \cite{algebraic examples}]\label{lemma: n+2 dim intro} Let $F=F(v^1,\dots,v^n)$, $E=E^\afa\p_\afa$ be the potential and the Euler vector field of $M$, and $d$ be the charge of $M$,
then the function
  \[
    \Ftil(v^0, v^1,\dots,v^n, v^{n+1}):=
    \frac12(v^0)^2v^{n+1}
   +\frac12 v^0 v^\afa v_\afa
   +F(v^1,\dots,v^n)
  \]
yields a Frobenius manifold structure of charge $d$ on
$\Mtil=\bbC\times M\times\bbC$ with flat coordinates $v^0, v^1,\dots,v^n, v^{n+1}$ and flat unity $\etil=\frac{\p}{\p v^0}$.
The flat metric and the Euler vector field are given by
  \begin{equation}\label{zh-25}
    \etatil=
    \begin{pmatrix}
        &   & 1 \\
        & \eta &   \\
      1 &   &
    \end{pmatrix}
  \end{equation}
  and
  \begin{equation}\label{n+2 Euler}
    \Etil = v^0\pp{v^0} + E +
    \left(
      (1-d)v^{n+1}+c_0
    \right) \pp{v^{n+1}},
  \end{equation}
  where the constant $c_0$ is defined in Lemma \ref{zh-17}, and $v^0, v^{n+1}$ are the coordinates of the first and the last components $\mathbb{C}$ of $\Mtil$.
\end{lem}

\begin{rmk}
The Frobenius manifold $\Mtil$ is not semisimple,
since the vector field $\pp{v^{n+1}}$ is nilpotent.
\end{rmk}

\begin{rmk}\label{injection n to n+2}
The map
\begin{align*}
  \imath \colon M &\hookrightarrow \Mtil,\quad
  (v^1,\dots,v^n) \mapsto (0, v^1,\dots,v^n, \theta_{0,0}(v^1,\dots,v^n))
\end{align*}
preserves the multiplication and the flat metric of the generalized Frobenius manifolds, so we can view $M$ as a generalized Frobenius submanifold of $\Mtil$.
\end{rmk}

The monodromy data $\mutil, \Rtil=\Rtil_1+\Rtil_2+\dots$ of the Frobenius manifold
$\Mtil$ at $z=0$ can be represented in terms of that of $M$ as follows:
\begin{equation}\label{n+2 mu R intro}
  \mutil = \begin{pmatrix}
              \mu_0 &  &  \\
               & \mu &  \\
               &   & \mu_{n+1}
            \end{pmatrix},\quad
  \Rtil_s =
  \begin{pmatrix}
    0 & 0 & 0 \\
    \bm{r}_s & R_s & 0 \\
    c_{s-1} & \bm{r}_s^\dag & 0
  \end{pmatrix}, \quad s\geq 1,
\end{equation}
where $\mu$ is defined in \eqref{zh-2},
\[
  \mu_0=-\frac d2,\quad
  \mu_{n+1}= \frac d2,
\]
and $\bm{r}_s$, $\bm{r}^\dag_s$ are given by the column and row vectors
\begin{equation}\label{bfr_s dag}
  \bm{r}_s=
(r_s^1,r_s^2,\dots,r_s^n)^{\transpose},\quad
  \bm{r}^\dag_s=(r_{s,1}\dots,r_{s,n})
\end{equation}
respectively with
\[r_{s,\al}:= (-1)^{s+1}  r_s^\beta \eta_{\afa\beta}.\]
It is easy to check that the matrices $\etatil$, $\mutil$ and $\Rtil_s$ satisfy the relations \eqref{zh-1}, \eqref{monodromy-R} and \eqref{zh-3}.
The following proposition justifies our assertion that $\mutil$, $\Rtil$
are indeed the monodromy data of the Frobenius manifold $\Mtil$ at $z=0$.

\begin{prop}[\cite{WYW}] \label{zh-20}
Let $\Mtil$ be the $(n+2)$-dimensional Frobenius manifold associated with $M$, and $\theta_{\afa,p}$, $\theta_{0,p}$ be the functions on $M$  defined in the previous sections. Introduce the following functions on
$\Mtil$:
\begin{align}\label{n+2 dim theta-1}
  \thetatil_{0,p}
  &=\sum_{k=0}^{p-1}
  \frac{(v^0)^k}{k!}\theta_{0,p-k}
  +\frac{(v^0)^p}{p!}v^{n+1},\quad  \thetatil_{n+1,p}
  =
    \frac{(v^0)^{p+1}}{(p+1)!},\quad p\ge 0;\\
  \thetatil_{\afa,p}
  &=
    \sum_{k=0}^p
    \frac{(v^0)^k}{k!}\theta_{\afa,p-k}
    ,\quad
  1\leq \afa\leq n,\, p\ge 0.
  \label{n+2 dim theta-2}
\end{align}
Then
\begin{equation}\label{n+2 deformed flat coordinate}
  (\thetatil_{0}(z),\dots,\thetatil_{n+1}(z))z^{\mutil}z^{\Rtil}
\end{equation}
yields a system of deformed flat coordinates of $\Mtil$, where
\begin{equation}\label{n+2 theta_i(z)}
\thetatil_i(z)=\sum_{p\geq 0}\thetatil_{i,p}z^p, \quad
0\leq i\leq n+1.
\end{equation}
Moreover, the functions $\thetatil_{i,p}$ satisfy the relations
\begin{equation}
  \thetatil_{i,0} =\sum_{j=0}^{n+1} \etatil_{ij} v^j,\quad
  \thetatil_{i,p} = \frac{\p\thetatil_{i,p+1}}{\p v^0},
  \quad 0\leq i\leq n+1,\quad  p\geq 0.
\end{equation}
\end{prop}

\begin{proof}
To prove that \eqref{n+2 deformed flat coordinate} gives a system of deformed flat coordinates of $\Mtil$, we need to show the validity of the following equations:
  \begin{align*}
    \frac{\p^2\thetatil_{\ell,p+1}}{\p v^i\p v^j}
    &=
     \sum_{k=0}^{n+1}\ctil_{ij}^{\,k}\frac{\p\thetatil_{\ell,p}}{\p v^k},\quad 0\le i, j, \ell\le n+1,\, p\ge 0; \\
   \p_{\Etil}\nablatil\thetatil_{i,p}
  &=(p+\mutil_i-\mutil)\nablatil\thetatil_{i,p}
  +\sum_{s=1}^p
    \nablatil\thetatil_{j,p-s}
    (\Rtil_s)^j_i,\quad 0\le i\le n+1,\, p\ge 0.
  \end{align*}
Here $\Etil$ is given by \eqref{n+2 Euler},
$\nablatil\thetatil_{i,p}$ are the gradients of $\thetatil_{i,p}$
with respect to the metric $\etatil$, and
\[\ctil_{ij}^{\,k}=\sum_{\ell=0}^{n+1}\etatil^{k\ell}\frac{\p^3\Ftil}{\p v^i\p v^j\p v^\ell},\quad 0\le i, j, k\le n+1.\]
These relations can be proved by a straightforward calculation, so we omit it here. The proposition is proved.
\end{proof}

\begin{rmk}
  Let $\ftil$ be a function on $\Mtil$. Define the restriction of $\ftil$ to $M$ by
\begin{equation}\label{restrict n+2 to n}
  \ftil|_{M} := \imath^*(\ftil),
\end{equation}
where the map $\imath\colon M\hookrightarrow\Mtil$
is introduced in Remark \ref{injection n to n+2}.
Then it is clear that
\begin{equation}\label{zh-28}
  v^0|_{M}=0,\quad
  v^\afa|_{M}=v^\afa,\quad
  v^{n+1}|_{M}=\theta_{0,0},
\end{equation}
and
\begin{equation}\label{qu-220801}
\thetatil_{0,p}|_{M} = \theta_{0,p},\quad
\thetatil_{\afa,p}|_{M} = \theta_{\afa,p},\quad
\thetatil_{n+1,p}|_{M} = 0,\quad p\ge 0.
\end{equation}
We note that the construction of $\thetatil_{i,p}$ and $\Rtil$ given above does not involve the functions $\theta_{0,-p}$ and the constants $c_{-p}$ for $p>0$, and we also have the following identities which can be easily proved:
\begin{align}
\frac{\p}{\p v^0}\left(
\thetatil_0(z),
\thetatil_\gamma(z),
\thetatil_{n+1}(z)
\right)\Big|_{M}
&= \left(z\theta_0(z),z\theta_\gamma(z),1\right), \label{zh-21-a}
\\
\frac{\p}{\p v^\al}\left(
\thetatil_0(z),
\thetatil_\gamma(z),
\thetatil_{n+1}(z)
\right)\Big|_{M} &=
\left(\frac{\p\theta_0(z)}{\p v^\al}-e_\afa,
\frac{\p\theta_\gamma(z)}{\p v^\al}, 0\right), \label{zh-21}
\end{align}
where $\theta_0(z)=\sum_{p=0}^{\infty}\theta_{0,p}z^p$.
\end{rmk}

\section{Periods of Generalized Frobenius manifolds
and Laplace-type integrals}

In this section we introduce the notion of periods of the generalized Frobenius manifold $M$, which is a slight modification of the one for a usual Frobenius manifold \cite{Dubrovin-2DTFT, Dubrovin-painleve, normal-form}. We will use a certain basis of periods of $M$ to represent the Virasoro symmetries of the tau-cover of the associated Principal Hierarchy in the next section.

For any complex parameter $\lmd\in\bbC$ we denote
\begin{equation}
  \Sigma_\lmd=
  \Bigset{v\in M}{\det (g^{\afa\beta}-\lmd\eta^{\afa\beta})|_v=0},
\end{equation}
here $g=(g^{\afa\beta})$ is the intersection form of $M$ defined in \eqref{zh-27}--\eqref{zh-19}.
When $|\lmd|$ is sufficiently large, the set $M\setminus\Sigma_\lmd$ is not empty,
and then the inverse
\[
  (g_{\afa\beta}(\lmd)):=(g^{\afa\beta}-\lmd\eta^{\afa\beta})^{-1}
\]
defines a flat metric $(\ ,\, )_\lm$ on $M\setminus\Sigma_\lmd$. In the flat coordinates $v^1,\dots, v^n$ of $M$, the contravariant components of the Levi-Civita connection $\nabla(\lm)$ of this metric are given by
\[\nabla(\lm)^\al \nd v^\beta=(g^{\al\gamma}-\lm \eta^{\al\gamma})\nabla(\lm)_\gamma\nd v^\beta=\Gamma^{\al\beta}_\gamma \nd v^\gamma,\]
where $\Gamma^{\al\beta}_\gamma$ are the contravariant coefficients of the Levi-Civita connection of $g$ given in \eqref{connection Gamma}. We can find a system of flat local coordinates
$p_1(v;\lm),\dots, p_n(v;\lm)$ of $(\ ,\,)_\lm$ on an open subset of $M\setminus \Sigma_\lm$ which satisfy the equations
\begin{equation}\label{flat 1-forms}
\nabla(\lm)\nd p_\al=0,\quad \al=1,\dots,n.
\end{equation}
These equations can be represented in the form
\begin{equation}\label{GM eqn 1}
  \p_\gamma\nabla p_\afa =
  -\left(\mcalU-\lmd I\right)^{-1}\mcalC_\gamma
  \left(
    \mu+\frac12
  \right)\nabla p_\afa,
\end{equation}
where $\nabla p_\al$ are the gradients of $p_\al$ with respect to the metric $\eta$, and
$\mcalU$, $\mcalC_\gamma$ are the operators of multiplication by the vector fields $E$ and $\p_\gamma$ respectively, i.e.,
\begin{equation}\label{zh-2024-4}
\mcalU^\afa_\beta=E^\gamma c^\afa_{\gamma\beta},\quad
(\mcalC_\gamma)^\afa_\beta=c_{\gamma\beta}^{\afa}.
\end{equation}
It can be verified directly that the functions $p_\afa$ can be chosen in such a way that
the following equations
\begin{equation}\label{GM eqn 2}
  \pp{\lmd}\nabla p_\afa =
  \left(\mcalU-\lmd I\right)^{-1}
  \left(
    \mu+\frac12
  \right)\nabla p_\afa
\end{equation}
also hold true.
  To show this, it suffices to verify the following compatibility condition
  of \eqref{GM eqn 1} and \eqref{GM eqn 2}:
\begin{equation}\label{zh-2024-1}
    \left[
      \pp{\lmd}, \pp{v^\gamma}
    \right] \nabla p_\afa = 0,\quad 1\leq\gamma\leq n.
  \end{equation}
In fact, by using the associativity equations \eqref{zh-5} we have
\[[\mcalU-\lmd I, \mcalC_\gamma]=0,\]
from which it follows that
\begin{align}
&\pp{\lmd}
  \left(
    \pp{v^\gamma}\nabla p_\afa
  \right)
=
  \pp{\lmd}
  \left(
    -\frac{1}{\mcalU-\lmd I}
    \mcalC_\gamma
    \left(
      \mu+\frac12
    \right)\nabla p_\afa
  \right)\notag\\
 &\quad=
  -\frac{\mcalC_\gamma}{\mcalU-\lmd I}
  \left(
    \mu+\frac 32
  \right)
  \frac{1}{\mcalU-\lmd I}
  \left(
    \mu+\frac12
  \right)\nabla p_\afa, \label{zh-2024-2}\\
&\pp{v^\gamma}
\left(
  \pp{\lmd}\nabla p_\afa
\right)
=
  \pp{v^\gamma}
  \left(
    \frac{1}{\mcalU-\lmd I}
    \left(\mu+\frac12\right)\nabla p_\afa
  \right) \notag\\
&\quad=
  -\frac{1}{\mcalU-\lmd I}
   \left(
     \pfrac{\mcalU}{v^\gamma}
   \right)
   \frac{1}{\mcalU-\lmd I}
   \left(\mu+\frac12\right)\nabla p_\afa \notag\\
&\qquad
  -\frac{1}{\mcalU-\lmd I}
  \left(\mu+\frac12\right)
  \frac{\mcalC_\gamma}{\mcalU-\lmd I}
  \left(\mu+\frac12\right)\nabla p_\afa.\label{zh-2024-3}
\end{align}
On the other hand, from \eqref{FM4-c} and \eqref{Euler field} we obtain the identity
\begin{align*}
  &\pp{v^\gamma} \left(\mcalU^\afa_\beta\right)
=
  \p_\gamma\left(E^\veps c^\afa_{\veps\beta}\right)
=
  \left(
    \frac{2-d}{2}-\mu_\gamma
  \right)c^\afa_{\gamma\beta}
  +E^\veps\p_\gamma c^\afa_{\veps\beta} \\
&\quad=
  \left(
    \frac{2-d}{2}-\mu_\gamma
  \right)c^\afa_{\gamma\beta}
  +\p_E c^\afa_{\gamma\beta}\\
&\quad=
  \left(
    \frac{2-d}{2}-\mu_\gamma
  \right)c^\afa_{\gamma\beta}
  +
    \left(
      \mu_\beta+\mu_\gamma-\mu_\afa+\frac d2
    \right)
  c^\afa_{\gamma\beta}\\
&\quad=
  (1+\mu_\beta-\mu_\afa)c^\afa_{\gamma\beta},
\end{align*}
which can be rewritten as
\begin{equation}\label{240818-1342}
  \pfrac{\mcalU}{v^\gamma} = \mcalC_\gamma + \mcalC_\gamma\mu - \mu\mcalC_\gamma.
\end{equation}
By using \eqref{zh-2024-2}, \eqref{zh-2024-3} and the above identity we arrive at the compatibility condition \eqref{zh-2024-1}.

The equations \eqref{GM eqn 1},\eqref{GM eqn 2} are called
the \textit{Gauss-Manin equations} of the generalized Frobenius manifold $M$,
and solutions of the Gauss-Manin equations are called \textit{periods}
of $M$.

Let $p_1(v;\lmd),\dots,p_n(v,\lmd)$ be a basis of periods of $M$,
then the entries
\[G^{\al\beta}=\left(\frac{\p}{\p p_\al},\frac{\p}{\p p_\beta}\right)_\lm\]
of the Gram matrix $(G^{\al\beta})$ are constants
which are also independent of $\lmd$ because of \eqref{GM eqn 2},
and that of the inverse $(G_{\al\beta})=(G^{\al\beta})^{-1}$ of the Gram matrix are given by
\begin{equation}\label{240813-0855}
  G_{\afa\beta}=\left(\nabla p_\afa\right)^{\transpose}\eta\left(\mcalU-\lmd I\right)\nabla p_\beta.
\end{equation}
We note that the Gauss-Manin equations also imply that the periods
satisfy the relations
\begin{equation}\label{240727-1957}
\pp\lmd \nabla p_\afa=-\p_e \nabla p_\afa,
\end{equation}
where $e$ is the unit vector field of $M$. For a
usual Frobenius manifold with flat unity,
the periods $p_\afa$ can be chosen in a way such that $\pp\lmd p_\afa=-\p_e p_\afa$;
however, this may not hold true for a generalized Frobenius manifold
because of the non-flatness of the unit vector field.

Let us introduce, following \cite{almost dual, normal-form}, the regularized periods
\begin{equation}\label{periods-Lapacle-2}
\left(
      p_1^{(\nu)}(v;\lmd),\dots,p_n^{(\nu)}(v;\lmd)
    \right) =\int_0^{\infty}
  \frac{\td z}{\sqrt{z}}
  \rme^{-\lmd z}
  (\theta_1(z),\dots,\theta_n(z))z^{\mu+\nu} z^R.
\end{equation}
%
Here the Laplace integral in the right-hand side of \eqref{periods-Lapacle-2}
is understood as a formal series of $\lmd$
via the following formula \cite{almost dual, normal-form}:
\begin{align}\label{basic Laplace formula}
 & \int_{0}^{\infty}
    \rme^{-\lmd z}
    z^{p+\mu+\nu-\frac12}
    z^R\td z\notag\\
  &\quad =
    \sum_{s\geq 0}
    \left[
      \rme^{R\p_\nu}
    \right]_s
    \Gamma\left(
      p+\mu+\nu+s+\frac12
    \right)
    \lmd^{-(p+\mu+\nu+s+\frac12)}
    \lmd^{-R}
\end{align}
for $p\in\bbZ_{\geq 0}$,
$\nu\in\bbC$,
with the components $[P(\mu, R)]_s$ for any polynomial $P(\mu, R)$ of $\mu, R$
are given by
\begin{equation}\label{zh-30}
  [P(\mu, R)]_s
 :=
   \sum_{\lm\in \textrm{Spec}\,\mu}\pi_{\lm+s}P(\mu, R)\pi_\lm,
\end{equation}
where $\pi_\lm\colon V\to V_\lm$ is the projection of
$V=\mathbb{C}^{n}$ to the eigenspace $V_\lm$ of the operator $\mu$.
The coefficients of the series are meromorphic functions of $\nu$.
It can be verified directly that $p^{(\nu)}_\afa$ satisfy the following \textit{twisted Gauss-Manin equations}:
\begin{align}
\p_\gamma P^{(\nu)} &=\, -(\mcalU-\lmd I)^{-1}\mcalC_\gamma\left(\mu+\nu+\frac12\right)P^{(\nu)},
\label{TGM-1}\\
\pp{\lmd}P^{(\nu)} &=\,
  (\mcalU-\lmd I)^{-1}\left(\mu+\nu+\frac12\right)P^{(\nu)},  \label{TGM-2}
\end{align}
where the matrix-valued function $P^{(\nu)}$ is defined by
\begin{equation}\label{def of P^(nu)}
P^{(\nu)}=P^{(\nu)}(v;\lmd):=
\left(
  \nabla p^{(\nu)}_1(v;\lmd),...,\nabla p^{(\nu)}_n(v;\lmd)
\right).
\end{equation}

We also introduce the $n\times n$ matrix
\begin{align}\label{def of G-nu}
  (G^{\afa\beta}(\nu))
&:=
 -\frac{1}{2\pi}
 \left(
   \rme^{\pi\rmi R}\rme^{\pi\rmi(\mu+\nu)}
  +\rme^{-\pi\rmi R}\rme^{-\pi\rmi(\mu+\nu)}
 \right)\eta^{-1}.
\end{align}
Note that all entries of $(G^{\afa\beta}(\nu))$ are independent of $v^\afa$ and $\lmd$.

\begin{thm}\label{thm:G-nu as Gram}
For a generalized Frobenius manifold the following identity holds true:
\begin{equation}\label{regularized Laplace-Gram matrix indenity}
P^{(-\nu)\transpose}\eta(\mcalU-\lmd I)P^{(\nu)}
=
\left(
  G^{\afa\beta}(\nu)
\right)^{-1},
\end{equation}
where $P^{(\nu)}$ is defined as in \eqref{def of P^(nu)}.
\end{thm}
Formula \eqref{regularized Laplace-Gram matrix indenity} also holds true for usual Frobenius manifolds \cite{normal-form}, a proof of which is given in \cite{almost dual} for semisimple Frobenius manifolds. We give a direct proof of this theorem in Appendix \ref{Appendix-A}.

In the non-resonant case where the spectrum of $\mu$ of $M$
contains no half integers,
the limits $\lim\limits_{\nu\to 0}p_\afa^{(\nu)}=p^{(0)}_\afa$ exist,
so from \eqref{240813-0855}, \eqref{regularized Laplace-Gram matrix indenity} it follows that
\begin{equation}\label{periods-Lapacle}
  (p_1(v;\lmd),\dots,p_n(v;\lmd))
:=
  \lim_{\nu\to 0}
    \left(
      p_1^{(\nu)}(v;\lmd),\dots,p_n^{(\nu)}(v;\lmd)
    \right)
    \end{equation}
 yields a basis of periods of the generalized Frobenius manifold $M$,
with associated Gram matrix
\begin{equation}
  (G^{\afa\beta})
:=
\lim_{\nu\to 0}(G^{\afa\beta}(\nu))=
 -\frac{1}{2\pi}
 \left(
   \rme^{\pi\rmi R}\rme^{\pi\rmi\mu}
  +\rme^{-\pi\rmi R}\rme^{-\pi\rmi\mu}
 \right)\eta^{-1}.
\end{equation}
However in general case, i.e. the spectrum of $\mu$ is allowed to contain half integers,
the limits $\lim\limits_{\nu\to 0}p_\alpha^{(\nu)}(v;\lmd)$ do not necessarily exist because of \eqref{basic Laplace formula}.
In fact, for any given basis of periods $p_1(v;\lmd),\dots, p_n(v;\lmd)$ of $M$,
from the basic theory of differential equations we know that
there locally exist functions
  \begin{equation}\label{240818-2116}
 p^{[\nu]}_1(v;\lmd),\dots, p^{[\nu]}_n(v;\lmd)
  \end{equation}
  which are analytic in $\nu$ near $\nu=0$
  such that
  \begin{equation}
    p_\afa^{[\nu]}\big|_{\nu=0} = p^{[0]}_\afa(v;\lmd) = p_\afa(v;\lmd),
  \end{equation}
  and
  \begin{equation}
    P^{[\nu]} = P^{[\nu]}(v;\lmd) :=
    \left(
      \nabla p^{[\nu]}_1(v;\lmd),...,\nabla p^{[\nu]}_n(v;\lmd)
    \right)
  \end{equation}
 satisfy the twisted Gauss-Manin equation \eqref{TGM-1}--\eqref{TGM-2}, i.e.
  \begin{align}
   \p_\gamma P^{[\nu]} &=\, -(\mcalU-\lmd I)^{-1}\mcalC_\gamma\left(\mu+\nu+\frac12\right)P^{[\nu]},
   \label{TGM-3}\\
   \pp{\lmd}P^{[\nu]} &=\,
  (\mcalU-\lmd I)^{-1}\left(\mu+\nu+\frac12\right)P^{[\nu]},
  \label{TGM-4}
\end{align}
see Lemma 4.1 of \cite{almost dual}.
Introduce matrix-valued functions
\begin{align}
  X &= X(v;\lmd;\nu) :=
  \left(P^{(\nu)}\right)^{-1}P^{[\nu]}, \label{X(nu)}\\
  H&= \left(H^{\afa\beta}(v;\lmd;\nu)\right) :=
    \left(
      {P^{[-\nu]}}^\transpose\eta(\mcalU-\lmd I)P^{[\nu]}
    \right)^{-1}. \label{H(nu)}
\end{align}
Since both $P^{(\nu)}$ and $P^{[\nu]}$ solve the linear equations \eqref{TGM-1}--\eqref{TGM-2},
the matrix $X$ only depends on the parameter $\nu$, and so we denote $X=X(\nu)$.
By using \eqref{TGM-3}--\eqref{TGM-4} and \eqref{240818-1342} we can verify that
\[
  \pfrac{H}{v^\gamma} = \pfrac H\lmd =0 ,\quad 1\leq\gamma\leq n,
\]
and so we can denote $H=H(\nu)$. Note that $G:=\lim\limits_{\nu\to 0} H(\nu) = H(0)$
is the Gram matrix with respect to the basis of periods $p_1,\dots, p_n$.
It follows from \eqref{regularized Laplace-Gram matrix indenity} and \eqref{X(nu)}--\eqref{H(nu)} that
the matrix $H(\nu)$ and $G(\nu)$ are related by
\begin{equation}\label{G-H relation}
  G(\nu) = X(\nu)H(\nu)X^\transpose(-\nu).
\end{equation}
  \begin{rmk}
    The functions $p^{[\nu]}_1(v;\lmd),..., p^{[\nu]}_n(v;\lmd)$ and their above-mentioned properties will be used in the proof of Theorem \ref{thm: loop-eqn n+2 version}.
  \end{rmk}

In the next section we also need to use the periods of the $(n+2)$-dimensional Frobenius manifold $\Mtil$ associated to $M$.
Recall that the monodromy data
$\mutil$ and $\Rtil$ of $\Mtil$ are defined in \eqref{n+2 mu R intro}, and the
functions $\thetatil_{0,p},\dots,\thetatil_{n+1,p}$ for $\Mtil$ are defined in \eqref{n+2 dim theta-1} and \eqref{n+2 dim theta-2}.
If the spectrum of $\mutil$ contains no half integers,
a basis of periods of $\Mtil$ can be chosen as follows:
\begin{equation}\label{n+2 twisted periods}
(\ptil_0,\ptil_1,\dots,\ptil_n,\ptil_{n+1})
= \lim_{\nu\to 0}
    \left(
      \ptil_0^{(\nu)},\ptil^{(\nu)}_1,\dots,\ptil^{(\nu)}_n,\ptil^{(\nu)}_{n+1}
    \right),
 \end{equation}
 where the regularized periods are defined as in \eqref{periods-Lapacle-2} by the formula
\begin{align}
  &\left(
      \ptil_0^{(\nu)},\ptil^{(\nu)}_1,\dots,\ptil^{(\nu)}_n,\ptil^{(\nu)}_{n+1}
    \right)\notag\\
=&\,
   \int_{0}^{+\infty}
    \frac{\td z}{\sqrt{z}}
    \rme^{-\lmd z}
    \left(
      \thetatil_0(z),
      \thetatil_1(z),\dots,\thetatil_n(z),
      \thetatil_{n+1}(z)
    \right)
    z^{\mutil+\nu}z^{\Rtil}.\label{n+2 twisted periods-2}
\end{align}
The associated Gram matrix is given by
\begin{equation}\label{n+2 Gram}
  (\Gtil^{ij})
:=\lim_{\nu\to 0}(\Gtil^{ij}(\nu))
=
 -\frac{1}{2\pi}
 \left(
   \rme^{\pi\rmi\Rtil}\rme^{\pi\rmi\mutil}
  +\rme^{-\pi\rmi\Rtil}\rme^{-\pi\rmi\mutil}
 \right)\etatil^{-1},
\end{equation}
here $\etatil$ is defined in \eqref{zh-25}.
For the $(n+2)$-dimensional Frobenius manifold $\widetilde{M}$
we can also introduce the functions $\tilde{p}^{[\nu]}_i$, $\widetilde{X}(\nu)$ and $\widetilde{H}(\nu)$ as we do above for the generalized Frobenius manifold $M$.

The following Lemma on Laplace-type integrals will be used in representing the Virasoro symmetries of the Principal Hierarchy of $M$ in terms of Virasoro operators.
\begin{lem}
\label{lemma: Laplace-Bilinear identity}
Let $\Bigset{a_p^{\afa}}{p\in\bbZ,\,\afa=1,2,\dots,n}$
be a family of formal variables in a certain $\bbC$-algebra,
and $\bfa_p:=(a_p^1,\dots,a_p^n)^{\transpose}$. Define the row vector-valued functions
\begin{equation} \label{240727-1107}
  \phi^{(\nu)}(\bfa;\lmd)
  :=
  \int_{0}^{\infty}
  \frac{\td z}{\sqrt{z}}\rme^{-\lmd z}
  \sum_{p\in\bbZ}
  \bfa_p^{\transpose}\eta
  z^{p+\mu+\nu}z^R,
\end{equation}
then the following identity holds true:
\begin{equation}\label{Laplace-bilinear identity}
  -\frac12
   \pfrac{\phi_\afa^{(\nu)}(\bfa;\lmd)}{\lmd}
   G^{\afa\beta}(\nu)
   \pfrac{\phi_\beta^{(-\nu)}(\bfa;\lmd)}{\lmd}
=
  \frac12
  \sum_{p,q\in\bbZ}
  \sum_{s\geq 0}
  \frac{1}{\lmd^{p+q+s+3}}
  \bfa_p^{\transpose}\eta
  \mathrm N_{p,q}(s;\nu)
  \bfa_q,
\end{equation}
where $G^{\afa\beta}(\nu)$ is defined in \eqref{def of G-nu}, and
\begin{align}
\mathrm N_{p,q}(s;\nu)
:=
  \frac{1}{\pi}
  \left[
    \rme^{R\p_\nu}
  \right]_s
  \left(
    \Gamma (p+\mu+\nu+s+\frac 32)
    \cos\left(\pi(\mu+\nu)\right)
    \Gamma (q-\mu-\nu+\frac 32)
  \right). \label{N pq}
\end{align}
\end{lem}

We shall give a direct proof of this lemma in Appendix \ref{Appendix-A}.
Note that the identity \eqref{Laplace-bilinear identity} in the above lemma is analogues to (3.10.31) of \cite{normal-form}.

\begin{rmk}\label{remark:polynomial-nu}
By using properties of the Gamma function we know that
  $\mathrm N_{p,q}(s;\nu)$ is a polynomial in $\nu$ if $p+q+s+3\geq 1$, see also \cite{Virasoro-Like}.
  As a consequence, the limit
  \[
    \lim_{\nu\to 0}
    \left[
      \pfrac{\phi_\afa^{(\nu)}(\bfa;\lmd)}{\lmd}
      G^{\afa\beta}(\nu)
      \pfrac{\phi_\beta^{(-\nu)}(\bfa;\lmd)}{\lmd}
    \right]_-
  \]
  is well-defined for any generalized Frobenius manifold,
  even though
  $\phi^{(\nu)}$ may be singular at $\nu=0$ in the resonant case when the spectrum of $\mu$ contains half integers.
  Here we denote $[\, \cdot\, ]_-$ the negative part of a formal series of $\lmd$, which is defined as follows:
  \begin{equation}\label{negative-part of lmd}
    \left[
      \sum_{m\in\bbZ}\frac{A_m}{\lmd^{m+2}}
    \right]_-
  :=
    \sum_{m\geq -1} \frac{A_m}{\lmd^{m+2}}.
  \end{equation}
\end{rmk}

We also introduce the following $(n+2)$-dimensional row vector-valued function for the $(n+2)$-dimensional Frobenius manifold $\Mtil$ associated to $M$:
\begin{equation}\label{n+2 formal phi-Laplace}
  \tilde\phi^{(\nu)}
  (\bfa;\lmd)
:=
  \int_{0}^{\infty}
    \frac{\td z}{\sqrt{z}}\rme^{-\lmd z}
    \sum_{p\in\bbZ}
    \bfa_p^{\transpose}
    \etatil
    z^{p+\mutil+\nu}
    z^{\Rtil},
\end{equation}
where $\bfa_p=(a_p^0,\dots,a_p^{n+1})^{\transpose}$,
and $\Bigset{a^i_p}{p\in\bbZ,\,0\leq i\leq n+1}$
is a family of formal variables in a certain $\bbC$-algebra.
The corresponding $(n+2)$-dimensional version of \eqref{Laplace-bilinear identity}
also holds true.

\section{Virasoro symmetries of the Principal Hierarchy}
Now let us study Virasoro symmetries of the Principal Hierarchy \eqref{principal hierarchy}.
Recall that a symmetry of the hierarchy \eqref{principal hierarchy}
is an evolutionary system
\[
  \pfrac vs=S(v, v_x,\dots;\bft)
\]
that commutes
with all the flows
$\pp{t^{\afa,p}}\,(1\leq\afa\leq n,\,p\geq 0)$ and
$\pp{t^{0,p}}\,(p\in\bbZ)$, i.e.,
\[\,
  \left[\pp s,\pp{t^{\afa,p}}\right]=\left[\pp{s},\pp{t^{0,q}}\right]=0,\quad
  1\leq\afa\leq n,\, p\geq 0,\,q\in\bbZ.
\]
All symmetries of the hierarchy form a Lie algebra
with respect to the commutator.

\begin{thm}
The Principal Hierarchy \eqref{principal hierarchy}
admits the symmetry
\begin{equation}\label{Galilean sym}
  \pfrac{v}{s_{-1}}
 =e+\sum_{(i,p)\in\mcalI}
  t^{i,p+1}\pfrac{v}{t^{i,p}},
\end{equation}
where $e=e^\afa\p_\afa$ is the unit vector field of $M$.
\end{thm}

\begin{proof}
Let us first show that
\[\left[\pp{s_{-1}}, \pp{t^{\beta,q}}\right]v^\lmd=0,\quad 1\leq\beta\leq n,\ q\geq 0,\ 1\leq\lmd\leq n.\]
From the relations
\begin{align*}
   & \left(
      \pp{s_{-1}}\circ\pp{t^{\beta,q}}
    \right)v^\lmd
  =
    \sum_{k\geq 0}
    \left(
      \pp{v^{\gamma,k}}
      \pfrac{v^\lmd}{t^{\beta,q}}
    \right)
    \p^k_x
    \left(
      e^\gamma
     +\sum_{\afa, p}
        t^{\afa,p+1}
        \pfrac{v^\gamma}{t^{\afa,p}}
     +\sum_{p\in\bbZ}
        t^{0,p}\pfrac{v^\gamma}{t^{0,p-1}}
    \right)\\
  &\qquad=
    e^\gamma\pp{v^\gamma}\pfrac{v^\lmd}{t^{\beta,q}}
   +e^\gamma_x\pp{v^{\gamma,1}}\pfrac{v^\lmd}{t^{\beta,q}}
   +\pfrac{v^\gamma}{t^{0,-1}}
    \left(
      \pp{v^{\gamma,1}}\pfrac{v^\lmd}{t^{\beta,q}}
    \right)\\
  &\qquad\quad
   +\sum_{\afa,p}
      t^{\afa,p+1}
      \frac{\p^2v^\lmd}{\p t^{\afa,p}\p t^{\beta,q}}
   +\sum_{p\in\bbZ}
     t^{0,p}\frac{\p^2v^\lmd}{\p t^{0,p-1}\p t^{\beta,q}}
  \\
&\qquad=
   \sum_{\afa,p}
      t^{\afa,p+1}
      \frac{\p^2v^\lmd}{\p t^{\afa,p}\p t^{\beta,q}}
   +\sum_{p\in\bbZ}
     t^{0,p}\frac{\p^2v^\lmd}{\p t^{0,p-1}\p t^{\beta,q}}
  \\
  &\qquad\quad
    +e^\gamma\pp{v^\gamma}\pfrac{v^\lmd}{t^{\beta,q}}
    +2e^\gamma_x\p_\gamma\p^\lmd\theta_{\beta,q+1},
\end{align*}
and
\begin{align*}
  \left(
    \pp{t^{\beta,q}}\circ\pp{s_{-1}}
  \right)v^\lmd
  =&\,
    \pfrac{e^\lmd}{t^{\beta,q}}
   +\pfrac{v^\lmd}{t^{\beta,q-1}}
   +\sum_{\afa,p}
      t^{\afa,p+1}
      \frac{\p^2 v^\lmd}{\p t^{\afa,p}\p t^{\beta,q}}
   +\sum_{p\in\bbZ}
      t^{0,p}\frac{\p^2 v^\lmd}{\p t^{0,p-1}\p t^{\beta,q}},
  \end{align*}
it follows that we only need to show the validity of the identity
  \begin{equation}\label{2021.10.2.1}
    e^\gamma\pp{v^\gamma}\pfrac{v^\lmd}{t^{\beta,q}}
   +2e^\gamma_x\p_\gamma\p^\lmd\theta_{\beta,q+1}
  =\pfrac{e^\lmd}{t^{\beta,q}}+\pfrac{v^\lmd}{t^{\beta,q-1}},
  \end{equation}
which holds true since
\begin{align*}
  &
    e^\gamma\pp{v^\gamma}\pfrac{v^\lmd}{t^{\beta,q}}
   +2e^\gamma_x\p_\gamma\p^\lmd\theta_{\beta,q+1}\\
  &\quad=
    e^\gamma\left(
      \p_\gamma\p^\lmd\theta_{\beta,q+1}
    \right)_x
   +2e^\gamma_x\p_\gamma\p^\lmd\theta_{\beta,q+1}
   =
   \p_x\left(
     e^\gamma\p_\gamma\p^\lmd\theta_{\beta,q+1}
   \right)
   +e^\gamma_x\p_\gamma\p^\lmd\theta_{\beta,q+1}\\
 &\quad=
   \p_x\left(
     e^\gamma c^\lmd_{\gamma\veps}\p^\veps\theta_{\beta,q}
   \right)
   +(\p_\delta e^\gamma)c^\lmd_{\gamma\veps}(\p^\veps\theta_{\beta,q})v^\delta_x
  =
   \p_x\p^\lmd\theta_{\beta,q}
   -e^\gamma c^\lmd_{\delta\gamma\veps}
    (\p^\veps\theta_{\beta,q})v^\delta_x\\
  &\quad=
    \pfrac{v^\lmd}{t^{\beta,q-1}}
  +(\p^\lmd e^\gamma)c_{\delta\gamma\veps}
    (\p^\veps\theta_{\beta,q})v^\delta_x
   =
    \pfrac{v^\lmd}{t^{\beta,q-1}}
  +(\p^\gamma e^\lmd)c_{\delta\gamma\veps}
    (\p^\veps\theta_{\beta,q})v^\delta_x\\
  &\quad=
    \pfrac{v^\lmd}{t^{\beta,q-1}}
   +\pfrac{e^\lmd}{t^{\beta,q}} .
  \end{align*}
By performing a similar calculation,
we can check that
\[\left[\pp {s_{-1}},\pp{t^{0,q}}\right]v^\lmd=0,\quad q\in\bbZ.\]
The theorem is proved.
\end{proof}

The following theorem shows that the symmetry \eqref{Galilean sym} of the Principal Hierarchy can be lifted to its tau-cover \eqref{tau-cover}.

\begin{thm}\label{zh-32}
The tau-cover \eqref{tau-cover} admits the following symmetry:
\begin{align}
  \pfrac f{s_{-1}} &=
  \sum_{(i,p)\in\mcalI}t^{i,p+1}f_{i,p}
  +\frac12\eta_{\afa\beta}t^{\afa,0}t^{\beta,0},\\
  \pfrac{f_{i,p}}{s_{-1}}&=
  f_{i,p-1}+\sum_{(j,q)\in\mcalI}t^{j,q+1}\Omg_{i,p; j,q}
  +\eta_{\afa\beta}t^{\afa,0}\delta^\beta_i\delta^0_p,\quad (i,p)\in\mcalI,\\
  \pfrac{v^\gamma}{s_{-1}}
  &=
  e^\gamma+\sum_{(i,p)\in\mcalI}t^{i,p+1}\pfrac{v^\gamma}{t^{i,p}},\quad \gamma=1,\dots,n.
\end{align}
Here $f_{i,p}:=0$ if $i\neq 0$ and $p<0$.
\end{thm}
\begin{proof}
  We need to check that\
\[\left[\pp{t^{j,q}},\pp{s_{-1}}\right]f=0,\quad
  \left[\pp{t^{j,q}},\pp{s_{-1}}\right]f_{i,p}=0,\quad (i,p),\ (j,q)\in\mcalI.\]
The first set of relations obviously hold true,
so we only need to check the validity of the second set of relations.
In fact, we have
\begin{align*}
   & \left(
      \pp{t^{j,q}}\circ\pp{s_{-1}}
    \right)f_{i,p}
  =
    \Omg_{i,p-1; j,q}+\Omg_{i,p;j,q-1}
    +\sum_{(k,r)\in\mcalI}
       t^{k,r+1}
       \pfrac{\Omg_{i,p; k,r}}{t^{j,q}}
    +\eta_{\afa\beta}\delta^\afa_i\delta^\beta_j\delta^0_p\delta^0_q,\\
    &\left(
      \pp{s_{-1}}\circ\pp{t^{j,q}}
    \right)f_{i,p}
  =
    \pp{s_{-1}}\Omg_{i,p;j,q}
   =\pfrac{v^\afa}{s_{-1}}\p_\afa\Omg_{i,p;j,q}\\
   &\qquad =
    \left(
      e^\afa+\sum_{(k,r)\in\mcalI}t^{k,r+1}\pfrac{v^\afa}{t^{k,r}}
    \right)
    \p_\afa\Omg_{i,p;j,q}
  =
    \p_e\Omg_{i,p; j,q}+\sum_{(k,r)\in\mcalI}t^{k,r+1}
    \pfrac{\Omg_{i,p;j,q}}{t^{k,r}},
  \end{align*}
here we assume that
\[\Omg_{i,p; j,q}:=0,\quad i\neq 0,\ p<0.\]
From the relations
\[\pfrac{\Omg_{i,p; k,r}}{t^{j,q}}=\frac{\p^3 f}{\p t^{i,p} \p t^{j,q}\p t^{k,r}}
=\pfrac{\Omg_{i,p;j,q}}{t^{k,r}}\]
it follows that we only need to check the validity of the identities
\begin{equation}\label{D e Omg}
\p_e\Omg_{i,p; j,q}
=
\Omg_{i,p-1;j,q}+\Omg_{i,p;j,q-1}+\eta_{\afa\beta}\delta^\afa_i\delta^\beta_j\delta^0_p\delta^0_q,\quad (i,p), (j,q)\in\mcalI.
\end{equation}
By using \eqref{zh-11}, \eqref{zh-7} and \eqref{theta0,-p-recur} we know that \[\p_e\nabla\theta_{i,p}=\nabla\theta_{i,p-1},\]
which, together with \eqref{Omg afa p beta q}--\eqref{Omg 0 p 0 q},
lead to the identities \eqref{D e Omg}. The theorem is proved.
\end{proof}

\begin{rmk}
If the flows $\frac{\p}{\p t^{0,p}}$, $p\in\mathbb{Z}$ are not included in the Principal Hierarchy, then we still have the symmetry
\[\frac{\p v^\gamma}{\p s_{-1}}=e^\gamma+x \frac{\p e^\gamma}{\p v^\xi} v^\xi_x+\sum_{p\ge 0} t^{\al,p+1}\frac{\p v^\gamma}{\p t^{\al,p}}.\]
However, in this case we are not able to lift this symmetry to the tau-cover of the Principal Hierarchy.
\end{rmk}

By applying the bihamiltonian recursion operator
\begin{equation}\label{real mcalR}
  \mcalR:=\mcalP_2\mcalP_1^{-1}=\mcalU+\mcalC_\gamma v^\gamma_x
  \left(\frac12+\mu\right)\p_x^{-1}
\end{equation}
to the symmetry $\pfrac{v}{s_{-1}}$ with $\mcalU$ and $\mcalC_\gamma$ defined in \eqref{zh-2024-4}, we obtain the following symmetry of the Principal Hierarchy:
\begin{align*}
  \pfrac{v}{s_0}:=&\,\mcalR \pfrac{v}{s_{-1}}\\
=&\,
  E+\sum_{p\geq 0}
  \left(
    p+\mu_\afa+\frac12
  \right)t^{\afa,p}\pfrac{v}{t^{\afa,p}}
 +\sum_{p\in\bbZ}
  \left(
    p-\frac d2+\frac12
  \right)t^{0,p}\pfrac{v}{t^{0,p}}
\\
&\,+
  \sum_{p\geq 0}\sum_{s\geq 1}
  (R_s)^\veps_\afa t^{\afa,p}
  \pfrac{v}{t^{\veps,p-s}}
 +\sum_{p\in\bbZ}
   \sum_{s\geq 1}
 r^\veps_s t^{0,p}\pfrac{v}{t^{\veps,p-s}}.
\end{align*}
Iterating this procedure we obtain
a series of symmetries
\[\pfrac{v}{s_{m+1}}:=\mcalR\pfrac{v}{s_m},\quad m\geq -1\]
of the Principal Hierarchy.
However, such symmetries involve non-local terms in general.
To eliminate such non-locality, we are to lift these symmetries to the tau-cover
\eqref{tau-cover} of the Principal Hierarchy. To this end,
we first introduce some notations.

Let us denote
\begin{align}
\Stil_i^{(\nu)}=&
\Stil_i^{(\nu)}(\bft, \pfrac{f}{\bft};\lmd)\notag\\
=&\,
\left(
  \int_{0}^{\infty}
    \frac{\td z}{\sqrt{z}}\rme^{-\lmd z}
    \left[
      \sum_{p\geq 0}
      \left(
        f_{0,p},\bm{f}_p,(-1)^{p+1}t^{-p-1}_{n+1}
      \right)z^p
    \right]
    z^{\mutil+\nu}z^{\Rtil}
\right)_i\notag\\
&\,
  +
\left(
\int_{0}^{\infty}
    \frac{\td z}{\sqrt{z}}\rme^{-\lmd z}
    \left[
     \sum_{p\geq 0}(-1)^p
     \left(
       (-1)^pf_{0,-p-1}, \bm{t}^p, t^p_{n+1}
     \right)z^{-p-1}
    \right]
    z^{\mutil+\nu}z^{\Rtil}
\right)_i \label{defn:Stil_i}
\end{align}
for $0\leq i\leq n+1$. Here $\lmd$ is a formal parameter, and the $n$-dimensional row vectors $\bm{f}_{p}$ and $\bm{t}^p$ are defined by
\[\bm{f}_p=\left(f_{1,p},\dots,f_{n,p}\right),\quad
\bm{t}^p=\left(t_1^p,t_2^p,\dots, t_n^p\right)\]
with $t_\afa^p:=\eta_{\afa\beta}t^{\beta,p}$
and $t_{n+1}^p:= t^{0,p}$.
We also introduce the $(n+2)$-dimensional row vector
\[\Stil^{(\nu)}=\left(\Stil_0^{(\nu)},
\Stil^{(\nu)}_1,\dots,\Stil^{(\nu)}_n,\Stil^{(\nu)}_{n+1}\right).\]
Notably, the notation $\Stil^{(\nu)}(\bft,\pfrac f{\bft};\lmd)$
is just obtained from
$\tilde\phi^{(\nu)}(\bfa;\lmd)$, which is defined in \eqref{n+2 formal phi-Laplace}, by
the following substitution:
\[
  a^i_p\mapsto
  \begin{cases}
    (-1)^{p+1}t^{0,-p-1}, & i=0;\\
    \eta^{i\beta}f_{\beta,p}, & 1\leq i\leq n,\, p\geq 0;\\
    (-1)^{p+1}t^{i,-p-1}, & 1\leq i\leq n,\, p<0;\\
    f_{0,p}, & i=n+1.
  \end{cases}
\]

Recall that the number of non-zero elements of the series $\{c_p\}_{p\in\bbZ}$ defined in Proposition \ref{zh-17}
is at most one, and $c_p\neq 0$ for some $p<0$ only if
the charge $d$ of the generalized Frobenius manifold $M$ is a negative odd integer
and in this case $p=d-1$.
Now let us introduce the notation $C_{p,q}(\lmd)$ for $p,q\in\bbZ$
as follows:
$C_{p,q}(\lmd)\neq 0$ \textit{only if}
the charge $d$ of $M$ is a negative odd integer and $p+q\leq d-1$.
In this case we set
\begin{align}
  C_{p,q}(\lmd)
&=
  \sum_{m\geq -1}
  \frac{C_{m;p,q}}{\lmd^{m+2}}\notag\\
&=
  \frac{
  (-1)^pc_{d-1}}{2\lmd^{d+1-p-q}}
  \sum_{k=0}^{d-1-p-q}
  (-1)^k
  \left(
    p-\frac d2+\frac12
  \right)^{[k]}
  \left(
    q-\frac d2+\frac12
  \right)^{[d-1-p-q-k]}; \label{C_p,q lmd}
\end{align}
otherwise we set $C_{p,q}(\lmd)=0$. Here $x^{[0]}=1$, and
\begin{equation}\label{perm symbol}
  x^{[k]}:=x(x+1)\cdots(x+k-1),\quad x\in\bbC,\, k\geq 1.
\end{equation}
It is straightforward to see that the constant
$C_{m;p,q}\neq 0$ only if $p+q+m=d-1$, $m\geq 0$ and $d$ is a negative odd integer.

\begin{thm}\label{thm: Vir sym on tau-cover}
The flows $\pp{s_m}$, $m\geq -1$ defined by the following formulae
are symmetries of the tau-cover \eqref{tau-cover} of the Principal Hierarchy of $M$:
\begin{align}
  \pp{s}=&\,
  \sum_{m\geq -1}\frac{1}{\lmd^{m+2}}
  \pp{s_m},\label{Vir sym on tau-cover 0}
\\ \label{Vir sym on tau-cover 1}
    \pfrac{f}{s}
  =&\,
    -\frac12
    \lim_{\nu\to 0}
    \left[
      \pfrac{\Stil_i^{(\nu)}}{\lmd}
      \Gtil^{ij}(\nu)
      \pfrac{\Stil_j^{(-\nu)}}{\lmd}
    \right]_-
  +
    \sum_{p,q\in\bbZ}
      C_{p,q}(\lmd)t^{0,p}t^{0,q},
    \\ \label{Vir sym on tau-cover 2}
    \pfrac{f_{i,p}}{s}
  =&\,
        \pp{t^{i,p}}
    \pfrac{f}{s},
  \\
\label{Vir sym on tau-cover 3}
    \pfrac{v^\afa}{s}
  =&\,
  \lim_{\nu\to 0}
    \left[
  \pfrac{\Stil_i^{(\nu)}}{\lmd}
  \Gtil^{ij}(\nu)
  \left(
    \p_x
    \p^\afa\ptil_j^{(-\nu)}
  \right)\Big|_{M}
\right]_-\notag\\
&\,
  +\lim_{\nu\to 0}
 \left[
   \p_x\p^\afa\xi^{(\nu)}(\lmd)
   \Gtil^{n+1,0}(\nu)
   \pfrac{\Stil_{n+1}^{(-\nu)}}{\lmd}
 \right]_-
-
 \left(
   \frac{1}{E-\lmd e}
 \right)^\afa.
  \end{align}
Here the restriction of functions on $\Mtil$ to $M$ is defined in \eqref{zh-28},
the negative part $[\,\cdot\,]_-$ is defined in \eqref{negative-part of lmd};
$\ptil_i^{(\nu)}$ are the regularized periods of $\Mtil$
defined by \eqref{n+2 twisted periods-2}, and
\begin{align}
  \xi^{(\nu)}(\lmd)=&\,
  \sum_{p\geq 0}
  \int_{0}^{\infty}
    \rme^{-\lmd z}\theta_{0,-p}z^{-p-\frac d2+\nu-\frac12}\td z\notag
\\
=&\,
   \sum_{p\geq 0}\theta_{0,-p}\lmd^{p+\frac d2-\nu-\frac12}
   \Gamma\left(-p-\frac d2+\nu+\frac12\right);\label{defn:xi(nu)(lmd)}
\end{align}
the vector field
\[\frac{1}{E-\lmd e}:=-\sum_{m\geq -1}\frac{1}{\lmd^{m+2}}E^{m+1}\]
is the inverse of $E-\lmd e$ with respect to the multiplication
of $M$.
\end{thm}

According to Remark \ref{remark:polynomial-nu},
all the limits ${\nu\to 0}$ in the right-hand side of
\eqref{Vir sym on tau-cover 1}--\eqref{Vir sym on tau-cover 3} are well defined.
In the non-resonant case that the spectrum of $\mutil$ does not contain half integers, the limits
  \begin{equation}\label{zh-40}
    \Stil_i=\lim_{\nu\to 0}\Stil^{(\nu)}_i,\quad
    \Gtil^{ij}=\lim_{\nu\to 0}\Gtil^{ij}(\nu),\quad
    \ptil_i=\lim_{\nu\to 0}\ptil^{(\nu)}_i,\quad
    \xi(\lmd)=\lim_{\nu\to0}\xi^{(\nu)}(\lmd)
  \end{equation}
exist.
For convenience, we will provide the detailed proof of the above theorem only for the non-resonant case.
The detail of the proof for the general case is similar, see Remark \ref{rmk: general case 240818}.
In order to prove the theorem, we need some lemmas.

\begin{lem}\label{zh-2023-1}
 If the spectrum of $\mutil$ contains no half integers,
then
\begin{align}
\p_{t^{\afa,0}}\pfrac{\Stil}{\lmd}
&=
  -\left.
  \p_\afa
  (\ptil_0,\bm{\ptil},\ptil_{n+1})
  \right|_{M}
  -\p_\afa(\xi(\lmd),\bm 0,0),\label{DtDlmdStil}
\\
\label{DxDlmdStil}
\p_x\pfrac{\Stil}{\lmd}
&=
  -\left.
  \p_0
  (\ptil_0,\bm{\ptil},\ptil_{n+1})
  \right|_{M}
  -(\psi(\lmd),\bm 0,0),
\end{align}
where $\bm{\ptil}=(\ptil_1,\dots,\ptil_n)$, and
\begin{align*}
  \xi(\lmd)&=
  \sum_{p\geq 0}
  \int_{0}^{\infty}
    \rme^{-\lmd z}\theta_{0,-p}z^{-p-\frac d2-\frac12}\td z
=
   \sum_{p\geq 0}\theta_{0,-p}\lmd^{p+\frac d2-\frac12}
   \Gamma\left(-p-\frac d2+\frac12\right),\\
  \psi(\lmd)&=
  \sum_{p\geq 0}
  \int_{0}^{\infty}
    \rme^{-\lmd z}\theta_{0,-p-1}
    z^{-p-\frac d2-\frac12}\td z
=
  \sum_{p\geq 0}\theta_{0,-p-1}\lmd^{p+\frac d2-\frac12}
  \Gamma\left(
    -p-\frac d2+\frac12
  \right).
\end{align*}
\end{lem}

\begin{proof}
From the relation between $\thetatil_i(z)$ and $\theta_\afa(z)$ that is given in \eqref{n+2 dim theta-1}, \eqref{n+2 dim theta-2}, \eqref{zh-21-a}, \eqref{zh-21}, and the relation
$\Omg_{\afa,0; i,p}=\p_\afa\theta_{i,p+1}$
for $(i,p)\in\mcalI$, it follows that
\begin{align*}
  \p_{t^{\afa,0}}\pfrac{\Stil}{\lmd}
  =&\,
    -\p_{t^{\afa,0}}
    \int_{0}^{\infty}
      \frac{\td z}{\sqrt{z}}\rme^{-\lmd z}
      \sum_{p\geq 0}
      \left(
        f_{0,p}, \bm{f}_p, (-1)^{p+1}t^{-p-1}_{n+1}
      \right)z^{p+1}z^{\mutil}z^{\Rtil}\\
  &
   -\p_{t^{\afa,0}}
   \int_{0}^{\infty}
     \frac{\td z}{\sqrt{z}}\rme^{-\lmd z}
     \sum_{p\geq 0}(-1)^p
     \left(
       (-1)^pf_{0,-p-1},
       \bm{t}^p,
       t_{n+1}^p
     \right)
     z^{-p}z^{\mutil}z^{\Rtil}\\
=&\,
  -
    \int_{0}^{\infty}
      \frac{\td z}{\sqrt{z}}\rme^{-\lmd z}
      \sum_{p\geq 0}
      \left(
        \Omg_{\afa,0;0,p}, \Omg_{\afa,0; \bullet,p}, 0
      \right)z^{p+1}z^{\mutil}z^{\Rtil}\\
  &
   -
   \int_{0}^{\infty}
     \frac{\td z}{\sqrt{z}}\rme^{-\lmd z}
     \sum_{p\geq 0}
     \left(
       \Omg_{\afa,0; 0,-p-1},
       \eta_{\al\bullet}\delta_{p,0},
       0
     \right)
     z^{-p}z^{\mutil}z^{\Rtil}\\
=&\,
  -
    \int_{0}^{\infty}
      \frac{\td z}{\sqrt{z}}\rme^{-\lmd z}
      \sum_{p\geq 0}
      \left(
        \p_\afa\theta_0(z)-e_\afa, \p_\afa\theta_\bullet(z), 0
      \right)z^{\mutil}z^{\Rtil}\\
  &
   -
   \int_{0}^{\infty}
   \rme^{-\lmd z}
     \sum_{p\geq 0}
     \left(
       \p_\afa\theta_{0,-p},
       \bm 0,
       0
     \right)
     \begin{pmatrix}
       z^{-p-\frac d2-\frac12} & \bm 0 & 0 \\
       * & * & * \\
       * & * & *
     \end{pmatrix}
     \td z
     \\
=&\,
  -\int_{0}^{\infty}
  \frac{\td z}{\sqrt{z}}\rme^{-\lmd z}
  \left(
    \p_\afa\thetatil_0(z),
    \p_\afa\thetatil_\bullet(z),
    \p_\afa\thetatil_{n+1}(z)
  \right)\Big|_{M}
  z^{\mutil}z^{\Rtil}
-
  \p_\afa(\xi(\lmd),\bm 0, 0)\\
=&\,
 -\left.
  \p_\afa
  (\ptil_0,\bm{\ptil},\ptil_{n+1})
  \right|_{M}
  -\p_\afa(\xi(\lmd),\bm 0,0),
\end{align*}
so the identity \eqref{DtDlmdStil} holds true. We can prove the  identity \eqref{DxDlmdStil} in a similar way.
The lemma is proved.
\end{proof}

\begin{rmk}
In the proof of the above lemma, we use $w_\bullet$ to denote
an $n$-dimensional row vector $(w_1,\dots, w_n)$. In some places of what follows we will also use such notations, and we will denote
by $w^\bullet$ an $n$-dimensional column vector $(w^1,\dots, w^n)^{\transpose}$.
\end{rmk}

The following lemma shows that the flows given in \eqref{Vir sym on tau-cover 0}--\eqref{Vir sym on tau-cover 3} are compatible with the relation $v^\afa=\eta^{\afa\beta}\p_x\p_{t^{\beta,0}}f$.

\begin{lem} If the spectrum of $\mutil$ contains no half integers, then the following relations hold true:
\begin{align}
&
-\frac12\eta^{\afa\beta}
\p_x\p_{t^{\beta,0}}
\left[
  \pfrac{\Stil_i}{\lmd}\Gtil^{ij}
  \pfrac{\Stil_j}{\lmd}
\right]_-\notag\\
&\qquad =
\left[
    \p_x(
      \p^\afa\ptil_i
    )\big|_{M}
    \Gtil^{ij}
    \pfrac{\Stil_j}{\lmd}
  \right]_-
+
 \left[
   \p_x\p^\afa\xi(\lmd)
   \Gtil^{n+1,0}
   \pfrac{\Stil_{n+1}}{\lmd}
 \right]_-
-
 \left(
   \frac{1}{E-\lmd e}
 \right)^\afa.\label{Qhn220721}
\end{align}
\end{lem}

\begin{proof}
From the definitions of $\mutil$ and $\Rtil$ given in \eqref{n+2 mu R intro} it follows that
\[
  \rme^{\pi\rmi\mutil}=\begin{pmatrix}
                    \rme^{-\frac{\pi\rmi d}{2}} &  &  \\
                     & \rme^{\pi\rmi\mu} &  \\
                     &  & \rme^{\frac{\pi\rmi d}{2}}
                  \end{pmatrix},\quad
  \rme^{\pi\rmi\Rtil} =
  \begin{pmatrix}
    1 & \bm 0 & 0 \\
    * & \rme^{\pi\rmi R} & \bm 0  \\
    * & * & 1
  \end{pmatrix},
\]
so the Gram matrix $(\Gtil^{ij})$ defined in \eqref{n+2 Gram} has the form
\[
  \Gtil =
  \begin{pmatrix}
    0 & \bm 0 & -\frac 1\pi\cos\frac{\pi d}{2} \\
    \bm 0 & G & * \\
    -\frac 1\pi\cos\frac{\pi d}{2} & * & *
  \end{pmatrix}.
\]
In particular, $\Gtil^{0,0}=\Gtil^{\afa,0}=\Gtil^{0,\afa}=0$,
and $\Gtil^{n+1,0}=\Gtil^{0,n+1}=-\frac 1\pi\cos\frac{\pi d}{2}$. Thus from Lemma \ref{zh-2023-1} it follows that
\begin{align*}
&
  -\frac12
  \p_x
  \p_{t^{\afa,0}}
  \left[
    \pfrac{\Stil_i}{\lmd}
    \Gtil^{ij}
    \pfrac{\Stil_j}{\lmd}
  \right]_-
\\
&\qquad =
  \p_x\left[
    (\p_\afa\ptil_i)\big|_{M}
    \Gtil^{ij}\pfrac{\Stil_j}{\lmd}
  \right]_-
 +
 \p_x\left[
   \p_\afa\xi(\lmd)\Gtil^{0,j}\pfrac{\Stil_j}{\lmd}
 \right]_-
\\
&\qquad =
  \left[
    \p_x(
      \p_\afa\ptil_i
   )\big|_{M}
    \Gtil^{ij}
    \pfrac{\Stil_j}{\lmd}
  \right]_-
 -\left[
   (
     \p_\afa\ptil_i
   )\big|_{M}
   \Gtil^{ij}
   \left(
     \p_0\ptil_j
   \right)\Big|_{M}
  \right]_-
\\
&\qquad\quad
  -\left[
    (
      \p_\afa\ptil_{n+1}
    )\big|_{M}
    \Gtil^{n+1,0}\psi(\lmd)
  \right]_-
+ \left[
    \p_x\p_\afa\xi(\lmd)
    \Gtil^{0,n+1}
    \pfrac{\Stil_{n+1}}{\lmd}
  \right]_-
\\
&\qquad\quad
-
  \left[
    \p_\afa\xi(\lmd)
    \Gtil^{0,n+1}
    (
      \p_0\ptil_{n+1}
    )\big|_{M}
  \right]_-.
\end{align*}
By using the relations
\begin{equation}\label{d0pn+1 to n-dim}
(
  \p_\afa\ptil_{n+1}
)\big|_{M}=0,\quad
  (
  \p_0\ptil_{n+1}
)\big|_{M}=
 \lmd^{-\frac d2-\frac 12}
 \Gamma\left(
   \frac d2+\frac12
 \right)
\end{equation}
we obtain
\begin{align*}
&
   -\left[
    (
      \p_\afa\ptil_{n+1}
    )\big|_{M}
    \Gtil^{n+1,0}\psi(\lmd)
  \right]_--\left[
    \p_\afa\xi(\lmd)
    \Gtil^{0,n+1}
    (
      \p_0\ptil_{n+1}
    )\big|_{M}
  \right]_-\\
&\qquad =
  \frac 1\pi\cos\frac{\pi d}{2}
  \left[
    \sum_{p\geq 0}
    \p_\afa\theta_{0,-p}
    \lmd^{p+\frac d2-\frac12}
    \Gamma\left(-p-\frac d2+\frac12\right)
    \lmd^{-\frac d2-\frac12}
    \Gamma\left(
      \frac d2+\frac12
    \right)
  \right]_-\\
&\qquad =
  \frac 1\pi\cos\frac{\pi d}{2}
  \cdot
    \p_\afa\theta_{0,0}
  \lmd^{-1}
  \Gamma\left(\frac{1-d}{2}\right)
  \Gamma\left(\frac{1+d}{2}\right)
 =
  \frac{e_\afa}{\lmd}.
\end{align*}
On the other hand, from the definition of periods $\ptil_i$ of the $(n+2)$-dimensional Frobenius manifold $\Mtil$
it follows that
\[\left(\p_\afa\ptil_i\right)\Gtil^{ij}
\left(\p_0\ptil_j\right)
=\left(\left(\gtil^{ij}-\lmd\etatil^{ij}\right)^{-1}\right)_{\afa,0},\]
where
\[
  (\gtil^{ij}) =
  \begin{pmatrix}
    0 & \bm 0 & v^0 \\
    \bm 0 & g^{\afa\beta}+v^0\eta^{\afa\beta} & E^\bullet \\
    v^0 & E_\bullet\eta^{-1} & (1-d)v^{n+1}+c_0
  \end{pmatrix}
\]
is the intersection form of $\Mtil$.
We observe that the operator $\mcalUtil=\left(\sum_{k=0}^{n+1}\gtil^{ik}\etatil_{kj}\right)$ of multiplication by the Euler vector field $\Etil$ of $\Mtil$
satisfies the relation
\[
  \left(
    \mcalUtil-\lmd\tilde{I}
  \right)^{-1}
  \Big|_{M}
=
  \begin{pmatrix}
    -\frac 1\lmd & \bm 0 & 0 \\[3pt]
    \frac 1\lmd\left(\mcalU-\lmd I\right)^{-1}E^\bullet & \left(\mcalU-\lmd I\right)^{-1}
    & \bm 0 \\[3pt]
    \frac{
      1
         }{\lmd^2}X_{n+1,0}
    &
    \frac1\lmd E_\bullet\left(\mcalU-\lmd I\right)^{-1} & -\frac{1}{\lmd}
  \end{pmatrix},
\]
where $\tilde{I}$ is the $(n+2)\times (n+2)$ identity matrix, and
\[X_{n+1,0}:= E_\bullet\left(\mcalU-\lmd I\right)^{-1}E^\bullet
      -(1-d)\theta_{0,0}-c_0.\]
Thus we arrive at
\begin{align*}
&
  \left[
    (
      \p_\afa\ptil_i
    )\big|_{M}
    \Gtil^{ij}
    (
      \p_0\ptil_j
    )\big|_{M}
  \right]_-
\\
&\quad =
  \left(
    \left(
      \gtil^{ij}-\lmd\etatil^{ij}
    \right)^{-1}\Big|_{M}
  \right)_{\afa,0}
=
 \left(
    \etatil
   \left(\mcalUtil-\lmd\tilde{I}\right)^{-1}
    \Big|_{M}
  \right)_{\afa,0}\\
&\quad =
 \frac1{\lmd} \left[\eta
   \left(\mcalU-\lmd I\right)^{-1} E^\bullet
  \right]^\afa
=
  \frac1\lmd \left[
    E_\bullet\left(\mcalU-\lmd I\right)^{-1}
  \right]_\afa
=
  \left(
    \frac e\lmd +\frac{1}{E-\lmd e}
  \right)_\afa.
\end{align*}
Here we use the identities
\begin{align*}
&\left(\mcalU-\lmd I\right)^{-1}=\eta^{-1} \left(\left(\mcalU-\lmd I\right)^{-1}\right)^{\transpose} \eta,\\
&\frac{1}{\lmd}E_\bullet\left(\mcalU-\lmd I\right)^{-1}=\frac1\lmd
  \left(
    \frac{E}{E-\lmd e}
  \right)_\bullet
=
  \left(
    \frac e\lmd +\frac{1}{E-\lmd e}
  \right)_\bullet,
\end{align*}
the second
one of which follows from
\[\left(\mcalU-\lmd I\right)\frac{E}{E-\lmd e}=\left(E-\lmd e\right)\cdot \frac{E}{E-\lmd e}=E.\]
Hence we have
\begin{align*}
&
  -\frac12\p_x\p_{t^{\afa,0}}
  \left[
    \pfrac{\Stil_i}{\lmd}
    \Gtil^{ij}
    \pfrac{\Stil_j}{\lmd}
  \right]_-\\
&\qquad =
  \left[
    \p_x(
      \p_\afa\ptil_i
    )\big|_{M}
    \Gtil^{ij}
    \pfrac{\Stil_j}{\lmd}
  \right]_-
 -\left(
   \frac e\lmd+\frac{1}{E-\lmd e}
 \right)_\afa \\
&\qquad\quad
  +\left[
    \p_x\p_\afa\xi(\lmd)
    \Gtil^{0,n+1}
    \pfrac{\Stil_{n+1}}{\lmd}
    \right]_-
  +\frac{e_\afa}{\lmd}\\
&\qquad =
  \left[
    \p_x (
      \p_\afa\ptil_i
    )\big|_{M}
    \Gtil^{ij}
    \pfrac{\Stil_j}{\lmd}
  \right]_-
 +\left[
    \p_x\p_\afa\xi(\lmd)
    \Gtil^{0,n+1}
    \pfrac{\Stil_{n+1}}{\lmd}
    \right]_-
 -\left(\frac{1}{E-\lmd e}\right)_\afa.
\end{align*}
The lemma is proved.
\end{proof}

  \begin{rmk}\label{rmk: general case 240818}
    In the general case when the spectrum of $\tilde\mu$ may contain half integers,
    by using the method that we used in the proof of Lemma \ref{zh-2023-1} we can verify the following analogues of the formulae \eqref{DtDlmdStil} and \eqref{DxDlmdStil}:
    \begin{align}
\p_{t^{\afa,0}}\pfrac{\Stil^{(\nu)}}{\lmd}
&=
  -\left.
  \p_\afa
  (\ptil_0^{(\nu)},\bm{\ptil}^{(\nu)},\ptil_{n+1}^{(\nu)})
  \right|_{M}
  -\p_\afa(\xi^{(\nu)}(\lmd),\bm 0,0),\label{DtDlmdStil-nu}
\\
\label{DxDlmdStil-nu}
\p_x\pfrac{\Stil^{(\nu)}}{\lmd}
&=
  -\left.
  \p_0
  (\ptil^{(\nu)}_0,\bm{\ptil}^{(\nu)},\ptil^{(\nu)}_{n+1})
  \right|_{M}
  -(\psi^{(\nu)}(\lmd),\bm 0,0),
\end{align}
here $\xi^{(\nu)}(\lmd)$ is defined as in \eqref{defn:xi(nu)(lmd)}, and
\begin{align*}
  \psi^{(\nu)}(\lmd)&:=
  \sum_{p\geq 0}
  \int_{0}^{\infty}
    \rme^{-\lmd z}\theta_{0,-p-1}
    z^{-p-\frac d2 + \nu -\frac12}\td z\\
&\phantom:=\,
  \sum_{p\geq 0}\theta_{0,-p-1}\lmd^{p+\frac d2 -\nu-\frac12}
  \Gamma\left(
    -p-\frac d2 + \nu +\frac12
  \right).
\end{align*}
Similarly, we can verify that
\begin{align}
&
-\frac12\eta^{\afa\beta}
\p_x\p_{t^{\beta,0}}
\left[
  \pfrac{\Stil_i^{(\nu)}}{\lmd}\Gtil^{ij}(\nu)
  \pfrac{\Stil_j^{(-\nu)}}{\lmd}
\right]_-\notag\\
&\qquad=
\left[
    \p_x(
      \p^\afa\ptil_i^{(\nu)}
    )\big|_{M}
    \Gtil^{ij}(\nu)
    \pfrac{\Stil_j^{(-\nu)}}{\lmd}
  \right]_-
+
 \left[
   \p_x\p^\afa\xi^{(\nu)}(\lmd)
   \Gtil^{n+1,0}(\nu)
   \pfrac{\Stil^{(-\nu)}_{n+1}}{\lmd}
 \right]_- \notag\\
&\qquad\quad
-
 \left(
   \frac{1}{E-\lmd e}
 \right)^\afa, \label{240818-2005}
\end{align}
which is an analogue of \eqref{Qhn220721}.
\end{rmk}

\begin{thm}\label{main-lemma-orig}
For $m\geq -1$ and $(i,p), (j,q)\in\mcalI$,
the following identities hold true:
\begin{align}\label{220720-D_E-Omg final form}
 & \p_{\frac{1}{E-\lmd e}}
  \Omg_{i,p;j,q}
:=
  -\sum_{m\geq -1}\frac{1}{\lmd^{m+2}}
  \p_{E^{m+1}}\Omg_{i,p;j,q}\notag\\
&\quad =
  \lim_{\nu\to 0}
  \left[
    \left(
    \pp{t^{i,p}}
    \pfrac{\Stil_{i'}^{(\nu)}}{\lmd}
    \right)
  \Gtil^{i'j'}(\nu)
    \left(
      \pp{t^{j,q}}
      \pfrac{\Stil_{j'}^{(-\nu)}}{\lmd}
    \right)
  \right]_-
-
 2\delta_{i,0}\delta_{j,0}C_{p,q}(\lmd),
\end{align}
where $C_{p,q}(\lmd)$ are defined by \eqref{C_p,q lmd}.
\end{thm}
The proof of this theorem is given in Appendix \ref{Appendix-B}.

Now we start to prove Theorem \ref{thm: Vir sym on tau-cover}
in the non-resonant case.

\begin{proof}[Proof of Theorem \ref{thm: Vir sym on tau-cover}]
  To show that \eqref{Vir sym on tau-cover 0}--\eqref{Vir sym on tau-cover 3}
  are symmetries of the tau-cover \eqref{tau-cover} of the Principle Hierarchy, we need to verify the following commutation relations:
  \begin{equation}\label{zh-29}
    \left[
      \pp{s},\pp{t^{k,\ell}}
    \right]f=0,\qquad
    \left[
      \pp{s},\pp{t^{k,\ell}}
    \right]f_{i,p}=0,\qquad
    \left[
      \pp{s},\pp{t^{k,\ell}}
    \right]v^\afa=0
  \end{equation}
  for all $(i,p), (k,\ell)\in\mcalI$ and $\afa=1,2,\dots,n$.
  From \eqref{Vir sym on tau-cover 2} we see that the first set of commutation relations $\left[
      \pp{s},\pp{t^{k,\ell}}
    \right]f=0$ hold true trivially.

In the non-resonant case we can take $\nu=0$, so
 from \eqref{Vir sym on tau-cover 3} we have
 \begin{align*}
    &\pp{s}\pfrac{f_{i,p}}{t^{k,\ell}}
  =
    \pp{s}\Omg_{i,p;k,\ell}
   =
    \p_\gamma\Omg_{i,p;k,\ell}
    \pfrac{v^\gamma}{s}
  \\
 &\qquad =
    \p_\gamma\Omg_{i,p;k,\ell}
    \left(
    \left[
      \pfrac{\Stil_{i'}}{\lmd}
      \Gtil^{i'j'}\p_x\p^\gamma\ptil_{j'}
    \right]_-
  +
    \left[
      \p_x\p^\gamma\xi(\lmd)
      \Gtil^{n+1,0}
      \pfrac{\Stil_{n+1}}{\lmd}
    \right]_-
  \right)\Bigg|_{M}\\
  &\qquad\quad-\p_{\frac{1}{E-\lmd e}}\Omg_{i,p;k,\ell}.
  \end{align*}
On the other hand, by using Theorem \ref{main-lemma-orig} we obtain
\begin{align*}
 & \pp{t^{k,\ell}}
  \pfrac{f_{i,p}}{s}
=
  \pp{t^{k,\ell}}
  \pp{t^{i,p}}
  \left(-\frac12
  \left[
    \pfrac{\Stil_{i'}}{\lmd}
    \Gtil^{i'j'}
    \pfrac{\Stil_{j'}}{\lmd}
  \right]_-
    +\sum_{s,q\in\bbZ}C_{s,q}(\lmd)t^{0,s}t^{0,q}
  \right)\\
&\qquad =
  -\pp{t^{k,\ell}}
  \left[
    \left(
      \pp{t^{i,p}}
      \pfrac{\Stil_{i'}}{\lmd}
    \right)
    \Gtil^{i'j'}
      \pfrac{\Stil_{j'}}{\lmd}
  \right]_-
+
  2\delta_{i,0}\delta_{k,0}C_{p,\ell}(\lmd)
\\
&\qquad =
  -
  \left[
    \left(
      \pp{t^{i,p}}
      \pfrac{\Stil_{i'}}{\lmd}
    \right)
    \Gtil^{i'j'}
    \left(
      \pp{t^{k,\ell}}
      \pfrac{\Stil_{j'}}{\lmd}
    \right)
  \right]_-
 +2\delta_{i,0}\delta_{k,0}C_{p,\ell}(\lmd)\\
&\qquad \quad
 -
  \left[
  \left(
      \frac{\p^2}{\p t^{i,p}\p t^{k,\ell}}
      \pfrac{\Stil_{i'}}{\lmd}
    \right)
    \Gtil^{i'j'}
      \pfrac{\Stil_{j'}}{\lmd}
  \right]_-\\
&\qquad =
  -
  \p_{\frac{1}{E-\lmd e}}\Omg_{i,p;k,\ell}
 -
  \left[
  \left(
      \frac{\p^2}{\p t^{i,p}\p t^{k,\ell}}
      \pfrac{\Stil_{i'}}{\lmd}
    \right)
    \Gtil^{i'j'}
      \pfrac{\Stil_{j'}}{\lmd}
  \right]_-.
\end{align*}
Thus by using the relation
\begin{align*}
&\frac{\p^2}{\p t^{i,p}\p t^{k,\ell}}
\pfrac{\Stil}{\lmd}
 = -\int_0^\infty
    \frac{\td z}{\sqrt{z}}\rme^{-\lmd z}
    \sum_{q\geq 0}
    \left(
      \frac{\p^2f_{0,q}}{\p t^{i,p}\p t^{k,\ell}},
      \frac{\p^2 \bm{f}_{q}}{\p t^{i,p}\p t^{k,\ell}}, 0
    \right)
    z^{q+1}z^{\mutil}z^{\Rtil}
\\
&\qquad\quad
  -\int_0^\infty
    \frac{\td z}{\sqrt{z}}\rme^{-\lmd z}
    \sum_{q\geq 0}
    \left(
      \frac{\p^2f_{0,-q-1}}{\p t^{i,p}\p t^{k,\ell}},
      \bm 0, 0
    \right)
    z^{-q}z^{\mutil}z^{\Rtil}
\\
&\qquad=
    -\int_0^\infty
    \frac{\td z}{\sqrt{z}}\rme^{-\lmd z}
    \sum_{q\geq 0}
    \left(
      \frac{\p \Omg_{i,p;k,\ell}}{\p t^{0,q}},
      \frac{\p \Omg_{i,p;k,\ell}}{\p \bm{t}^q}, 0
    \right)
    z^{q+1}z^{\mutil}z^{\Rtil}
\\
&\qquad\quad
  -\int_0^\infty
    \frac{\td z}{\sqrt{z}}\rme^{-\lmd z}
    \sum_{q\geq 0}
    \left(
      \frac{\p \Omg_{i,p;k,\ell}}{\p t^{0,-q-1}},
      \bm 0, 0
    \right)
    z^{-q}z^{\mutil}z^{\Rtil}\\
&\qquad=
  -\p_\gamma\Omg_{i,p;k,\ell}
  \p_x
  \left(
    (\p^\gamma\ptil_0,\p^\gamma\bm{\ptil},\p^\gamma\ptil_{n+1})
   +(\p^\gamma\xi(\lmd),\bm 0, 0)
  \right)\big|_{M},
\end{align*}
we arrive at the validity of the second set of relations of \eqref{zh-29}. In particular, we have
\begin{equation}\label{cor of 2nd vir sym}
  \left[\pp{s},\pp{t^{i,p}}\right]f_{\beta,0}=0,\quad
 \forall\, (i,p)\in\mcalI,\,\beta=1,2,\dots,n.
\end{equation}
Therefore, from \eqref{Qhn220721} and \eqref{cor of 2nd vir sym} we obtain
\begin{align*}
  &\pp{t^{i,p}}
  \pfrac{v^\afa}{s}
=
  \eta^{\afa\beta}
  \pp{t^{i,p}}
  \left(
  \p_x\pfrac{f_{\beta,0}}{s}
  \right)
=
  \eta^{\afa\beta}
  \p_x\left(
    \pp{t^{i,p}}
    \pp{s}
    f_{\beta,0}
  \right)\\
&\qquad =
  \eta^{\afa\beta}
  \p_x\left(
    \pp{s}
    \pp{t^{i,p}}
    f_{\beta,0}
  \right)
=
  \pp{s}
  \left(
    \eta^{\afa\beta}
    \p_x\Omg_{i,p;\beta,0}
  \right)
=
  \pp{s}\pfrac{v^\afa}{t^{i,p}},
\end{align*}
which leads to $\left[\pp{s},\pp{t^{k,\ell}}\right]v^\afa=0$.

In the general case when the spectrum of $\tilde\mu$ may contain half integer,
we can prove the theorem in a similar way by using \eqref{DtDlmdStil-nu}--\eqref{240818-2005}.
The theorem is proved.
\end{proof}

It is easy to see from Lemma \ref{lemma: Laplace-Bilinear identity} that
the symmetries $\pfrac{f}{s_m}$, $m\geq -1$ given in Theorem
\ref{thm: Vir sym on tau-cover} can be represented in the form
  \begin{align}
      \pfrac{f}{s_m}
    =&\,
      \sum_{(i,p),(j,q)\in\mcalI}
        a^{i,p;j,q}_m
        \pfrac{f}{t^{i,p}}
        \pfrac{f}{t^{j,q}}
     +\sum_{(i,p),(j,q)\in\mcalI}
        b^{i,p}_{m;j,q}
        t^{j,q}
        \pfrac{f}{t^{i,p}}\notag
    \\
    &\,
     +\sum_{(i,p),(j,q)\in\mcalI}
        c_{m;i,p;j,q}
        t^{i,p}t^{j,q}
     +\sum_{p,q\in\bbZ}
      C_{m;p,q}t^{0,p}t^{0,q}, \label{ABC of f_sm}
  \end{align}
  for some constants $a^{i,p;j,q}_m$, $b^{i,p}_{m;j,q}$,
  $c_{m;i,p;j,q}\in\bbC$ which satisfy the condition
  \[a^{i,p;j,q}_m=a^{j,q;i,p}_m,\quad c_{m;i,p;j,q}=c_{m;j,q;i,p}.\]
  We will also use the following generating series:
  \begin{equation}
  a^{i,p;j,q}(\lmd):=
    \sum_{m\geq -1}
    \frac{a^{i,p;j,q}_m}{\lmd^{m+2}},\
  b^{i,p}_{j,q}(\lmd):=
    \sum_{m\geq -1}
    \frac{b^{i,p}_{m;j,q}}{\lmd^{m+2}},\
  c_{i,p;j,q}(\lmd):=
    \sum_{m\geq -1}
    \frac{c_{m;i,p;j,q}}{\lmd^{m+2}}.
  \end{equation}
In general, let $\Bigset{\Phi^{i,p},\,\Psi_{j,q}}{(i,p),(j,q)\in\mcalI}$
be a family of formal variables in some $\bbC$-algebra, and let us denote
\[
  \Stil^{(\nu)}(\Phi,\Psi;\lmd):=
  \Stil^{(\nu)}(\bft,\pfrac{f}{\bft};\lmd)
  \Big|_{t^{i,p}\mapsto\Phi^{i,p},\, f_{j,q}\mapsto\Psi_{j,q}},
\]
then the following identity holds true:
\begin{align}\label{formal Laplace bilinear identity}
&
  -\frac12\lim_{\nu\to 0}
  \left[
    \pfrac{\Stil_i^{(\nu)}(\Phi,\Psi;\lmd)}{\lmd}
    \Gtil^{ij}(\nu)
    \pfrac{\Stil_j^{(-\nu)}(\Phi,\Psi;\lmd)}{\lmd}
  \right]_-\notag\\
&\qquad=
  \sum_{(i,p),(j,q)\in\mcalI}
        a^{i,p;j,q}(\lmd)
        \Psi_{i,p}\Psi_{j,q}
     +\sum_{(i,p),(j,q)\in\mcalI}
        b^{i,p}_{j,q}(\lmd)
        \Phi^{j,q}
        \Psi_{i,p}\notag\\
    &\qquad\quad +\sum_{(i,p),(j,q)\in\mcalI}
        c_{i,p;j,q}(\lmd)
        \Phi^{i,p}\Phi^{j,q}.
\end{align}

\begin{defn}\label{defn:ext Vira coef}
We call the constants $a^{i,p;j,q}_m$, $b^{i,p}_{m;j,q}$,
  $c_{m;i,p;j,q}$ and $C_{m;p,q}$ that are given by \eqref{ABC of f_sm} and \eqref{C_p,q lmd}
the extended Virasoro coefficients of $M$.
\end{defn}

The extended Virasoro coefficients
can be calculated explicitly by using the formulae given in
  \eqref{Laplace-bilinear identity}--\eqref{N pq}.
The following properties of these coefficients
can be verified by straightforward calculation:
\begin{align}
  a_m^{i,p;j,q}&=0,\qquad \textrm{if} \ i=0\ \textrm{or}\  j=0, \label{Vira-coef relation-1}\\
  b^{0,q}_{m;i,p}&\neq 0, \qquad\textrm{only if}\ i=0\ \textrm{and}\ q=p+m, \\
b^{i,q}_{m;i,p}&=
  \delta^q_{p+m}
  \left(
    p+\mu_i+\frac12
  \right)^{[m+1]},
\label{Vira-coef relation-3}
\end{align}
here $(i,p),(j,q)\in\mcalI$ and $m\geq -1$, and
the symbol $x^{[m]}$ is defined in \eqref{perm symbol}.

\section{The Virasoro operators}
We are to construct in this section a collection of Virasoro operators in terms of the monodromy data $\eta, \mu, R$ of $M$ at $z=0$, the constant vectors
$\bm{r}_s=(r^1_s,\dots,r^n_s)^\transpose$, $s\ge 1$ and the constants $c_p$, $p\in\mathbb{Z}$ that are specified in lemma
\ref{lemma-existence of theta-0-p} and Proposition \ref{zh-17}.

Our construction of the Virasoro operators for $M$ is based on the one given in \cite{Frob mfd and Vir const, normal-form} for the $(n+2)$-dimensional Frobenius manifold $\Mtil$ that is introduced in Lemma \ref{lemma: n+2 dim intro}.
Let $a_p^i$, $0\le i\le n+1$, $p\in\mathbb{Z}$ and $\mathbf 1$ be the generators of a Heisenberg algebra $\mathcal{H}$, which satisfy the commutation relations
\begin{equation}
[\mathbf{1}, a^i_p]=0,\quad [a^i_p, a^j_q]=(-1)^p \tilde{\eta}^{ij}\delta_{p+q+1,0}\cdot\mathbf{1}.
\end{equation}
Introduce the normal ordering as follows:
\[:a^i_p a^j_q:=\begin{cases} a^j_q a^i_p, &\textrm{if}\ p\ge 0,\ q<0,\\
a^i_p a^j_q,&\textrm{other cases},\end{cases}\]
then the Virasoro operators $\Ltil_m, m\ge -1$ of $\Mtil$ are defined by
\begin{align}
  \Ltil(\lmd)&=
\sum_{m\geq -1}
  \frac{\Ltil_m}{\lmd^{m+2}}\notag\\
&=
  -\frac12\lim_{\nu\to 0}
  :
  \left[
    \pfrac{\tilde\phi^{(\nu)}_i}{\lmd}
    \Gtil^{ij}(\nu)
    \pfrac{\tilde\phi^{(-\nu)}_j}{\lmd}
  \right]_-
  :
 +\frac{1}{4\lmd^2}\mathrm{tr}
  \left(
    \frac14-\mutil^2
  \right), \label{Lmtil(lmd)}
\end{align}
where $\tilde\phi^{(\nu)}=\tilde\phi^{(\nu)}(\bfa;\lmd)$
is given in \eqref{n+2 formal phi-Laplace}, and the negative part $[\,\cdot\,]_-$  is introduced in \eqref{negative-part of lmd}.
From Remark \ref{remark:polynomial-nu} we know that the above limit is well defined.
These operators satisfy the commutation relations
\[[\Ltil_m, \Ltil_n] = (m-n)\Ltil_{m+n},\quad m,n\ge -1.\]
By putting
\begin{equation}\label{Heisenberg a_p}
  a^i_p :=
  \begin{cases}
   \etatil^{ij}\pp{t^{j,p}}, & p\geq 0,\\[3pt]
    (-1)^{p+1}t^{i,-p-1}, & p<0,
  \end{cases}\qquad
  \end{equation}
for $0\leq i\leq n+1$, we obtain a realization of the Heisenberg algebra in the space of functions of the variables $t^{i,p}$, $0\le i\le n+1$, $p\ge 0$, and a collection of linear
differential operators $\Ltil_m$ acting on this space.

Let us proceed to define the Virasoro operators for the generalized Frobenius manifold $M$ by modifying the expression \eqref{Lmtil(lmd)} of $\Ltil(\lm)$ and by using a different realization of the Heisenberg algebra $\mathcal{H}$. This realization of the Heisenberg algebra is given by
\begin{equation}\label{Heisenberg b_p}
  a^i_p :=
  \begin{cases}
    (-1)^{p+1}t^{0,-p-1}, & i=0,\,p\in\bbZ,\\
    \eta^{i\beta}\pp{t^{\beta,p}}, & i=1,\dots,n,\, p\geq 0,\\
    (-1)^{p+1}t^{i,-p-1}, & i=1,\dots,n,\, p<0,\\
    \pp{t^{0,p}}, & i=n+1,\, p\in\bbZ.
  \end{cases}
\end{equation}
\begin{defn}\label{zh-23-08-1}
The Virasoro operators $L_m, m\ge -1$ of the generalized Frobenius manifold $M$ are defined by the following formula:
\begin{align}
 & L(\bft,\pp\bft;\lmd)
=
  \sum_{m\geq -1}
  \frac{L_m(\bft,\pp\bft)}{\lmd^{m+2}}\notag\\
&\qquad=
  -\frac12\lim_{\nu\to 0}
  :\left[
    \pfrac{\tilde\phi^{(\nu)}_i}{\lmd}
    \Gtil^{ij}(\nu)
    \pfrac{\tilde\phi^{(-\nu)}_j}{\lmd}
    \right]_-
  :
   +\sum_{p,q\in\bbZ}C_{p,q}(\lmd)t^{0,p}t^{0,q}\notag\\
  &\qquad\quad+\frac{1}{4\lmd^2}\mathrm{tr}
  \left(
    \frac14-\mu^2
  \right), \label{Lm final}
\end{align}
where $\tilde\phi^{(\nu)}=\tilde\phi^{(\nu)}(\bfa;\lmd)$
is defined in \eqref{n+2 formal phi-Laplace} with
$\mathbf{a}_p=(a^0_p,\dots,a^{n+1}_p)^{\transpose}$ having the realization
\eqref{Heisenberg b_p},
the series $C_{p,q}(\lmd)$ are defined in \eqref{C_p,q lmd},
and the
normal ordering $:\!\cdot\!:$ means to put the differential operator terms on the right.
\end{defn}

It is important to note that the $n\times n$ matrix $\mu$ that appears in \eqref{Lm final} is not the $(n+2)\times(n+2)$ matrix $\mutil$.
From the above definition it follows that
the Virasoro operators $L_m$, $m\geq -1$ have the form
\begin{align}
&L_m(\bft,\pp{\bft})
=
  \sum_{(i,p),(j,q)\in\mcalI}
    a_m^{i,p;j,q}
    \frac{\p^2}{\p t^{i,p}\p t^{j,q}}
 +\sum_{(i,p),(j,q)\in\mcalI}
   b_{m; i,p}^{j,q}
   t^{i,p}\pp{t^{j,q}}
\notag\\
&\quad +
  \sum_{(i,p),(j,q)\in\mcalI}
   c_{m;i,p;j,q}t^{i,p}t^{j,q}
 +
  \sum_{p,q\in\bbZ}
   C_{m;p,q}t^{0,p}t^{0,q}\notag\\
&\quad
 +\frac14\delta_{m,0}
 \mathrm{tr}
 \left(
   \frac14-\mu^2
 \right)
 \bm 1,  \label{ABC of Lm}
\end{align}
where the extended Virasoro coefficients
$a_m^{i,p;j,q}, b_{m; i,p}^{j,q}, c_{m;i,p;j,q}$
and $C_{m;p,q}$ are specified in Definition \ref{defn:ext Vira coef}.
By using \eqref{Laplace-bilinear identity} and
\eqref{N pq},
we can also obtain the following formulae for $L_m(\bft,\pp\bft)$ (see \cite{Frob mfd and Vir const}):
\begin{align*}
&  L_m(\bft,\pp\bft)=\frac12\sum_{p,q\in\bbZ}
    (-1)^{p+1}
    :
      \bfa^\transpose_q\tilde{\eta}
      \left[
        P_m(\tilde{\mu}-p,\tilde{R})
      \right]_{m-1-q-p}
      \bfa_p
    :\\
&\quad
  +\sum_{p,q\in\bbZ}C_{m;p,q}t^{0,p}t^{0,q}
+\frac14\delta_{m,0}\mathrm{tr}
\left(\frac14-\mu^2\right)
\bm 1,\quad m\ge -1.
\end{align*}
Here $\mathbf{a}_p=(a^0_p,\dots,a^{n+1}_p)^{\transpose}$
have the realization
\eqref{Heisenberg b_p},
and $\etatil,\mutil,\Rtil$ are the monodromy data of $\Mtil$ at $z=0$
given by \eqref{zh-25} and \eqref{n+2 mu R intro}. The
$(n+2)\times(n+2)$ matrices $P_m$ have the expressions
\begin{equation}
  P_m(\tilde{\mu}, \tilde{R}):=
  \begin{cases}
  \left.
    \rme^{\tilde{R}\p_x}
    \prod_{j=0}^{m}
    \left(
      x+\tilde{\mu}+j-\frac12
    \right)
  \right|_{x=0},
  & m\geq 0,
  \\
  1, & m=-1,
  \end{cases}
\end{equation}
and their components $[P_m]_k$ are defined as in \eqref{zh-30}.

\begin{prop}
The Virasoro operators $L_m, m\ge -1$
of the generalized Frobenius manifold $M$ satisfy the following commutation relations:
\begin{equation}\label{vira commu final}
  [L_m, L_k] = (m-k)L_{m+k},\quad
  m, k\geq -1.
\end{equation}
\end{prop}

\begin{proof}
 We can prove the commutation relations of the Virasoro operators in the same way as it is done in \cite{Frob mfd and Vir const, normal-form} for the Virasoro operators of a usual Frobenius manifold.
As for the additional term $\sum_{p,q\in\bbZ}C_{m;p,q}t^{0,p}t^{0,q}$ in $L_m$,
we note, due \eqref{Vira-coef relation-1}-\eqref{Vira-coef relation-3}, that the only term in $L_{k}$
 which has non-trivial commutator with $\sum_{p,q\in\bbZ}C_{m;p,q}t^{0,p}t^{0,q}$ is
 \[
   \sum_{p\in\bbZ}b_{k;0,p}^{0,p+k}
   t^{0,p}\pp{t^{0,p+k}}.
 \]
 The commutator
 \[\left[
   \sum_{p,q\in\bbZ}C_{m;p,q}t^{0,p}t^{0,q} ,
   \sum_{p\in\bbZ}b_{k;0,p}^{0,p+k}
   t^{0,p}\pp{t^{0,p+k}}
 \right]\]
 can be calculated directly
 by using the explicit formulae \eqref{C_p,q lmd} and \eqref{Vira-coef relation-3}.
 We omit the details here. The proposition is proved.
\end{proof}

At the end of this section,
we provide the explicit formulae of $L_{-1}$, $L_0$, $L_1$ and $L_2$.
Recall that
\[\Rtil_{s;1}:=\Rtil_s=
\begin{pmatrix}
  0 &   &   \\
  \bm{r}_{s} & R_s &  \\
  c_{s-1} & \bm{r}^\dag_s & 0
\end{pmatrix}.\]
It is easy to check that
\begin{equation}\label{zh-39}
\tilde{R}_{s;k}:=
[\tilde{R}^k]_s
=
    \begin{pmatrix}
      0 &   &   \\
      \langle R\bm{r}\rangle_{s;k} & R_{s;k} &   \\
      \langle \bm{r}^\dag R\bm{r}\rangle_{s;k} & \langle \bm{r}^\dag R\rangle_{s;k} &  0
    \end{pmatrix},\quad k\geq 2,
\end{equation}
where $R_{s;k}:=[R^k]_s$,
and
\begin{align*}
\langle R\bm{r}\rangle_{s;k}&=\sum_{\ell\geq 1}R_{\ell;k-1}\bm{r}_{s-\ell},\\
\langle \bm{r}^\dag R\rangle_{s;k}&=\sum_{\ell\geq 1}\bm{r}^\dag_{s-\ell} R_{\ell;k-1},\\
\langle \bm{r}^\dag R\bm{r}\rangle_{s;k}&=
\sum_{p+q+t=s}\bm{r}^\dag_{p} R_{q;k-2}\bm{r}_{t}.
\end{align*}
By using the above notations and our assumption that
$\pp{t^{\afa, p}}= 0$ if $\afa\neq 0$ and $p<0$,
we can write down the explicit expressions of
the operators $L_m$ for $m=-1,0,1,2$ as follows:
\begin{align}
  L_{-1}=&\,
  \sum_{p\geq 0}
      t^{\afa,p+1}\pp{t^{\afa,p}}
  +\sum_{p\in\bbZ}
      t^{0,p+1}\pp{t^{0,p}}
  +\frac12\eta_{\afa\beta}t^{\afa,0}t^{\beta,0},
\label{explicit L-1}
\\
  L_0=&\,
    \sum_{p\geq 0}
      \left(
        p+\mu_\afa+\frac12
      \right)
      t^{\afa,p}\pp{t^{\afa,p}}
    +\sum_{p\in\bbZ}
      \left(
        p-\frac d2+\frac12
      \right)
      t^{0,p}\pp{t^{0,p}}\notag
\\
&
    +\sum_{p\geq 0}
     \sum_{s\geq 1}
       (R_s)^\veps_\afa
       t^{\afa,p}
       \pp{t^{\veps, p-s}}
    +\sum_{p\in\bbZ}
     \sum_{s\geq 1}
      r^\veps_s t^{0, p}
      \pp{t^{\veps,p-s}}
\notag\\
  &
    +\frac12\sum_{p,q\geq 0}
     (-1)^p\eta_{\afa\veps}
     (R_{p+q+1})^\veps_\beta
     t^{\afa,p}t^{\beta,q}
    +\sum_{p\geq 0}
     \sum_{s\geq 1}
     (-1)^{p+s+1}(r_s)_\afa t^{0,s-1-p}t^{\afa,p}\notag\\
 &   +\frac12\sum_{p,q\in\bbZ}
     (-1)^p c_{p+q}t^{0,p}t^{0,q}
     +\frac14\mathrm{tr}
     \left(
       \frac14-\mu^2
     \right)\bm 1,
\\
  L_1=&\,
    \frac12\left(\frac14-\mu^2\right)_\veps^\afa
    \eta^{\veps\beta}
    \frac{\p^2}{\p t^{\afa,0}\p t^{\beta,0}}
    \notag\\
   &
    +\sum_{p\geq 0}
       \left(
         p+\mu_\afa+\frac12
       \right)
       \left(
         p+\mu_\afa+\frac32
       \right)
       t^{\afa,p}\pp{t^{\afa,p+1}}
\notag\\
  &
    +\sum_{p\in\bbZ}
       \left(
         p-\frac d2+\frac12
       \right)
       \left(
         p-\frac d2+\frac32
       \right)
       t^{0,p}\pp{t^{0,p+1}}
    \notag\\
  &+2\sum_{p\geq 0,\, s\ge 1}
        (R_s)^\veps_\afa
        (p+\mu_\afa+1)
        t^{\afa,p}\pp{t^{\veps,p+1-s}}
        \notag\\
        &
        +2\sum_{p\in\bbZ,\,s\ge 1}
         r^\veps_s
        \left(p-\frac d2+1\right)
        t^{0,p}\pp{t^{\veps,p+1-s}}\notag\\
   &
    +\sum_{p\geq 0,\,s\ge 2}
       (R_{s;2})^\veps_\afa
       t^{\afa,p}\pp{t^{\veps, p+1-s}}
    +\sum_{p\in\bbZ,\,s\ge 2}
         \langle R\bm{r}\rangle_{s;2}^\veps
       t^{0,p}\pp{t^{\veps, p+1-s}}
\notag\\
  &
    +\sum_{p,q\geq 0}
       (-1)^q(p+\mu_\afa+1)
       (R_{p+q+2})^\veps_\afa\eta_{\veps\beta}
       t^{\afa,p}t^{\beta,q}\notag\\
   & +\frac12
     \sum_{p,q\geq 0}
       (-1)^q
       (R_{p+q+2;2})^\veps_\afa\eta_{\veps\beta}
       t^{\afa,p}t^{\beta,q}
       \notag\\
       &
        +\sum_{p\geq 0,\,s\ge 2}
       (-1)^{p+s}
       \left(
         \langle \bm{r}^\dag R\rangle_{s;2}
       \right)_\afa
       t^{0,s-2-p}t^{\afa,p}
\notag\\
  &
    +2\sum_{p\geq 0,\,s\ge 1}
         (-1)^{p+s}
       (p+\mu_\afa+1)(r_s)_\afa
       t^{0,s-2-p}t^{\afa,p}
\notag\\
  &
    +\frac12
     \sum_{p\in\mathbb{Z},\,s\geq 2}(-1)^p
      \langle \bm{r}^\dag R\bm{r}\rangle_{s;2}
     t^{0,s-2-p}t^{0,p}
   +\frac12
     \sum_{p,q\in\bbZ}
       (-1)^p
       c_{p+q+1}
       (q-p)
       t^{0,p}t^{0,q},
\\
  L_2=&\,
    \frac12
    \left(-3\mu_\afa^2+3\mu_\afa+\frac14\right)
    (R_1)^\afa_\lmd\eta^{\lmd\beta}
    \frac{\p^2}{\p t^{\afa,0}\p t^{\beta,0}}
\notag\\
  &
    +
      \left(
        \frac12-\mu_\afa
      \right)
      \left(
        \frac32-\mu_\afa
      \right)
      \eta^{\afa\beta}
      \left(
        \frac12-\mu_\beta
      \right)
      \frac{\p^2}{\p t^{\afa,0}\p t^{\beta,1}}
\notag\\
  &
    +\sum_{p\geq 0}
       \left(
         p+\mu_\afa+\frac12
       \right)
       \left(
         p+\mu_\afa+\frac32
       \right)
       \left(
         p+\mu_\afa+\frac52
       \right)
       t^{\afa,p}\pp{t^{\afa,p+2}}
\notag\\
  &
    +\sum_{p\in\bbZ}
       \left(
         p-\frac d2+\frac12
       \right)
       \left(
         p-\frac d2+\frac32
       \right)
       \left(
         p-\frac d2+\frac52
       \right)
       t^{0,p}\pp{t^{0,p+2}}
\notag\\
  &
    +\sum_{p\geq 0,\,s\ge 1}
\left(3\left(p+\mu_\al+\frac32\right)^2-1\right)
        (R_s)^\veps_\afa
        t^{\afa,p}\pp{t^{\veps,p+2-s}}
\notag\\
  &
    +\sum_{p\in\bbZ,\,s\ge 1}
\left(3\left(p-\frac{d}2+\frac32\right)^2-1\right)
        r_s^\veps
        t^{0,p}\pp{t^{\veps,p+2-s}}
\notag\\
  &
    +\sum_{p\geq 0,\,s\ge 2}
       3\left(p+\mu_\afa+\frac32\right)
       (R_{s;2})^\veps_\afa
       t^{\afa,p}\pp{t^{\veps, p+2-s}}
\notag\\
  &
    +\sum_{p\in\bbZ,\,s\ge 2}
       3\left(p-\frac d2+\frac32\right)
       \langle
         R\bm{r}
       \rangle_{s;2}^\veps
       t^{0,p}\pp{t^{\veps, p+2-s}}
\notag\\
  &
    +\sum_{p\geq 0,\,s\ge 3}
        (R_{s;3})^\veps_\afa t^{\afa,p}\pp{t^{\veps,p+2-s}}
    +\sum_{p\in\bbZ}
     \sum_{s\geq 3}
       \langle
         R\bm{r}
       \rangle_{s;3}^\veps
       t^{0,p}\pp{t^{\veps,p+2-s}}
\notag\\
  &
    +\frac12
     \sum_{p,q\geq 0}(-1)^q
       \Bigg[
         (R_{p+q+3;3})^\veps_\afa
        +3\left(
          p+\mu_\afa+\frac32
        \right)
         (R_{p+q+3;2})^\veps_\afa\notag\\
  &+
    \left(3\left(p+\mu_\al+\frac32\right)^2-1\right)
    (R_{p+q+3})^\veps_\afa
    \Bigg]\eta_{\veps\beta}
    t^{\afa,p}t^{\beta,q}
\notag\\
  &
    +\sum_{p\geq 0,\,s\ge 1}
      (-1)^{s+p+1}
\left(3\left(p+\mu_\al+\frac32\right)^2-1\right)
(r_s)_\afa
      t^{0,s-3-p}t^{\afa,p}
\notag\\
  &
    +\sum_{p\geq 0,\,s\ge 2}
      (-1)^{s+p+1}
      3\left(p+\mu_\afa+\frac32\right)
      \left(
        \langle
          \bm{r}^\dag R
        \rangle_{s;2}
      \right)_\afa
      t^{0,s-3-p}t^{\afa,p}
\notag\\
  &
    +\sum_{p\geq 0,\,s\ge 2}
       (-1)^{s+p+1}
       \left(
         \langle \bm{r}^\dag R \rangle_{s;3}
       \right)_\afa
       t^{0,s-3-p}t^{\afa,p}
\notag\\
  &
    +\frac34
    \sum_{p,q\in\bbZ}
      (-1)^q(p-q)
      \langle
        \bm{r}^\dag R\bm{r}
      \rangle_{p+q+3;2}
    t^{0,p}t^{0,q}
  +\frac12\sum_{p,q\in\bbZ}
    (-1)^p
    \langle
        \bm{r}^\dag R\bm{r}
      \rangle_{p+q+3;3}
    t^{0,p}t^{0,q}
\notag\\
  &
    +\frac12
     \sum_{p,q\in\bbZ}
     (-1)^pc_{p+q+2}
     \left[
       \left(
         p+\frac12-\frac d2
       \right)
       \left(
         p+\frac32-\frac d2
       \right)\right.\notag\\
       &
     \left. + \left(
         q+\frac12-\frac d2
       \right)
       \left(
         q+\frac32-\frac d2
       \right)
     -\left(p+\frac12-\frac d2\right)
      \left(q+\frac12-\frac d2\right)
     \right]
     t^{0,p}t^{0,q}.\label{explicit L2}
\end{align}

\section{Linearization of the Virasoro symmetries and the loop equation}
Let us consider the integrable hierarchy
\begin{equation}\label{deformed PH}
  \pfrac{w^\afa}{t^{i,p}}
  =K^\al_{i,p}(w;w_x, w_{xx},\dots),\quad
  (i,p)\in\mcalI,\, \afa=1,2,\dots,n
\end{equation}
that is obtained from the Principal Hierarchy \eqref{principal hierarchy} by using a certain quasi-Miura transformation of the form
\begin{equation}\label{quasi-Miura transf}
  v^\afa\mapsto w^\afa =v^\afa+\eta^{\al\gamma}\p_x\p_{t^{\gamma,0}}
  \sum_{k\geq 1}
  \veps^{k}\mcalF^{[k]},
\end{equation}
where $\mcalF^{[k]}=\mcalF^{[k]}(v;v_x,\dots,v^{(m_k)})$
are functions on the jet space $J^\infty(M^n)$.

Denote
\begin{equation}\label{full-genera tau function}
  \mcalF=\veps^{-2}f+\Delta\mcalF,\quad
  \Delta\mcalF=\sum_{k\geq 1}
  \veps^{k-2}\mcalF^{[k]},
\end{equation}
then the quasi-Miura transformation \eqref{quasi-Miura transf}  yields a deformation of the tau-cover \eqref{tau-cover} of the Principal Hierarchy of $M$ which has the form
\[\veps\frac{\p\mcalF}{\p t^{j,q}}=\mcalF_{j,q},\quad \veps\frac{\p\mcalF_{i,p}}{\p t^{j,q}}=\hat{\Omega}_{i,p;j,q},\quad \frac{\p w^\al}{\p t^{j,q}}=\eta^{\al\gamma}\p_x\hat{\Omega}_{\gamma,0;j,q},\]
here the functions $\hat{\Omega}_{i,p;j,q}=\hat{\Omega}_{i,p;j,q}(w;w_x,\dots)$ are given by
\[\hat{\Omega}_{i,p;j,q}=\left.\left(\Omega_{i,p;j,q}(v)+\veps^2\frac{\p^2\Delta\mcalF}{\p t^{i,p}\p t^{j,q}}\right)\right|_{v\to v(w,w_x,\dots)}.\]
This deformation of the tau-cover of the Principal Hierarchy also possesses the following Virasoro symmetries:
\begin{align*}
  \pfrac{\mcalF}{s}
&=
  \veps^{-2}
  \pfrac{f}{s}
 +\sum_{k\geq 1}\veps^{k-2}
  \sum_{r=0}^{m_k}
    \pfrac{\mcalF^{[k]}}{v^{\beta,r}}
    \p_x^r\pfrac{v^\beta}{s},
\\
  \pfrac{\mcalF_{i,p}}{s}
&=
 \veps \pp{t^{i,p}}
  \pfrac{\mcalF}{s},
\\
  \pfrac{w^\afa}{s}
&=\veps\eta^{\afa\beta}\p_x\frac{
  \p\mcalF_{\beta,0}}{\p s},
\end{align*}
where $\pp{s}=\sum_{m\geq -1}\frac{1}{\lmd^{m+2}}\pp{s_m}$.

We say that the quasi-Miura transformation
\eqref{quasi-Miura transf} linearizes the Virasoro symmetries of the Principal Hierarchy of $M$ if the actions of these symmetries on the
tau function
\begin{equation}\label{zh-51}
\tau=\exp{(\mcalF)}
\end{equation}
of the deformed Principal Hierarchy \eqref{deformed PH} are given by
\begin{equation}\label{Linearized Virasoro sym}
\pfrac{\tau}{s_m}
=
  L_m(\veps^{-1}\bft, \veps\pp{\bft})\tau,\quad m\ge -1,
\end{equation}
where the Virasoro operators $L_m(\veps^{-1}\bft, \veps\pp{\bft})$ are introduced in Definiton \ref{zh-23-08-1}.
We note that the linearization condition
\eqref{Linearized Virasoro sym} can be rewritten as
\begin{align}
 & \pfrac{\mcalF}{s}
=
  \veps^2
  \sum_{(i,p),(j,q)\in\mcalI}
  a^{i,p;j,q}(\lmd)
  \left(
    \frac{\p^2\mcalF}{\p t^{i,p}\p t^{j,q}}
   +\pfrac{\mcalF}{t^{i,p}}
    \pfrac{\mcalF}{t^{j,q}}
  \right)\notag
\\
&\quad
  +\sum_{(i,p),(j,q)\in\mcalI}
   b^{i,p}_{j,q}(\lmd)t^{j,q}
   \pfrac{\mcalF}{t^{i,p}}
  +
   \veps^{-2}\sum_{(i,p),(j,q)\in\mcalI}
   c_{i,p;j,q}(\lmd)t^{i,p}t^{j,q}\notag
\\
&\quad
  +
   \veps^{-2}
   \sum_{p,q\in\bbZ}
     C_{p,q}(\lmd)t^{0,p}t^{0,q}
  +
  \frac{1}{4\lmd^2}
  \mathrm{tr}\left(
    \frac14-\mu^2
  \right),
\end{align}
which leads to the following lemma.
\begin{lem}\label{Loop eqn lemma1}
The quasi-Miura transformation \eqref{quasi-Miura transf}
linearizes the Virasoro symmetries of the Principal Hierarchy of $M$ if and only if
the function $\Delta\mcalF=\sum_{k\geq 1}\veps^{k-2}\mcalF^{[k]}$
defined on the jet space of $M$ satisfies the equation
\begin{align}
  &\pfrac{\Delta\mcalF}{s}
  =
    \mcalD(\Delta\mcalF)
  +
    \sum_{(i,p),(j,q)\in\mcalI}
      a^{i,p;j,q}(\lmd)\frac{\p^2 f}{\p t^{i,p}\p t^{j,q}}\notag\\
  &\quad
   +\veps^2\sum_{(i,p),(j,q)\in\mcalI}
    a^{i,p;j,q}(\lmd)
    \left(
      \frac{\p^2\Delta\mcalF}{\p t^{i,p}\p t^{j,q}}
     +\pfrac{\Delta\mcalF}{t^{i,p}}
      \pfrac{\Delta\mcalF}{t^{j,q}}
    \right)
   +\frac{1}{4\lmd^2}
    \mathrm{tr}\left(
      \frac14-\mu^2\right),
\end{align}
where the linear operator $\mcalD$ is defined by
\begin{equation} \label{defn:operator mcalD}
\mcalD=
  \sum_{(i,p),(j,q)\in\mcalI}
    2a^{i,p;j,q}(\lmd)
    \pfrac{f}{t^{i,p}}
    \pfrac{}{t^{j,q}}
   +\sum_{(i,p),(j,q)\in\mcalI}
    b^{j,q}_{i,p}(\lmd)t^{i,p}
    \pfrac{}{t^{j,q}},
\end{equation}
and
\[\pfrac{\Delta\mcalF}{s}=\sum_{r\geq 0}
\pfrac{\Delta\mcalF}{v^{\gamma,r}}\p_x^r\pfrac{v^\gamma}{s}.\]
\end{lem}

The following lemma gives a more explicit formula for the linear operator $\mcalD$.

\begin{lem}\label{Loop eqn lemma2}
Let $\Mtil$ be the $(n+2)$-dimensional Frobenius manifold associated with $M$,
$\tilde p_i^{(\nu)}$ are defined by the Laplace-type integral \eqref{n+2 twisted periods-2},
and $(\Gtil^{ij}(\nu))$ is given by \eqref{n+2 Gram}.
Then for any function $\Delta\mcalF$ defined on the jet space of $M$ we have
\begin{align}
&\mcalD(\Delta\mcalF)
=
  \pfrac{\Delta\mcalF}{s}
 +
  \sum_{\ell\geq 0}
    \pfrac{\Delta\mcalF}{v^{\gamma,\ell}}
    \p_x^\ell
    \left(
      \frac{1}{E-\lmd e}
    \right)^\gamma
   -\frac{1}{\lmd}
    \sum_{\ell\geq 1}
    \pfrac{\Delta\mcalF}{v^{\gamma,\ell}}\ell\p_x^\ell e^\gamma\notag
\\
&\quad
+
  \sum_{\ell\geq 1}
   \pfrac{\Delta\mcalF}{v^{\gamma,\ell}}
   \sum_{k=1}^{\ell}
   \binom{\ell}{k}
   \lim_{\nu\to 0}
   \left(
   (\p_x^{k-1}\p_0\ptil_i^{(\nu)})\Gtil^{ij}(\nu)
   \left(
     \p_x^{\ell+1-k}\p^\gamma\ptil_j^{(-\nu)}
   \right)\right)\Big|_{M},\label{lemma:mcalD(Delta mcaF)}
\end{align}
here $\p_0$ is the unit vector field $\frac{\p}{\p v^0}$ on $\Mtil$.
\end{lem}

\begin{proof}
Let $\Stil^{(\nu)}_i=\Stil^{(\nu)}_i(\bft,\pfrac{f}{\bft};\lmd)$
be defined in \eqref{defn:Stil_i}.
From the relation \eqref{formal Laplace bilinear identity} between Laplace-type integrals and the extended Virasoro coefficients it follows that
\begin{align}
&  \mcalD(\Delta\mcalF)
=
  \lim_{\nu\to 0}
  \left[
    \pfrac{\Stil_i^{(\nu)}}{\lmd}
    \Gtil^{ij}(\nu)
    \left(
      \int_{0}^{\infty}
        \frac{\td z}{\sqrt{z}}\rme^{-\lmd z}
        \sum_{p\geq 0}
        \left(
          \pfrac{\Delta\mcalF}{t^{0,p}},
          \pfrac{\Delta\mcalF}{\bm{t}^p}, 0
        \right)
    z^{p+1}
    z^{\mutil-\nu}z^{\Rtil}
    \right)_j
  \right]_-\notag\\
&\quad+\lim_{\nu\to 0}
  \left[
    \pfrac{\Stil_i^{(\nu)}}{\lmd}
    \Gtil^{ij}(\nu)
    \left(
      \int_{0}^{\infty}
        \frac{\td z}{\sqrt{z}}\rme^{-\lmd z}
        \sum_{p\geq 0}
        \left(
          \pfrac{\Delta\mcalF}{t^{0,-p-1}},
          \bm 0, 0
        \right)
    z^{-p}
    z^{\mutil-\nu}z^{\Rtil}
    \right)_j
  \right]_-.\label{DDeltaF pre}
\end{align}
To simplify the presentation of the proof of this lemma, we only consider the non-resonant case.
In this case we have the limits given in \eqref{zh-40}.
The proof of the general case is similar.
By using \eqref{qu-220801}--\eqref{zh-21} and \eqref{DxDlmdStil},
we can represent the first term of the right-hand side of \eqref{DDeltaF pre} as follows:
\begin{align*}
&
  \left[
    \pfrac{\Stil_i}{\lmd}
    \Gtil^{ij}
    \left(
      \int_{0}^{\infty}
        \frac{\td z}{\sqrt{z}}\rme^{-\lmd z}
        \sum_{p\geq 0}
        \left(
          \pfrac{\Delta\mcalF}{t^{0,p}},
          \pfrac{\Delta\mcalF}{\bm{t}^p}, 0
        \right)
    z^{p+1}
    z^{\mutil}z^{\Rtil}
    \right)_j
  \right]_-
\\
&\quad=
  \sum_{s\geq 0}
  \pfrac{\Delta\mcalF}{v^{\gamma,s}}
  \left[
    \pfrac{\Stil_i}{\lmd}
    \Gtil^{ij}
    \left(
      \int_{0}^{+\infty}
        \frac{\td z}{\sqrt{z}}\rme^{-\lmd z}
        \sum_{p\geq 0}
        \p_x^s
        \left(
          \pfrac{v^\gamma}{t^{0,p}},
          \pfrac{v^\gamma}{\bm{t}^p}, 0
        \right)z^{p+1}
      z^{\mutil}z^{\Rtil}
    \right)_j
  \right]_-\\
&\quad=
  \sum_{s\geq 0}
  \pfrac{\Delta\mcalF}{v^{\gamma,s}}
  \left[
    \pfrac{\Stil_i}{\lmd}
    \Gtil^{ij}
    \left(
      \int_{0}^{+\infty}
        \frac{\td z}{\sqrt{z}}\rme^{-\lmd z}
        \p_x^{s+1}
        \left(
          \p^\gamma\theta_0(z)-e^\gamma,
          \p^\gamma\theta_\bullet(z), 0
        \right)
        z^{\mutil}z^{\Rtil}
    \right)_j
  \right]_-\\
&\quad=
  \sum_{s\geq 0}
  \pfrac{\Delta\mcalF}{v^{\gamma,s}}
  \left[
    \pfrac{\Stil_i}{\lmd}
    \Gtil^{ij}
    \p_x^{s+1}
    \left(
      \p^\gamma\ptil_j
    \right)\Big|_{M}
  \right]_-\\
&\quad=
  \sum_{s\geq 0}
  \pfrac{\Delta\mcalF}{v^{\gamma,s}}
  \p_x^s
  \left[
    \pfrac{\Stil_i}{\lmd}
    \Gtil^{ij}
    \p_x
    \left(
      \p^\gamma\ptil_j
    \right)\Big|_{M}
  \right]_-\\
  &\quad\quad-\sum_{s\geq 1}
  \pfrac{\Delta\mcalF}{v^{\gamma,s}}
  \sum_{k=1}^{s}
 \binom{s}{k}
  \left[
    \left(
    \p_x^k
    \pfrac{\Stil_i}{\lmd}
    \right)\Gtil^{ij}
    \p_x^{s+1-k}
    \left(
      \p^\gamma\ptil_j
    \right)\Big|_{M}
  \right]_-
\\
&\quad=
  \sum_{s\geq 0}
  \pfrac{\Delta\mcalF}{v^{\gamma,s}}
  \p_x^s
  \left[
    \pfrac{\Stil_i}{\lmd}
    \Gtil^{ij}
    \p_x
    \left(
      \p^\gamma\ptil_j
    \right)\Big|_{M}
  \right]_-\\
&\quad\quad+ \sum_{s\geq 1}
  \pfrac{\Delta\mcalF}{v^{\gamma,s}}
  \sum_{k=1}^{s}
 \binom{s}{k}
  \left(
    \left(
    \p_x^{k-1}
    \p_0\ptil_i
    \right)\Gtil^{ij}
    \left(
    \p_x^{s+1-k}
      \p^\gamma\ptil_j
    \right)
    \right)\Big|_{M}.
\end{align*}
Similarly, by using \eqref{defn:Stil_i} and \eqref{defn:xi(nu)(lmd)} we can represent
the second term of the right-hand side of \eqref{DDeltaF pre}
in the form
\begin{align*}
&
\left[
    \pfrac{\Stil_i}{\lmd}
    \Gtil^{ij}
    \left(
      \int_{0}^{\infty}
        \frac{\rme^{-\lmd z}}{\sqrt{z}}
        \sum_{p\geq 0}
        \left(
          \pfrac{\Delta\mcalF}{t^{0,-p-1}};
          \bm 0; 0
        \right)
    z^{-p}
    z^{\mutil}z^{\Rtil}
    \right)_j
  \right]_-\\
&\quad=
\left[
    \pfrac{\Stil_{n+1}}{\lmd}
    \Gtil^{n+1,0}
      \int_{0}^{\infty}
        \rme^{-\lmd z}
        \sum_{p\geq 0}
          \pfrac{\Delta\mcalF}{t^{0,-p-1}}
    z^{-p-\frac d2-\frac12}\td z
  \right]_-\\
&\quad=
\sum_{s\geq 0}
  \pfrac{\Delta\mcalF}{v^{\gamma,s}}
  \left[
    \pfrac{\Stil_{n+1}}{\lmd}
    \Gtil^{n+1,0}
    \int_{0}^{\infty} \rme^{-\lmd z}
    \p_x^s
    \left(
      \p_x\p^\gamma\theta_{0,-p}
    \right)z^{-p-\frac d2-\frac12}\td z
  \right]_-\\
&\quad=
  \sum_{s\geq 0}
  \pfrac{\Delta\mcalF}{v^{\gamma,s}}
  \p_x^s
  \left[
    \pfrac{\Stil_{n+1}}{\lmd}
    \Gtil^{n+1,0}
    \p_x\p^\gamma
    \xi(\lmd)
  \right]_-\\
&\quad\quad
-
  \sum_{s\geq 1}
  \pfrac{\Delta\mcalF}{v^{\gamma,s}}
  \sum_{k=1}^{s}
 \binom{s}{k}
  \left[
    \left(
      \p_x^k
      \pfrac{\Stil_{n+1}}{\lmd}
      \Gtil^{n+1,0}
      \left(
        \p_x^{s+1-k}\p^\gamma\xi(\lmd)
      \right)
    \right)
  \right]_-
\\
&\quad=
\sum_{s\geq 0}
  \pfrac{\Delta\mcalF}{v^{\gamma,s}}
  \p_x^s
  \left[
    \pfrac{\Stil_{n+1}}{\lmd}
    \Gtil^{n+1,0}
    \p_x\p^\gamma
    \xi(\lmd)
  \right]_-\\
&\quad\quad
-
  \sum_{s\geq 1}
  \pfrac{\Delta\mcalF}{v^{\gamma,s}}
 \binom{s}{1}
  \left[
   -\lmd^{-\frac d2-\frac12}
   \Gamma\left(\frac d2+\frac12\right)
   \left(-\frac{1}{\pi}\right)\cos\frac{\pi d}{2}\cdot
   \p_x^s\p^\gamma\xi(\lmd)\right]_-
\\
&\quad=
  \sum_{s\geq 0}
  \pfrac{\Delta\mcalF}{v^{\gamma,s}}
  \p_x^s
  \left[
    \pfrac{\Stil_{n+1}}{\lmd}
    \Gtil^{n+1,0}
    \p_x\p^\gamma
    \xi(\lmd)
  \right]_-
-\frac 1\lmd\sum_{s\geq 1}
\pfrac{\Delta\mcalF}{v^{\gamma,s}}s\p_x^se^\gamma.
\end{align*}
Hence by using \eqref{Vir sym on tau-cover 3} we arrive at
\begin{align*}
&\mcalD(\Delta\mcalF)\\
=&\,
  \sum_{s\geq 0}
  \pfrac{\Delta\mcalF}{v^{\gamma,s}}
  \p_x^s
  \left[
    \pfrac{\Stil_i}{\lmd}
    \Gtil^{ij}
    \p_x
    \left(
      \p^\gamma\ptil_j
    \right)\Big|_{M}
  \right]_-
+
\sum_{s\geq 0}
  \pfrac{\Delta\mcalF}{v^{\gamma,s}}
  \p_x^s
  \left[
    \pfrac{\Stil_{n+1}}{\lmd}
    \Gtil^{n+1,0}
    \p_x\p^\gamma
    \xi(\lmd)
  \right]_-\\
&\,
+ \sum_{s\geq 1}
  \pfrac{\Delta\mcalF}{v^{\gamma,s}}
  \sum_{k=1}^{s}
 \binom{s}{k}
  \left(
    \left(
    \p_x^{k-1}
    \p_0\ptil_i
    \right)\Gtil^{ij}
    \left(
    \p_x^{s+1-k}
      \p^\gamma\ptil_j
    \right)
    \right)\Big|_{M}
-\frac 1\lmd\sum_{s\geq 1}
\pfrac{\Delta\mcalF}{v^{\gamma,s}}s\p_x^se^\gamma\\
=&\,
  \sum_{s\geq 0}
  \pfrac{\Delta\mcalF}{v^{\gamma,s}}
  \p_x^s
  \left[
  \pfrac{v^\gamma}{s}
  +
  \left(
    \frac{1}{E-\lmd e}
  \right)^\gamma
  \right]
-\frac 1\lmd\sum_{s\geq 1}
\pfrac{\Delta\mcalF}{v^{\gamma,s}}s\p_x^se^\gamma\\
&\,
+
\sum_{s\geq 1}
  \pfrac{\Delta\mcalF}{v^{\gamma,s}}
  \sum_{k=1}^{s}
 \binom{s}{k}
  \left(
    \left(
    \p_x^{k-1}
    \p_0\ptil_i
    \right)\Gtil^{ij}
    \left(
    \p_x^{s+1-k}
      \p^\gamma\ptil_j
    \right)
    \right)\Big|_{M}\\
=&\,
  \pfrac{\Delta\mcalF}{s}
 +
  \sum_{s\geq 0}
    \pfrac{\Delta\mcalF}{v^{\gamma,s}}
    \p_x^s
    \left(
      \frac{1}{E-\lmd e}
    \right)^\gamma
   -\frac{1}{\lmd}
    \sum_{s\geq 1}
    \pfrac{\Delta\mcalF}{v^{\gamma,s}}s\p_x^se^\gamma
\\
&\,
+
  \sum_{s\geq 1}
   \pfrac{\Delta\mcalF}{v^{\gamma,s}}
   \sum_{k=1}^{s}
   \binom{s}{k}
   \left(
   (\p_x^{k-1}\p_0\ptil_i)\Gtil^{ij}
   \left(
     \p_x^{s+1-k}\p^\gamma\ptil_j
   \right)\right)\Big|_{M}.
\end{align*}
The lemma is proved.
\end{proof}

We introduce, following \cite{normal-form}, the star product map $*$ as follows:
\begin{equation}\label{defn: star product}
\theta_{i,p} * \theta_{j,q}=\Omg_{i,p;j,q},\quad
1 * \theta_{i,p} = \theta_{i,p} * 1 =0.
\end{equation}
Since the regularized periods
$p_\afa^{(\nu)}(\lmd)$ defined in \eqref{periods-Lapacle-2}
are series of $\lmd$ with coefficients depending linearly on $\{\theta_{i,p}\}$,
the star products $\pfrac{p^{(\nu)}_\afa}{\lmd} * \pfrac{p^{(-\nu)}_\beta}{\lmd}$
are well defined.

\begin{thm}\label{thm: loop-eqn n+2 version}
The linearization condition \eqref{Linearized Virasoro sym} holds true
if and only if $\Delta\mcalF$ satisfies the
differential equation
\begin{align}
&
    \sum_{s\geq 0}
    \pfrac{\Delta\mcalF}{v^{\gamma,s}}
    \p_x^s
    \left(
      \frac{1}{E-\lmd e}
    \right)^\gamma
   -\frac{1}{\lmd}
    \sum_{s\geq 1}
    \pfrac{\Delta\mcalF}{v^{\gamma,s}}s\p_x^se^\gamma
\notag\\
&\quad\quad
+
  \sum_{s\geq 1}
   \pfrac{\Delta\mcalF}{v^{\gamma,s}}
   \sum_{k=1}^{s}
\binom{s}{k}
   \left(
   \left(\p_x^{k-1}\p_0\ptil_i
   \right)\Gtil^{ij}
   \left(
     \p_x^{s+1-k}\p^\gamma\ptil_j
   \right)
   \right)\Big|_{M}
\notag\\
&\quad=
 \frac{\veps^2}{2}
 \sum_{k,\ell\geq 0}
 \left(
   \pfrac{\Delta\mcalF}{v^{\afa,k}}
   \pfrac{\Delta\mcalF}{v^{\beta,\ell}}
  +\frac{\p^2\Delta\mcalF}{\p v^{\afa,k}\p v^{\beta,\ell}}
 \right)
 (\p_x^{k+1}\p^\afa p_\sigma)
 G^{\sigma\rho}
 (\p_x^{\ell+1}\p^\beta p_\rho)\notag
\\
&\quad\quad
 +\frac{\veps^2}{2}
  \sum_{k\geq 0}
  \pfrac{\Delta\mcalF}{v^{\afa,k}}
  \p_x^{k+1}
  \left[
    \nabla\pfrac{p_\sigma}{\lmd}
    \cdot
    \nabla\pfrac{p_\rho}{\lmd}
    \cdot v_x
  \right]^\afa
  G^{\sigma\rho}
\notag
\\
&\quad\quad
+
 \frac12 G^{\afa\beta}
 \pfrac{p_\afa}{\lmd}*\pfrac{p_\beta}{\lmd}-
 \frac{1}{4\lmd^2}
 \mathrm{tr}
 \left(
   \frac14-\mu^2
 \right),
\label{loop equation 1}
\end{align}
which is called the loop equation of $M$.
Here the star product map $*$ is defined as in \eqref{defn: star product},
$\ptil_0,\ptil_1,\dots,\ptil_{n+1}$ is any basis of periods of $\Mtil$ with Gram matrix $\Gtil=(\Gtil^{ij})$,
and $p_1,\dots, p_n$ is any basis of periods of $M$ with Gram matrix $G=(G^{\afa\beta})$.
\end{thm}

Let us note that the equation \eqref{loop equation 1}
is independent of the choice of the periods
$\{p_\afa\}$ and $\{\ptil_i\}$,
so in the non-resonant case we can choose
$p_\afa$ and $\ptil_i$ as in
\eqref{periods-Lapacle}, \eqref{n+2 twisted periods} respectively.

\begin{proof}
From Lemma \ref{Loop eqn lemma1} and Lemma \ref{Loop eqn lemma2} we know that
the linearization condition \eqref{Linearized Virasoro sym} is equivalent to the following equation:
\begin{align*}
0=&\,
\sum_{s\geq 0}
    \pfrac{\Delta\mcalF}{v^{\gamma,s}}
    \p_x^s
    \left(
      \frac{1}{E-\lmd e}
    \right)^\gamma
   -\frac{1}{\lmd}
    \sum_{s\geq 1}
    \pfrac{\Delta\mcalF}{v^{\gamma,s}}s\p_x^se^\gamma
\\
&
+
  \sum_{s\geq 1}
   \pfrac{\Delta\mcalF}{v^{\gamma,s}}
   \sum_{k=1}^{s}
  \binom{s}{k}
  \lim_{\nu\to 0}
   \left(
   (\p_x^{k-1}\p_0\ptil_i^{(\nu)})\Gtil^{ij}(\nu)
   \left(
     \p_x^{s+1-k}\p^\gamma\ptil_j^{(-\nu)}
   \right)\right)\Big|_{M}
\\
&
  +\veps^2
  \sum_{(i,p),(j,q)\in\mcalI}
  a^{i,p;j,q}(\lmd)
  \left(
    \pfrac{\Delta\mcalF}{t^{i,p}}
    \pfrac{\Delta\mcalF}{t^{j,q}}
   +
    \frac{\p^2\Delta\mcalF}{\p t^{i,p}\p t^{j,q}}
  \right)\\
&
  +\sum_{(i,p),(j,q)\in\mcalI}
    a^{i,p;j,q}(\lmd)
    \frac{\p^2 f}{\p t^{i,p}\p t^{j,q}}
  +\frac1{4\lmd^2}
  \mathrm{tr}
   \left(
     \frac14-\mu^2
   \right).
\end{align*}
From \eqref{formal Laplace bilinear identity} it follows that
\begin{align*}
&
  \veps^2
  \sum_{(i,p),(j,q)\in\mcalI}
  a^{i,p;j,q}(\lmd)
  \pfrac{\Delta\mcalF}{t^{i,p}}
  \pfrac{\Delta\mcalF}{t^{j,q}}\\
&\quad =
  -\frac{\veps^2}{2}
  \lim_{\nu\to 0}
  \left[
    \left(
    \int_{0}^{\infty}
    \frac{\td z}{\sqrt{z}}\rme^{-\lmd z}
    \sum_{p\geq 0}
    \left(
      \pfrac{\Delta\mcalF}{t^{0,p}},
      \pfrac{\Delta\mcalF}{\bm{t}^p}, 0
    \right)
    z^{p+1}z^{\mutil+\nu}z^{\Rtil}
    \right)_i\Gtil^{ij}(\nu)
  \right.\\
&\quad\quad
  \times
  \left.
  \left(
    \int_{0}^{\infty}
      \frac{\td z}{\sqrt{z}}\rme^{-\lmd z}
      \sum_{q\geq 0}
      \left(
        \pfrac{\Delta\mcalF}{t^{0,q}},
        \pfrac{\Delta\mcalF}{\bm{t}^q}, 0
      \right)z^{q+1}z^{\mutil-\nu}z^{\Rtil}
  \right)_j
  \right]_-\\
&\quad=
  -\frac{\veps^2}{2}
  \sum_{k,\ell\geq 0}
  \pfrac{\Delta\mcalF}{v^{\afa,k}}
  \pfrac{\Delta\mcalF}{v^{\beta,\ell}}
  \lim_{\nu\to 0}
  \left(
  \left(
    \p_x^{k+1}
    \p^\afa\ptil_i^{(\nu)}
  \right)
  \Gtil^{ij}(\nu)
  \left(
    \p_x^{\ell+1}
    \p^\beta\ptil_j^{(-\nu)}
  \right)
  \right)\Big|_{M}\\
&\quad=
  -\frac{\veps^2}{2}
  \sum_{k,\ell\geq 0}
  \pfrac{\Delta\mcalF}{v^{\afa,k}}
  \pfrac{\Delta\mcalF}{v^{\beta,\ell}}
  \lim_{\nu\to 0}
  \left(
    \p_x^{k+1}
    \p^\afa p_{\sigma}^{(\nu)}
  \right)
  G^{\sigma\rho}(\nu)
  \left(
    \p_x^{\ell+1}
    \p^\beta p_\rho^{(-\nu)}
  \right).
\end{align*}
In a similar way, we can also verify that
\begin{align*}
&\veps^2
\sum_{(i,p),(j,q)\in\mcalI}
a^{i,p;j,q}(\lmd)
\frac{\p^2\Delta\mcalF}{\p t^{i,p}\p t^{j,q}}\\
&\quad=
 -\frac{\veps^2}{2}
 \sum_{k,\ell\geq 0}
 \frac{\p^2\Delta\mcalF}{\p v^{\afa,k}\p v^{\beta,\ell}}
 \lim_{\nu\to 0}
 \left(
   \p_x^{k+1}\p^\afa p_\sigma^{(\nu)}
 \right)
 G^{\sigma\rho}(\nu)
 \left(
   \p_x^{\ell+1}
   \p^\beta p_\rho^{(-\nu)}
 \right)\\
&\quad\quad
 -\frac{\veps^2}{2}
  \sum_{k\geq 0}
  \pfrac{\Delta\mcalF}{v^{\afa,k}}
  \lim_{\nu\to 0}
  \p_x^{k+1}
  \left[
    \nabla\pfrac{p_\sigma^{(\nu)}}{\lmd}
    \cdot
    \nabla\pfrac{p_\rho^{(-\nu)}}{\lmd}
    \cdot v_x
  \right]^\afa
  G^{\sigma\rho}(\nu),
  \end{align*}
 and
 \begin{align*}
&\sum_{(i,p),(j,q)\in\mcalI}
  a^{i,p;j,q}(\lmd)
  \frac{\p^2 f}{\p t^{i,p}\p t^{j,q}}
=
  \sum_{(i,p),(j,q)\in\mcalI}
  a^{i,p;j,q}\theta_{i,p} * \theta_{j,q} \\
&\qquad =
  -\frac12
  \lim_{\nu\to 0}
  G^{\afa\beta}(\nu)
  \pfrac{p_\afa^{(\nu)}}{\lmd}
  *\pfrac{p_\beta^{(-\nu)}}{\lmd}.
\end{align*}
Therefore the linearization condition \eqref{Linearized Virasoro sym}
is equivalent to the following equation for $\Delta\mcalF$:
\begin{align}
&
    \sum_{s\geq 0}
    \pfrac{\Delta\mcalF}{v^{\gamma,s}}
    \p_x^s
    \left(
      \frac{1}{E-\lmd e}
    \right)^\gamma
   -\frac{1}{\lmd}
    \sum_{s\geq 1}
    \pfrac{\Delta\mcalF}{v^{\gamma,s}}s\p_x^se^\gamma
\notag\\
&\quad\quad
+
  \sum_{s\geq 1}
   \pfrac{\Delta\mcalF}{v^{\gamma,s}}
   \sum_{k=1}^{s}
\binom{s}{k}
\lim_{\nu\to 0}
   \left(
   \left(\p_x^{k-1}\p_0\ptil_i^{(\nu)}
   \right)\Gtil^{ij}(\nu)
   \left(
     \p_x^{s+1-k}\p^\gamma\ptil_j^{(-\nu)}
   \right)
   \right)\Big|_{M}
\notag\\
&\quad=
 \frac{\veps^2}{2}
 \sum_{k,\ell\geq 0}
 \left(
   \pfrac{\Delta\mcalF}{v^{\afa,k}}
   \pfrac{\Delta\mcalF}{v^{\beta,\ell}}
  +\frac{\p^2\Delta\mcalF}{\p v^{\afa,k}\p v^{\beta,\ell}}
 \right)
 \lim_{\nu\to 0}
 (\p_x^{k+1}\p^\afa p_\sigma^{(\nu)})
 G^{\sigma\rho}(\nu)
 (\p_x^{\ell+1}\p^\beta p_\rho^{(-\nu)})\notag
\\
&\quad\quad
 +\frac{\veps^2}{2}
  \sum_{k\geq 0}
  \pfrac{\Delta\mcalF}{v^{\afa,k}}
  \p_x^{k+1}
  \lim_{\nu\to 0}
  \left[
    \nabla\pfrac{p_\sigma^{(\nu)}}{\lmd}
    \cdot
    \nabla\pfrac{p_\rho^{(-\nu)}}{\lmd}
    \cdot v_x
  \right]^\afa
  G^{\sigma\rho}(\nu)
\notag
\\
&\quad\quad
+
 \frac12 \lim_{\nu\to 0}G^{\afa\beta}(\nu)
 \pfrac{p_\afa^{(\nu)}}{\lmd}*\pfrac{p_\beta^{(-\nu)}}{\lmd}-
 \frac{1}{4\lmd^2}
 \mathrm{tr}
 \left(
   \frac14-\mu^2
 \right).
\end{align}
Thus, in order to prove the theorem it suffices to verify that the identities
\begin{align}
    \lim_{\nu\to 0}
   \left(\p_x^{k-1}\p_0\ptil_i^{(\nu)}
   \right)\Gtil^{ij}(\nu)
   \left(
     \p_x^{s+1-k}\p^\gamma\ptil_j^{(-\nu)}
   \right)
   &=
   \left(\p_x^{k-1}\p_0\ptil_i
   \right)\Gtil^{ij}
   \left(
     \p_x^{s+1-k}\p^\gamma\ptil_j
   \right),
\label{limit eqn 1}\\
  \lim_{\nu\to 0}
 (\p_x^{k+1}\p^\afa p_\sigma^{(\nu)})
 G^{\sigma\rho}(\nu)
 (\p_x^{\ell+1}\p^\beta p_\rho^{(-\nu)})
&=
 (\p_x^{k+1}\p^\afa p_\sigma)
 G^{\sigma\rho}
 (\p_x^{\ell+1}\p^\beta p_\rho),
\label{limit eqn 2}\\
\lim_{\nu\to 0}
  \left[
    \nabla\pfrac{p_\sigma^{(\nu)}}{\lmd}
    \cdot
    \nabla\pfrac{p_\rho^{(-\nu)}}{\lmd}
    \cdot v_x
  \right]^\afa
  G^{\sigma\rho}(\nu)
&=
  \left[
    \nabla\pfrac{p_\sigma}{\lmd}
    \cdot
    \nabla\pfrac{p_\rho}{\lmd}
    \cdot v_x
  \right]^\afa
  G^{\sigma\rho},
\label{limit eqn 3}\\
\lim_{\nu\to 0}G^{\afa\beta}(\nu)
 \pfrac{p_\afa^{(\nu)}}{\lmd}*\pfrac{p_\beta^{(-\nu)}}{\lmd}
&=
  G^{\afa\beta}
 \pfrac{p_\afa}{\lmd}*\pfrac{p_\beta}{\lmd}
\label{limit eqn 4}
  \end{align}
hold true for any basis of periods $p_1(v;\lmd),\dots, p_n(v;\lmd)$ of $M$ with Gram matrix $G=(G^{\afa\beta})$,
and for any basis of periods $\tilde p_0(v;\lmd),\dots, \tilde p_{n+1}(v;\lmd)$ of $\widetilde{M}$ with Gram matrix $\widetilde{G}=(\widetilde{G}^{ij})$.
In fact, from the definition of the functions $p_\al^{[\nu]}$ and their properties given in \eqref{240818-2116}--\eqref{G-H relation}, it follows that the left-hand side of \eqref{limit eqn 2} can be written as
\begin{align*}
  &
  \lim_{\nu\to 0}
  \left(
    \p_x^{k+1}\p^\afa p_\sigma^{(\nu)}
  \right) G^{\sigma\rho}(\nu)
  \left(
    \p_x^{\ell+1}\p^\beta p_\rho^{(-\nu)}
  \right) \\
&\quad =
  \lim_{\nu\to 0}
  \left(
    \p_x^{k+1}\p^\afa p_\sigma^{(\nu)}
  \right)
    X^\sigma_\xi(\nu)
    H^{\xi\zeta}(\nu)
    X^\rho_\zeta(-\nu)
  \left(
    \p_x^{\ell+1}\p^\beta p_\rho^{(-\nu)}
  \right)
\\
&\quad =
  \lim_{\nu\to 0}
  \left[
    \p_x^{k+1}\left(X^\sigma_\xi(\nu)\p^\afa p_\sigma^{(\nu)}\right)
  \right]
    H^{\xi\zeta}(\nu)
  \left[
    \p_x^{\ell+1}\left(X^\rho_\zeta(-\nu)\p^\beta p_\rho^{(-\nu)}\right)
  \right]
\\
&\quad =
  \lim_{\nu\to 0}
  \left(
    \p_x^{k+1}\p^\afa p^{[\nu]}_\xi
  \right)
    H^{\xi\zeta}(\nu)
  \left(
    \p_x^{\ell+1}\p^\beta p^{[-\nu]}_\zeta
  \right) \\
&\quad =
  \left(
    \p_x^{k+1}\p^\afa p_\sigma
  \right) G^{\sigma\rho}
  \left(
    \p_x^{\ell+1}\p^\beta p_\rho
  \right),
\end{align*}
therefore the identity \eqref{limit eqn 2} holds true. The validity of \eqref{limit eqn 1}, \eqref{limit eqn 3} can be verified in a similar way by
introducing $\tilde p_\afa^{[\nu]}$, $\widetilde{X}(\nu)$ and $\widetilde{H}(\nu)$ for the associated $(n+2)$-dimensional Frobenius manifold $\widetilde{M}$, which are analogues of $p_\afa^{[\nu]}$, ${X}(\nu)$ and ${H}(\nu)$.

Finally,  let us verify the validity of \eqref{limit eqn 4}. From \eqref{X(nu)} we obtain
\[
  \p^\gamma \left(
    p_\afa^{[\nu]} - p_\beta^{(\nu)}X_\afa^\beta(\nu)
  \right)=0, \quad 1\leq\afa,\beta,\gamma\leq n,
\]
thus there exist functions $\{w_\afa(\lmd;\nu)\}$, which are independent of $v^\afa$, such that
\begin{equation}\label{psi(lmd nu)}
  \left(p_1^{[\nu]},...,p_n^{[\nu]}\right)
 =
  \left(p_1^{(\nu)},...,p_n^{(\nu)}\right)X(\nu)
 +\left(w_1,...,w_n\right).
\end{equation}
Then the left-hand side of \eqref{limit eqn 4} can be represented as
\begin{align*}
&
  \lim_{\nu\to 0}
  G^{\afa\beta}(\nu)
  \pfrac{p_\afa^{(\nu)}}{\lmd}
  *
  \pfrac{p_\beta^{(-\nu)}}{\lmd}
\\
&\quad=
  \lim_{\nu\to 0}
  H^{\afa\beta}(\nu)
  \pp{\lmd}
  \left(
    p_\afa^{[\nu]}-w_\afa(\lmd;\nu)
  \right)
  *
  \pp{\lmd}\left(
    p_\beta^{[-\nu]}-w_\beta(\lmd;-\nu)
  \right).
\end{align*}
Since $w_\afa(\lmd;\nu)$ are independent of $v^\afa$,
it follows from the definition of the star product given in \eqref{defn: star product} that
\[
  \pfrac{w_\afa(\lmd;\nu)}{\lmd}
  *
  \pfrac{p_\beta^{[-\nu]}}{\lmd}
=
  \pfrac{w_\afa(\lmd;\nu)}{\lmd}
  *
  \pfrac{w_\beta(\lmd;-\nu)}{\lmd}
= 0,
\]
therefore
\begin{align*}
&\,
  \lim_{\nu\to 0}
  G^{\afa\beta}(\nu)
  \pfrac{p_\afa^{(\nu)}}{\lmd}
  *
  \pfrac{p_\beta^{(-\nu)}}{\lmd}
=
  \lim_{\nu\to 0}
  H^{\afa\beta}(\nu)
  \pfrac{
    p_\afa^{[\nu]}
  }{\lmd}
  *
  \pfrac{
    p_\beta^{[-\nu]}
  }{\lmd}
=
  G^{\afa\beta}
  \pfrac{p_\afa}{\lmd}
  *
  \pfrac{p_\beta}{\lmd}.
\end{align*}
The theorem is proved.
\end{proof}

\section{Quasi-periods and the simplified loop equation}

The goal of this section
is to simplify the term
\[\left(
   \left((\p_x^{k-1}\p_0\ptil_i
   \right)\Gtil^{ij}
   \left(
     \p_x^{s+1-k}\p^\gamma\ptil_j
   \right)
   \right)\Big|_{M}\]
in the left-hand side of the loop equation
\eqref{loop equation 1}. To this end, we are to choose the periods of $\Mtil$ and of $M$ in a certain special way,
since the loop equation is independent of the choice of these periods.

\begin{prop}\label{regularized periods}
There exists a basis $p_1(v;\lmd),\dots,p_n(v;\lmd)$ of periods of $M$
which satisfy the quasi-homogeneous condition
\begin{align}
 & \left(
    \p_E+\lmd\pp{\lmd}
  \right)\nabla p_\afa
+
  \left(\mu+\frac12\right)\nabla p_\afa
=
0,
\label{GM eqn 3}
\\
&\left(
    \p_E+\lmd\pp{\lmd}
  \right)\pfrac{p_\afa}{\lmd}
+ \frac{1+d}{2}\pfrac{p_\afa}{\lmd}
=
0.
\label{GM eqn 4}
\end{align}
Such periods of $M$ are said to be quasi-homogeneous.
\end{prop}

\begin{proof}
Suppose $p(v;\lmd)$ is a solution of
the Gauss-Manin equations \eqref{GM eqn 1} and \eqref{GM eqn 2}, then
\begin{align*}
\p_E\nabla p
&=
  E^\afa\p_\afa\nabla p
=
  -\mcalU\left(\mcalU-\lmd I\right)^{-1}
  \left(
    \mu+\frac12
  \right)\nabla p\\
&=
  -\left(\mu+\frac12\right)\nabla p
  -\lmd\left(\mcalU-\lmd I\right)^{-1}
  \left(\mu+\frac12\right)\nabla p\\
&=
  -\lmd\pp{\lmd}\nabla p
  -\left(\mu+\frac12\right)\nabla p,
\end{align*}
thus the equation \eqref{GM eqn 3} holds true automatically for $p(v;\lmd)$. From this equation it follows, for $\al=1,\dots,n$, that
\begin{align*}
&
  \p^\afa
  \left[
    \left(
      \p_E+\lmd\pp{\lmd}
    \right)\pfrac{p}{\lmd}
  \right]
=
\left(
  \frac{2-d}{2}+\mu_\afa
\right)
\pp{\lmd}\p^\afa p
+
  \left(
    \p_E+\lmd\pp{\lmd}
  \right)\pp{\lmd}\p^\afa p\\
&\quad=
  \left(-\frac d2+\mu_\afa\right)
  \pp\lmd\p^\afa p
 +\pp\lmd\left[
   \left(
     \p_E+\lmd\pp\lmd
   \right)\p^\afa p
  \right]
=
  \p^\afa\left(
    -\frac{d+1}{2}
    \pfrac p\lmd
  \right),
\end{align*}
therefore, there exists a function $c=c(\lmd)$ such that
\[
  \left(
    \p_E+\lmd\pp\lmd
  \right)\pfrac p\lmd
 +\frac{1+d}{2}\pfrac p\lmd = c(\lmd).
\]
Let $y=y(\lmd)$ be a solution of the following ODE of Euler-type:
\[\lmd y'(\lmd)+\frac{1+d}{2}y(\lmd)=c(\lmd).\]
Consider the substitution
$p\mapsto \hat p:=p-\int y(\lmd)\td\lmd$,
then \eqref{GM eqn 4} holds true for $\hat p$ which
still satisfies the Gauss-Manin equations
\eqref{GM eqn 1} and \eqref{GM eqn 2}.
The proposition is proved.
\end{proof}

We note that solutions of the Gauss-Manin equations
\eqref{GM eqn 1}, \eqref{GM eqn 2} and the equation \eqref{GM eqn 4} have the following ambiguity
\begin{equation}\label{ambiguity of reg-period}
  p(v;\lmd)\mapsto
  p(v;\lmd)+a\lmd^{\frac{1-d}{2}}+b,
\end{equation}
where $a,b$ are any constants.
We also note that the periods given by the Laplace-type integrals
\eqref{periods-Lapacle} are not necessarily quasi-homogeneous.

\begin{rmk}\label{rmk: reg period with flat unity}
If $M$ has flat unity $e=\pp{v^1}$, then $\mu_1=-\frac d2$.
It is easy to show that any period $p(v;\lmd)$ of $M$ which
satisfies the condition $\pp\lmd p=-\p_ep$ is quasi-homogeneous.
Thus the quasi-homogeneous  conditions
  \eqref{GM eqn 3} and \eqref{GM eqn 4}
  generalize the condition $\pp\lmd p=-\p_ep$ for periods of a usual Frobenius manifold.
\end{rmk}

\begin{rmk}
If a period $p(v;\lmd)$ of $M$ is quasi-homogeneous,
then for $\afa=1,2,\dots,n$ we have
\begin{equation}\label{Qhnsi-homog 220814}
  \p^\afa
  \left[
    \left(
      \p_E+\lmd\pp\lmd+\frac{d-1}{2}
    \right)p
  \right]
=
  \pp\lmd
  \left[
    \left(
      \p_E+\lmd\pp\lmd+\frac{d-1}{2}
    \right)p
  \right]=0,
\end{equation}
which yields
\begin{equation}\label{D_E periods}
  \left(
      \p_E+\lmd\pp\lmd+\frac{d-1}{2}
    \right)p=\mathrm{const.}
\end{equation}
If the charge $d\neq 1$,
then we can adjust the parameter $b$ in \eqref{ambiguity of reg-period}
such that the right-hand side of \eqref{D_E periods} equals zero.
\end{rmk}

Now we are to construct a basis
$\ptil_0,\dots,\ptil_{n+1}$ of periods of $\Mtil$ from a basis of
quasi-homogeneous periods of $M$ such that $\pp\lmd \ptil_i=-\p_{\tilde e}\ptil_i$ for $i=0,1,\dots,n+1$.

\begin{lem}
  Let $p_1(v;\lmd),\dots,p_n(v;\lmd)$ be a basis of periods of $M$,
  and $Q=Q(v;\lmd)$ be a function on $M\times\mathbb{C}$. Then the functions
  $\ptil_0,\dots,\ptil_{n+1}$ defined by
 \begin{align}
      \ptil_0(v^0,v,v^{n+1};\lmd)
   &=
      Q(v;\lmd-v^0)+v^{n+1}(\lmd-v^0)^{\frac{d-1}{2}},
  \label{n+2 periods-1}
  \\
    \ptil_\afa(v^0,v,v^{n+1};\lmd)
    &=
      p_\afa(v;\lmd-v^0),
  \label{n+2 periods-2}
  \\
    \ptil_{n+1}(v^0,v,v^{n+1};\lmd)
    &=
      \begin{cases}
        (\lmd-v^0)^{\frac{1-d}{2}}, & \text{if}\quad d\neq 1\\
        \log(\lmd-v^0), & \text{if}\quad d=1
      \end{cases}
  \label{n+2 periods-3}
\end{align}
form a basis of periods of $\Mtil$ if and only if
the periods $p_1,\dots,p_n$ of $M$ are quasi-homogeneous,
and the function $Q(v;\lmd)$ satisfies the following system of equations:
\begin{align}
&  \left(\mcalU-\lmd I\right)
  \p_\gamma\nabla Q
=
  -\mcalC_\gamma
  \left(
    \mu+\frac12
  \right)\nabla Q
  +
    \frac{d-1}{2}\lmd^{\frac{d-1}{2}}
    \p_\gamma,
\label{nGM-1}
\\
 & \left(\mcalU-\lmd I\right)
  \pp\lmd\nabla Q
=
  \left(\mu+\frac12\right)\nabla Q
  -\frac{d-1}{2}\lmd^{\frac{d-3}{2}}E,
\label{nGM-2}\\
 & \left(
    \p_E+\lmd\pp\lmd
  \right)\nabla Q
  +\left(\mu+\frac12\right)\nabla Q
= 0,
\label{nGM-3}
\\
&  \left(\p_E+\lmd\pp\lmd\right)\pfrac Q\lmd
 +\frac{d+1}{2}\pfrac Q\lmd
= 0.
\label{nGM-4}
\end{align}
\end{lem}

\begin{proof}
It is straightforward to prove the lemma by checking that $\ptil_0,\dots,\ptil_{n+1}$
satisfy the Gauss-Manin equations of $\Mtil$, we omit the details here.
\end{proof}

Since the homogeneous part of the system of equations \eqref{nGM-1} and \eqref{nGM-2}
coincides with the system of Gauss-Manin equations \eqref{GM eqn 1} and \eqref{GM eqn 2},
if $Q(v;\lmd)$ is a solution of the equations \eqref{nGM-1} and \eqref{nGM-2}
then $Q+p$ is also a solution of these equations
for any period $p$ of $M$.
Notably, the system of equations
\eqref{nGM-3} and \eqref{nGM-4} coincides with that of the equations  \eqref{GM eqn 3} and \eqref{GM eqn 4}. Therefore, for the particular case when $d=1$,
the function $Q(v,\lmd)$ can be chosen as any quasi-homogeneous period of $M$.

We proceed
to prove the existence of a function $Q(v,\lmd)$ satisfying the equations \eqref{nGM-1}--\eqref{nGM-4}.

\begin{lem}
There exists a vector field $Y(v;\lmd)=Y^\afa(v;\lmd)\p_\afa$
on $M$, such that
\begin{align}
  \left(\mcalU-\lmd I\right)\p_\gamma Y
&=
  -\mcalC_\gamma\left(\mu+\frac12\right)Y
  +\frac{d-1}{2}\lmd^{\frac{d-1}2}\p_\gamma,
\label{YGM-1}
\\
  \left(\mcalU-\lmd I\right)\pp\lmd Y
&=
  \left(\mu+\frac12\right)Y
 -\frac{d-1}{2}\lmd^{\frac{d-3}{2}}E
\label{YGM-2}
\end{align}
hold true for $\gamma=1,2,\dots,n$.
\end{lem}

\begin{proof}
  Let $p_1(v;\lmd),\dots,p_n(v;\lmd)$ be a basis of periods of $M$ with Gram matrix
  $G=(G^{\afa\beta})$, and
  \begin{equation}\label{Qhn Y Vect}
    Y=-\lmd^{\frac{d-1}{2}}PG
    \left(\p_E+\lmd\pp\lmd\right)\bm{p},
  \end{equation}
where
\[P=(\nabla p_1,\dots,\nabla p_n),\quad
 \bm{p}=(p_1,p_2,\dots,p_n)^{\transpose}.\]
Then by using \eqref{Qhnsi-homog 220814} and properties of $P, G$ it is straightforward to check that the vector field $Y$ satisfies the equations \eqref{YGM-1} and \eqref{YGM-2}.
  The lemma is proved.
  \end{proof}

\begin{thm}\label{thm: ass Q function}
Let $p_1(v;\lmd),\dots,p_n(v,\lmd)$ be a basis of quasi-homogeneous periods of $M$,
and the vector field $Y=Y(v;\lmd)$ be given by \eqref{Qhn Y Vect}.
Then there exists a function $Q=Q(v;\lmd)$ on $M\times\mathbb{C}$ such that
$\nabla Q=Y$ and \eqref{nGM-3}, \eqref{nGM-4} hold true.
This function together with \eqref{n+2 periods-1}--\eqref{n+2 periods-3} yields
a basis of periods of $\Mtil$ which satisfy the condition $\pp\lmd \ptil_i=-\p_{\tilde e}\ptil_i$
and moreover, the associated Gram matrix $\Gtil=(\Gtil^{ij})$ has the form
\begin{align}
  \Gtil &=
  \begin{pmatrix}
    0 & \bm 0 & \frac{2}{1-d} \\[3pt]
    \bm 0 & G & \bm 0 \\[3pt]
    \frac{2}{1-d} & \bm 0 & *
  \end{pmatrix},
  \quad
  \textrm{if}\quad
  d\neq 1, \label{ass n+2 G form 1}
\\[3pt]
  \Gtil &=
  \begin{pmatrix}
    0 & \bm 0 & 1 \\
    \bm 0 & G & \bm 0 \\
    1 & \bm 0 & *
  \end{pmatrix},
  \quad
  \textrm{if}\quad
  d = 1. \label{ass n+2 G form 2}
\end{align}
\end{thm}

\begin{proof}
  By using \eqref{D_E periods} it is easy to see that the vector field
  $Y$ is a gradient field for a certain function $Q_0(v;\lmd)$,
 which satisfies the equations \eqref{nGM-1} and \eqref{nGM-2}.
  Then we can adjust $Q_0$ by adding a certain function $c=c(\lmd)$ so that
  $Q=Q_0+c(\lmd)$ satisfies the equations \eqref{nGM-1}--\eqref{nGM-4}.
It is easy to check that the associated Gram matrix $\Gtil$ has the
form \eqref{ass n+2 G form 1} or \eqref{ass n+2 G form 2}.
The theorem is proved.
\end{proof}

\begin{defn}
Let $p_1(v;\lmd),\dots,p_n(v;\lmd)$
be a basis of quasi-homogeneous periods of $M$,
then we call the function $Q(v;\lmd)$
determined by Theorem \ref{thm: ass Q function} the
\textbf{quasi-period} of $M$ associated with $p_1,\dots,p_n$.
\end{defn}

\begin{rmk}\label{rmk: quasi-period with flat unity}
If $M$ has flat unity $e=\pp{v^1}$,
then for any basis of  periods $p_1,\dots,p_n$ of $M$
satisfying $\p_e p_\afa=-\pfrac{p_\afa}{\lmd}$ we have
\begin{align*}
  Y &= -\lmd^{\frac{d-1}{2}}PG
    \left(\p_E+\lmd\pp\lmd\right)\bm{p}
  =
    -\lmd^{\frac{d-1}{2}}PG\p_{E-\lmd e}\bm{p}\\
  &=
    -\lmd^{\frac{d-1}{2}}PGP^{\transpose}\eta(E-\lmd e)
  =
    -\lmd^{\frac{d-1}{2}}(\mcalU-\lmd)^{-1}(E-\lmd e)\\
 & =
    -\lmd^{\frac{d-1}{2}} e,
\end{align*}
therefore the associated quasi-period $Q(v;\lmd)$ can be chosen as
\[
  Q=-\lmd^{\frac{d-1}{2}}v_1+c(\lmd)
\]
for a certain function $c(\lmd)$.
\end{rmk}

Now we are ready to simplify the loop equation \eqref{loop equation 1}.
\begin{thm}
The loop equation \eqref{loop equation 1}
of $M$ can be represented in the form
\begin{align}
&
    \sum_{s\geq 0}
    \pfrac{\Delta\mcalF}{v^{\gamma,s}}
    \p_x^s
    \left(
      \frac{1}{E-\lmd e}
    \right)^\gamma
+
  \sum_{s\geq 1}
  \pfrac{\Delta\mcalF}{v^{\gamma,s}}
  s\p_x^s
  \left[
    \left(\p^\gamma p_\afa\right)
    G^{\afa\beta}
    \left(
      \p_e+\pp\lmd
    \right)p_\beta
  \right]
\notag
\\
&\quad\quad
-
  \sum_{s\geq 1}
   \pfrac{\Delta\mcalF}{v^{\gamma,s}}
   \sum_{k=1}^{s}
 \binom{s}{k}
   \left(\p_x^{k-1}\pfrac{p_\afa}\lmd
   \right)G^{\afa\beta}
   \left(
     \p_x^{s+1-k}\p^\gamma p_\beta
   \right)
\notag\\
&\quad=
 \frac{\veps^2}{2}
 \sum_{k,\ell\geq 0}
 \left(
   \pfrac{\Delta\mcalF}{v^{\afa,k}}
   \pfrac{\Delta\mcalF}{v^{\beta,\ell}}
  +\frac{\p^2\Delta\mcalF}{\p v^{\afa,k}\p v^{\beta,\ell}}
 \right)
 (\p_x^{k+1}\p^\afa p_\sigma)
 G^{\sigma\rho}
 (\p_x^{\ell+1}\p^\beta p_\rho)
\notag\\
&\quad\quad
 +\frac{\veps^2}{2}
  \sum_{k\geq 0}
  \pfrac{\Delta\mcalF}{v^{\afa,k}}
  \p_x^{k+1}
  \left[
    \nabla\pfrac{p_\sigma}{\lmd}
    \cdot
    \nabla\pfrac{p_\rho}{\lmd}
    \cdot v_x
  \right]^\afa
  G^{\sigma\rho}
\notag
\\
&\quad\quad
+
 \frac12 G^{\afa\beta}
 \pfrac{p_\afa}{\lmd}*\pfrac{p_\beta}{\lmd}-
 \frac{1}{4\lmd^2}
 \mathrm{tr}
 \left(
   \frac14-\mu^2
 \right),\label{reduced loop equation}
\end{align}
where $p_1,\dots,p_n$ is any basis of periods of $M$
and $(G^{\afa\beta})$ is the associated Gram matrix.
\end{thm}
\begin{proof}
Since the loop equation \eqref{loop equation 1} is independent of
the choice of periods $\{p_\afa\}$ and $\{\ptil_i\}$,
we first fix a basis of quasi-homogeneous periods $\{p_\afa\}$ of $M$,
and then choose the basis of periods $\{\ptil_i\}$ of $\Mtil$ by using the associated quasi-period $Q(v;\lmd)$ and the formulae \eqref{n+2 periods-1}--\eqref{n+2 periods-3}.
From the form \eqref{ass n+2 G form 1} or \eqref{ass n+2 G form 2} of the Gram matrix $(\Gtil^{ij})$ associated with $\{\ptil_i\}$, it follows that the left-hand side of the loop equation \eqref{loop equation 1} can be represented as
\begin{align*}
&
    \sum_{s\geq 0}
    \pfrac{\Delta\mcalF}{v^{\gamma,s}}
    \p_x^s
    \left(
      \frac{1}{E-\lmd e}
    \right)^\gamma
   -\frac{1}{\lmd}
    \sum_{s\geq 1}
    \pfrac{\Delta\mcalF}{v^{\gamma,s}}s\p_x^se^\gamma
-
  \frac{1}{\lmd^{\frac{d+1}{2}}}
  \sum_{s\geq 1}
  \pfrac{\Delta\mcalF}{v^{\gamma,s}}
  s\p_x^s\p^\gamma Q\notag
\\
&\quad
-
  \sum_{s\geq 1}
   \pfrac{\Delta\mcalF}{v^{\gamma,s}}
   \sum_{k=1}^{s}
  \binom{s}{k}
   \left(\p_x^{k-1}\pfrac{p_\afa}\lmd
   \right)G^{\afa\beta}
   \left(
     \p_x^{s+1-k}\p^\gamma p_\beta
   \right).
\end{align*}
Since $Y=\nabla Q$ is given by \eqref{Qhn Y Vect},
and $PG\p_{E-\lmd e}\bm{p}=e$, we have
\begin{align*}
&
  -\frac 1\lmd
   \sum_{s\geq 1}
   \pfrac{\Delta\mcalF}{v^{\gamma,s}}
   s\p_x^s e^\gamma
  -
   \frac{1}{\lmd^{\frac{d+1}{2}}}
  \sum_{s\geq 1}
  \pfrac{\Delta\mcalF}{v^{\gamma,s}}
  s\p_x^s\p^\gamma Q
\\
&\qquad =
   -\frac 1\lmd
   \sum_{s\geq 1}
   \pfrac{\Delta\mcalF}{v^{\gamma,s}}
   s\p_x^s e^\gamma
  +
  \frac 1\lmd
  \sum_{s\geq 1}
  \pfrac{\Delta\mcalF}{v^{\gamma,s}}
  s\p_x^s
  \left[
    PG\left(\p_E+\lmd\pp\lmd\right)\bm{p}
  \right]^\gamma
\\
&\qquad =
  \frac 1\lmd
  \sum_{s\geq 1}
  \pfrac{\Delta\mcalF}{v^{\gamma,s}}
  s\p_x^s
  \left[
    PG\p_{E-\lmd e}\bm{p}-e
    +\lmd PG\left(\p_e+\pp\lmd\right)\bm{p}
  \right]^\gamma
\\
&\qquad =
  \sum_{s\geq 1}
  \pfrac{\Delta\mcalF}{v^{\gamma,s}}
  s\p_x^s
  \left[
    PG\left(\p_e+\pp\lmd\right)\bm{p}
  \right]^\gamma
\\
&\qquad =
  \sum_{s\geq 1}
  \pfrac{\Delta\mcalF}{v^{\gamma,s}}
  s\p_x^s
  \left[
    \left(\p^\gamma p_\afa\right)
    G^{\afa\beta}
    \left(
      \p_e+\pp\lmd
    \right)p_\beta
  \right],
\end{align*}
hence the equation \eqref{loop equation 1}
is equivalent to \eqref{reduced loop equation}
if $\{p_\afa\}$ are quasi-homogeneous. On the other hand, from the expression of the equation \eqref{reduced loop equation} we see that this equation is independent of the choice of the basis of periods of $M$.
The theorem is proved.
\end{proof}

\begin{rmk}
If $M$ has a flat unity $e=\pp{v^1}$,
then according to \eqref{240727-1957} we can choose a basis of periods $p_1,\dots,p_n$ of
$M$ such that
\[\p_ep_\afa = \p_1p_\afa=-\pp\lmd p_\afa,\]
therefore the second term in the left-hand side of \eqref{reduced loop equation} vanishes,
and the loop equation \eqref{reduced loop equation} coincides with the one given in (3.10.95) of \cite{normal-form},
which is the loop equation of a usual Frobenius manifolds.
\end{rmk}

\begin{cor}
The loop equation \eqref{reduced loop equation}
of $M$ can also be represented in the form
\begin{align}
&
  \sum_{s\geq 0}
    \pfrac{\Delta\mcalF}{v^{\gamma,s}}
    (s+1)\p_x^s\left(\frac{1}{E-\lmd e}\right)^\gamma\notag\\
   &\quad\quad   +
      \sum_{s\geq 0}
    \pfrac{\Delta\mcalF}{v^{\gamma,s}}
    \sum_{k=1}^{s}
     k\binom{s+1}{k+1}
        \left(
          \p_x^{s-k}\p^\gamma p_\afa
        \right)
        G^{\afa\beta}
        \left(
          \pp{\lmd}
          \p_x^k p_\beta
        \right)
\notag\\
&\quad=
 \frac{\veps^2}{2}
 \sum_{k,\ell\geq 0}
 \left(
   \pfrac{\Delta\mcalF}{v^{\afa,k}}
   \pfrac{\Delta\mcalF}{v^{\beta,\ell}}
  +\frac{\p^2\Delta\mcalF}{\p v^{\afa,k}\p v^{\beta,\ell}}
 \right)
 (\p_x^{k+1}\p^\afa p_\sigma)
 G^{\sigma\rho}
 (\p_x^{\ell+1}\p^\beta p_\rho)
\notag\\
&\quad\quad
 +\frac{\veps^2}{2}
  \sum_{k\geq 0}
  \pfrac{\Delta\mcalF}{v^{\afa,k}}
  \p_x^{k+1}
  \left[
    \nabla\pfrac{p_\sigma}{\lmd}
    \cdot
    \nabla\pfrac{p_\rho}{\lmd}
    \cdot v_x
  \right]^\afa
  G^{\sigma\rho}
\notag
\\
&\quad\quad
+
 \frac12 G^{\afa\beta}
 \pfrac{p_\afa}{\lmd}*\pfrac{p_\beta}{\lmd}-
 \frac{1}{4\lmd^2}
 \mathrm{tr}
 \left(
   \frac14-\mu^2
 \right),\label{loop equation-2308}
\end{align}
where $p_1,\dots,p_n$ is any basis of periods of $M$
and $(G^{\afa\beta})$ is the associated Gram matrix.
\end{cor}
\begin{proof}
We have the following simple relation:
\begin{align*}
 & s\p_x^s
  \left[
    \left(\p^\gamma p_\afa\right)
    G^{\afa\beta}
    \left(
      \p_e+\pp\lmd
    \right)p_\beta
  \right]
\notag
-
   \sum_{k=1}^{s}
  \binom{s}{k}
   \left(\p_x^{k-1}\pfrac{p_\afa}\lmd
   \right)G^{\afa\beta}
   \left(
     \p_x^{s+1-k}\p^\gamma p_\beta
   \right)\\
&\quad=
s\p_x^s\left[(\p^\gamma p_\afa)G^{\afa\beta}(\p_e p_\beta)\right]
  + s\sum_{k=1}^{s}\binom{s}{k}
      (\p_x^{s-k}\p^\gamma p_\afa)G^{\afa\beta}
      \left(\pp\lmd\p_x^k p_\beta\right) \\
&\quad\quad
  -\sum_{k=2}^{s}
   \binom{s}{k}
    \left(\pp\lmd\p_x^{k-1}p_\afa\right)G^{\afa\beta}
    (\p_x^{s+1-k}\p^\gamma p_\beta)\\
&\quad=
s\p_x^s\left[(\p^\gamma p_\afa)G^{\afa\beta}(\p_e p_\beta)\right]
  +\sum_{k=1}^{s}s\binom{s}{k}
    (\p_x^{s-k}\p^\gamma p_\afa)G^{\afa\beta}
    \left(\pp\lmd\p_x^kp_\beta\right) \\
&\quad\quad
  -\sum_{k=1}^{s-1}
  \binom{s}{k+1}
   (\p_x^{s-k}\p^\gamma p_\afa)G^{\afa\beta}
   \left(\pp\lmd\p_x^kp_\beta\right) \\
&\quad=
 s\p_x^s\left[(\p^\gamma p_\afa)G^{\afa\beta}(\p_e p_\beta)\right]
  +\sum_{k=1}^{s}
   \left[s\binom{s}{k}-\binom{s}{k+1}\right]
   (\p_x^{s-k}\p^\gamma p_\afa)G^{\afa\beta}
   \left(\pp\lmd\p_x^kp_\beta\right) \\
&\quad=
  s\p_x^s\left[(\p^\gamma p_\afa)G^{\afa\beta}(\p_e p_\beta)\right]
  +\sum_{k=1}^{s} k\binom{s+1}{k+1}
  (\p_x^{s-k}\p^\gamma p_\afa)G^{\afa\beta}
   \left(\pp\lmd\p_x^kp_\beta\right).
\end{align*}
Then by using the fact that
\begin{equation}
  (\p^\gamma p_\afa)G^{\afa\beta}(\p_e p_\beta)
=
  \left(\frac{1}{E-\lmd e}\cdot e\right)^\gamma =
  \left(\frac{1}{E-\lmd e}\right)^\gamma
\end{equation}
we prove the corollary.
\end{proof}
We note that the loop equation of a usual Frobenius manifold with flat unity can also be represented in precisely the same form as the one given in the above corollary.

The existence and uniqueness (up to the addition of constants) of solutions of the loop equation for a semisimple generalized Frobenius manifold with non-flat unity are proved in \cite{LWZ-2}. Such a solution of the loop equation yields a quasi-Miura transformation \eqref{quasi-Miura transf} which gives a deformation \eqref{deformed PH} of the Principal Hierarchy of the semisimple generalized Frobenius manifold. In the next section, we will present two examples of semisimple generalized Frobenius manifolds, and show that the associated deformed
Principal Hierarchies are closely related to the Hodge integrals and the equivariant Gromov-Witten invariants of the resolved conifold with anti-diagonal action, so we call such deformations of the Principal Hierarchy the topological deformations.

\section{Two examples}

In this section we present two examples of
generalized Frobenius manifolds of dimension $n=1$ and $2$ respectively, and
establish their relations with the well-known integrable hierarchies: the Volterra hierarchy, the \textit{q}-deformed KdV hierarchy and the Ablowitz-Ladik hierarchy.

\begin{ex}
Let us consider the 1-dimensional generalized Frobenius manifold $M$ with potential
\begin{equation}\label{zh-55}
F=\frac1{12} v^4,
\end{equation}
here we denote the flat coordinate $v^1$ of $M$ by $v$.
The flat metric, the unity and the Euler vector field are given respectively by
\begin{equation}\label{zh-56}
\eta=1,\quad e=\frac1{2 v}\p_v,\quad E=\frac12 v\p_v.
\end{equation}
This generalized Frobenius manifold has charge $d=1$, and $\mu=R=0$.

The Hamiltonian densities $\theta_{\al,p}$ for the Principal Hierarchy are given by
\begin{align*}
&\theta_{0,0}=\frac12 \log v,\\
&\theta_{0,p}=\frac{2^{p-1} (2p-2)!!}{(2p)!} v^{2p},\quad
\theta_{0,-p}=(-1)^{p+1}\frac{(2p-1)!}{2^{p+1} (2p)!!} v^{-2p},\quad p\ge 1,\\
&\theta_{1,p}=\frac{2^{p} (2p-1)!!}{(2p+1)!} v^{2p+1},\quad p\ge 0.
\end{align*}
They satisfy the quasi-homogeneity conditions \eqref{real D-E-theta}
and \eqref{def of c-p} with
\[r^\veps_p=0,\quad c_p=\frac14\delta_{p,0}.\]

The first few flows of the Principal Hierarchy \eqref{principal hierarchy}
have the expressions
\begin{align*}
&\frac{\p v}{\p t^{0,-3}}=-\frac{15}{8 v^6} v_x,\quad \frac{\p v}{\p t^{0,-2}}=\frac3{4 v^4} v_x,\quad  \frac{\p v}{\p t^{0,-1}}=-\frac1{2 v^2}v_x,\\
&\frac{\p v}{\p t^{0,0}}=v_x,\quad \frac{\p v}{\p t^{0,1}}=2 v^2 v_x,
\quad \frac{\p v}{\p t^{0,2}}=\frac43 v^4 v_x,\\
&\frac{\p v}{\p t^{1,0}}=2 v v_x,\quad \frac{\p v}{\p t^{1,1}}=2 v^3 v_x,\quad \frac{\p v}{\p t^{1,2}}=v^5 v_x.
\end{align*}
The tau function $\tau^{[0]}=\rme^f$ of the Principal Hierarchy is defined by
\[ \frac{\p^2\log\tau^{[0]}}{\p t^{\al,p}\p t^{\beta,q}}=\Omega_{\al,p;\beta,q}=
\p_x^{-1}\frac{\p\theta_{\al,p}}{\p t^{\beta,q}}.\]
In particular, we have
\[\log v=2 \frac{\p^2\log\tau^{[0]}}{\p x^2},\quad v=\frac{\p^2\log\tau^{[0]}}{\p x\p t^{1,0}}.\]
We have the Virasoro operators
\begin{align*}
  L_{-1}
=&\,
  \sum_{p\in\bbZ}t^{0,p+1}\pp{t^{0,p}}
 +\sum_{p\geq 0}
 t^{1,p+1}\pp{t^{1,p}}
 +\frac12 t^{1,0}t^{1,0},\\
L_0
=&\,
  \sum_{p\in\bbZ}pt^{0,p}\pp{t^{0,p}}
  +\sum_{p\geq 0}
  \left(
    p+\frac12
  \right)
  t^{1,p}\pp{t^{1,p}}
+\frac18\sum_{p\in\bbZ}
(-1)^p t^{0,p}t^{0,-p}+\frac{1}{16},\\
L_m
=&\,
\frac12\sum_{p=0}^{m-1}
\frac{(2p+1)!!(2m-2p-1)!!}{2^{m+1}}
\frac{\p^2}{\p t^{1,p}\p t^{1,m-1-p}}\\
&
+
\sum_{p\geq 1}
\frac{(p+m)!}{(p-1)!}
\left(
  t^{0,p}\pp{t^{0,p+m}}
  +(-1)^{m+1}
  t^{0,-p-m}
  \pp{t^{0,-p}}
\right)\\
&
+\sum_{p\geq 0}
  \frac{(2p+2m+1)!!}{2^{m+1}(2p-1)!!}
  t^{1,p}\pp{t^{1,p+m}}\\
&
+
  \frac14\sum_{p\geq 1}
  (-1)^{p+m}
  \frac{(p+m)!}{(p-1)!}
  \left(
      \sum_{k=p}^{p+m}
      \frac{1}{k}
  \right)
  t^{0,p}t^{0,-m-p}\\
&
+
  \frac18\sum_{p=0}^{m}
  (-1)^m
  p!(m-p)!
  t^{0,-p}t^{0,p-m},\quad  m\geq 1.
\end{align*}

The generalized Frobenius manifold $M$ has a quasi-homogeneous  period
\[
  p(v,\lmd)=\log\left(v+\sqrt{v^2-\lmd}\right)
 \]
 with the associated Gram matrix $ (G^{\afa\beta})=1$ and
the star product
\[
  \pfrac{p}{\lmd}*\pfrac{p}{\lmd} =
  \frac{1}{8\lmd^2}-\frac{1}{8}\frac{1}{(v^2-\lmd)^2}.
\]
Thus the loop equation \eqref{reduced loop equation} of $M$ reads
\begin{align*}
&
   \sum_{s\geq 0}
   \pfrac{\Delta\mcalF}{v^{(s)}}
   \p_x^s\frac{1}{2v(v^2-\lmd)}
  +
   \sum_{s\geq 1}
   \pfrac{\Delta\mcalF}{v^{(s)}}
   s\p_x^s
   \left[
     \frac{1}{2\lmd}
     \left(
       \frac{1}{\sqrt{v^2-\lmd}}-\frac 1v
     \right)
   \right]\\
&\quad\quad
  +\sum_{s\geq 1}
   \pfrac{\Delta\mcalF}{v^{(s)}}
   \sum_{k=1}^{s}
   \binom{s}{k}
   \left[
     \p_x^{k-1}
     \frac{1}{2\lmd}
     \left(
       \frac{v}{\sqrt{v^2-\lmd}}-1
     \right)
   \right]
   \left(
     \p_x^{s+1-k}
     \frac{1}{\sqrt{v^2-\lmd}}
   \right)\\
&\quad=
  \frac12\veps^2
  \sum_{k,\ell\geq 0}
  \left(
    \pfrac{\Delta\mcalF}{v^{(k)}}
    \pfrac{\Delta\mcalF}{v^{(\ell)}}
   +\frac{\p^2\Delta\mcalF}{\p v^{(k)}\p v^{(\ell)}}
  \right)
  \left(
    \p_x^{k+1}
    \frac{1}{\sqrt{v^2-\lmd}}
  \right)
  \left(
    \p_x^{\ell+1}
    \frac{1}{\sqrt{v^2-\lmd}}
  \right)\\
&\quad\quad
 +
   \frac12\veps^2
     \sum_{k\geq 0}
     \pfrac{\Delta\mcalF}{v^{(k)}}
     \p_x^{k+1}
     \frac{v^2v_x}{(v^2-\lmd)^3}
 -
   \frac{1}{16}\frac{1}{(v^2-\lmd)^2}.
\end{align*}
Since the generalized Frobenius manifold $M$ is semisimple, from \cite{LWZ-2} we know that its loop equation has a unique solution
\[\Delta\mcalF=\sum_{k\ge 1}\veps^{k-2}\mcalF^{[k]}
=\sum_{g\ge 1}\veps^{2g-2} \mcalF_g(v, v_x,\dots, v^{(3g-2)})\]
up to the addition of constant series $a_1\veps^{-1}+a_2+a_3\veps+\dots$.
The coefficient of $\veps^{0}$ of the loop equation yields the equation
\[
  \frac{1}{(v^2-\lmd)^2}
  \left(
  \frac{1}{16}-\frac 32v_x\pfrac{\mcalF_1}{v_x}
  \right)
 +\frac{1}{2v^2(v^2-\lmd)}
  \left(
    v\pfrac{\mcalF_1}{v}
   -2v_x\pfrac{\mcalF_1}{v_x}
  \right) =0 ,
\]
which leads to
\[
  \mcalF_1=\frac{1}{24}\log v_x+\frac{1}{12}\log v.
\]
One can solve the loop equation recursively to obtain $\mcalF_g, g\ge 2$. For example, we have
\begin{align*}
\mcalF_2 =&\,
 \frac{v_{xx}}{120 v}
-\frac{v_x^2}{120 v^2}
+\frac{v v^{(4)}}{576 v_x^2}
+\frac{37 v^{(3)}}{2880 v_x}
+\frac{vv_{xx}^3}{180 v_x^4}
-\frac{11 v_{xx}^2}{960 v_x^2}
-\frac{7 vv_{xx}v^{(3)}}{960 v_x^3},
\\
\mcalF_3 =&\,
  \frac{v^{(4)}}{1512 v}
 -\frac{v_{xx}^2}{504 v^2}
 -\frac{v_x^4}{252 v^4}
 +\frac{v^2 v^{(7)}}{20736 v_x^3}
 +\frac{91 v v^{(6)}}{103680 v_x^2}
 +\frac{913 v^{(5)}}{241920 v}
\\
&\,
 -\frac{103 v^2 (v^{(4)})^2}{120960 v_x^4}
 +\frac{59 v^2 (v^{(3)})^3}{16128 v_x^5}
 -\frac{v_x v^{(3)} }{378 v^2}
 -\frac{1669 (v^{(3)})^2}{145152 v_x^2}
 -\frac{5 v^2 v_{xx}^6}{162 v_x^8}
 +\frac{13 v v_{xx}^5}{252 v_x^6}
\\
&\,
 -\frac{193 v_{xx}^4}{8064 v_x^4}
 +\frac{v_x^2 v_{xx}}{126 v^3}
 -\frac{7 v^2 v_{xx}v^{(6)}}{11520 v_x^4}
 -\frac{53 v^2 v^{(3)} v^{(5)}}{40320 v_x^4}
 +\frac{353 v^2 v_{xx}^2 v^{(5)}}{80640 v_x^5}
\\
&\,
 -\frac{419 v v_{xx}v^{(5)}}{60480 v_x^3}
 -\frac{9169 v v^{(3)} v^{(4)}}{725760 v_x^3}
 -\frac{83 v^2 v_{xx}^3 v^{(4)} }{3780 v_x^6}
 +\frac{545 v v_{xx}^2 v^{(4)}}{16128 v_x^4}
 -\frac{3727 v_{xx}v^{(4)}}{241920 v_x^2}
\\
&\,
 +\frac{59 v^2 v_{xx}^4 v^{(3)} }{756 v_x^7}
 -\frac{5555 v v_{xx}^3 v^{(3)} }{48384 v_x^5}
 +\frac{325 v_{xx}^2 v^{(3)} }{6912 v_x^3}
 -\frac{83 v^2 v_{xx}^2 (v^{(3)})^2}{1792 v_x^6}
\\
&\,
 +\frac{97 vv_{xx} (v^{(3)})^2}{2016 v_x^4}
 +\frac{1273 v^2 v_{xx}v^{(3)} v^{(4)}}{80640 v_x^5}.
\end{align*}
Let us note that after the replacement
\[v\to \rme^{-\frac12 v},\quad \ve \to -\frac{\rmi}{\sqrt{2}}\epsilon,\]
the function $\mcalF_1(v,v_x)+\ve^2\mcalF_2(v,v_x,\dots,v^{(4)})$ coincides, up to the addition of a constant, with
the function $H_1(v,v_x)+\epsilon^2 H_2(v_x,v_{xx},v^{(3)},v^{(4)})$
given in (26), (27) of \cite{DD-CNTP} as the generating function for the genus one and genus two special cubic Hodge integrals, see also \cite{DLYZ-1, DLYZ-2}.

Now let us consider the topological deformation \eqref{deformed PH} of the Principal Hierarchy \eqref{principal hierarchy} obtained via the quasi-Miura transformation \eqref{quasi-Miura transf}, i.e.,
\[w=v+\ve^2\frac{\p^2 \mcalF_1}{\p x\p t^{1,0}}+\ve^4\frac{\p^2 \mcalF_2}{\p x\p t^{1,0}}+\dots.\]
In order to represent the deformed integrable hierarchy in a simple form, we use the unknown function
\begin{align}\label{zh-31}
U&=\frac12 \frac{\Lambda-\Lambda^{-1}}{\sqrt{2}\,\rmi\ve \p_x}
\left(v+\ve^2\frac{\p^2 \mcalF_1}{\p x\p t^{1,0}}+\ve^4\frac{\p^2 \mcalF_2}{\p x\p t^{1,0}}+\dots\right)\notag\\
&=w-\frac{\ve^2}3 w_{xx}+\frac{\ve^4}{30} w^{(4)}-\frac{\ve^6}{630} w^{(6)}+\dots,
\end{align}
where
\[\Lambda=\rme^{\sqrt{2}\,\rmi\ve\p_x}=\rme^{\epsilon\p_x},\quad \epsilon=\sqrt{2}\,\rmi\ve.\]
Then, at the approximation up to $\ve^6$, we can verify that $U$ satisfies the following equation:
\begin{equation}
\frac{\p U}{\p t^{1,0}}=\frac{4 U}{\sqrt{2}\,\rmi \ve}\frac{\Lambda-1}{\Lambda+1} U
=\frac{4 U}{\epsilon}\frac{\Lambda-1}{\Lambda+1} U.
\end{equation}
This equation is just the \textit{q}-deformed KdV equation introduced in
\cite{Frenkel-1}, and it
can be represented as the following Lax equation:
\[\frac{\p L}{\p t^{1,0}}=-\frac{4}{\epsilon}\left[\left(L^{\frac12}\right)_+,L\right],\]
here the Lax operator is given by
\begin{equation}\label{zh-57}
L=\Lambda^2+U\Lambda+1.
\end{equation}

We can verify that the $t^{1,1}$-flow and the $t^{1,2}$-flow of the deformed Principal Hierarchy can be represented,  also at the approximation up to $\ve^6$, by the Lax equations
\begin{align*}
\frac{\p L}{\p t^{1,1}}&=\frac{32}{3\epsilon}\left[\left(L^{\frac32}\right)_+-\frac32 \left(L^{\frac12}\right)_+,L\right],\\
\frac{\p L}{\p t^{1,2}}&=-\frac{256}{15\epsilon}\left[\left(L^{\frac52}\right)_+-\frac52 \left(L^{\frac32}\right)_++\frac{15}8\left(L^{\frac12}\right)_+,L\right].
\end{align*}

Let us proceed to consider the flows $\frac{\p}{\p t^{0,p}},\,p\in\mathbb{Z}$ of the deformed Principal Hierarchy.
Introduce the unknown function
\begin{align}
W&=-\left(\Lambda^{\frac12}+\Lambda^{-\frac12}\right)\log U=-2\log v+\ve^2 w_1(v, v_x,\dots)+\dots,\label{zh-53}
\end{align}
then at the approximation up to $\epsilon^6$ we know that the
$\frac{\p}{\p t^{0,-1}}$-flow can be represented in the form
\[\frac{\p W}{\p t^{0,-1}}=-\frac1{4\epsilon}\left(\Lambda-\Lambda^{-1}\right)\rme^W.\]
This is just the Volterra equation (also called the discrete KdV equation)
which is a well-known discrete integrable system \cite{Kac-Moerbecke, Manakov}
originally introduced for the description of the population
dynamics \cite{Lotka, Lotka-2, Volterra}. The relation of this equation to the special
cubic Hodge integrals is given in \cite{DLYZ-1, DLYZ-2}, which is in accordance with the afore mentioned relation between the function $\mcalF_1+\ve^2\mcalF_2$ and the generating function of the special cubic Hodge integrals.
In terms of the Lax operator
\begin{equation}\label{zh-58}
\mathcal{L}=\Lambda+\rme^W\Lambda^{-1},
\end{equation}
the Volterra equation can be represented by the Lax equation
\[\frac{\p\mathcal{L}}{\p t^{0,-1}}=-\frac1{4\epsilon}\left[\left(\mathcal{L}^2\right)_+,\mathcal{L}\right].\]
The $t^{0,-2}$-flow and the $t^{0,-3}$-flow can also be represented,  at the approximation up to $\ve^6$, by the Lax equations
\[\frac{\p\mathcal{L}}{\p t^{0,-2}}=\frac{1}{16\epsilon}\left[\left(\mathcal{L}^4\right)_+,\mathcal{L}\right],\quad
\frac{\p\mathcal{L}}{\p t^{0,-3}}=-\frac{1}{32\epsilon}\left[\left(\mathcal{L}^6\right)_+,\mathcal{L}\right].\]

We have the following conjecture which relates the topological deformation of the Principal Hierarchy with the \textit{q}-deformed KdV hierarchy
\begin{equation}\label{zh-2023-2}
\frac{\p L}{\p \tau_k}=\frac1{\epsilon}\left[\left(L^{k+\frac12}\right)_+,L\right],\quad k\ge 0
\end{equation}
and the Volterra hierarchy
\begin{equation}\label{zh-2023-3}
\frac{\p\mathcal{L}}{\p \sigma_m}=\frac1{\epsilon}\left[\left(\mathcal{L}^{2m}\right)_+,\mathcal{L}\right],\quad m\ge 1.
\end{equation}
\begin{conj}
The $\frac{\p}{\p t^{1,k}}$-flows and the $\frac{\p}{\p t^{0,-m}}$-flows  of the topological deformation of the Principal Hierarchy of the generalized Frobenius manifold defined by \eqref{zh-55}, \eqref{zh-56},
for $k\ge 0$ and $m\ge 1$ can be represented by the following Lax equations:
\begin{align*}
\frac{\p L}{\p t^{1,k}}&
=\sum_{\ell=0}^k (-1)^{k+\ell+1} \frac{2^{3k-\ell+2}}{\ell! (2k-2\ell+1)!!} \frac{\p L}{\p \tau_{k-\ell}},\quad k\ge 0\\
\frac{\p \mathcal{L}}{\p t^{0,-m}}&
=(-1)^m \frac{(m-1)!}{2^{2m}}\frac{\p \mathcal{L}}{\p\sigma_{m}},\quad m\ge 1.
\end{align*}
Here the Lax operators $L$, $\mathcal{L}$ are defined by \eqref{zh-31},
\eqref{zh-57} and \eqref{zh-53},
\eqref{zh-58}.
\end{conj}

For the flows $\frac{\p}{\p t^{0,p}}$, $p\ge 0$ we have not found their relation with known integrable hierarchies yet. A proof of this conjecture is recently given in \cite{LWZ-2}.

\begin{rmk}
The \textit{q}-deformed KdV hierarchy \eqref{zh-2023-2}
is bihamiltonian with respect to the bihamiltonian structure $(\mathcal{P}_1, \mathcal{P}_2)$
given by the Hamiltonian operators (cf. \cite{Frenkel-1})
\[\mathcal{P}_1=\frac1{2\epsilon}\left(\Lambda-\Lambda^{-1}\right),\quad  \mcalP_2=\frac2{\epsilon}U\frac{\Lmd-1}{\Lmd+1}U.\]
The hydrodynamic limits
\[\mathcal{P}_1^{[0]}=\p_x,\quad
\mathcal{P}_2^{[0]}=U^2\p_x+ U U_x \]
of these Hamiltonian operators yield the bihamiltonian structure $(\mathcal{P}^{[0]}_1, \mathcal{P}^{[0]}_2)$ for the Principal Hierarchy of $M$. It is easy to check, after representing $\epsilon$ in terms of the dispersion parameter $\varepsilon$ with $\epsilon=\sqrt{2} i\varepsilon$, the bihamiltonian structure $(\mathcal{P}_1, \mathcal{P}_2)$ has central invariant $c=\frac1{24}$. The notion of central invariants
for deformations of a semisimple bihamiltonian structure of hydrodynamic type is introduced in \cite{DLZ-06,liu2005deformations}, which are functions of  a single variable and characterize equivalence classes of deformations of a semisimple bihamiltonian structure of hydrodynamic type under Miura type transformations \cite{carlet2018deformations,DLZ-06,liu2005deformations,liu2013bihamiltonian, lorenzoni2002}.

The flows of the \textit{q}-deformed KdV hierarchy \eqref{zh-2023-2} can be represented as Hamiltonian systems with respect to $\mathcal{P}_1$ and $\mathcal{P}_2$ as follows:
\[\pfrac{U}{\tau_k}
=\mcalP_1\dtafrac{H_k}{U}=\mcalP_2\dtafrac{G_k}{U},\quad k\ge 0,\]
here the Hamiltonians are defined by
\[
  H_k=\frac{4}{2k+3}\int \res L^{\frac{2k+3}{2}}\nd x,\quad
  G_k=-\frac14\sum_{j=0}^k H_{j-1}.
\]

The Volterra hierarchy \eqref{zh-2023-3} is also bihamiltonian with respect to the bihamiltonian structure $(\mathcal{P}_1, \mathcal{P}_2)$ given above \cite{Adler, DLYZ-1, FT}.
In terms of the unknown function $W$ the Hamiltonian operators $\mathcal{P}_1$ and $\mathcal{P}_2$ can be represented in the form
\begin{align*}
\hat{\mathcal{P}}_1&=\frac1{2\epsilon}\left( \left(\Lmd+1\right)\rme^W\left(\Lmd+1\right)
 -\left(1+\Lmd^{-1}\right)\rme^W\left(1+\Lmd^{-1}\right)\right),\\
 \hat{\mathcal{P}}_2&=\frac2{\epsilon}\left(\Lmd-\Lmd^{-1}\right),
 \end{align*}
and the flows of the Volterra hierarchy can be represented as bihamiltonian systems as follows:
\[\frac{\p W}{\p\sigma_m}=\hat{\mathcal{P}}_1\frac{\delta \hat{H}_m}{\delta W}=\hat{\mathcal{P}}_2\frac{\delta \hat{G}_m}{\delta W},\quad m\ge 1.\]
Here the Hamiltonians are given by
\[\hat{H}_1=\int W(x)\nd x,\quad \hat{H}_{m+1}=\frac1{m}\int\res \mathcal{L}^{2m}\nd x,\quad  \hat{G}_m=\frac1{4m}\int\res \mathcal{L}^{2m}\nd x.\]
\end{rmk}
\end{ex}

\begin{ex}
We consider the 2-dimensional generalized Frobenius manifold with potential
\begin{equation}\label{zh-42}
   F =\frac12(v^1)^2 v^2+v^1\rme^{v^2}+\frac12(v^1)^2\log v^1.
\end{equation}
The flat metric,  the unity and the Euler vector field are given respectively by
\begin{equation}\label{zh-43}
\eta=\begin{pmatrix}0 & 1 \\ 1 & 0\end{pmatrix},
\quad  e = \frac{v^1\p_{v^1}-\p_{v^2}}{v^1-\rme^{v^2}},\quad E = v^1\p_{v^1}+\p_{v^2}.
\end{equation}
It has charge $d=1$ and  monodromy data
\[
    \mu=\begin{pmatrix}
          -\frac12 &  0 \\
       0     & \frac12
        \end{pmatrix},\quad
    R=R_1=\begin{pmatrix}
        0 & 0 \\
        2 & 0
      \end{pmatrix}.
  \]
This generalized Frobenius manifold and its relation to the dispersionless   Ablowitz-Ladik hierarchy are presented in \cite{Brini-2}, see also \cite{AL-intro-SXL}.

The Hamiltonian densities $\theta_{2,p},\, p\ge 0$ and $\theta_{0,-p},\, p\ge 1$ for the Principal Hierarchy are given by
  \begin{align*}
    \theta_{2,p}
   &= \frac{1}{(p+1)!}
    \sum_{k=0}^{p+1}
   \binom{p+1}{k}\binom{p+k}{k}
    \rme^{k v^2}(v^1-\rme^{v^2})^{p+1-k} \\
   \theta_{0,0}
   &= v^2-\log(v^1-\rme^{v^2})\\
   \theta_{0,-p}
   &=
   \frac{(-1)^p(p-1)!p!}{(v^1-\rme^{v^2})^{2p}}\theta_{2,p-1}
   ,\qquad p\geq 1,
  \end{align*}
and the first few $\theta_{0,p}$ and $\theta_{1,p}$ for $p\geq 0$ have the expressions
\begin{align*}
  \theta_{0,1}
&=
  v^1 v^2, \\
  \theta_{0,2}
&=
  \frac{1}{2} (v^2+1) (v^1)^2+(v^2-1) \rme^{v^2} v^1,\\
\theta_{0,3}
&=
  \frac{1}{12} v^1
  \left[(2 v^2+3) (v^1)^2+6 \rme^{v^2} (2 v^2-1) v^1+\rme^{2 v^2} (6 v^2-9)\right],\\
\theta_{1,0}
&= v^2,\\
\theta_{1,1}
&=
  (v^2+\log v^1)v^1 +(\rme^{v^2}-v^1),\\
\theta_{1,2}
&=
  (v^2+\log v^1)\theta_{2,1}+
  \frac{1}{4} \left(\rme^{2 v^2}-4 \rme^{v^2} v^1-(v^1)^2\right),\\
\theta_{1,3}
&=
  (v^2+\log v^1)\theta_{2,2}+
  \frac{1}{18} \left(\rme^{3 v^2}-9 \rme^{2 v^2} v^1-27 \rme^{v^2} (v^1)^2-(v^1)^3\right).
\end{align*}

We have the Virasoro operators
\begin{align*}
L_{-1}
=&\,
  \sum_{p\geq 0}
    \left(
    t^{1,p+1}\pp{t^{1,p}}
   +t^{2,p+1}\pp{t^{2,p}}
    \right)
 +\sum_{p\in\bbZ}
  t^{0,p+1}\pp{t^{0,p}}
 +t^{1,0}t^{2,0},\\
L_0 =&\,
  \sum_{p\in\bbZ}
    pt^{0,p}\pp{t^{0,p}}
 +\sum_{p\geq 1}
   p\left(
     t^{1,p}\pp{t^{1,p}}
    +t^{2,p-1}\pp{t^{2,p-1}}
   \right) \\
&\,
  +\sum_{p\geq 1}
  \left(
    t^{0,p}\pp{t^{2,p-1}}
   +2t^{1,p}\pp{t^{2,p-1}}
  \right)
  +
   \sum_{p\geq 0}(-1)^pt^{0,-p}t^{1,p}
  +t^{1,0}t^{1,0},
\\
L_{m} =&\,
  \sum_{p=1}^{m-1}
    p!(m-p)!
    \frac{\p^2}{\p t^{2,p-1}\p t^{2,m-p-1}}\\
&\,
  +
  \sum_{p\geq 1}
  \frac{(p+m)!}{(p-1)!}
  \left(
    t^{0,p}\pp{t^{0,p+m}}
   +(-1)^{m+1}t^{0,-p-m}\pp{t^{0,-p}}
  \right.\\
&
  \quad
  \left.
   +t^{1,p}\pp{t^{1,p+m}}
   +t^{2,p-1}\pp{t^{2,p+m-1}}
  \right)\\
&\,
   +\sum_{p=0}^{m-1}
     (-1)^pp!(m-p)!t^{0,-p}\pp{t^{2,-p+m-1}}
\\
&\,
 +\sum_{p\geq 1}
  \frac{(p+m)!}{(p-1)!}
  \left(
    \sum_{k=p}^{p+m}\frac 1k
  \right)
  t^{0,p}\pp{t^{2,p+m-1}}\\
&\,
 +2m!t^{1,0}\pp{t^{2,m-1}}
 +2\sum_{p\geq 1}
  \frac{(p+m)!}{(p-1)!}
  \left(
    \sum_{k=p}^{p+m}\frac 1k
  \right)
  t^{1,p}\pp{t^{2,p+m-1}}\\
&\,
 +(-1)^mm! t^{0,-m}t^{1,0}
 +\sum_{p\geq 1}
  (-1)^{p+m}
  \frac{(p+m)!}{(p-1)!}
  \left(
    \sum_{k=p}^{p+m}\frac 1k
  \right)
  t^{0,-p-m}t^{1,p},
\end{align*}
here $m\ge 1$. The intersection form of the generalized Frobenius manifold has the expression
\[
  (g^{\afa\beta})
 =\begin{pmatrix}
    2v^1\rme^{v^2} & v^1+\rme^{v^2} \\
    v^1+\rme^{v^2} & 2
  \end{pmatrix}.
\]
We have the following basis of quasi-homogeneous periods
\[p_1=\log\left(
      v^1-\lmd-\rme^{v^2}+\sqrt{D}
    \right)-\frac {v^2}2,\quad
    p_2=v^2
  \]
with the associated Gram matrix
\[
  (G^{\afa\beta})
  =\begin{pmatrix}
     -2 & 0 \\
     0 & \frac12
   \end{pmatrix}.
\]
It can be verified that
\begin{align*}
&\frac{1}{E-\lmd e}
=
\frac{v^1(3\rme^{v^2}+v^1-\lmd)\p_{v^1}-(\rme^{v^2}+3v^1-\lmd)\p_{v^2}}{(v^1-\rme^{v^2})D},\\
&\frac12G^{\afa\beta}
\pfrac{p_\afa}{\lmd}*\pfrac{p_\beta}{\lmd}
=
-\frac{v^1\rme^{v^2}}{D^2},
\end{align*}
where $D=(\lmd-v^1-\rme^{v^2})^2-4v^1\rme^{v^2}$.
Hence the loop equation \eqref{reduced loop equation}
reads
\begin{align*}
&
  \sum_{s\geq 0}
  \left(
    \pfrac{\Delta\mcalF}{v^{1,s}}
    \p_x^s\frac{v^1(3\rme^{v^2}+v^1-\lmd)}{(v^1-\rme^{v^2})D}
   -\pfrac{\Delta\mcalF}{v^{2,s}}
    \p_x^s\frac{\rme^{v^2}+3v^1-\lmd}{(v^1-\rme^{v^2})D}
  \right)
\\
&\quad\quad
   -\frac{1}{\lmd}
    \sum_{s\geq 1}
    s\left(
      \pfrac{\Delta\mcalF}{v^{1,s}}
      \p_x^s\frac{v^1}{v^1-\rme^{v^2}}
     -\pfrac{\Delta\mcalF}{v^{2,s}}
      \p_x^s\frac{1}{v^1-\rme^{v^2}}
    \right)
\\
&\quad\quad
   -\frac{1}{2\lmd}
   \sum_{s\geq 1}
   \sum_{k=1}^{s}
    \binom{s}{k}
     \left(
       \p_x^{k-1}\frac{\lmd+v^1-\rme^{v^2}}{\sqrt{D}}
     \right)
     \left(
       \pfrac{\Delta\mcalF}{v^{1,s}}
       \p_x^{s+1-k}
       \frac{\lmd-v^1-\rme^{v^2}}{\sqrt{D}}
      +\pfrac{\Delta\mcalF}{v^{2,s}}
       \p_x^{s+1-k}\frac{2}{\sqrt{D}}
     \right)\\
&\quad=
  -\frac{1}{4}\veps^2
  \sum_{k,\ell\geq 0}
  \left[
    \left(
      \pfrac{\Delta\mcalF}{v^{1,k}}
      \pfrac{\Delta\mcalF}{v^{1,\ell}}
     +\frac{\p^2\Delta\mcalF}{\p v^{1,k}\p v^{1,\ell}}
    \right)
    \p_x^{k+1}\frac{\lmd-v^1-\rme^{v^2}}{\sqrt{D}}
    \cdot
    \p_x^{\ell+1}\frac{\lmd-v^1-\rme^{v^2}}{\sqrt{D}}
  \right.
\\
&\quad\quad
  \left.+
    4\left(
      \pfrac{\Delta\mcalF}{v^{1,k}}
      \pfrac{\Delta\mcalF}{v^{2,\ell}}
     +\frac{\p^2\Delta\mcalF}{\p v^{1,k}\p v^{2,\ell}}
    \right)
    \p_x^{k+1}\frac{\lmd-v^1-\rme^{v^2}}{\sqrt{D}}
    \cdot
    \p_x^{\ell+1}\frac{1}{\sqrt{D}}
  \right.
\\
&\quad\quad
  \left.+
    4\left(
      \pfrac{\Delta\mcalF}{v^{2,k}}
      \pfrac{\Delta\mcalF}{v^{2,\ell}}
     +\frac{\p^2\Delta\mcalF}{\p v^{2,k}\p v^{2,\ell}}
    \right)
    \p_x^{k+1}\frac{1}{\sqrt{D}}
    \cdot
    \p_x^{\ell+1}\frac{1}{\sqrt{D}}
   \right]
\\
&\quad\quad
  -\veps^2
  \sum_{k\geq 0}
  \left[
    \pfrac{\Delta\mcalF}{v^{1,k}}
    \p_x^{k+1}
    \frac{e^{v^2}\left(
      Kv^1_x-2v^1\rme^{v^2}
      \left(
        (\rme^{v^2}-v^1)^2-\lmd^2
      \right)v^2_x
    \right)}
    {D^3}
  \right.
\\
&\quad\quad
  \left.
    -\pfrac{\Delta\mcalF}{v^{2,k}}
    \p_x^{k+1}
    \frac{e^{v^2}\left(
      2\left(
        (\rme^{v^2}-v^1)^2-\lmd^2
      \right)v^1_x-Kv^2_x
    \right)}{D^3}
  \right]
  -\frac{v^1\rme^{v^2}}{D^2},
\end{align*}
where
\[
   K=
  (v^1+\rme^{v^2})D+8\lmd v^1\rme^{v^2}.\]
The loop equation has a unique solution
\[\Delta\mcalF=\sum_{k\ge 1}\veps^{k-2}\mcalF^{[k]}
=\sum_{g\ge 1}\veps^{2g-2} \mcalF_g(v^1,v^2, v^1_x, v^2_x,\dots, v^{1,3g-2}, v^{2,3g-2})\]
up to the addition of a constant series $b_1\veps^{-1}+b_2+b_3\veps+\dots$.
The coefficient of $\veps^{0}$ of the loop equation yields the following equation for $\mcalF_1$:
\begin{align*}
 & \pfrac{\mcalF_1}{v^1}
  \frac{v^1(3\rme^{v^2}+v^1-\lmd)}{(v^1-\rme^{v^2})D}
 -\pfrac{\mcalF_1}{v^2}
  \frac{\rme^{v^2}+3v^1-\lmd}{(v^1-\rme^{v^2})D}
+
  \pfrac{\mcalF_1}{v^1_x}
  \p_x \frac{v^1(3\rme^{v^2}+v^1-\lmd)}{(v^1-\rme^{v^2})D}\\
  &\quad
-
  \pfrac{\mcalF_1}{v^2_x}
  \p_x\frac{\rme^{v^2}+3v^1-\lmd}{(v^1-\rme^{v^2})D}
-\frac1\lmd
 \left(
   \pfrac{\mcalF_1}{v^1_x}
   \p_x\frac{v^1}{v^1-\rme^{v^2}}
  -\pfrac{\mcalF_1}{v^2_x}
   \p_x\frac{1}{v^1-\rme^{v^2}}
 \right)
\\
&\quad
-\frac{\lmd+v^1-\rme^{v^2}}{2\lmd\sqrt{D}}
\left(
  \pfrac{\mcalF_1}{v^1_x}
  \p_x\frac{\lmd-v^1-\rme^{v^2}}{\sqrt{D}}
 +\pfrac{\mcalF_1}{v^2_x}
  \p_x\frac{2}{\sqrt{D}}
\right)
+\frac{v^1\rme^{v^2}}{D^2}=0,
\end{align*}
from which it follows that
\[
  \mcalF_1=
  \frac{1}{24}\log
  \left(
    (v^1_x)^2-v^1\rme^{v^2}(v^2_x)^2
  \right)
 +\frac{1}{12}
  \log\left(
    v^1-\rme^{v^2}
  \right)
 -\frac18\log v^1-\frac{1}{24}v^2.
\]
This function coincides, up to the addition of a constant, with the following genus one free energy for the equivariant Gromov-Witten invariants of the resolved conifold with anti-diagonal action \cite{Brini-1}:
\[\hat{\mathcal{F}}_1=\frac1{24}\log\left(v'(x)^2+\frac{\lmd^2 \rme^{w(x)}}{1-\rme^{w(x)}} w'(x)^2\right)+\frac1{12}\mathrm{Li}_1\left(\rme^{w(x)}\right)-\frac{w(x)}{24}\]
after putting $\lmd=1$ and taking the change of variables
\[v^1=\rme^v (\rme^w-1),\quad v^2=v+w,\]
which relates the flat coordinates of the Frobenius manifold associated with the Ablowitz-Ladik hierarchy and its almost dual \cite{Brini-2}.

Now let us consider the relation of the deformed Principal Hierarchy with the positive flows of the Ablowitz-Ladik hierarchy \cite{AL-95, AL-96, Brini-2, AL-intro-SXL, Suris} defined by
\begin{equation}\label{zh-41}
\frac{\p L}{\p t_k}=\frac1{(k+1)!}[(L^{k+1})_+, L],\quad k\ge 0,
\end{equation}
where the Lax operator $L$ has the expression
\begin{align*}
L&=(1-Q\clm^{-1})^{-1}(\clm-P)\\
&=\clm+Q-P+Q(\clm-Q)^{-1}(Q-P)
\end{align*}
with $\Lambda=e^{\ve\p_x}$.
The first and the second positive flows of the Ablowitz-Ladik hierarchy have the expressions \cite{AL-intro-SXL}
\begin{align}
\ve P_{t_0}&=P(Q^+-Q),\quad \ve Q_{t_0}=Q (Q^+-Q^--P+P^-). \label{RTL-flow}\\
\ve P_{t_1}&=\frac12 P(PQ-PQ^++P^-Q-P^+Q^++Q^+Q^{++}+Q^+Q^+\notag\\
&\quad-QQ^--Q^2),\\
\ve Q_{t_1}&=\frac12Q(P^2-P^-P^--P^+Q^+-2PQ^+-PQ+P^-Q+2P^-Q^-+P^{--}Q^-\notag\\
&\quad
 +Q^+Q^{++}+Q^+Q^++QQ^+-Q^-Q^{--}-QQ^--Q^-Q^-).
\end{align}
Here we use the notion $P^\pm=\Lambda^{\pm} P$, $Q^\pm=\Lambda^{\pm} Q$.

\begin{rmk}
  The system of equations given in \eqref{RTL-flow} is called the relativistic Toda lattice in the literature \cite{lqp-RTL,Suris}.
We thank Qing Ping Liu for pointing out to us this fact, and informing us the existence of a tri-Hamiltonian structure for this system given by Ovel, Fuchssteiner, Zhang and Ragnisco in \cite{lqp-RTL} after we published the paper \cite{AL-intro-SXL}, in which we re-derived the tri-Hamiltonian structure for the positive and negative flows of the Ablowitz-Ladik hierarchy.
The Ablowitz-Ladik hierarchy is bihamiltonian with respect to the bihamiltonian structure $(\mathcal{P}_1, \mathcal{P}_2)$ given by
\[\mathcal{P}_1=\begin{pmatrix} Q\Lambda^{-1}-\Lambda Q& (1-\Lambda)Q\\[5pt] Q(\Lambda^{-1}-1)& 0\end{pmatrix},\quad \mathcal P_2=\begin{pmatrix}0 & P(\clm-1) Q\\[5pt] Q(1-\clm^{-1}) P & Q(\clm-\clm^{-1}) Q\end{pmatrix}.\]
The positive flows of the Ablowitz-Ladik hierarchy can be represented
as bihamiltonian systems as follows:
\[
\begin{pmatrix} \frac{\p P}{\p {t_k}}\\[5pt] \frac{\p Q}{\p {t_k}}\end{pmatrix}=\mathcal{P}_1\begin{pmatrix} \frac{\delta H_{k}}{\delta P}\\[5pt] \frac{\delta H_{k}}{\delta Q}\end{pmatrix}=\frac1{k+1}\mathcal{P}_2\begin{pmatrix} \frac{\delta H_{k-1}}{\delta P}\\[5pt] \frac{\delta H_{k-1}}{\delta Q}\end{pmatrix},\quad k\ge 0.\]
Here the Hamiltonians are given by
\[H_k=\frac{1}{(k+2)!}\int \res L^{k+2}\nd x,\quad k\ge -1.\]
If we introduce the new unknown functions
\begin{equation}\label{zh-2023-5}
U=Q-P,\quad W=\log Q,
\end{equation}
then after representing the Hamiltonian operators $\mathcal{P}_1$ and $\mathcal{P}_2$ in terms of $U, W$ and taking the hydrodynamic limit
we obtain
\[\mathcal{P}_1^{[0]}=\begin{pmatrix}0& \p_x\\ \p_x&0\end{pmatrix},\quad
\mathcal{P}_2^{[0]}=\begin{pmatrix}2 v^1 \rme^{v^2}\p_x+(v^1 \rme^{v^2})'& (v^1+\rme^{v^2})\p_x\\
(v^1+\rme^{v^2})\p_x+(v^1+\rme^{v^2})'& 2\p_x\end{pmatrix},
\]
where we replace $U, W$ by $v^1, v^2$ respectively. We note that
$\bigl(\mathcal{P}_1^{[0]}, \mathcal{P}_2^{[0]}\bigr)$ is the bihamiltonian structure
for the Principal Hierarchy of the generalized Frobenius manifold $M$, and $(\mathcal{P}_1, \mathcal{P}_2)$ is a deformation of $\bigl(\mathcal{P}_1^{[0]}, \mathcal{P}_2^{[0]}\bigr)$ with central invariants $c_1=c_2=\frac1{24}$, as it is shown in \cite{AL-intro-SXL}.
\end{rmk}

Now let us denote
\begin{align}
w^1=&\,U-\frac{\ve}2 U_x+\frac{\ve^2}{12} U_{xx},\label{zh-45}\\
w^2=&\,W-\frac{\ve}{2}\p_x\log U+\frac{\ve^2}{12}\frac{\p}{\p x}\left(\frac{2 U_x-(2 \rme^W+U) W_x}{U}\right)\notag\\
&\,+\frac{\ve^3}{12}\frac{\p}{\p x}\left(\frac{(U U_{xx}-U_x^2)\rme^W}{U^3}\right),\label{zh-46}
\end{align}
then the first positive flow of the Ablowitz-Ladik hierarchy can be represented in the form
\begin{align*}
 \frac{\p w^1}{\p t_0}
=&\,
  \rme^{w^2}w^1_x+\rme^{w^2}w^1 w^2_x
\\
&\,
  +\frac{\ve^2 \rme^{w^2}}{24 (w^1)^2}
  \left(
    5 (w^1_x)^3+4 \rme^{w^2} (w^1_x)^2 w^2_x-5 w^1 (w^1_x)^2 w^2_x
    -8 \rme^{w^2} w^1 w^1_x (w^2_x)^2
  \right.\\
&\,
  +8 \rme^{w^2} (w^1)^2 (w^2_x)^3-10 w^1 w^1_x w^1_{xx}
  -4 \rme^{w^2} w^1 w^2_x w^1_{xx}+6 (w^1)^2 w^2_x w^1_{xx}\\
&\,
  -4 \rme^{w^2} w^1 w^1_x w^2_{xx}+2 (w^1)^2 w^1_x w^2_{xx}
+16 \rme^{w^2} (w^1)^2 w^2_x w^2_{xx}+2 (w^1)^3 w^2_x w^2_{xx}\\
&\,
  \left.+6 (w^1)^2 w^1_{xxx}
  +4 \rme^{w^2} (w^1)^2 w^2_{xxx}+2 (w^1)^3 w^2_{xxx}\right)+O(\ve^4),\\
\frac{\p w^2}{\p t_0}
=&\,
  w^1_x+\rme^{w^2} w^2_x\\
&\,
  +\frac{\ve^2 \rme^{w^2}}{24 (w^1)^3}
  \left(6 (w^1_x)^3-3 w^1 (w^1_x)^2 w^2_x-8 w^1 w^1_x w^1_{xx}+2 (w^1)^2 w^2_x w^1_{xx}\right.\\
&\,\left.+2 (w^1)^3 w^2_x w^2_{xx}+2 (w^1)^2 w^1_{xxx}+2 (w^1)^3 w^2_{xxx}\right)+O(\ve^4).
\end{align*}
This flow coincides, at the approximation up to $\ve^3$, with the
$\frac{\p}{\p t^{2,0}}$-flow of the deformed Principal Hierarchy \eqref{deformed PH}, where the unknown functions $w^1, w^2$ are related to the unknown functions $v^1, v^2$ of the Principal Hierarchy by the quasi-Miura transformation \eqref{quasi-Miura transf}, i.e.,
\begin{equation}\label{zh-52}
w^1=\ve^2\frac{\p^2\log\tau}{\p x\p t^{2,0}}=v^1+\ve^2\frac{\p^2\Delta\mcalF}{\p x\p t^{2,0}},\quad
w^2=\ve^2\frac{\p^2\log\tau}{\p x\p t^{1,0}}=v^2+\ve^2\frac{\p^2\Delta\mcalF}{\p x\p t^{1,0}},
\end{equation}
where the $\tau$ function is defined by
\[\log\tau=\ve^{-2} f+\sum_{g\ge 1}\ve^{2g-2}\mcalF_g=\ve^{-2} f+\Delta\mcalF.\]
In a similar way, we can verify such a relation of the second positive flow $\frac{\p}{\p t_1}$ of the Ablowitz-Ladik hierarchy with the $\frac{\p}{\p t^{2,1}}$-flow of the deformed Principal Hierarchy.

We also obtain the explicit expression of the function $\mcalF_2$ which is quite long, so we do not present here. By using the expression of $\mcalF_2$ and by adding certain $\ve^4$-terms at the right-hand sides of \eqref{zh-45}, \eqref{zh-46}, we check the validity of the above-mentioned relation of the first two positive flows of the Ablowitz-Ladik hierarchy and the $\frac{\p}{\p t^{2,p}}$-flows ($p=0,1$) of the topological deformation of the Principal Hierarchy at the approximation up to $\ve^4$. Actually, we conjecture that under the Miura-type transformation given by the relation
\begin{align}
U&=\sum_{k=0}^\infty \frac{\ve^k}{(k+1)!} \p_x^k w^1,\label{zh-49}\\
W&=\left(\sum_{k=0}^\infty (-1)^k \frac{\ve^k}{(k+1)!}\left(\frac{\p}{\p x}\right)^k\right) \left(\sum_{\ell=0}^\infty \frac{\ve^\ell}{(\ell+1)!}\left(\frac{\p}{\p t^{1,0}}\right)^\ell\right) w^2,\label{zh-50}
\end{align}
the positive flows $\frac{\p}{\p t_k}$ of the Ablowitz-Ladik hierarchy \eqref{zh-41} coincide with the flows $\frac{\p}{\p t^{2,k}}$ of the topological deformation of the Principal Hierarchy. In other words, we have the following conjecture.
\begin{conj}
Let $\tau$ be a tau function of the topological deformation of the Principal Hierarchy of the generalized Frobenius manifold defined by \eqref{zh-42}, \eqref{zh-43}, then the following special two-point functions
\begin{equation}\label{zh-47}
U=\ve(\Lambda-1)\frac{\p\log\tau}{\p t^{2,0}},\quad
W=\left(1-\Lambda^{-1}\right)\left(\rme^{\ve\frac{\p}{\p t^{1,0}}}-1\right)\log\tau
\end{equation}
satisfy the positive flows of the Ablowitz-Ladik hierarchy \eqref{zh-41}
if we identify $t^{2,k}$ with $t_k$ for $k\ge 0$. Here $\Lambda=\rme^{\ve \frac{\p}{\p x}}$, and $U, W$ are related with $P, Q$ by \eqref{zh-2023-5}.
\end{conj}

From \eqref{zh-52} we see that the relations \eqref{zh-49}, \eqref{zh-50} follow from \eqref{zh-47}. This representation of the unknown functions $U, W$ in terms the tau function was obtained by D. Yang and C. Zhou in their study of the relation between the equivariant Gromov-Witten invariants of the resolved conifold with anti-diagonal action and the Ablowitz-Ladik hierarchy \cite{Yang-Zhou}. A proof of this conjecture is recently given in \cite{LWZ-3}.
\end{ex}

\section{Conclusion}
In this paper we study the relationship between a class of generalized Frobenius manifolds with non-flat unit vector fields and integrable hierarchies. For any such generalized Frobenius manifold, we construct an analogue of the Principal Hierarchy of a usual Frobenius manifold which contains infinitely many additional flows. We show that this Principal Hierarchy has a tau-cover, and possesses Virasoro symmetries which can be lifted to this tau-cover. The condition of linearization of the actions of the Virasoro symmetries on the tau function enables us to derive the loop equation for the generalized Frobenius manifold, which has a unique solution in the semsisimple case as it is shown in \cite{LWZ-2}. This solution of the loop equation yields a quasi-Miura transformation which gives the topological deformation of the Principal Hierarchy
of the semisimple generalized Frobenius manifold.

We will study the bihamiltonian structure of the topological deformation of the Principal Hierarchy of a semisimple generalized Frobenius manifold in subsequent publications.
We know that the quasi-Miura transformation that relates the Principal Hierarchy and its topological deformation transforms the bihamiltonian structure of the Principal Hierarchy
to that of its topological deformation,
the main problem is whether this deformed bihamiltonian structure possesses the polynomiality property, i.e., whether it can be represented in terms of differential polynomials of the unknown functions of the deformed integrable hierarchy. For a usual semisimple Frobenius manifold with flat unity the polynomiality of the first deformed Hamiltonian structure is proved in \cite{shadrin-1, shadrin-2}; the polynomiality of the deformed bihamiltonian structure is studied in \cite{shadrin-3, LWZ-1} and is proved
in \cite{LWZ-1}. Since this proof depends essentially on the flatness of the unit vector field of the Frobenius manifold, it can not be applied directly to the proof of the polynomiality of the deformed bihamiltonian structure for a semisimple generalized Frobenius manifold with non-flat unity. Nonetheless, we conjecture that the deformed bihamiltonian structure of a semisimple generalized Frobenius manifold also possesses the polynomiality property, and its central invariants (see the definition given in \cite{DLZ-06}) are all equal to $\frac1{24}$.

We will also study more examples of generalized Frobenius manifolds with non-flat unity and their relations to integrable hierarchies. Notably, even in the 1-dimensional case we have other examples apart from the one that is given in the present paper, which may have close relations to some important integrable hierarchies. For example, we have
the $1$-dimensional generalized Frobenius manifold with
\[
  F=\rme^v,\quad
  E=\p_v,\quad
  e=\rme^{-v}\p_v,
\]
its charge and flat metric are given by $d=2$ and $\eta=1$, and
its monodromy data at $z=0$ are given by $\mu=R=0$.
The Hamiltonian densities $\theta_{i,p}$ of the Principal Hierarchy have the expressions
\begin{align*}
  \theta_{0,p}
&=
  \begin{cases}
    \frac12 v^2, & p=1,\\
    \frac{1}{(p-1)!(p-1)}(v-H_{p-1})\rme^{(p-1)v}, & p\geq 2;
  \end{cases}\\[3 pt]
\theta_{0,-p}
&=
  \frac{(-1)^{p+1}p!}{p+1}\rme^{-(p+1)v},\qquad p\geq 0;\\[3 pt]
\theta_{1,p}
&=
  \begin{cases}
    v, & p=0,\\
    \frac{1}{p!p}\rme^{pv}, & p\geq 1.
  \end{cases}
\end{align*}
where $H_p=1+\frac12+\cdots+\frac 1p$. The loop equation has a solution
$\Delta\mcalF=\mcalF_1+\veps^2\mcalF_2+\cdots$ with
\begin{align*}
  \mcalF_1&=\frac{1}{24}\log v_x
  +\frac{1}{12}v,\\
\mcalF_2&=
 \frac{49 \rme^v v_x^2}{5760}
+\frac{\rme^v v_{xx}}{64}
-\frac{11 \rme^v v_{xx}^2}{1920 v_x^2}
+\frac{\rme^v v_{xx}^3}{360 v_x^4}
+\frac{37 \rme^v v_{xxx}}{5760 v_x}
-\frac{7 \rme^v v_{xx} v_{xxx}}{1920 v_x^3}
+\frac{\rme^v v_{xxxx}}{1152 v_x^2}.
\end{align*}
We will study the topological deformation of the corresponding Principal Hierarchy in subsequent works.

Before we list other examples of 1-dimensional generalized Frobenius manifolds, let us first introduce the following notion.
\begin{defn}
Two generalized Frobenius manifolds $M$ and $M'$ with non-flat unity are called equivalent if there exists a smooth or analytic homeomorphism $h\colon M\to M'$
  such that
  \[
    h_* (E)=E',\quad
    h^*(\eta')=k_1\eta, \quad
    h^*(c')=k_2 c
  \]
  for some nonzero constants $k_1, k_2$,
  where $E,\eta$ and $E',\eta'$
  are the Euler vector fields, the flat metrics of $M$ and $M'$ respectively, and the 3-tensors $c, c'$ of $M$ and $M'$ are defined as in Definition \ref{zh-38}.
  \end{defn}

From the above definition, it follows that if
$M, M'$ are equivalent, then their charges are equal, i.e., $d=d'$, and
$h_*(e)=\frac{k_2}{k_1}e'$. We can classify the 1-dimensional generalized Frobenius manifolds with non-flat unity and get the following result.
\begin{prop}
There are four equivalence classes of 1-dimensional generalized Frobenius manifolds with non-flat unity, and their representatives are given in the following table:
\[
  \begin{tabular}{m{1.8cm} m{3.2cm} m{2.1cm} m{4.0cm}}
    \toprule
   \textrm{charge}  & \textrm{Euler vector field}
    & \textrm{potential}  & \textrm{unit vector field}  \\[3pt]
    \toprule
    $d=2$ & $E=\p_v$ & $F=\mathrm e^{v}$ & $e=\mathrm e^{-v}\p_v$ \\[3pt]
    \midrule
    $d=3$ & $E=-\frac12v\p_v$ & $F=\log v$ & $e=\frac12v^3\p_v$ \\[3pt]
    \midrule
    $d=4$ & $E=-v\p_v$ & $F=v\log v$ & $e=-v^2\p_v$ \\[3pt]
    \midrule
    $d\neq 2,3,4$ & $E=\frac{2-d}{2}v\p_v$ & $F=v^{\frac{6-2d}{2-d}}$ &     $e=-\frac14\frac{(d-2)^3}{(d-3)(d-4)}v^{\frac{d}{d-2}}\p_v$ \\[3pt]
    \bottomrule
  \end{tabular}
\]
The flat metrics of these generalized Frobenius manifolds are given by $\eta=1$.
\end{prop}

We remark that the 1-dimensional generalized Frobenius manifold with $d=3$ and $F=\log v$ arises in the study of properties of a negative spin version of Witten's $r$-spin class, and its relation to integrable hierarchies \cite{CGG}.

Finally, we remark that the genus zero free energy $\mcalF_0$ defined by \eqref{genus 0 free energy} satisfies the genus zero Virasoro constraints
\begin{align*}
 & \sum_{(i,p),(j,q)\in\mcalI}
  a^{i,p;j,q}_m
  \pfrac{\mcalF_0}{t^{i,p}}
  \pfrac{\mcalF_0}{t^{j,q}}
 +\sum_{(i,p),(j,q)\in\mcalI}
  b^{j,q}_{m;i,p}\ttil^{i,p}\pfrac{\mcalF_0}{t^{j,q}}
\\
&\qquad
 +
  \sum_{(i,p),(j,q)\in\mcalI}
    c_{m;i,p;j,q}\ttil^{i,p}\ttil^{j,q}
 +
  \sum_{(i,p),(j,q)\in\mcalI}
    \delta_{i,0}
    \delta_{j,0}
    C_{m;p,q}\ttil^{i,p}\ttil^{j,q}=0
\end{align*}
for $m\geq -1$, this fact can be easily proved by using
the formulae \eqref{D Em+ Omg} and the Euler-Lagrange equation \eqref{EL equation}. For a semisimple generalized Frobenius manifold $M$, we have a unique solution
\[\Delta\mcalF=\sum_{k\ge 1}\veps^{k-2}\mcalF^{[k]}
=\sum_{g\ge 1}\veps^{2g-2} \mcalF_g(v, v_x,\dots, v^{(3g-2)}),\]
up to the addition of a constant series $\sum_{g\ge 1} a_g \ve^{2g-2}$, of the loop equation. Then from the condition of linearization of the Virasoro symmetries of the Principal Hierarchy of $M$ it follows that
the tau function
\[\tau=\exp{(\ve^{-2}\mcalF_0+\Delta\mcalF)}\]
satisfies the Virasoro constraints
\[
  L_m(\veps^{-1}\tilde\bft,\veps\pp{\bft})\tau=0,\quad m\geq -1.
\]

\medskip

\bmhead{Acknowledgments}
This work was supported by NSFC No.\,12171268 and No.\,11725104. The authors would like to thank Yuewei Wang for pointing out to them the construction of the $(n+2)$-dimensional Frobenius manifold from an $n$-dimensional generalized Frobenius manifold, and sharing with them her result that is given in Proposition \ref{zh-20}. They would also like to thank Di Yang, Chunhui Zhou and Zhe Wang for very helpful discussions on this work.
\vskip 0.3truecm

\noindent\textbf{Data availability statement} Data sharing not applicable to this article as no datasets were generated or analysed during the current study.
\vskip 0.3truecm

\noindent\textbf{Declarations}
\vskip 0.3truecm
\noindent\textbf{Conflict of interest} The authors have no competing interests to declare that are relevant to the content of this article.

\medskip
\appendix
\section{Proof of Theorem \ref{thm:G-nu as Gram} and Lemma \ref{lemma: Laplace-Bilinear identity}}
\label{Appendix-A}

In this appendix we are to prove Theorem \ref{thm:G-nu as Gram}
and Lemma \ref{lemma: Laplace-Bilinear identity} respectively.

Firstly, Theorem \ref{thm:G-nu as Gram} is equivalent to the validity of the following formula
\begin{equation}\label{Qu230811-1258}
  P^{(-\nu)\transpose}\eta(\mcalU-\lmd I)P^{(\nu)}
=
\left(
  -\frac{1}{2\pi}
  \left(
    \rme^{\pi\rmi R}
    \rme^{\pi\rmi (\mu+\nu)}
 +
    \rme^{-\pi\rmi R}
    \rme^{-\pi\rmi (\mu+\nu)}
  \right)\eta^{-1}
\right)^{-1},
\end{equation}
where $P^{(\nu)}=P^{(\nu)}(v;\lmd)$ is defined as in \eqref{def of P^(nu)}.

To simplify the right-hand side of \eqref{Qu230811-1258},
we introduce the following matrix $R^*$, which is called the $\eta$-transpose of $R$:
\begin{equation} \label{R-star}
  R^*:=R_1-R_2+R_3-R_4+\cdots = \sum_{s\geq 1}(-1)^{s+1}R_s,
\end{equation}
then \eqref{zh-3} can be rewritten as $R^{\transpose}\eta = \eta R^*$.
\begin{lem} \label{lemma:240727-1453}
For any meromorphic function $f(z)$ which satisfies $f(z+1)=-f(z)$, the following identity holds true:
\[
  \left(\rme^{R\p_\nu}f(\mu+\nu)\right)
  \left(\rme^{-R^*\p_\nu}\frac{1}{f(\mu+\nu)}\right)
=I,
\]
where $I$ is the identity matrix.
\end{lem}

\begin{proof}
 It is easy to verify (or see \cite{normal-form, Virasoro-Like}) the following identities satisfied by $\mu$ and $R$:
  \begin{align}
    \left[(R^*)^p\right]_\ell &=\,(-1)^{p+\ell}(R^p)_\ell,  \label{Qu230811-1423-1}\\
    f(\mu+\nu)(R^p)_\ell&=\,(R^p)_\ell f(\mu+\nu+\ell)  \label{Qu230811-1423-2}
  \end{align}
  for $p,\ell\geq 0$, therefore we have
 \begin{align*}
  &
    \left(
      \rme^{R\p_\nu}f(\mu+\nu)
    \right)
    \left(
      \rme^{-R^*\p_\nu}\frac{1}{f(\mu+\nu)}
    \right)\\
 &\quad =
    \sum_{k,\ell\geq 0}
      \left[\rme^{R\p_\nu}\right]_kf(\mu+\nu)
      \cdot
      \left[\rme^{-R^*\p_\nu}\right]_\ell\frac{1}{f(\mu+\nu)}\\
 &\quad =
    \sum_{k,\ell\geq 0}
    \sum_{p,q\geq 0}
      \frac{(R^p)_k}{p!}\p^p_\nu f(\mu+\nu)
      \cdot
      \frac{\left[(-R^*)^q\right]_\ell}{q!}\p_\nu^q \frac{1}{f(\mu+\nu)}\\
 &\quad =
    \sum_{p,q\geq 0}\frac{1}{p!q!}
      \sum_{k,\ell\geq 0}
        (-1)^{\ell}(R^p)_k\p_\nu^pf(\mu+\nu)\cdot
        (R^q)_\ell\p^q_\nu\frac{1}{f(\mu+\nu)}\\
  &\quad =
    \sum_{p,q\geq 0}\frac{1}{p!q!}
      \left(
      \sum_{k,\ell\geq 0}
        (R^p)_k(R^q)_\ell\right)
        \p_\nu^pf(\mu+\nu)\cdot
        \p^q_\nu\frac{1}{f(\mu+\nu)}\\
  &\quad =
    \sum_{s\geq 0}\frac{R^s}{s!}\p^s_\nu
    \left(f(\mu+\nu)\cdot\frac{1}{f(\mu+\nu)}\right)
  =
    \rme^{R\p_\nu}I=I,
  \end{align*}
the lemma is proved.
\end{proof}

In particular, by taking $f(z)=\cos \pi z$ we obtain
\begin{equation}\label{230329-0032}
\left(\rme^{R\p_\nu}\cos\left(\pi(\mu+\nu)\right)\right)^{-1}=
  \rme^{-R^*\p_\nu}\frac{1}{\cos\left(\pi(\mu+\nu)\right)}.
\end{equation}
So we can get a simplified expression of the right-hand side of \eqref{Qu230811-1258} by using the identity
\begin{equation} \label{230329-0031}
    \rme^{\pi\rmi R}
    \rme^{\pi\rmi (\mu+\nu)}
 +
    \rme^{-\pi\rmi R}
    \rme^{-\pi\rmi (\mu+\nu)}
=
  2\rme^{R\p_\nu}\cos\left(\pi(\mu+\nu)\right).
\end{equation}

Let us proceed to simplify the left-hand side of \eqref{Qu230811-1258}.
We first introduce the following notations:
\begin{align}\label{zh-2024-5}
\Gamma_k(z+\mu+\nu)&:=\left[\rme^{R\p_\nu}\right]_k
      \Gamma( z+\mu+\nu),\\
       \tilde\Gamma_k(z+\mu-\nu)&:=\Gamma\left(z+\mu-\nu\right)
      \left[\rme^{-R^{\transpose}\overleftarrow{\p_\nu}}\right]_k , \label{zh-2024-6}
      \end{align}
where the action of the right differential operator $\overleftarrow{\p_\nu}$ on $f(\nu)$
is defined by
\[f(\nu)\overleftarrow{\p_\nu}:=\p_\nu f(\nu).\]
Then by using \eqref{periods-Lapacle-2},
  \eqref{basic Laplace formula} and \eqref{def of P^(nu)} we arrive at
\begin{align}
     P^{(\nu)}
  &=
    \sum_{p\geq 0}
    \int \rme^{-\lmd z}\Theta_p
    z^{p+\mu+\nu-\frac12}z^R\td z \notag\\
  &=
    \sum_{p\geq 0}\Theta_p
    \sum_{s\geq 0}
      \Gamma_s\left( p+\mu+\nu+s+\frac12\right)
       \lmd^{-\left(p+\mu+\nu+s+\frac12\right)}
      \lmd^{-R} \notag \\
  &=
    \sum_{p\geq 0}
              \sum_{k=0}^{p}
          \Theta_{p-k}
         \Gamma_k\left(p+\mu+\nu+\frac12\right)
         \lmd^{-\left( p+\mu+\nu+\frac12\right)}
      \lmd^{-R}, \label{P^nu}
\end{align}
here $\Theta_p:=(\nabla\theta_{1,p},...,\nabla\theta_{n,p})$ for $p\geq 0$.
\begin{lem}The following identity holds true:
\begin{align}\label{(U-lmdI)P}
 & (\mcalU-\lmd I)P^{(\nu)}
=
  -\left(\mu+\nu-\frac12\right)
   \sum_{p\geq 0}
   \sum_{k=0}^{p}
   \Theta_{p-k}
   \Gamma_k\left(p+\mu+\nu-\frac12\right)
   \notag\\
&\qquad
  \times \lmd^{-\left(p+\mu+\nu-\frac12\right)}\lmd^{-R}.
\end{align}
\end{lem}
\begin{proof}
By using \eqref{zh-12} we obtain the identity
\begin{align}
  &  (\mcalU-\lmd I)\Theta_p
  =
    \p_E\Theta_{p+1}-\lmd\Theta_p \notag\\
  &\quad=
    (p+1-\mu)\Theta_{p+1}
   +\sum_{s=0}^{p+1}\Theta_{p+1-s}B_s-\lmd\Theta_p,  \label{Qu230811-1429}
  \end{align}
  where
  \[B_s:=\begin{cases}
  \mu, & s=0,\\
  R_s, & s>0.
\end{cases}\]
We also obtain from \eqref{Qu230811-1423-1} and \eqref{Qu230811-1423-2} the following identity:
\begin{align}
 & \Gamma_k\left(p+\mu+\nu+\frac12\right)
=
  \left(p+\nu-k-\frac12\right)
    \Gamma_k\left(p+\mu+\nu-\frac12\right) \notag\\
&\qquad
 +\sum_{s=0}^k
    B_s
    \Gamma_{k-s}\left(p+\mu+\nu-\frac12\right).  \label{e^RGamma}
\end{align}
Then by using \eqref{P^nu}, \eqref{Qu230811-1429} and \eqref{e^RGamma} we obtain
\begin{align*}
  &
    (\mcalU-\lmd I)P^{(\nu)}\\
&\quad=
    \sum_{p\geq 0}
      \sum_{k=0}^{p}
        \left(
          (p+1-k-\mu)\Theta_{p+1-k}
         +\sum_{s=0}^{p+1-k}\Theta_{p+1-k-s}B_s
         \right)
  \\
  &\quad\quad
     \times \Gamma_k\left(p+\mu+\nu+\frac12\right)
        \lmd^{-\left(p+\mu+\nu+\frac12\right)}\lmd^{-R}
  \\
  &\quad\quad
    -\sum_{p\geq 0}\sum_{k=0}^p\Theta_{p-k}
     \Gamma_k\left(p+\mu+\nu+\frac12\right)
     \lmd^{-\left(p+\mu+\nu-\frac12\right)}\lmd^{-R}
  \\
&\quad=
  \sum_{p\geq 0}
      \sum_{k=0}^{p}
        \left(
          (p+1-k-\mu)\Theta_{p+1-k}
         +\sum_{s=0}^{p+1-k}\Theta_{p+1-k-s}B_s
         \right)
  \\
  &\quad\quad
     \times\Gamma_k\left(p+\mu+\nu+\frac12\right)
       \lmd^{-\left(p+\mu+\nu+\frac12\right)}\lmd^{-R}
  \\
  &\quad\quad
    -
  \left(
     \sum_{p\geq 0}\sum_{k=0}^p\Theta_{p-k}
       \left(p+\nu-k-\frac12\right)
          \Gamma_k\left(p+\mu+\nu-\frac12\right)\right.
  \\
  &\quad \quad
    +
     \left.
     \sum_{p\geq 0}\sum_{k=0}^{p}\sum_{s=0}^{k}\Theta_{p-k}B_s
     \Gamma_{k-s}\left(p+\mu+\nu-\frac12\right)
     \right)
      \lmd^{-\left(p+\mu+\nu-\frac12\right)}\lmd^{-R}
  \\
&\quad=
  -\left(\mu+\nu-\frac12\right)
   \sum_{p\geq 0}
   \sum_{k=0}^{p}
   \Theta_{p-k}
   \Gamma_k\left(p+\mu+\nu-\frac12\right)
  \lmd^{-\left(p+\mu+\nu-\frac12\right)}\lmd^{-R}.
  \end{align*}
The lemma is proved.
\end{proof}

\begin{lem}\label{lemma:Qu-A3}
We have the following formula:
\begin{equation}
   P^{(-\nu)\transpose}\eta(\mcalU-\lmd I)P^{(\nu)} = W_1+W_2,
\end{equation}
where
\begin{align}
   & W_1 :=\,
      -
  \sum_{p,p'\geq 0}
    \lmd^{-R^{\transpose}}
      \tilde\Gamma_p\left(p+\mu-\nu+\frac12\right)
       \notag \\
&\quad\times
    \eta
    \left(\mu+\nu-\frac12\right)
  \Gamma_{p'}\left(p'+\mu+\nu-\frac12\right)
  \lmd^{-R}\\
&W_2:=
-\sum_{p,p'\geq 0}
     \sum_{k=0}^{p-1}
     \sum_{k'=0}^{p'-1}
       \lmd^{-R^{\transpose}}
      \tilde \Gamma_k\left(p+\mu-\nu+\frac12\right)
\notag\\
  &\quad
    \times
    (-1)^{p-k+1}\eta R_{p-k+p'-k'}
        \Gamma_{k'}\left(p'+\mu+\nu-\frac12\right)
         \lmd^{-R}.
\end{align}
\end{lem}
\begin{proof}
From \eqref{P^nu} and \eqref{(U-lmdI)P} it follows that
\begin{align}
  &
    P^{(-\nu)\transpose}\eta(\mcalU-\lmd I)P^{(\nu)} \notag\\
  &\quad=
    -\sum_{p,p'\geq 0}
    \frac{1}{\lmd^{p+p'}}
     \sum_{k=0}^{p}
     \sum_{k'=0}^{p'}
       \lmd^{-R^{\transpose}}\lmd^{-\mu}
       \tilde\Gamma_k\left(p+\mu-\nu+\frac12\right)
  \notag\\
  &\quad\quad
    \times \Theta^{\transpose}_{p-k}\eta\left(\mu+\nu-\frac12\right)
    \Theta_{p'-k'}
       \Gamma_{k'}\left(p'+\mu+\nu-\frac12\right)
       \lmd^{-\mu}\lmd^{-R}.\label{zh-23-08-2}
\end{align}
For $p,q\geq 0$, the following formulae for $\Theta_p$ and $\Omg_{\afa,p;\beta,q}$
are well known
(see details in \cite{Frob mfd and Vir const}, or \eqref{nabla-theta pair}, \eqref{Theta eta-mu lemma} in Appendix \ref{Appendix-B}):
\begin{align}
  \Theta_p^{\transpose}\eta\Theta_q
&=\,
  \Omg_{p-1,q}+\Omg_{p,q-1}+\eta\delta_{p,0}\delta_{q,0},\\
  \Theta_p^{\transpose}\eta\mu\Theta_q
&=\,
  -p\Omg_{p,q-1}+q\Omg_{p-1,q}
  -\sum_{s=0}^{p}B_s^{\transpose}\Omg_{p-s,q-1}
  +\sum_{s=0}^{q}\Omg_{p-1,q-s}B_s
\notag\\
&\quad
  +(-1)^{p+1}\eta R_{p+q}\chi_{p\geq 1,q\geq 1}
  +\eta\mu\delta_{p,0}\delta_{q,0},\label{zh-23-08-3}
\end{align}
here we use the matrix-valued function
\[\Omg_{p,q}=(\Omg_{\afa,p;\beta,q})_{1\leq\afa,\beta\leq n},\quad \Omg_{-1,p}=\Omg_{p,-1}=0,\quad p,q\geq 0,\]
and the symbol
\[
  \chi_{p\geq 1, q\geq 1}:=
    \begin{cases}
      1, &  \text{if $p\geq 1$ and $q\geq 1$},\\
      0, &  \text{other cases}.
    \end{cases}
\]
Then by using \eqref{zh-23-08-2}, \eqref{zh-23-08-3} we obtain
\begin{align}
&
   P^{(-\nu)\transpose}\eta(\mcalU-\lmd I)P^{(\nu)}  \notag\\
&\quad=
  \sum_{p,p'\geq 0}\frac{1}{\lmd^{p+p'}}
  \sum_{k=0}^{p}
  \sum_{k'=0}^{p'}
    \lmd^{-R^{\transpose}}\lmd^{-\mu}
      \tilde\Gamma_k\left(p+\mu-\nu+\frac12\right)  \notag \\
&\quad\quad \times
  \left(
    \left(p-k-\nu+\frac12\right)
      \Omg_{p-k,p'-k'-1}
    +\sum_{s=0}^{p-k}
       B_s^{\transpose}\Omg_{p-k-s,p'-k'-1}
  \right.  \notag\\
&\quad\quad
   \left.
    -\left(p'-k'+\nu-\frac12\right)\Omg_{p-k-1,p'-k'}
    -\sum_{s=0}^{p'-k'}\Omg_{p-k-1,p'-k'-s}B_s
   \right)
\notag\\
&\quad\quad\times
  \Gamma_{k'}\left(p'+\mu+\nu-\frac12\right)\lmd^{-\mu}\lmd^{-R}
\notag\\
&\quad\quad -
  \sum_{p,p'\geq 0}\frac{1}{\lmd^{p+p'}}
    \lmd^{-R^{\transpose}}\lmd^{-\mu}
      \tilde\Gamma_p\left(p+\mu-\nu+\frac12\right)  \notag\\
&\quad\quad\times
    \eta
    \left(\mu+\nu-\frac12\right)
  \Gamma_{p'}\left(p'+\mu+\nu-\frac12\right)\lmd^{-\mu}\lmd^{-R}
\notag \\
&\quad\quad
-\sum_{p,p'\geq 0}
    \frac{1}{\lmd^{p+p'}}
     \sum_{k=0}^{p-1}
     \sum_{k'=0}^{p'-1}
       \lmd^{-R^{\transpose}}\lmd^{-\mu}
       \tilde\Gamma_k\left(p+\mu-\nu+\frac12\right)
\notag\\
  &\quad\quad
    \times (-1)^{p-k+1}\eta R_{p-k+p'-k'}
       \Gamma_{k'}\left(p'+\mu+\nu-\frac12\right)
       \lmd^{-\mu}\lmd^{-R}
\notag \\
&\quad=
    \sum_{p,p'\geq 0}\frac{1}{\lmd^{p+p'}}
  \sum_{k=0}^{p}
  \sum_{k'=0}^{p'}
    \lmd^{-R^{\transpose}}\lmd^{-\mu}
\notag\\
&\quad\quad
    \left(
     \tilde\Gamma_k\left(p+\mu-\nu+\frac32\right)
           \Omg_{p-k,p'-k'-1}
         \Gamma_{k'}\left(p'+\mu+\nu-\frac12\right)
    \right.
    \notag\\
&\quad\quad \left.-
       \tilde\Gamma_k\left(p+\mu-\nu+\frac12\right)
            \Omg_{p-k-1,p'-k'}
       \Gamma_{k'}\left(p'+\mu+\nu+\frac12\right) \right)
\lmd^{-\mu}\lmd^{-R}
\notag\\
&\quad \quad-
  \sum_{p,p'\geq 0}\frac{1}{\lmd^{p+p'}}
    \lmd^{-R^{\transpose}}\lmd^{-\mu}\times
      \tilde\Gamma_p\left(p+\mu-\nu+\frac12\right)
   \notag\\
&\quad\quad\times
    \eta
    \left(\mu+\nu-\frac12\right)
  \Gamma_{p'}\left(p'+\mu+\nu-\frac12\right)\lmd^{-\mu}\lmd^{-R}
\notag \\
&\quad\quad
-\sum_{p,p'\geq 0}
    \frac{1}{\lmd^{p+p'}}
     \sum_{k=0}^{p-1}
     \sum_{k'=0}^{p'-1}
       \lmd^{-R^{\transpose}}\lmd^{-\mu}
       \tilde\Gamma_k\left(p+\mu-\nu+\frac12\right)\notag\\
   &\quad\quad
    \times    (-1)^{p-k+1}\eta R_{p-k+p'-k'}
           \Gamma_{k'}\left(p'+\mu+\nu-\frac12\right)
       \lmd^{-\mu}\lmd^{-R}
\notag \\
&\quad=\, W_1+W_2.
\label{230328-1335}
\end{align}
The lemma is proved.
\end{proof}

\begin{lem}
The following identity holds true:
\begin{equation}\label{230329-0034}
  P^{(-\nu)\transpose}\eta(\mcalU-\lmd I)P^{(\nu)}
=
  -\pi\lmd^{-R^{\transpose}}\eta
  \left(
    \rme^{-R^*\p_\nu}
    \frac{1}{\cos\left(\pi(\mu+\nu)\right)}
  \right)\lmd^{-R},
\end{equation}
here $R^*$ is defined in \eqref{R-star}.
\end{lem}
\begin{proof}
It is straightforward to verify the validity of the identity
\begin{align}
      &\,\p_\nu^\ell\Gamma\left(-\mu-\nu-p+\frac12\right)\cdot\p^{\ell'}_\nu
      \Gamma\left(p+\mu+\nu+\frac12\right)  \notag\\
    &\quad =
      (-1)^p\p^\ell_\nu\Gamma\left(-\mu-\nu+\frac12\right)
      \cdot\p^{\ell'}_\nu\Gamma\left(\mu+\nu+\frac12\right) \notag\\
    &\quad\quad
      +\ell\sum_{k=1}^{p}(-1)^k
        \p^{\ell-1}_\nu\Gamma\left(-\mu-\nu-p+k-\frac12\right)
        \cdot
        \p^{\ell'}_\nu\Gamma\left(p+\mu+\nu-k+\frac12\right) \notag\\
    &\quad\quad
      -\ell'\sum_{k=1}^{p}(-1)^k
        \p^{\ell}_\nu\Gamma\left(-\mu-\nu-p+k-\frac12\right)
        \cdot
        \p^{\ell'-1}_\nu\Gamma\left(p+\mu+\nu-k+\frac12\right)  \label{230328-1349}
\end{align}
for all non-negative integers $\ell,\ell'$ and $p$.
Introduce the notations
\begin{equation}
  R_{k;\ell}:=(R^\ell)_k,\quad (-R^*)_{k;\ell}:=\left[(-R^*)^\ell\right]_k=(-1)^{k}R_{k;\ell},
\end{equation}
then by using \eqref{230328-1349} we can rewrite the term $W_1$ which is introduced in Lemma \ref{lemma:Qu-A3} as follows:
\begin{align*}
W_1 =&\,
  -\sum_{p,p'\geq 0}\lmd^{-R^{\transpose}}\eta
  \left[\rme^{-R^*\p_\nu}\right]_p
  \Gamma\left(-\mu-\nu+\frac12\right)
  \\
&\,
  \times\left(\mu+\nu+\frac12\right)
    \Gamma_{p'}\left(p'+\mu+\nu-\frac12\right)\lmd^{-R}
\\
=&\,
  -\lmd^{-R^{\transpose}}\eta
   \sum_{p,p'\geq 0}
   \sum_{\ell,\ell'\geq 0}
     \frac{(-1)^{p'}}{\ell!\ell'!}
     (-R^*)_{p;\ell}(-R^*)_{p';\ell'}
\\
&\,
  \times\p^\ell_{\nu}\Gamma\left(-\mu-\nu-p'+\frac12\right)
  \cdot \left(\mu+\nu+p'-\frac12\right)
  \cdot \p_\nu^{\ell'}\Gamma\left(p'+\mu+\nu-\frac12\right)\lmd^{-R}
\\
=&\,
  -\lmd^{-R^{\transpose}}\eta
   \sum_{p,p'\geq 0}
   \sum_{\ell,\ell'\geq 0}
     \frac{(-1)^{p'}}{\ell!\ell'!}
     (-R^*)_{p;\ell}(-R^*)_{p';\ell'}
\\
&\,
  \times
  \p^\ell_{\nu}\Gamma\left(-\mu-\nu-p'+\frac12\right)
  \cdot \p_\nu^{\ell'}\Gamma\left(p'+\mu+\nu+\frac12\right)
  \lmd^{-R}
\\
&\,  +\lmd^{-R^{\transpose}}\eta
   \sum_{p,p'\geq 0}
  \sum_{{\ell\geq 0, \ell'\geq 1}}
     \frac{(-1)^{p'}}{\ell!(\ell'-1)!}
     (-R^*)_{p;\ell}(-R^*)_{p';\ell'}
\\
&\,
  \times \p^\ell_{\nu}\Gamma\left(-\mu-\nu-p'+\frac12\right)
  \cdot \p_\nu^{\ell'-1}\Gamma\left(p'+\mu+\nu-\frac12\right)\lmd^{-R}
\\
=&\,
  -\lmd^{-R^{\transpose}}\eta
  \sum_{p,p'\geq 0}
    \sum_{\ell,\ell'\geq 0}
      \frac{1}{\ell!\ell'!}
        (-R^*)_{p;\ell}(-R^*)_{p';\ell'}
\\
&\,
  \times
  \p_\nu^\ell\Gamma\left(-\mu-\nu+\frac12\right)
  \cdot \p_\nu^{\ell'}\Gamma\left(\mu+\nu+\frac12\right)
  \lmd^{-R}
\\
&\,
   -\lmd^{-R^{\transpose}}\eta
  \sum_{p,p'\geq 0}
   \sum_{{\ell\geq 1, \ell'\geq 0}}
      \frac{(-1)^{p'}}{(\ell-1)!\ell'!}
        (-R^*)_{p;\ell}(-R^*)_{p';\ell'}
\\
&\,
  \times \sum_{m=1}^{p'}
    (-1)^m
    \p_\nu^{\ell-1}\Gamma\left(-\mu-\nu-p'+m-\frac12\right)
    \cdot \p_\nu^{\ell'}\Gamma\left(p'+\mu+\nu-m+\frac12\right)
    \lmd^{-R}
\\
&\,
  +\lmd^{-R^{\transpose}}\eta
  \sum_{p,p'\geq 0}
   \sum_{{\ell\geq 0, \ell'\geq 1}}
      \frac{(-1)^{p'}}{\ell!(\ell'-1)!}
        (-R^*)_{p;\ell}(-R^*)_{p';\ell'}
\\
&\,
  \times\sum_{m=1}^{p'}
    (-1)^m
    \p_\nu^{\ell}\Gamma\left(-\mu-\nu-p'+m-\frac12\right)
    \cdot \p_\nu^{\ell'-1}\Gamma\left(p'+\mu+\nu-m+\frac12\right)
    \lmd^{-R}
\\
&\,
  +\lmd^{-R^{\transpose}}\eta
  \sum_{p,p'\geq 0}
   \sum_{{\ell\geq 0, \ell'\geq 1}}
      \frac{(-1)^{p'}}{\ell!(\ell'-1)!}
        (-R^*)_{p;\ell}(-R^*)_{p';\ell'}
\\
&\,
  \times
    \p_\nu^{\ell}\Gamma\left(-\mu-\nu-p'+\frac12\right)
    \cdot \p_\nu^{\ell'-1}\Gamma\left(p'+\mu+\nu-\frac12\right)
    \lmd^{-R}
\\
=&\,
  -\lmd^{-R^{\transpose}}\eta
  \left(
  \rme^{-R^*\p_\nu}\frac{\pi}{\cos\left(\pi(\mu+\nu)\right)}
  \right)
  \lmd^{-R}
\\
&\,
  +\lmd^{-R^{\transpose}}\eta
   \sum_{\ell,\ell'\geq 0}
     \frac{1}{\ell!\ell'!}
     \sum_{p,p'\geq 0}
         (-R^*)_{p+1;\ell+1}
         (-R^*)_{p'+1;\ell'}
\\
&\,
  \times\sum_{m=0}^{p'}
    (-1)^{m+1}
      \p_\nu^\ell\Gamma\left(-\mu-\nu-m-\frac12\right)
      \cdot
      \p_\nu^{\ell'}\Gamma\left(\mu+\nu+m+\frac12\right)
  \lmd^{-R}
\\
&\,
   +\lmd^{-R^{\transpose}}\eta
   \sum_{\ell,\ell'\geq 0}
     \frac{1}{\ell!\ell'!}
     \sum_{p,p'\geq 0}
         (-R^*)_{p;\ell}
         (-R^*)_{p'+2;\ell'+1}
\\
&\,
  \times\sum_{m=0}^{p'}
    (-1)^{m}
      \p_\nu^\ell\Gamma\left(-\mu-\nu-m-\frac12\right)
      \cdot
      \p_\nu^{\ell'}\Gamma\left(\mu+\nu+m+\frac12\right)
      \lmd^{-R}
\\
=&\,
  -\lmd^{-R^{\transpose}}\eta
  \left(
  \rme^{-R^*\p_\nu}\frac{\pi}{\cos\left(\pi(\mu+\nu)\right)}
  \right)
  \lmd^{-R}
\\
&\,
  +\lmd^{-R^{\transpose}}\eta
  \sum_{\ell,\ell'\geq 0}
   \sum_{m\geq 0}
     \frac{(-1)^{m+1}}{\ell!\ell'!}
\\
&\,\times
     \sum_{p,p'\geq 0}
       \Big(
         (-R^*)_{p+1;\ell+1}(-R^*)_{p'+m+1;\ell'}
        -(-R^*)_{p;\ell}(-R^*)_{p'+m+2;\ell'+1}
       \Big)
\\
&\,\times
  \p^\ell_\nu\Gamma\left(-\mu-\nu-m-\frac12\right)
  \cdot\p^{\ell'}_\nu\Gamma\left(\mu+\nu+m+\frac12\right)
  \lmd^{-R}.
\end{align*}
In a similar way, we can rewrite the term $W_2$ introduced in Lemma \ref{lemma:Qu-A3} as follows:
\begin{align*}
W_2=&\,
\lmd^{-R^{\transpose}}\eta
  \sum_{\ell,\ell'\geq 0}
   \sum_{m\geq 0}
     \frac{(-1)^{m+1}}{\ell!\ell'!}
       \left(
         \sum_{p\geq 0}
           \sum_{k\geq 0}
             \sum_{k'=0}^{m}
               (-R^*)_{k;\ell}
               (-R^*)_{p+m+2-k'}
               (-R^*)_{k';\ell'}
       \right)
\\
&\,\times
  \p^\ell_\nu\Gamma\left(-\mu-\nu-m-\frac12\right)
  \cdot\p^{\ell'}_\nu\Gamma\left(\mu+\nu+m+\frac12\right)
  \lmd^{-R},
\end{align*}
Therefore, in order to prove the lemma we only need to show the validity of the following identities for all $\ell,\ell'\geq 0$ and $m\geq 0$:
\begin{align}
0=&\,
  \sum_{p,p'\geq 0}
       \Big(
         (-R^*)_{p+1;\ell+1}(-R^*)_{p'+m+1;\ell'}
        -(-R^*)_{p;\ell}(-R^*)_{p'+m+2;\ell'+1}
       \Big)
\notag\\
&\,
  +
  \sum_{p\geq 0}
           \sum_{k\geq 0}
             \sum_{k'=0}^{m}
               (-R^*)_{k;\ell}
               (-R^*)_{p+m+2-k'}
               (-R^*)_{k';\ell'}, \label{230329-0000}
\end{align}
which can be verified by a straightforward calculation.
The lemma is proved.
\end{proof}

We are now ready to prove Theorem \ref{thm:G-nu as Gram}.
\begin{proof}[Proof of Theorem \ref{thm:G-nu as Gram}]
From the relation $[\mu, R_k]=kR_k$ for $k\geq 1$ it follows the identities
\begin{align}
  \rme^{\pi\rmi\mu}\lmd^R
&=
  \left(
    \rme^{\pi\rmi\mu}\lmd^R\rme^{-\pi\rmi\mu}
  \right)\rme^{\pi\rmi\mu}
=
 \left(\lmd^{\exp\left(\mathrm{ad}\,\pi\rmi\mu\right)\sum_{k\geq 1}R_k}\right)
 \rme^{\pi\rmi\mu} \notag\\
&=
 \left( \lmd^{\sum_{k\geq 1}\rme^{\pi\rmi k}R_k}\right)\rme^{\pi\rmi\mu}
=
  \lmd^{-R^*}\rme^{\pi\rmi\mu} \label{240727-1400-1}
\end{align}
and
\[\rme^{-\pi\rmi\mu}\lmd^R=\lmd^{-R^*}\rme^{-\pi\rmi\mu}.\]
Then by using \eqref{230329-0031}, \eqref{230329-0032} and \eqref{230329-0034} we arrive at
\begin{align}
&\,
  -\frac{1}{2\pi}
  \left(
    \rme^{\pi\rmi R}\rme^{\pi\rmi(\mu+\nu)}
   +\rme^{-\pi\rmi R}\rme^{-\pi\rmi(\mu+\nu)}
  \right)\eta^{-1}
  P^{(-\nu)\transpose}\eta(\mcalU-\lmd I)P^{(\nu)}
\notag\\
&\qquad =
  \frac12
  \left(
    \rme^{\pi\rmi R}\rme^{\pi\rmi(\mu+\nu)}
   +\rme^{-\pi\rmi R}\rme^{-\pi\rmi(\mu+\nu)}
  \right)\eta^{-1}
  \lmd^{-R^{\transpose}}\eta\left(
  \rme^{-R^*\p_\nu}
    \frac{1}{\cos\left(\pi(\mu+\nu)\right)}
  \right)\lmd^{-R}  \notag\\
&\qquad =
  \frac12\lmd^R
    \left(
    \rme^{\pi\rmi R}\rme^{\pi\rmi(\mu+\nu)}
   +\rme^{-\pi\rmi R}\rme^{-\pi\rmi(\mu+\nu)}
  \right)
  \left(\rme^{-R^*\p_\nu}
    \frac{1}{\cos\left(\pi(\mu+\nu)\right)}
  \right)\lmd^{-R} \notag\\
&\qquad =
  \lmd^R
  \left(
    \rme^{R\p_\nu}\cos\left(\pi(\mu+\nu)\right)
    \cdot
    \rme^{-R^*\p_\nu}\frac{1}{\cos\left(\pi(\mu+\nu)\right)}
  \right)
  \lmd^{-R}
=
  \lmd^R\lmd^{-R}=I, \label{240727-1400-2}
\end{align}
therefore the formula \eqref{Qu230811-1258} holds true.
Theorem \ref{thm:G-nu as Gram} is proved.
\end{proof}

Now, let us give a proof of Lemma \ref{lemma: Laplace-Bilinear identity}.
\begin{proof}[Proof of Lemma \ref{lemma: Laplace-Bilinear identity}]
By using the definition \eqref{240727-1107} of $\phi^{(\nu)}(\bfa;\lmd)$ and the integral formula \eqref{basic Laplace formula} we obtain
\begin{align*}
  &\pfrac{\phi^{(\nu)}(\bfa;\lmd)}{\lmd}
=
  -\int_{0}^{\infty}
    \frac{\td z}{\sqrt{z}}
    \sum_{p\in\bbZ}
      \bfa_p^\transpose \eta z^{p+\mu+\nu+1}z^R \\
&\quad=
  -\sum_{p\in\bbZ}\bfa_p^\transpose\eta
    \sum_{s\geq 0}
    \Gamma_s\left(p+\mu+\nu+s+\frac 32\right)
    \lmd^{\left(p+\mu+\nu+s+\frac 32\right)}\lmd^{-R}.
\end{align*}
Similarly, by using \eqref{zh-1}, \eqref{zh-3} and \eqref{R-star} we arrive at
\begin{align*}
  &\pfrac{\phi^{(-\nu)}(\bfa;\lmd)}{\lmd}
=
  -\sum_{q\in\bbZ}\bfa_q^\transpose\eta
   \sum_{s\geq 0}
     \left[\rme^{-R\p_{\nu}}\right]_s
   \Gamma\left(q+\mu-\nu+s+\frac 32\right)
   \cdot
   \lmd^{-\left(q+\mu-\nu+s+\frac 32\right)}
   \lmd^{-R} \\
&\quad=
  -\sum_{q\in\bbZ}\bfa_q^\transpose
   \sum_{s\geq 0}
     \left[\rme^{-(R^*)^\transpose\p_{\nu}}\right]_s
   \Gamma\left(q-\mu-\nu+s+\frac 32\right)
   \cdot
   \lmd^{-\left(q-\mu-\nu+s+\frac 32\right)}
   \lmd^{-(R^*)^\transpose}\eta .
\end{align*}
Similar to the derivation of \eqref{240727-1400-1}--\eqref{240727-1400-2}, we have
\begin{align*}
  &
  \lmd^{-R}\rme^{R\p_\nu}\cos\left(\pi(\mu+\nu)\right)\lmd^{-R^*}
=
  \lmd^{-R}
  \frac{\rme^{\pi\rmi R}\rme^{\pi\rmi(\mu+\nu)}+\rme^{-\pi\rmi R}\rme^{-\pi\rmi(\mu+\nu)}}
      {2}\lmd^{-R^*} \\
&\quad=
  \frac{\rme^{\pi\rmi R}\rme^{\pi\rmi(\mu+\nu)}+\rme^{-\pi\rmi R}\rme^{-\pi\rmi(\mu+\nu)}}
      {2}\lmd^{R^*} \lmd^{-R^*}
=
  \rme^{R\p_\nu}\cos\left(\pi(\mu+\nu)\right).
\end{align*}
Thus, the left-hand side of \eqref{Laplace-bilinear identity} reads
\begin{align*}
& -\frac12
  \pfrac{\phi_\afa^{(\nu)}(\bfa;\lmd)}{\lmd}
  G^{\afa\beta}(\nu)
  \pfrac{\phi_\beta^{(-\nu)}(\bfa;\lmd)}{\lmd} \\
&\quad =
  \frac{1}{2\pi}
  \pfrac{\phi^{(\nu)}(\bfa;\lmd)}{\lmd}
  \left(
    \rme^{R\p_\nu}\cos\left(\pi(\mu+\nu)\right)
  \right)\eta^{-1}
  \left(
    \pfrac{\phi^{(-\nu)}(\bfa;\lmd)}{\lmd}
  \right)^\transpose \\
&\quad =
  \frac{1}{2\pi}
  \sum_{p,q\in\bbZ}\sum_{s,k\geq 0}
    \bfa_p^\transpose\eta
       \Gamma_s\left(p+\mu+\nu+s+\frac 32\right)
    \lmd^{-\left(p+\mu+\nu+s+\frac 32\right)}\lmd^{-R} \\
&\qquad
   \times \left(\rme^{R\p_\nu}\cos\left(\pi(\mu+\nu)\right)\right) \eta^{-1}
   \cdot \eta \lmd^{-R^*}
   \lmd^{-\left(q-\mu-\nu+k+\frac 32\right)} \\
&\qquad
   \times \Gamma\left(q-\mu-\nu+k+\frac 32\right)
   \left[
     \rme^{-R^*\overleftarrow{\p_\nu}}
   \right]_k \bfa_q \\
&\quad =
  \frac{1}{2\pi}
  \sum_{p,q\in\bbZ}\sum_{s,k\geq 0}
  \frac{1}{\lmd^{p+q+s+k+3}}
  \bfa_p^\transpose\eta
      \Gamma_s\left(p+\mu+\nu+s+\frac 32\right) \\
&\qquad
  \times \lmd^{-\mu}
  \left(
    \rme^{R\p_\nu}\cos\left(\pi(\mu+\nu)\right)
  \right) \lmd^\mu
  \cdot
  \left[
    \rme^{-R^*\p_\nu}
  \right]_k
  \Gamma\left(q-\mu-\nu+\frac 32\right) \bfa_q \\
&\quad =
  \frac{1}{2\pi}
  \sum_{p,q\in\bbZ}\sum_{s,k\geq 0}
  \frac{1}{\lmd^{p+q+s+k+3}}
  \bfa_p^\transpose\eta
      \Gamma_s\left(p+\mu+\nu+s+\frac 32\right) \\
&\qquad
  \times\sum_{\ell\geq 0}
  \left[
    \rme^{R\p_\nu}
  \right]_\ell\cos\left(\pi(\mu+\nu)\right)\cdot
  \lmd^{-(\mu+\ell)}\cdot
   \lmd^\mu
  \cdot
  \left[
    \rme^{-R^*\p_\nu}
  \right]_k
  \Gamma\left(q-\mu-\nu+\frac 32\right) \bfa_q \\
&\quad =
  \frac{1}{2\pi}
  \sum_{p,q\in\bbZ}\sum_{s\geq 0}
    \frac{1}{\lmd^{p+q+s+3}}
    \bfa_p^\transpose\eta
        \Gamma_s\left(p+\mu+\nu+s+\frac 32\right) \\
&\qquad
  \times\left(
    \sum_{k,\ell\geq 0}
    \frac{1}{\lmd^{k+\ell}}
    \left[\rme^{R\p_\nu}\right]_\ell
    \cos\left(\pi(\mu+\nu)\right)\cdot
    \left[
      \rme^{-R^*\p_\nu}
    \right]_k
    \Gamma\left(q-\mu-\nu+\frac 32\right)
  \right) \bfa_q.
\end{align*}
By using the same method as we do in the proof of Lemma \ref{lemma:240727-1453}, we can verify that
\begin{align*}
  &
    \sum_{k,\ell\geq 0}
    \frac{1}{\lmd^{k+\ell}}
    \left[\rme^{R\p_\nu}\right]_\ell
    \cos\left(\pi(\mu+\nu)\right)\cdot
    \left[
      \rme^{-R^*\p_\nu}
    \right]_k
    \Gamma\left(q-\mu-\nu+\frac 32\right) \\
&\quad=
  \sum_{k\geq 0}
    \frac{1}{\lmd^k}
    \left[\rme^{R\p_\nu}\right]_k
    \left(
      \cos\left(\pi(\mu+\nu)\right)
      \Gamma\left(
        q-\mu-\nu+\frac 32
      \right)
    \right).
\end{align*}
Therefore we obtain
\begin{align*}
&
  -\frac12
  \pfrac{\phi_\afa^{(\nu)}(\bfa;\lmd)}{\lmd}
  G^{\afa\beta}(\nu)
  \pfrac{\phi_\beta^{(-\nu)}(\bfa;\lmd)}{\lmd} \\
&\quad=
  \frac{1}{2\pi}\sum_{p,q\in\bbZ}
  \frac{1}{\lmd^{p+q+3}}
  \sum_{s,k\geq 0}
  \frac{1}{\lmd^{s+k}}
    \bfa_p^\transpose\eta
    \left[
      \rme^{R\p_\nu}
    \right]_s
    \Gamma\left(q-\mu-\nu+\frac 32\right) \\
&\qquad
  \times \left[\rme^{R\p_\nu}\right]_k
  \left(
    \cos\left(\pi(\mu+\nu)\right)
      \Gamma\left(
        q-\mu-\nu+\frac 32
      \right)
  \right) \bfa_q \\
&\quad=
  \frac{1}{2\pi}
  \sum_{p,q\in\bbZ}\sum_{s\geq 0}
  \frac{1}{\lmd^{p+q+s+3}}
  \bfa_p^\transpose\eta \\
&\qquad\times
  \left[\rme^{R\p_\nu}\right]_s
  \left(
    \Gamma\left(p+\mu+\nu+s+\frac 32\right)
    \cos\left(\pi(\mu+\nu)\right)
    \Gamma\left(q-\mu-\nu+\frac 32\right)
  \right)\bfa_q \\
&\quad=
  \frac12\sum_{p,q\in\bbZ}\sum_{s\geq 0}
  \frac{1}{\lmd^{p+q+s+3}}
  \bfa_p^\transpose\eta
  \mathrm{N}_{p,q}(s;\nu)\bfa_q,
\end{align*}
hence Lemma \ref{lemma: Laplace-Bilinear identity} is proved.
\end{proof}

\section{Proof of Theorem \ref{main-lemma-orig}}\label{Appendix-B}

In this appendix, we prove Theorem \ref{main-lemma-orig}.
From the definition \eqref{defn:Stil_i} of $\Stil^{(\nu)}_i$
and the relation \eqref{formal Laplace bilinear identity}
between the Laplace-type integrals and the extended Virasoro coefficients,
it follows that
\begin{align*}
&
  \lim_{\nu\to 0}
  \left[
    \left(
    \pp{t^{i,p}}
    \pfrac{\Stil^{(\nu)}_{i'}}{\lmd}
    \right)
  \Gtil^{i'j'}(\nu)
    \left(
      \pp{t^{j,q}}
      \pfrac{\Stil_{j'}^{(-\nu)}}{\lmd}
    \right)
  \right]_-
 -
  2\delta_{i,0}\delta_{j,0}C_{p,q}(\lmd)
\\
&\quad=
  -2\sum_{(i',p'),(j',q')\in\mcalI}
  a^{i',p';j',q'}(\lmd)
  \Omg_{i,p;j',q'}\Omg_{i',p';j,q}\\
&\qquad
  -\sum_{(i',p'),(j',q')\in\mcalI}
  \left(
    b^{i',p'}_{j',q'}(\lmd)
   \Omg_{i,p;i',p'}\delta^{j',q'}_{j,q}
  + b_{i',p'}^{j',q'}(\lmd)
   \Omg_{j',q';j,q}\delta^{i',p'}_{i,p}
  \right)\\
&\qquad
  -2\sum_{(i',p'),(j',q')\in\mcalI}
  c_{i',p';j',q'}(\lmd)
  \delta^{i',p'}_{i,p}\delta^{j',q'}_{j,q}
 -
  2\delta_{i,0}\delta_{j,0}C_{p,q}(\lmd),
\\
&\quad=
  -2\sum_{(k,r),(l,s)\in\mcalI}
  a^{k,r;l,s}(\lmd)
  \Omg_{i,p;k,r}\Omg_{l,s;j,q}
\\
&\qquad
-
 \sum_{(k,r)\in\mcalI}
 \Big(
   b_{i,p}^{k,r}(\lmd)
   \Omg_{k,r;j,q}
  +b^{k,r}_{j,q}(\lmd)
   \Omg_{i,p;k,r}
 \Big)
-
 2c_{i,p;j,q}(\lmd)
-
  2\delta_{i,0}\delta_{j,0}C_{p,q}(\lmd),
\end{align*}
so the identity \eqref{220720-D_E-Omg final form} of Theorem \ref{main-lemma-orig}
is equivalent to the following identities:
\begin{align}
 & \p_{E^{m+1}}
  \Omg_{i,p;j,q}
=
  2\sum_{(k,r),(l,s)\in\mcalI}
  a^{k,r;l,s}_m
  \Omg_{i,p;k,r}\Omg_{l,s;j,q}\notag
\\
&\qquad
+
 \sum_{(k,r)\in\mcalI}
 \Big(
   b_{m;i,p}^{k,r}
   \Omg_{k,r;j,q}
  +b^{k,r}_{m;j,q}\Omg_{i,p;k,r}\notag
 \Big)
\\
&\qquad
+
 2c_{m;i,p;j,q}
+2\delta_{i,0}\delta_{j,0}C_{m;p,q} \label{D Em+ Omg}
\end{align}
for $(i,p), (j,q)\in\mcalI$ and $m\geq -1$,
where $a^{i,p;j,q}_m$, $b^{j,q}_{m;i,p}$,
$c_{m;i,p;j,q}$ and $C_{m,p,q}$ are the extended Virasoro coefficients
given by Definition \ref{defn:ext Vira coef}, and they are explicitly given for $m=-1,0,1,2$ by the coefficients of $L_m$ that are presented in \eqref{explicit L-1}--\eqref{explicit L2}.
From the commutation relations
\[
  [E^{m+1}, E^{k+1}]=(k-m)E^{m+k+1},\quad
  \forall\, m,k\geq -1
\]
that is proved in \cite{Manin-Hertling} and \eqref{vira commu final}, it follows that we only need to show the validity of the identities
\eqref{D Em+ Omg} for $m=-1,0,1,2$.
We can prove this fact in a similar way as it is done in \cite{Frob mfd and Vir const} for a usual Frobenius manifold.
The main subtlety that we need to deal with for a generalized Frobenius manifold is in the computation of the derivatives of the functions $\Omg_{0,p;\afa,q}$ and $\Omg_{0,p;0,q}$ along powers of the Euler vector field.

To simplify the presentation of the proof of \eqref{D Em+ Omg}, we first introduce some notations. Define the
$(n+1)\times (n+1)$ matrices
\[
  \muhat=
  \begin{pmatrix}
    \mu_0 &  \\
      & \mu
  \end{pmatrix},\quad
  \Rhat_s=
  \begin{pmatrix}
    0 &  \bm{0} \\
    \bm{r}_s & R_s
  \end{pmatrix},\quad s\geq 1,
\]
where $\mu_0=-\frac d2$, and denote
$\Rhat=\sum_{s\geq 1}\Rhat_s$.
We also introduce the $n\times (n+1)$ matrix-valued functions
\[
  \Thetahat_p(z)=
  \sum_{k\geq 0}
  \Big(
  \nabla\theta_{0,p+k}, \nabla\theta_{1,p+k},\dots,\nabla\theta_{n,p+k}
  \Big)z^k,\quad p\in\bbZ,
\]
here $\theta_{i,p}=0$ if $i\neq 0$ and $p<0$, and the $n\times n$ matrix-valued function
\[
  \Theta(w)=
  \sum_{k\geq 0}
  \Big(
    \nabla\theta_{1,k},\dots,\nabla\theta_{n,k}
  \Big)w^k.
\]
By using these notations, we can represent the quasi-homogenous conditions
\eqref{zh-12}, \eqref{theta0p-quasi-homog} and \eqref{theta0,-p-quasi-homog} as follows:
\begin{align}\label{D-E Theta}
\p_E\Theta(w)&=
\Theta(w)\rmD_w-\mu\Theta(w),\\
\label{D-E Thetahat}
  \p_E\Thetahat_p(w)
&=\left(
  p-\mu+w\frac{\td}{\td w}
 \right)
 \Thetahat_p(w)
+
 \sum_{l\geq 0}
 \Thetahat_{p-l}(w)\Bhat_l,
\end{align}
where
\[
\Bhat_0=\muhat,\quad \Bhat_l=\Rhat_l \quad\text{for $l\geq 1$},
\]
and the action of the right operator $\rmD_w$
on an $n\times n$ matrix function $X(w)=\sum_{k\geq 0}X_kw^k$ is defined by
\[
  X(w)\rmD_w=w\frac{\td X(w)}{\td w}+X(w)B(w),\quad
  B(w)=\mu+\sum_{s\geq 1}R_sw^s.
\]

On the other hand, from the relations \eqref{zh-11}, \eqref{zh-7} and \eqref{theta0,-p-recur} it follows that
\begin{align}\label{126-1}
\p_Y\Theta(w)&=w\mcalC(Y)\Theta(w),\\
\label{126-2}
  \p_Y\Thetahat_p(w)&=\mcalC(Y)\Thetahat_{p-1}(w),
\end{align}
for any vector field $Y=Y^\afa\p_\afa$,
where $(\mcalC(Y))^\afa_\beta=Y^\veps c_{\veps\beta}^\afa$.
In particular, let $Y$ be the Euler vector field $E$, then we have
\begin{align}\label{D-E Theta to mcalU}
\p_E\Theta(w)&=w\mcalU\Theta(w),\\
\label{D-E Theta-p to mcalU}
  \p_E\Thetahat_p(w)&=\mcalU\Thetahat_{p-1}(w),
\end{align}
where $\mcalU=(\mcalU^\afa_\beta)=\bigl(E^\gamma c^\afa_{\beta\gamma}\bigr)$
is the operator of multiplication by the Euler vector field $E$.

Now we introduce the $(n+1)\times n$ matrix-valued functions
$\Omghat_p(w)$, $p\in\bbZ$, $w\in\bbC$ as follows:
\[
\left(
\Omghat_p(w)
\right)_{i,\beta}=
  \sum_{q\geq 0}\Omg_{i,p;\beta,q}w^q,\quad
  0\leq i\leq n,\, 1\leq \beta\leq n,
\]
here $\Omg_{i,p;\beta,q}= 0$ if $i\neq 0$ and $p<0$.
By using the relation \eqref{nabla-theta-ortho}
we know that the functions $\Omg_{i,p;\beta,q}$ defined in
\eqref{Omg afa p beta q}--\eqref{Omg 0 p beta q} can be represented in the form
\begin{equation}\label{main prop of Omghat-p(w)}
  \Omghat_p(w)=\Thetahat^\transpose_{p+1}(-w)\eta\Theta(w)
  +(-1)^p\chi_{p<0}
  \begin{pmatrix}
    0 \\
    \eta
  \end{pmatrix}w^{-p-1},\quad p\in\bbZ,
\end{equation}
where we use the function
\[
  \chi_{\textrm P}:=
  \begin{cases}
    1, & \textrm{if P is true}, \\
    0, & \textrm{if P is false},
  \end{cases}
\]
for any proposition $\textrm{P}$.
We also denote
\[
\Omgtilhat_p(w)=\Thetahat^\transpose_{p+1}(-w)\eta\Theta(w),\quad p\in\bbZ,
\]
then we have
\[\Omghat_p(w)=\Omgtilhat_p(w)+(-1)^p\chi_{p<0}
  \begin{pmatrix}
    0 \\
    \eta
  \end{pmatrix}w^{-p-1}.\]
Here we emphasize that the wide tilde
used in the above notation does not mean that this notation is for the $(n+2)$-dimensional Frobenius manifold.

In order to prove Theorem \ref{main-lemma-orig}, we need to compute
the following derivatives:
\[
  \p_{E^s}\Omghat_p(w),\quad
  \p_{E^s}\Omg_{0,p;0,q},\quad
  p,q\in\bbZ,\, s=0,1,2,3.
\]
Recall that the derivatives $\p_e\Omg_{i,p; j,q}$ are already
computed in \eqref{D e Omg}.
From the relations \eqref{Omg prop 4} it follows that
\[
  \p_e\Omg_{i,p; j,q}
=
\pair{e}{\nabla\Omg_{i,p; j,q}}
=
\pair{e}{\nabla\theta_{i,p}\cdot\nabla\theta_{j,q}}
=\pair{e\cdot\nabla\theta_{i,p}}{\nabla\theta_{j,q}}
=
\pair{\nabla\theta_{i,p}}{\nabla\theta_{j,q}},
\]
hence we have the identities
\begin{equation}\label{nabla-theta pair}
  \pair{\nabla\theta_{i,p}}{\nabla\theta_{j,q}}
=\Omg_{i,p-1; j,q}+\Omg_{i,p; j,q-1}+
 \eta_{\afa\beta}\delta^\afa_i\delta^\beta_j\delta_{p,0}\delta_{q,0},\quad (i,p), (j,q)\in\mcalI.
\end{equation}
In fact, the above identities hold true for $(i,p),(j,q)\in\{0,1,\dots,n\}\times\bbZ$, since in our conventions we have $\Omg_{\afa,p;j,q}=0$ if $\afa\neq 0$ and $p<0$.

Now let us start to compute $\p_E\Omghat_p(w)$.

\begin{prop} \label{thm D-E Omghat-p}
The following identities hold true:
\[
 \p_E\Omghat_p(w)
=
  \left(
    p+\muhat+\frac12
  \right)
  \Omgtilhat_p(w)
+
  \sum_{s\geq 1}
  \Rhat_s^\transpose
  \Omgtilhat_{p-s}(w)
 +
  \Omgtilhat_p(w)
  \left(
    \rmD_w+\frac12
  \right),\quad p\in\bbZ. \label{D-E Omghat-p}
\]
\end{prop}

\begin{proof}
By using the identities
\eqref{D-E Theta}, \eqref{D-E Thetahat}, \eqref{main prop of Omghat-p(w)} and the relation \eqref{zh-1}, we obtain
\begin{align*}
  &\p_E\Omghat_p(w)
=
  \left(
    \p_E\Thetahat^\transpose_{p+1}(-w)
  \right)\eta\Theta(w)
 +
  \Thetahat^\transpose_{p+1}(-w)
  \eta\p_E\Theta(w)
\\
&\quad=
  \left[
  \left(p+1-w\frac{\td}{\td w}\right)
  \Thetahat^\transpose_{p+1}(-w)
  \right]\eta\Theta(w)
 -
  \Thetahat^\transpose_{p+1}(-w)\mu\eta\Theta(w)
\\
&\qquad
  +\sum_{l\geq 0}
    \Bhat^\transpose_l
    \Thetahat^\transpose_{p+1-l}(-w)\eta\Theta(w)
  +
   \Thetahat_{p+1}^\transpose(-w)
   \eta\Big(
     \Theta(w)\rmD_w-\mu\Theta(w)
   \Big)
\\
&\quad=
  \Big(p+\muhat+1\Big)
  \Theta^\transpose_{p+1}(-w)\eta\Theta(w)
 +\sum_{s\geq 1}
  \Rhat^\transpose_s \Theta^\transpose_{p+1-s}(-w)\eta\Theta(w)
\\
&\qquad
  +\Big[
  \Theta^\transpose_{p+1}(-w)\eta\Theta(w)\Big]\rmD_w
\\
&\quad=
\left(
    p+\muhat+\frac12
  \right)
  \Omgtilhat_p(w)
+
  \sum_{s\geq 1}
  \Rhat_s^\transpose
  \Omgtilhat_{p-s}(w)
+
  \Omgtilhat_p(w)
  \left(
    \rmD_w+\frac12
  \right).
\end{align*}
The proposition is proved.
\end{proof}

We proceed to compute $\p_{E^2}\Omghat_p(w)$.
Denote
\[
  \Thetahat_p=
  \Thetahat_p(w)\big|_{w=0}
 =\Big(\nabla\theta_{0,p}, \nabla\theta_{1,p},\dots,\nabla\theta_{n,p}
  \Big),\quad p\in\bbZ.
\]
Then it is clear that
\begin{align}\label{Theta-p and Theta-p-w}
  &\frac 1w
  \Thetahat^\transpose_{p-1}(-w)
 +\Thetahat^\transpose_p(-w)
 =\frac1w\Thetahat^\transpose_{p-1},\\
\label{D-E Thetahat-p}
&\p_E\Thetahat^\transpose_{p+1}
=
(p+1)\Thetahat^\transpose_{p+1}
-\Thetahat^\transpose_{p+1}\mu
+\sum_{l\geq 0}
\Bhat^\transpose_l\Thetahat^\transpose_{p+1-l},
\end{align}
where $\Bhat_0=\muhat$, $\Bhat_s=\Rhat_s$, $s\geq 1$.
We first prove a useful lemma.
\begin{lem} For $p\in\bbZ$ we have the following identity:
\begin{equation}\label{Theta eta-mu lemma}
  \frac 1w\Thetahat^\transpose_p\eta\mu\Theta(w)
=
  -\Big(p+\muhat\Big)
  \Omgtilhat_p(w)
 -\sum_{s\geq 1}
  \Rhat^\transpose_s\Omgtilhat_{p-s}(w)
 +\left(
  \Omgtilhat_{p-1}(w)
  \frac1w
  \right)\big(\rmD_w+1\big).
\end{equation}
In particular, for $p,q\in\bbZ$, $\afa, \beta\in\{1,\dots,n\}$, we have
\begin{align}
  &\nabla\theta_{0,p+1}^\transpose\eta\mu\nabla\theta_{\beta,q}
=
    \left(q+\mu_{\beta}\right)\Omega_{0,p;\beta,q}
   -\left(
    p+1-\frac d2
    \right) \Omega_{0,p+1;\beta,q-1}\notag\\
  &\quad
    +\sum_{s\geq 1}\Omega_{0,p;\veps,q-s}(R_s)^\veps_\beta
    -\sum_{s\geq 1}
    r^\veps_s\Omega_{\veps,p+1-s;\beta,q-1}\notag\\
  &\quad
    +(-1)^q\chi_{q\geq 1}\chi_{p+q\geq 0}r^\veps_{p+q+1}\eta_{\veps\beta}, \label{0,p eta-mu beta,q}\\
&\nabla\theta_{\afa,p+1}^\transpose\eta\mu\nabla\theta_{\beta,q}
=
    (q+\mu_\beta)\Omega_{\afa,p;\beta,q}
   -(p+1+\mu_\afa)\Omega_{\afa,p+1;\beta,q-1}\notag\\
  &\quad
    +\sum_{s\geq 1}
      (R_s)^\veps_\beta\Omega_{\afa,p;\veps,q-s}
    -\sum_{s\geq 1}
      (R_s)^\veps_\afa\Omega_{\veps,p+1-s;\beta,q-1}\notag\\
  &\quad
    +(-1)^p\eta_{\afa\veps}
     (R_{p+q+1})^\veps_\beta
     \chi_{p\geq 0}
     \chi_{q\geq 1}
     +(\eta\mu)_{\al\beta}\delta_{p,-1}\delta_{q,0}.
    \label{afa,p eta-mu beta,q}
\end{align}
\end{lem}

\begin{proof}
From the identities \eqref{D-E Theta to mcalU}, \eqref{D-E Theta-p to mcalU}, \eqref{Theta-p and Theta-p-w} and the relation
$\mcalU^\transpose\eta=\eta\mcalU$ it follows that
 \begin{align*}
  &\p_E\Omghat_p(w)
  =
  \Thetahat^\transpose_p(-w)
  \mcalU^\transpose\eta\Theta(w)
 +\Thetahat^\transpose_{p+1}(-w)\eta\p_E\Theta(w)
  \\
  &\quad=
  \left[
    \frac1w\Thetahat_p^\transpose(-w)
    +\Thetahat^\transpose_{p+1}(-w)
  \right]\eta
  \p_E\Theta(w)
\\
&\quad=
  \frac1w\Thetahat_p^\transpose
  \eta
  \Big(
    \Theta(w)\rmD_w-\mu\Theta(w)
  \Big)
\\
&\quad=
  -\frac1w\Theta^\transpose_p\eta\mu\Theta(w)
  +\left(\frac1w\Theta_p^\transpose\eta\Theta(w)\right)
   (\rmD_w+1),
  \end{align*}
which yields
\begin{align*}
&\frac1w\Thetahat^\transpose_p\eta\mu\Theta(w)
=
  \left(
    \frac1w\Theta_p^\transpose\eta\Theta(w)
  \right)(\rmD_w+1)
-\p_E\Omghat_p(w)
\\
&\quad=
  \left(
    \frac1w\Omgtilhat_{p-1}(w)+\Omgtilhat_p(w)
  \right) (\rmD_w+1)
-\p_E\Omghat_p(w).
\end{align*}
Then by comparing the above identity with \eqref{D-E Omghat-p} we arrive at \eqref{Theta eta-mu lemma}.
The lemma is proved.
\end{proof}

\begin{prop} \label{thm D-E2 Omghat p}
We have the following identities:
\begin{align}
 & \p_{E^2}\Omghat_p(w)
=
  \Omgtilhat_p(0)
  \left(
    \frac14-\mu^2
  \right)\eta^{-1}\Omg(0,w)\notag
\\
&\quad
  +\left(
    p+\muhat+\frac12
  \right)
  \left(
    p+\muhat+\frac32
  \right)
  \Omgtilhat_{p+1}(w)\notag
\\
&\quad
  +\sum_{l\geq 1}
  2\Big(p+\muhat+1\Big)\Rhat^\transpose_l
  \Omgtilhat_{p+1-l}(w)
 +\sum_{l\geq 2}
  \Rhat^\transpose_{l;2}
  \Omgtilhat_{p+1-l}(w)\notag
\\
&\quad
  +
  \left[
    \Omgtilhat_p(w)
    \frac 1w
    \left(
      \rmD_w+\frac12
    \right)
    \left(
      \rmD_w+\frac32
    \right)
  \right]_+,\quad p\in\bbZ.\label{D-E2 Omghat p}
\end{align}
Here $\Rhat_{l;2}=[\Rhat^2]_l$,
and the definition of component $[- ]_l$ is obtained from
\eqref{zh-30} by replacing the $n\times n$ matrices $\mu, R$ with the
$(n+1)\times(n+1)$ matrices $\muhat,\Rhat$. The $n\times n$ matrix-valued formal power series
$\Omg(z,w)$ in $z,w$ is given by
\begin{equation}\label{def of Omg zw}
  \Big(\Omg(z,w)\Big)_{\afa\beta}
  =\sum_{p,q\geq 0}
   \Omg_{\afa,p; \beta,q}z^pw^q,
\end{equation}
and for any Laurent series $A=\sum_{k\geq s}A_kw^k$ we denote
$[A]_+=\sum_{k\geq 0}A_kw^k$.
\end{prop}

\begin{proof}
By using the identities \eqref{126-1}--\eqref{D-E Theta-p to mcalU},
\eqref{Theta-p and Theta-p-w}, \eqref{D-E Thetahat-p} and the relation
$\mcalU^\transpose\eta=\eta\mcalU$ we obtain
\begin{align*}
  &\p_{E^2}\Omghat_p(w)
=
  \p_{E^2}
  \Big(
    \Thetahat^\transpose_{p+1}(-w)\eta\Theta(w)
  \Big)\\
&\quad=
  \Thetahat^\transpose_p(-w)(\mcalU^\transpose)^2\eta\Theta(w)
 +w\Thetahat^\transpose_{p+1}(-w)
  \eta\mcalU^2\Theta(w)
\\
&\quad=
   \Thetahat^\transpose_p(-w)\mcalU^\transpose\eta\mcalU\Theta(w)
 +w\Thetahat^\transpose_{p+1}(-w)
  \mcalU^\transpose\eta\mcalU\Theta(w)
\\
&\quad=
  \left(
    \frac1w\p_E\Thetahat^\transpose_{p+1}(-w)
   +\p_E\Thetahat^\transpose_{p+2}(-w)
  \right)\eta
  \Big(\p_E\Theta(w)\Big)
\\
&\quad=
  \left(
    \frac1w\p_E\Thetahat^\transpose_{p+1}
  \right)\eta
  \Big(\p_E\Theta(w)\Big)
\\
&\quad=
\frac1w
  \Big[
    (p+1)\Thetahat^\transpose_{p+1}
   -\Thetahat^\transpose_{p+1}\mu
   +\sum_{l\geq 0}
    \Bhat_l^\transpose\Thetahat^\transpose_{p+1-l}
  \Big]
  \eta
  \Big(
    \Theta(w)\rmD_w-\mu\Theta(w)
  \Big).
\end{align*}
Then from \eqref{Theta eta-mu lemma} it follows that
\begin{align*}
&
\p_{E^2}\Omghat_p(w)
 =
  \frac1w\Thetahat^\transpose_{p+1}\mu\eta\mu\Theta(w)
  +(p+1)^2\Omgtilhat_{p+1}(w)
\\
&\qquad
  +\sum_{l\geq 0}
  \Bhat^\transpose_l(2p+2-l)\Omgtilhat_{p+1-l}(w)
  +\sum_{l\geq 0}
  \Bhat^\transpose_{l;2}
  \Omgtilhat_{p+1-l}(w)
  +
  \Omgtilhat_p(w)\frac1w
  (\rmD_w+1)^2
\\
&\quad=
  \frac1w\Thetahat^\transpose_{p+1}\mu\eta\mu\Theta(w)
  +\Big(p+\muhat+1\Big)^2\Omgtilhat_{p+1}(w)
\\
&\qquad
  +2\sum_{s\geq 1}
  \Big(p+\muhat+1\Big)\Rhat^\transpose_s\Omgtilhat_{p+1-s}(w)
  +\sum_{s\geq 2}
  \Rhat^\transpose_{s;2}
  \Omgtilhat_{p+1-s}(w)
  +
  \Omgtilhat_p(w)\frac1w
  (\rmD_w+1)^2
\\
&\quad=
  \frac1w\Thetahat^\transpose_{p+1}\mu\eta\mu\Theta(w)
  +\left(p+\muhat+\frac12\right)
  \left(p+\muhat+\frac32\right)
  \Omgtilhat_{p+1}(w)
\\
&\qquad
  +2\sum_{s\geq 1}
  \Big(p+\muhat+1\Big)\Rhat^\transpose_s\Omgtilhat_{p+1-s}(w)
  +\sum_{s\geq 2}
  \Rhat^\transpose_{s;2}
  \Omgtilhat_{p+1-s}(w)
\\
&\qquad
  +
  \Omgtilhat_p(w)\frac1w
  \left(
    \rmD_w+\frac12
  \right)
  \left(
    \rmD_w+\frac32
  \right)
  +\frac14
  \left(
    \frac1w\Omgtilhat_p(w)+\Omgtilhat_{p+1}(w)
  \right),
\end{align*}
where $\Bhat_{l;2}:=[\Bhat^2]_l=[(\muhat+\Rhat)^2]_l$.
On the other hand, from the relations
\begin{align*}
  \Thetahat^\transpose_p
&=
  \Omgtilhat_{p-1}(0)\eta^{-1},
\\
  \Theta(w)
&=
  I+w\eta^{-1}\Omg(0,w)
\end{align*}
for $p\in\bbZ$ and \eqref{nabla-theta pair}, it follows that
\[
  \frac1w
  \Big(
    \Omgtilhat_p(w)-\Omgtilhat_p(0)
  \Big)+\Omgtilhat_{p+1}(w)
=
  \frac1w\Thetahat^\transpose_{p+1}\eta
  \Big(\Theta(w)-I\Big)
=
  \Omgtilhat_p(0)\eta^{-1}\Omg(0,w),
\]
and
\begin{align*}
&\frac1w\Theta^\transpose_{p+1}\mu\eta\mu\Theta(w)
=
  \frac1w\Omgtilhat_p(0)\eta^{-1}\mu\eta\mu
  \Big(
    w\eta^{-1}\Omg(0,w)+I
  \Big)\notag\\
&\quad=
  -\Omgtilhat_p(0)\mu^2\eta^{-1}\Omg(0,w)
  -\frac1w\Omgtilhat_p(0)\mu^2.
\end{align*}
By using these relations we get
\begin{align*}
&
\p_{E^2}\Omghat_p(w)\\
&\quad=
  \Omgtilhat_p(0)
  \left(
    \frac14-\mu^2
  \right)\eta^{-1}\Omg(0,w)
  +\left(
    p+\muhat+\frac12
  \right)
  \left(
    p+\muhat+\frac32
  \right)
  \Omgtilhat_{p+1}(w)
\\
&\qquad
  +\sum_{l\geq 1}
  2\Big(p+\muhat+1\Big)\Rhat^\transpose_l
  \Omgtilhat_{p+1-l}(w)
  +\sum_{l\geq 2}
  \Rhat^\transpose_{l;2}
  \Omgtilhat_{p+1-l}(w)
\\
&\qquad
  +
  \left[
  \frac1w\Omgtilhat_p(0)
  \left(\frac14-\mu^2\right)+
    \Omgtilhat_p(w)
    \frac 1w
    \left(
      \rmD_w+\frac12
    \right)
    \left(
      \rmD_w+\frac32
    \right)
  \right]
\\
&\quad=
  \Omgtilhat_p(0)
  \left(
    \frac14-\mu^2
  \right)\eta^{-1}\Omg(0,w)
  +\left(
    p+\muhat+\frac12
  \right)
  \left(
    p+\muhat+\frac32
  \right)
  \Omgtilhat_{p+1}(w)
\\
&\qquad
  +\sum_{l\geq 1}
  2\Big(p+\muhat+1\Big)\Rhat^\transpose_l
  \Omgtilhat_{p+1-l}(w)
  +\sum_{l\geq 2}
  \Rhat^\transpose_{l;2}
  \Omgtilhat_{p+1-l}(w)
\\
&\qquad
  +
  \left[
    \Omgtilhat_p(w)
    \frac 1w
    \left(
      \rmD_w+\frac12
    \right)
    \left(
      \rmD_w+\frac32
    \right)
  \right]_+.
\end{align*}
The proposition is proved.
\end{proof}

Now let us start to prepare the computation of $\p_{E^3}\Omghat_p(w)$. Introduce the
$n\times n$ matrix
\begin{equation}\label{def of C}
  \mcalC:=\left.
  \frac{\td\Theta(w)}{\td w}
  \right|_{w=0}
  =
    \eta^{-1}\Omg(0,0)
  =
  \big(
    \nabla\theta_{1,1}, \dots, \nabla\theta_{n,1}
  \big),
\end{equation}
then it is clear that
$
  (\p_\afa\mcalC)^\beta_\gamma=c^\beta_{\afa\gamma}
$, so by using \eqref{zh-12} we arrive at
\begin{equation}\label{mcalU mcalC mu R1}
  \mcalU=(1-\mu)\mcalC+\mcalC\mu+R_1.
\end{equation}
We also have, for any vector field $Y=Y^\afa\p_\afa$, the relation
\begin{equation}\label{D-E mcalC}
  \p_Y\mcalC=\mcalC(Y),
\end{equation}
where $\mcalC(Y)$ is defined as in \eqref{126-1} and \eqref{126-2}.
In particular, we have
$\p_E\mcalC=\mcalU$, $\p_{E^2}\mcalC=\mcalU^2$.

\begin{lem}\label{mcalC Theta lemma}
The following relations hold true for $i=0,1,\dots,n$ and $p\in\bbZ$:
\begin{equation}\label{128}
\mcalC\nabla\theta_{i,p}
=
  \nabla\theta_{i,p+1}
  +\eta^{-1}\Omg_{\bullet,1;i,p-1}
  -\delta_{p+1,0}\delta^\beta_i\p_\beta,
\end{equation}
where the $n\times 1$ matrix $\Omg_{\bullet,1;i,p-1}$ is given by
$(\Omg_{1,1 ;i,p-1},\dots,\Omg_{n,1 ;i,p-1})^\transpose$,
and the vector field $\p_\beta$ is regarded as the column vector
$(0,\dots,1,\dots,0)^\transpose$,
here the number $1$ is the $\beta^{\text{th}}$ component.
\end{lem}

\begin{proof}
From the identity \eqref{nabla-theta-ortho} it follows that
\[
    \p_\afa\theta_{\beta,1}=\p_\beta\theta_{\afa,1},\quad
    1\leq \afa,\beta\leq n.
  \]
Then, for any fixed $1\leq\gamma\leq n$,
by using the relation \eqref{nabla-theta pair} and the definition of $\Omg_{i,p;j,q}$ we obtain
\begin{align*}
   & \big(\mcalC\nabla\theta_{i,p}\big)^\gamma
   =
    \mcalC^\gamma_\veps\p^\veps\theta_{i,p}
   =
    \eta^{\gamma\beta}\p_\beta\theta_{\veps,1}
    \p^\veps\theta_{i,p}
   =
    \eta^{\gamma\beta}
      \p_\veps\theta_{\beta,1}\p^\veps\theta_{i,p}
  \\
  &\quad=
  \eta^{\gamma\beta}
  \pair{\nabla\theta_{\beta,1}}{\nabla\theta_{i,p}}
  =
    \eta^{\gamma\beta}
    \big(
      \Omg_{\beta,0; i,p}
     +\Omg_{\beta,1; i,p-1}
    \big) \\
  &\quad=
    \eta^{\gamma\beta}
      \big(
        \p_\beta\theta_{i,p+1}
       -\delta_{p+1,0}
        \delta_i^\gamma
        \eta_{\beta\gamma}
       +\Omg_{\beta,1; i,p-1}
      \big)\\
  &\quad=
    \p^\gamma\theta_{i,p+1}
    +\eta^{\gamma\beta}\Omg_{\beta,1;i,p-1}
    -\delta_{p+1,0}\delta^\gamma_i,
  \end{align*}
hence the relation \eqref{128} holds true. The lemma is proved.
\end{proof}

We can rewrite the relation \eqref{128} in matrix form
\begin{align}
\label{mcalC Theta-1}
  \Thetahat^\transpose_p\mcalC^\transpose
 &= \Thetahat^\transpose_{p+1}
 +\p_w\Omgtilhat_{p-1}(0)\eta^{-1},
\\
\label{mcalC Theta-2}
 \Thetahat^\transpose_p(w)\mcalC^\transpose
&=
 \Thetahat^\transpose_{p+1}(w)
 +\sum_{k\geq 0}
 \big(\p_w\Omgtilhat_{p+k-1}(0)\big)\eta^{-1}w^k,
\\
\label{mcalC Theta-3}
  \mcalC\Theta(w)
  &=
  \frac1w(\Theta(w)-I)+w\eta^{-1}\p_z\Omg(0,w)
\end{align}
here the operators $\p_w$, $\p_z$ are defined by
\[
  \p_w\Omgtilhat_{p-1}(0)
=
  \left.
  \frac{\td}{\td w}\Omgtilhat_{p-1}(w)
  \right|_{w=0},\quad
  \p_z\Omg(0,w)
=
  \left.
  \frac{\td}{\td z}\Omg(z,w)
  \right|_{z=0},
\]
and $\Omg(z,w)$ is defined in \eqref{def of Omg zw}.

\begin{lem}\label{eta mu mcalC lemma}
For $p\in\bbZ$ the following relations hold true:
\begin{align}
  \Thetahat_p^\transpose\eta\mu\mcalC
=&\,
  -(p+\muhat)\Omgtilhat_p(0)
  -\sum_{s\geq 1}
  \Rhat^\transpose_s\Omgtilhat_{p-s}(0)+\p_w\Omgtilhat_{p-1}(0)(\mu+1)
  \notag\\
&\,
  +\Omgtilhat_{p-1}(0)R_1,\label{Theta-p eta mu mcalC}\\
\label{mcalC mu eta Theta}
\mcalC^\transpose\mu\eta\Theta(w)
=&\,
  -\eta\Theta(w)
  \big(\rmD_w-1\big)\frac1w
  +w(1+\mu)\p_z\Omg(0,w)+
  R_1^\transpose\eta\Theta(w)
\notag\\
&\,
-(1+\mu)\eta\cdot\frac 1w.
\end{align}
\end{lem}
\begin{proof}
By using the identities \eqref{Theta eta-mu lemma}, we have
\[
  \Thetahat_p^\transpose\eta\mu\Theta(w)
=-(p+\muhat)\Omgtilhat_p(w)w
 -\sum_{s\geq 1}\Rhat^\transpose_s\Omgtilhat_{p-s}(w)w
 +\Omgtilhat_{p-1}(w)\rmD_w
\]
which, together with the definition of $\mcalC$ given in \eqref{def of C}, yield the result of the lemma.
\end{proof}

\begin{lem}
We have the following relations for $p\in\bbZ$:
\begin{align}\label{D-E2 Theta p w}
&\p_{E^2}\Thetahat_p^\transpose(w)
=
  \left(
    p+\muhat-\frac12+w\frac{\td}{\td w}
  \right)
  \left(
    p+\muhat+\frac12+w\frac{\td}{\td w}
  \right)
\Thetahat_{p+1}^\transpose(w)\notag
\\
&\qquad
  +2\sum_{s\geq 1}
  \Big(p+\muhat+\frac{\td}{\td w}\Big)
  \Rhat^\transpose_s
  \Thetahat^\transpose_{p+1-s}(w)
  +
   \sum_{s\geq2}
   \Rhat_{s;2}^\transpose
   \Thetahat_{p+1-s}^\transpose(w)\notag
\\
&\qquad
  +\sum_{k\geq 0}
   \p_w\Omgtilhat_{p-k+1}(0)\eta^{-1}
   \left(
     \mu-\frac12
   \right)
   \left(
     \mu-\frac32
   \right)w^k\notag
\\
&\qquad
  -2\Thetahat_p^\transpose(w)R_1^\transpose(\mu-1)
  +\Thetahat_p^\transpose(w)
  \left(
    \frac14-\mu^2
  \right)\mcalC^\transpose,\\
\label{D-E2 Theta w}
  &\p_{E^2}\Theta(w)
=\Theta(w)
  \left(
    \rmD_w-\frac12
  \right)
  \left(
    \rmD_w-\frac32
  \right)\frac1w
 -
  \frac1w\left(\frac12-\mu\right)\left(\frac32-\mu\right)\notag\\
&\qquad+
  w\left(\frac12-\mu\right)\left(\frac32-\mu\right)
  \eta^{-1}\p_z\Omg(0,w)\notag\\
&\qquad-
  2(\mu-1)R_1\Theta(w)
 +
  \mcalC\left(\frac14-\mu^2\right)\Theta(w).
\end{align}
\end{lem}

\begin{proof}
Let us prove this lemma by computing $\p_E\p_E\Thetahat_p(w)$
in two different ways. On the one hand, by using \eqref{126-2},
  \eqref{mcalU mcalC mu R1} and \eqref{D-E mcalC} we obtain
\begin{align*}
  &
    \p_E\p_E\Theta_p(w)\\
  &\quad=
    \p_E\big(\mcalU\Thetahat_{p-1}(w)\big)
   =
    \big[(1-\mu)\mcalU+\mcalU\mu\big]
    \Thetahat_{p-1}(w)
   +\mcalU^2\Thetahat_{p-2}(w)\\
 &\quad=
    \p_{E^2}\Thetahat_{p-1}(w)
    +2(1-\mu)\p_E\Thetahat_p(w)
    -(1-\mu)\mcalU\Thetahat_{p-1}(w)
    +\mcalU\mu\Thetahat_{p-1}(w)
  \\
  &\quad=
     \p_{E^2}\Thetahat_{p-1}(w)
    +2(1-\mu)\p_E\Thetahat_p(w)\\
  &\qquad
    -(1-\mu)
    \big[
      (1-\mu)\mcalC+\mcalC\mu+R_1
    \big]\Thetahat_{p-1}(w)\\
  &\qquad
   +\big[
      (1-\mu)\mcalC+\mcalC\mu+R_1
    \big]\mu\Thetahat_{p-1}(w)
  \\
  &\quad=
    \p_{E^2}\Thetahat_{p-1}(w)
   +2(1-\mu)\p_E\Thetahat_p(w)
   -(1-\mu)^2\mcalC\Thetahat_{p-1}(w)\\
  &\qquad
   +2(\mu-1)R_1\Thetahat_{p-1}(w)
   +\mcalC\mu^2\Thetahat_{p-1}(w),
  \end{align*}
from which it follows that
\begin{align}\label{1282}
   & \p_{E^2}\Thetahat_{p-1}(w)
  =
    \p_E\p_E\Thetahat_p(w)
    -2(1-\mu)\p_E\Thetahat_p(w)
    +(1-\mu)^2\mcalC\Thetahat_{p-1}(w)\notag\\
  &\qquad-
   2(\mu-1)R_1\Thetahat_{p-1}(w)-\mcalC\mu^2\Thetahat_{p-1}(w).
\end{align}
On the other hand, by using the relation \eqref{D-E Thetahat} we have
\begin{align*}
&
  \p_E\p_E\Thetahat^\transpose_p(w)\\
&\quad=
  \p_E\left[
    \left(
      p+\muhat+w\frac{\td}{\td w}
    \right)\Thetahat^\transpose_p(w)
     +\sum_{s\geq 1}
      \Rhat^\transpose_s
      \Thetahat^\transpose_{p-s}(w)
     -\Thetahat^\transpose_p(w)\mu
  \right]\\
&\quad=
  \left(
    p+\muhat+w\frac{\td}{\td w}
  \right)
  \left[
    \left(
      p+\muhat+w\frac{\td}{\td w}
    \right)\Thetahat^\transpose_p(w)
     +\sum_{s\geq 1}
      \Rhat^\transpose_s
      \Thetahat^\transpose_{p-s}(w)
     -\Thetahat^\transpose_p(w)\mu
  \right]\\
&\qquad
  +
  \sum_{s\geq 1}
  \Rhat^\transpose_s
  \left[
    \left(
      p-s+\muhat+w\frac{\td}{\td w}
    \right)\Thetahat^\transpose_{p-s}(w)
     +\sum_{l\geq 1}
      \Rhat^\transpose_l
      \Thetahat^\transpose_{p-s-l}(w)
     -\Thetahat^\transpose_{p-s}(w)\mu
  \right]\\
&\qquad
  -
  \left[
    \left(
      p+\muhat+w\frac{\td}{\td w}
    \right)\Thetahat^\transpose_p(w)
     +\sum_{s\geq 1}
      \Rhat^\transpose_s
      \Thetahat^\transpose_{p-s}(w)
     -\Thetahat^\transpose_p(w)\mu
  \right]\mu \\
&\quad=
  \left(
    p+\muhat+w\frac{\td}{\td w}
  \right)^2
  \Thetahat^\transpose_p(w)
 +2\sum_{s\geq 1}
   \left(
    p+\muhat+w\frac{\td}{\td w}
  \right)\Rhat^\transpose_s\Thetahat^\transpose_{p-s}(w)\\
&\qquad
  -2\left(
    p+\muhat+w\frac{\td}{\td w}
  \right)\Thetahat^\transpose_p(w)\mu
  -2\sum_{s\geq 1}
   \Rhat^\transpose_s\Thetahat^\transpose_{p-s}(w)\mu\\
&\qquad
  +\sum_{s\geq 2}
   \Rhat^\transpose_{s;2}\Thetahat^\transpose_{p-s}(w)
  +\Thetahat^\transpose_p(w)\mu^2.
\end{align*}
Combining this relation with \eqref{1282} and \eqref{mcalC Theta-2}, we arrive at the relation \eqref{D-E2 Theta p w}.
The relation \eqref{D-E2 Theta w} can be proved in a similar way, and we omit the details here. The lemma is proved.
\end{proof}

\begin{prop} For $p\in\bbZ$, we have
\begin{align}
  &\p_{E^3}\Omghat_p(w)
=
  \Omgtilhat_p(0)
  \left(
    \frac12-\mu
  \right)
  \left(
    \frac 32-\mu
  \right)
  \eta^{-1}
  \left(
    \frac12-\mu
  \right)
  \p_z\Omg(0,w)\notag
\\
&\qquad
  +\p_w\Omgtilhat_{p}(0)
  \left(
    \frac12-\mu
  \right)
  \eta^{-1}
  \left(
    \frac12-\mu
  \right)
  \left(
    \frac 32-\mu
  \right)
  \Omg(0,w)
\notag\\
&\qquad
  +\Omgtilhat_p(0)
  \left(
    -3\mu^2+3\mu+\frac14
  \right)R_1\eta^{-1}\Omg(0,w)
\notag\\
&\qquad
  +
  \left(
    p+\muhat+\frac12
  \right)
  \left(
    p+\muhat+\frac32
  \right)
  \left(
    p+\muhat+\frac52
  \right)
  \Omgtilhat_{p+2}(w)
\notag\\
&\qquad
  +
  \sum_{s\geq 1}
\left(3\left(p+\muhat+\frac32\right)^2-1\right)
  \Rhat_s^\transpose\Omgtilhat_{p+2-s}(w)
\notag\\
&\qquad
  +
  \sum_{s\geq 2}
  3\left(p+\muhat+\frac32\right)
  \Rhat^\transpose_{s;2}
  \Omgtilhat_{p+2-s}(w)
  +\sum_{s\geq 3}
  \Rhat^\transpose_{s;3}\Omgtilhat_{p+2-s}(w)
\notag\\
&\qquad
  +\left[
    \Omgtilhat_p(w)
    \frac{1}{w^2}
    \left(
      \rmD_w+\frac12
    \right)
    \left(
      \rmD_w+\frac32
    \right)
    \left(
      \rmD_w+\frac52
    \right)
  \right]_+. \label{D-E3 Omghat}
\end{align}
\end{prop}

\begin{proof} From the proof of Propositions \ref{thm D-E Omghat-p},
\ref{thm D-E2 Omghat p} and the relations
\eqref{126-1}--\eqref{126-2}, \eqref{Theta-p and Theta-p-w},
it follows that
\begin{align}
  \p_E\Omghat_p(w)
=&\,
  \Thetahat_p^\transpose(-w)\eta\mcalU\Theta(w)
 +w\Thetahat^\transpose_{p+1}(-w)\eta\mcalU\Theta(w)
\notag\\
=&\,
  \Thetahat^\transpose_p\eta\mcalU\Theta(w),
\label{eta mcalU}
\\
  \p_{E^2}\Omghat_p(w)
=&\,
  \frac1w\Big(\p_E\Thetahat^\transpose_{p+1}\Big)
  \eta
  \big(
    \p_E\Theta(w)
  \big)
\notag\\
=&\,
  \Thetahat^\transpose_p\mcalU^\transpose\eta\mcalU\Theta(w)
 =
  \Thetahat^\transpose_p\eta\mcalU^2\Theta(w).
\end{align}
On the other hand, by using \eqref{mcalU mcalC mu R1}--\eqref{D-E mcalC} we obtain
\[
  \p_{E^2}\mcalU=(1-\mu)\mcalU^2+\mcalU^2\mu.
\]
Thus we have
\begin{align*}
  &\p_{E^3}\Omghat_p(w)\\
&\quad=
  \Thetahat^\transpose_p(-w)(\mcalU^\transpose)^3\eta\Theta(w)
 +w\Thetahat^\transpose_{p+1}(-w)\eta\mcalU^3\Theta(w)\\
&\quad=
   \Thetahat^\transpose_p(-w)(\mcalU^\transpose)^2\eta\mcalU\Theta(w)
 +w\Thetahat^\transpose_{p+1}(-w)\eta\mcalU\cdot\mcalU^2\Theta(w)\\
&\quad=
  \p_{E^2}
  \Big(
    \Thetahat^\transpose_{p+1}(-w)\eta\mcalU\Theta(w)
  \Big)
 -\Thetahat^\transpose_{p+1}(-w)\eta
 \big(\p_{E^2}\mcalU\big)\Theta(w)
\\
&\quad=
  \sum_{k\geq 0}
    (-1)^k\p_{E^2}\p_E
    \Omghat_{p+1+k}(w)\cdot w^k
  -\Thetahat^\transpose_{p+1}(-w)
  \eta
  \big[
    (1-\mu)\mcalU^2+\mcalU^2\mu
  \big]\Theta(w)
\\
&\quad=
  \sum_{k\geq 0}(-1)^k
  \p_{E^2}\p_E\Omghat_{p+1+k}(w)\cdot w^k
 -2\Thetahat^\transpose_{p+1}(-w)\eta\mcalU^2\Theta(w)
\\
&\qquad
  +\Thetahat^\transpose_{p+1}(-w)\eta
  \left(
    \mu+\frac12
  \right)\mcalU^2\Theta(w)
 +
   \Thetahat^\transpose_{p+1}(-w)(\mcalU^\transpose)^2\eta
   \left(
     \frac12-\mu
   \right)\Theta(w)
\\
&\quad=
  \sum_{k\geq 0}
  \p_{E^2}(\p_E-2)
  \Omghat_{p+1+k}(w)\cdot (-w)^k
  +\frac1w\Thetahat^\transpose_{p+1}(-w)
  \eta\left(\mu+\frac12\right)
  \p_{E^2}\Theta(w)\\
  &\qquad
  +\p_{E^2}\Thetahat^\transpose_{p+2}(-w)
   \left(\mu+\frac12\right)\eta\Theta(w).
\end{align*}
Then the relation \eqref{D-E3 Omghat} follows from
\eqref{D-E Omghat-p}, \eqref{Theta eta-mu lemma}, \eqref{D-E2 Omghat p}, \eqref{mcalC Theta-1}--\eqref{D-E2 Theta w} and a straightforward calculation.
The theorem is proved
\end{proof}

Finally, let us compute $\p_{E^s}\Omg_{0,p;0,q}$
for $s=0,1,2,3$ and $p,q\in\bbZ$.
\begin{prop}\label{thm D-Es Omg 0p0q}
For $p,q\in\bbZ$ we have the following identities:
\begin{align}
& \p_e\Omg_{0,p;0,q}
=
  \Omg_{0,p-1;0,q}+\Omg_{0,p;0,q-1}.\label{zh-33}
\\
&\p_E\Omg_{0,p;0,q}
=
  \Big(
    p+q+1-d
  \Big)\Omg_{0,p;0,q}+
  \sum_{s\geq 1}
  r^\veps_s\Omg_{\veps,p-s; 0,q}\notag\\
&\qquad
+ \sum_{s\geq 1}
  r^\veps_s\Omg_{0,p;\veps, q-s}
+(-1)^pc_{p+q}.
\label{D-E Omg 0p0q}\\
&\p_{E^2}\Omg_{0,p;0,q}
=
  \Omg_{0,p;\bullet,0}
  \left(
    \frac14-\mu^2
  \right)\eta^{-1}
  \Omg_{\bullet,0;0,q}\notag\\
&\qquad +
  \left(
    p-\frac d2+\frac12
  \right)
  \left(
    p-\frac d2+\frac32
  \right)\Omg_{0,p+1;0,q}
\notag\\
&\qquad
 +\left(
    q-\frac d2+\frac12
  \right)
  \left(
    q-\frac d2+\frac32
  \right)\Omg_{0,p;0,q+1}
\notag\\
&\qquad
  +2\left(p-\frac d2+1\right)
  \sum_{s\geq 1}
  r^\veps_s
  \Omg_{\veps,p+1-s;0,q}
  +2\left(q-\frac d2+1\right)
  \sum_{s\geq 1}
  r^\veps_s
  \Omg_{0,p;\veps,q+1-s}
\notag\\
&\qquad
  +\sum_{s\geq 2}
    \langle
      R\bm{r}
    \rangle_{s;2}^\veps
    \Omg_{\veps,p+1-s;0,q}
  +
   \sum_{s\geq 2}
    \langle
      R\bm{r}
    \rangle_{s;2}^\veps
    \Omg_{0,p;\veps,q+1-s}
\notag\\
&\qquad
  +(-1)^p\chi_{p+q\geq 0}
  \langle
    \bm{r}^\dag R\bm{r}
  \rangle_{p+q+2;2}
  +(-1)^p(q-p)c_{p+q+1}.\label{D-E2 Omg 0p0q}
\end{align}
We also have
\begin{align}
&\p_{E^3}\Omg_{0,p;0,q}\notag\\
&\quad=
  \Omg_{0,p;\bullet,0}
  \left(\frac12-\mu\right)
  \left(\frac32-\mu\right)
  \eta^{-1}
  \left(\frac12-\mu\right)
  \Omg_{\bullet,1;0,q}
\notag
\\
&\qquad
  +\Omg_{0,p;\bullet,1}
  \left(\frac12-\mu\right)
  \eta^{-1}
  \left(\frac12-\mu\right)
  \left(\frac32-\mu\right)
  \Omg_{\bullet,0;0,q}
\notag\\
&\qquad
  +\Omg_{0,p;\bullet 0}
  \left(
    -3\mu^2+3\mu+\frac14
  \right)R_1\eta^{-1}
  \Omg_{\bullet,0;0,q}
\notag\\
&\qquad
  +
  \left(
    p-\frac d2+\frac12
  \right)
  \left(
    p-\frac d2+\frac32
  \right)
  \left(
    p-\frac d2+\frac52
  \right)
  \Omg_{0,p+2;0,q}
\notag\\
&\qquad
  +
  \left(
    q-\frac d2+\frac12
  \right)
  \left(
    q-\frac d2+\frac32
  \right)
  \left(
    q-\frac d2+\frac52
  \right)
  \Omg_{0,p;0,q+2}
\notag\\
&\qquad
  +
  \left(
    3\left(
    p-\frac d2+\frac32
  \right)^2
  -1
  \right)
  \sum_{s\geq 1}
  r^\veps_s
  \Omg_{\veps,p+2-s;0,q}
\notag\\
&\qquad
  +
  \left(
    3\left(
    q-\frac d2+\frac32
  \right)^2
  -1
  \right)
  \sum_{s\geq 1}
  r^\veps_s
  \Omg_{0,p;\veps,q+2-s}
\notag\\
&\qquad
  +\sum_{s\geq 2}
   3\left(p-\frac d2+\frac32\right)
   \langle
     R\bm{r}
   \rangle_{s;2}^\veps
   \Omg_{\veps,p+2-s; 0,q}\notag\\
 &\qquad +\sum_{s\geq 2}
   3\left(q-\frac d2+\frac32\right)
   \langle
     R\bm{r}
   \rangle_{s;2}^\veps
   \Omg_{0,p;\veps,q+2-s}
\notag\\
&\qquad
  +\sum_{s\geq 3}
   \langle
     R\bm{r}
   \rangle_{s;3}^\veps
   \Omg_{\veps,p+2-s; 0,q}
  +\sum_{s\geq 3}
   \langle
     R\bm{r}
   \rangle_{s;3}^\veps
   \Omg_{0,p;\veps,q+2-s}
\notag\\
&\qquad
  +(-1)^q\cdot\frac 32(p-q)
  \langle
    \bm{r}^\dag R\bm{r}
  \rangle_{p+q+3;2}
  +(-1)^p
  \langle
    \bm{r}^\dag R\bm{r}
  \rangle_{p+q+3;3}
\notag\\
&\qquad
  +(-1)^p c_{p+q+2}
  \left[
  \left(
    p-\frac d2+\frac12
  \right)
  \left(
    p-\frac d2+\frac32
  \right)\right.\notag\\
  &\qquad
  \left.+
  \left(
    q-\frac d2+\frac12
  \right)
  \left(
    q-\frac d2+\frac32
  \right)
  -
  \left(
    p-\frac d2+\frac12
  \right)
  \left(
    q-\frac d2+\frac32
  \right)\right],\label{D-E3 Omg 0p0q}
\end{align}
where we denote the $n\times 1$ matrix $(\Omg_{i,p; 1,q},\dots,\Omg_{i,p; n,q})$ by $\Omg_{i,p; \bullet,q}$, and denote the $1\times n$ matrix $(\Omg_{1,p;j,q},\dots, \Omg_{n,p;j,q})^{\transpose}$ by $\Omg_{\bullet,p; j,q}$.
\end{prop}

\begin{proof}
The relation \eqref{zh-33} is given by \eqref{D e Omg}, and the relation \eqref{D-E Omg 0p0q} can be easily verified by
   using \eqref{Omg 0 p 0 q},
   \eqref{def of c-p},  \eqref{theta0p-quasi-homog} and \eqref{nabla-theta pair}, so we are left to compute the derivatives
$\p_{E^2}\Omg_{0,p;0,q}$ and $\p_{E^3}\Omg_{0,p;0,q}$.

From the relations given in \eqref{Omg prop 4} it follows that
   \begin{align} \label{0p eta mcalU 0q}
        \p_E\Omg_{0,p;0,q}
    &=
     \pair{E}{\nabla\Omg_{0,p;0,q}}
    =
     \pair{E}{\nabla\theta_{0,p}\cdot\nabla\theta_{0,q}}
    =
     \pair{E\cdot\nabla\theta_{0,p}}{\nabla\theta_{0,q}}\notag\\
    &=
     \nabla\theta_{0,p}^\transpose
     \mcalU^\transpose\eta\nabla\theta_{0,q}
    =
     \big(
       \p_E\nabla\theta_{0,p+1}^\transpose
     \big)\eta\nabla\theta_{0,q},
      \end{align}
   then by using the relation \eqref{theta0p-quasi-homog}
   we get another way to compute $\p_E\Omg_{0,p;0,q}$.
   Comparing the resulting formula with \eqref{D-E Omg 0p0q},
   we obtain
\begin{align}\label{theta 0p eta mu theta 0q}
 & \nabla\theta_{0,p+1}^T\eta\mu\nabla\theta_{0,q}
  =
    \left(q-\frac d2\right)\Omega_{0,p;0,q}
   -\left(
    p+1-\frac d2
    \right) \Omega_{0,p+1;0,q-1}\notag\\
  &\qquad
    +\sum_{s\geq 1}
    r^\veps_s\Omega_{0,p;\veps,q-s}
    -\sum_{s\geq 1}
    r^\veps_s\Omega_{\veps,p+1-s;0,q-1}
    +(-1)^pc_{p+q}.
\end{align}
Then, by using \eqref{Omg prop 4},
\eqref{theta0p-quasi-homog} and \eqref{theta0,-p-quasi-homog}
we have
\begin{align*}
&
  \p_{E^2}\Omg_{0,p;0,q}\\
&\quad=
  \pair{E^2}{\nabla\theta_{0,p}\cdot\nabla\theta_{0,q}}
 =
  \pair{E\cdot\nabla\theta_{0,p}}{E\cdot\nabla\theta_{0,q}}
 =
  \pair{\p_E\nabla\theta_{0,p+1}}{\p_E\nabla\theta_{0,q+1}}\\
&\quad=
  \left[
    \nabla\theta_{0,p+1}^\transpose
    \left(
      p+1-\frac d2-\mu
    \right)
   +\sum_{s\geq 1}
   r^\veps_s\nabla\theta^\transpose_{\veps,p+1-s}
  \right]\\
  &\qquad\times \eta
  \left[
    \left(
      q+1-\frac d2-\mu
    \right)\nabla\theta_{0,q+1}
   +\sum_{s\geq 1}
   r^\lmd_s\nabla\theta_{\lmd,q+1-s}
  \right]\\
&\quad=
  \left(
    p+1-\frac d2
  \right)
  \left(
    q+1-\frac d2
  \right)
  \pair{\nabla\theta_{0,p+1}}{\nabla\theta_{0,q+1}}
\\
&\qquad +
 \left[
   -\left(p+1-\frac d2\right)
   +\left(q+1-\frac d2\right)
 \right]
 \nabla\theta^\transpose_{0,p+1}\eta\mu\nabla\theta_{0,q+1}
+
  \nabla\theta_{0,p+1}^\transpose\mu\eta\mu\nabla\theta_{0,q+1}
\\
&\qquad+
  \left(
    q+1-\frac d2
  \right)
  \sum_{s\geq 1}
  r^\veps_s\pair{\nabla\theta_{\veps,p+1-s}}{\nabla\theta_{0,q+1}}
 +\left(
    p+1-\frac d2
  \right)
  \sum_{s\geq 1}
  r^\veps_s
  \pair{\nabla\theta_{0,p+1}}{\nabla\theta_{\veps,q+1-s}}
\\
&\qquad
 -\sum_{s\geq 1}
  r^\veps_s
  \nabla\theta^\transpose_{\veps,p+1-s}
  \eta\mu\nabla\theta_{0,q+1}
 +\sum_{s\geq 1}
  r^\lmd_s\nabla\theta^\transpose_{0,p+1}
  \eta\mu\nabla\theta_{\lmd,q+1-s}\\
  &\qquad
 +\sum_{k,l\geq 1}
  r^\veps_k r^\lmd_l
  \pair{\nabla\theta_{\veps,p+1-k}}{\nabla\theta_{\lmd,q+1-l}},
\end{align*}
from this and the relations given in
\eqref{nabla-theta pair},
\eqref{0,p eta-mu beta,q}
and \eqref{theta 0p eta mu theta 0q}
we arrive at the formula \eqref{D-E2 Omg 0p0q}.

Finally, let us compute $\p_{E^3}\Omg_{0,p;0,q}$.
By using Lemma \ref{mcalC Theta lemma} and the way that is used to prove Lemma
\ref{eta mu mcalC lemma}
it is easy to show that
\begin{align}
  \mcalC\nabla\theta_{0,p}
=&\, \nabla\theta_{0,p+1}+\eta^{-1}\Omega_{\bullet,1;0,p-1}
\label{00 mcalC 1}\\
  \nabla\theta^\transpose_{0,p}\eta\mcalC
=&\,
  \nabla\theta_{0,p+1}^\transpose\eta+\Omega_{0,p-1;\bullet,1},
\label{00 mcalC 2}\\
  \nabla\theta_{0,p}^\transpose\eta\mu\mcalC
=&\,
  \Omega_{0,p-1;\bullet,1}
  (1+\mu)
 -\left(p-\frac d2\right)
  \nabla\theta_{0,p+1}^\transpose\eta+\nabla\theta_{0,p}^\transpose\eta R_1\notag
 \\
&\,
  -\sum_{s\geq 1}
   r^\veps_s
   \nabla\theta_{\veps,p+1-s}^\transpose\eta
  -(-1)^p\chi_{p\geq 0}(r_{p+1})_\bullet.
\label{00 mcalC 3}
\end{align}
Then by using the formulae given in \eqref{theta0p-quasi-homog},
\eqref{theta0,-p-quasi-homog}, \eqref{Omg prop 4} and the relations \eqref{mcalU mcalC mu R1},
\eqref{eta mcalU}, \eqref{0p eta mcalU 0q},
we get
\begin{align}
&
  \p_{E^3}\Omg_{0,p;0,q}\notag\\
&\quad=
  \pair{E^3}{\nabla\theta_{0,p}\cdot\nabla\theta_{0,q}}
 =
  \nabla\theta^\transpose_{0,p}
  \eta\mcalU^3\nabla\theta_{0,q}
 =
  \big(\p_E\nabla\theta_{0,p+1}^\transpose\big)
  \eta\mcalU
  \big(\p_E\nabla\theta_{0,q+1}\big)
\notag\\
&\quad=
  \left[
    \nabla\theta^\transpose_{0,p+1}
    \left(
      p+1-\frac d2-\mu
    \right)
   +\sum_{s\geq 1}
    r^\veps_s\nabla\theta_{\veps,p+1-s}^\transpose
  \right]
  \eta\mcalU
  \notag \\
 &\qquad
  \times
  \left[
    \left(
      q+1-\frac d2-\mu
    \right)\nabla\theta_{0,q+1}
   +\sum_{s\geq 1}
    r^\veps_s\nabla\theta_{\veps,q+1-s}
  \right]
\notag\\
 &\quad=
  \left(
    p+1-\frac d2
  \right)
  \left(
    q+1-\frac d2
  \right)
  \nabla\theta^\transpose_{0,p+1}
  \eta\mcalU\nabla\theta_{0,q+1}
\notag\\
&\qquad
  -\left(q+1-\frac d2\right)
   \nabla\theta^\transpose_{0,p+1}
   \mu\eta\mcalU
   \nabla\theta_{0,q+1}
  -\left(p+1-\frac d2\right)
   \nabla\theta^\transpose_{0,p+1}
   \eta\mcalU\mu
   \nabla\theta_{0,q+1}
\notag\\
&\qquad
 -\sum_{s\geq 1}
  r^\veps_s\nabla\theta^\transpose_{0,p+1}
  \mu\eta\mcalU
  \nabla\theta_{\veps,q+1-s}
 -\sum_{s\geq 1}
  r^\veps_s\nabla\theta^\transpose_{\veps,p+1-s}
  \eta\mcalU\mu\nabla\theta_{0,q+1}
 +\nabla\theta^\transpose_{0,p+1}
  \mu\eta\mcalU\mu
  \nabla\theta_{0,q+1}
\notag\\
&\qquad
  +\left(
    q+1-\frac d2
   \right)
   \sum_{s\geq 1}r^\veps_s
   \nabla\theta_{\veps,p+1-s}^\transpose
   \eta\mcalU\nabla\theta_{0,q+1}
  +\left(
    p+1-\frac d2
   \right)
   \sum_{s\geq 1}
   r^\veps_s
   \nabla\theta_{0,p+1}^\transpose
   \eta\mcalU\nabla\theta_{\veps,q+1-s}
\notag\\
&\qquad
  +\sum_{k,l\geq 1}
   r^\veps_k r^\lmd_l
   \nabla\theta^\transpose_{\veps ,p+1-k}
   \eta\mcalU\nabla\theta_{\lmd,q+1-l}.
\notag\\
 &\quad=
  \left(
    p+1-\frac d2
  \right)
  \left(
    q+1-\frac d2
  \right)
  \p_E\Omg_{0,p+1;0,q+1}
\notag\\
&\qquad
  +\left(q+1-\frac d2\right)
   \nabla\theta^\transpose_{0,p+1}
   \eta\mu\mcalU
   \nabla\theta_{0,q+1}
  -\left(p+1-\frac d2\right)
   \nabla\theta^\transpose_{0,p+1}
   \eta\mcalU\mu
   \nabla\theta_{0,q+1}
\notag\\
&\qquad
  +\sum_{s\geq 1}
   r^\veps_s
   \nabla^\transpose_{0,p+1}\eta\mu
   \big(
     \p_E\nabla\theta_{\veps,q+2-s}
   \big)
  -\sum_{s\geq 1}
   r^\veps_s
   \big(
     \p_E\nabla\theta^\transpose_{\veps,p+2-s}
   \big)\eta\mu\nabla\theta_{0,q+1}
\notag\\
&\qquad
  +\nabla\theta^\transpose_{0,p+1}
   \mu\eta
   \big[
     (1-\mu)\mcalC+\mcalC\mu+R_1
   \big]\mu\nabla\theta_{0,q+1}
\notag\\
&\qquad
  +
  \left(
    q+1-\frac d2
  \right)
  \sum_{s\geq 1}
  r^\veps_s\p_E\Omg_{\veps,p+1-s; 0,q+1}
  +\left(
    p+1-\frac d2
  \right)
  \sum_{s\geq 1}
  r^\veps_s
  \p_E\Omg_{0,p+1;\veps,q+1-s}\notag\\
&\qquad
  +\sum_{k,l\geq 1}
  r^\veps_k r^\lmd_l
  \p_E\Omg_{\veps,p+1-k;\lmd,q+1-l}.
\label{zh-36}
\end{align}
On the other hand, for any parameter $x,y\in\bbC$ we have
\begin{align*}
&
  \left(
    q+1-\frac d2
  \right)
  \nabla\theta_{0,p+1}^\transpose\eta\mu\mcalU
  \nabla\theta_{0,q+1}
\\
 &\quad=
  \left(
    q-\frac d2+x
  \right)
  \nabla\theta_{0,p+1}^\transpose
  \eta\mu\big(
  \p_E\nabla\theta_{0,q+2}\big)\\
 &\qquad +
  \big(
    1-x
  \big)
  \nabla\theta_{0,p+1}^\transpose
  \eta\mu
  \big[
    (1-\mu)\mcalC+\mcalC\mu+R_1
  \big]
  \nabla\theta_{0,q+1}
\\
 &\quad=
  \left(
    q-\frac d2+x
  \right)
  \nabla\theta_{0,p+1}^\transpose
  \eta\mu\big(
  \p_E\nabla\theta_{0,q+2}\big)
+
  \big(1-x\big)
  \nabla\theta_{0,p+1}^\transpose
  \eta\mu\mcalC\nabla\theta_{0,q+1}
\\
&\quad= -
  \big(1-x\big)
  \nabla\theta^\transpose_{0,p+1}
  \eta\mu^2
  \big(
    \mcalC\nabla\theta_{0,q+1}
  \big)
+
  \big(1-x\big)
  \big(
    \nabla\theta_{0,p+1}^\transpose
    \eta\mu\mcalC
  \big)\mu\nabla\theta_{0,q+1}
\\
&\qquad +
  \big(1-x\big)
  \nabla\theta_{0,p+1}^\transpose
  \eta\mu R_1\nabla\theta_{0,q+1}\\
 &\quad=
  \left(
    q-\frac d2+x
  \right)
  \nabla\theta_{0,p+1}^\transpose
  \eta\mu\big(
  \p_E\nabla\theta_{0,q+2}\big)\\
&\qquad+  (1-x-y)
  \nabla\theta^\transpose_{0,p+1}
  \eta\mu
  \big(
    \mcalC\nabla\theta_{0,q+1}
  \big)
+
  y\big(
    \nabla\theta_{0,p+1}^\transpose
    \eta\mu\mcalC
  \big)\nabla\theta_{0,q+1}\\
&\qquad-
  \big(1-x\big)
  \nabla\theta^\transpose_{0,p+1}
  \eta\mu^2
  \big(
    \mcalC\nabla\theta_{0,q+1}
  \big)+
  \big(1-x\big)
  \big(
    \nabla\theta_{0,p+1}^\transpose
    \eta\mu\mcalC
  \big)\mu\nabla\theta_{0,q+1}\\
&\qquad+
  \big(1-x\big)
  \nabla\theta_{0,p+1}^\transpose
  \eta\mu R_1\nabla\theta_{0,q+1}.
\end{align*}
We take $x=\frac 54$, $y=\frac 38$,
and replace the second term after the last equality of \eqref{zh-36} by the expression that appears after the last equality of the above formula.
Then by using the relations given in
\eqref{zh-12},
\eqref{D-E Omghat-p},
\eqref{0,p eta-mu beta,q},
\eqref{D-E Omg 0p0q},
\eqref{theta 0p eta mu theta 0q},
and \eqref{00 mcalC 1}--\eqref{00 mcalC 3} we can verify the validity of the formula \eqref{D-E3 Omg 0p0q}.
The proposition is proved.
\end{proof}

\end{document}